\documentclass[12pt]{book}
\usepackage{times,euler}
\usepackage{amsmath,a4wide,array,axodraw,epsfig,floatfig,minitoc,multirow,theorem}
\usepackage[english,dutch]{babel}
\usepackage{fancyheadings}
\newcommand{\clearemptydoublepage}{\newpage{\pagestyle{empty}\cleardoublepage}}
\renewcommand{\baselinestretch}{1.1}
\theoremstyle{break}\theorembodyfont{\rmfamily}
                    \newtheorem{Alg}{Algorithm}[section]
\newcounter{appendix}[chapter]
\renewcommand{\theappendix}{\thechapter\Alph{appendix}}
\newcommand{\Appendix}[1]{%
   \refstepcounter{appendix}%
   \vspace{1.2\baselineskip}\noindent%
   {\large\bfseries Appendix \theappendix #1}%
   \vspace{0.5\baselineskip}\\}
\addtocounter{secnumdepth}{1}
\begin{document}

%
\newcommand{\eqn}[1]{Eq.\!~(\ref{#1})}
\newcommand{\fig}[1]{Fig.\!~\ref{#1}}
\newcommand{\tab}[1]{Tab.\!~\ref{#1}}
\newcommand{\Sec}[1]{Section~\ref{#1}}
\newcommand{\Chp}[1]{Chapter~\ref{#1}}
\newcommand{\App}[1]{Appendix~\ref{#1}}
%
\newcommand{\al}{\alpha}
\newcommand{\be}{\beta}
\newcommand{\om}{\omega}
\newcommand{\vt}{\vartheta}
\newcommand{\ep}{\epsilon}
\newcommand{\ve}{\varepsilon}
\newcommand{\vp}{\varepsilon}
\newcommand{\ga}{\gamma}
\newcommand{\Ga}{\Gamma}
\newcommand{\vGa}{\varGamma}
\newcommand{\vDe}{\varDelta}
\newcommand{\la}{\lambda}
\newcommand{\si}{\sigma}
\newcommand{\Si}{\Sigma}
\newcommand{\vSi}{\varSigma}
\newcommand{\de}{\delta}
\newcommand{\vhi}{\varphi}
\newcommand{\phiv}{\phi^{\textrm{v}}}
\newcommand{\vhiv}{\vhi^{\textrm{v}}}
\newcommand{\phib}{\overline{\phi}}
\newcommand{\psib}{\bar{\psi}}
\newcommand{\hphi}{\widehat{\phi}}
\newcommand{\vLa}{\varLambda}
\newcommand{\vTh}{\varTheta}
%
\newcommand{\Kube}{{\bf K}}
\newcommand{\Aset}{{\bf A}}
\newcommand{\Bset}{{\bf B}}
\newcommand{\Fset}{{\bf F}}
\newcommand{\Mset}{{\bf M}}
\newcommand{\Pset}{{\bf P}}
\newcommand{\Pol}{{\bf P}}
\newcommand{\XN}{\mathscr{X}_N}
\newcommand{\Natu}{{\bf N}}
\newcommand{\Qrat}{{\bf Q}}
\newcommand{\Zatu}{{\bf Z}}
\newcommand{\Fatu}{{\bf F}}
\newcommand{\Real}{{\bf R}}
\newcommand{\Comp}{{\bf C}}
\newcommand{\Inat}{{\bf I}}
\newcommand{\Afam}{A}
\newcommand{\Cfam}{C}
\newcommand{\Ffam}{\mathscr{F}}
\newcommand{\Mfam}{M}
\newcommand{\Hilb}{H}
\newcommand{\bs}{\mathbf}
%
\newcommand{\intk}{\int_{\Kube}}
\newcommand{\intkk}{\int_{\Kube^2}}
\newcommand{\Dp}{d\mu[\phi]}
\newcommand{\Dpath}{[d\phi]}
\newcommand{\lK}{\langle}
\newcommand{\rK}{\rangle}
\newcommand{\leb}[1]{\langle #1\rangle}
\newcommand{\inp}[2]{\langle #1\,#2\rangle}
\newcommand{\qnm}[1]{\langle\,(#1)^2\,\rangle}
\newcommand{\norm}[1]{\|#1\|}
\newcommand{\lebM}[3]{\langle #1\rangle_{#2}^{#3}}
\newcommand{\Vol}{\textrm{Vol}}
%
\newcommand{\Int}{J}
\newcommand{\Est}{\mathscr{E}}
\newcommand{\Tran}{\mathscr{T}}
%
\newcommand{\twoC}{\mathscr{C}}
\newcommand{\twoB}{\mathscr{B}}
\newcommand{\twoI}{\mathscr{I}}
\newcommand{\twoJ}{\mathscr{J}}
\newcommand{\twoK}{\mathscr{K}}
\newcommand{\prop}{\mathscr{G}}
\newcommand{\Prim}{\mathscr{P}}
\newcommand{\Pspc}{\hspace{1pt}}
\newcommand{\Prid}{\mathscr{P}^{\dagger}}
\newcommand{\Prit}{\tilde{\mathscr{P}}}
\newcommand{\Quad}{\mathscr{Q}}
\newcommand{\Npnt}{\mathscr{N}}
%
\newcommand{\df}{:=}
\newcommand{\ra}{\rightarrow}
\newcommand{\lar}{\leftarrow}
\newcommand{\da}{\downarrow}
\newcommand{\vn}{\vec{n}}
\newcommand{\vm}{\vec{m}}
\newcommand{\vk}{\vec{k}}
\newcommand{\Smath}{\mathscr{S}}
\newcommand{\rmin}{\textrm{min}}
\newcommand{\half}{{\textstyle\frac{1}{2}}}
\newcommand{\shalf}{{\scriptstyle\frac{1}{2}}}
\newcommand{\sfrac}[2]{{\textstyle\frac{#1}{#2}}}
\newcommand{\sss}[1]{\scriptscriptstyle{#1}}
\newcommand{\Ord}{{\mathscr O}}
\newcommand{\one}{{\bf 1}}
\newcommand{\Tr}{\textrm{Tr}}
\newcommand{\sgn}{\textrm{sgn}}
\newcommand{\RE}{\textrm{Re}}
\newcommand{\Exp}{\mathsf{E}}
\newcommand{\Var}{\mathsf{V}}
\newcommand{\Pro}{\mathsf{P}}
\newcommand{\VAR}{\textrm{V}}
\newcommand{\Qw}{Q_{\scriptscriptstyle\textrm{W}}}
\newcommand{\Qfp}{Q_{\scriptscriptstyle\textrm{F}}^{\scriptscriptstyle\prod}}
\newcommand{\Qfs}{Q_{\scriptscriptstyle\textrm{F}}^{\scriptscriptstyle\sum}}
\newcommand{\brfr}[2]{\left({{#1}\over{#2}}\right)}
\newcommand{\dcon}{\overset{\textrm{d}}{\longrightarrow}}
\newcommand{\wcon}{\overset{\textrm{w}}{\longrightarrow}}
\newcommand{\ccon}{\overset{\textrm{c}}{\longrightarrow}}
\newcommand{\Wcon}{\Rightarrow}
\newcommand{\pcon}{\overset{\textrm{p}}{\longrightarrow}}
\newcommand{\acon}{\overset{\textrm{a.s.}}{\longrightarrow}}
\newcommand{\Lcon}{\overset{\textrm{$\mathscr{L}_p$}}{\longrightarrow}}
\newcommand{\alasym}{\alpha^{\textrm{asym}}}
\newcommand{\rambo}{\texttt{RAMBO}}
\newcommand{\sarge}{\texttt{SARGE}}
\newcommand{\np}{n}
\newcommand{\wcm}{\sqrt{s}\,}
\newcommand{\scm}{s}
\newcommand{\ip}[2]{#1\!\cdot\!#2}
\newcommand{\ipb}[2]{(#1#2)}
\newcommand{\inip}[2]{\{#1|#2\}_\de}
\newcommand{\Bo}{\mathscr{H}}
\newcommand{\Ro}{\mathscr{R}}
\newcommand{\Lo}{\mathscr{L}}
\newcommand{\Aqcd}{A^{\textrm{QCD}}}
\newcommand{\Ant}{A^{\scriptscriptstyle\textrm{QCD}}_{\scm}}
\newcommand{\pin}{0}
\newcommand{\nip}{\tilde{0}}
\newcommand{\ppin}{p_0}
\newcommand{\pnip}{\tilde{p}_{0}}
\newcommand{\xim}{\xi_{\textrm{m}}}
\newcommand{\xilow}{\xi_{\textrm{low}}}
\newcommand{\enul}{e_0}
\newcommand{\ethr}{e_3}
\newcommand{\dQ}{d^4Q}        
\newcommand{\dfp}{d^4p}       
\newcommand{\dnp}{d^{4n}p}    
\newcommand{\invs}[2]{{#1}^{2}_{#2}}
\newcommand{\invm}[1]{m_{#1}}
\newcommand{\tcpu}{t_{\textrm{cpu}}}
\newcommand{\reject}{{\tt REJECT}}
\newcommand{\ouralg}{{\tt OURALG}}
\newcommand{\dt}{d\tau}
\newcommand{\oPhi}{\overline{\Phi}}
\newcommand{\diagram}[3]{\raisebox{-#3pt}{\epsfig{figure=figures/#1.eps,width=#2pt}}}     
%
%
\selectlanguage{english}
\thispagestyle{empty}
\vspace*{8mm}
\begin{center}
{\bf\huge Loaded Dice in Monte Carlo}

\vspace{\baselineskip}

{\bf\large importance sampling in phase space integration}\\
\vspace{0.2\baselineskip}
{\bf\large and probability distributions for discrepancies}\\

\vspace{\baselineskip}

\vspace{\baselineskip}

{\large Andr\'e van Hameren}

\vspace{2pt}

{\large University of Nijmegen, Nijmegen, the Netherlands}

\vspace{2pt}

{\tt andrevh@sci.kun.nl}



\vspace{3\baselineskip}

PhD thesis

\vspace{3\baselineskip}

{\bf Abstract}

\vspace{\baselineskip}

\parbox{0.8\linewidth}
{\hspace*{15pt}Discrepancies play an important r\^ole in the study of
uniformity properties of point sets. Their probability distributions are a help
in the analysis of the efficiency of the Quasi Monte Carlo method of numerical
integration, which uses point sets that are distributed more uniformly than
sets of independently uniformly distributed random points. In this thesis,
generating functions of probability distributions of quadratic discrepancies
are calculated using techniques borrowed from quantum field theory.

\hspace*{15pt}The second part of this manuscript deals with the application of 
the Monte Carlo method to phase space integration, and in particular with an
explicit example of importance sampling.  It concerns the integration of
differential cross sections of multi-parton QCD-processes, which contain the
so-called kinematical antenna pole structures.  The algorithm is presented and
compared with \rambo, showing a substantial reduction in computing time.

\hspace*{15pt}In behalf of completeness of the thesis, short introductions to
probability theory, Feynman diagrams and the Monte Carlo method of numerical 
integration are included.}

\end{center}

\newpage

\thispagestyle{empty}
\vspace*{\fill}
\noindent
This thesis was typeset with \LaTeXe.

%
\selectlanguage{dutch}
\thispagestyle{empty}
\vspace*{20mm}
\begin{center}
{\Huge Loaded Dice in Monte Carlo}

\vspace{\baselineskip}

{\large importance sampling in phase space integration}\\
\vspace{0.2\baselineskip}
{\large and probability distributions for discrepancies}

\vspace{\baselineskip}

\vspace{\baselineskip}

\vspace{\baselineskip}

\vspace{\baselineskip}

\newcommand{\spd}{\hspace{4pt}}
\parbox{0.53\linewidth}{een wetenschappelijke proeve op het gebied van de 
     Natuurwetenschappen, \spd Wiskunde \spd en \spd Informatica}
     
\vspace{\baselineskip}

\vspace{\baselineskip}

\vspace{\baselineskip}

\vspace{\baselineskip}

\newcommand{\spe}{\hspace{1pt}}
{\large P{\spe}r{\spe}o{\spe}e{\spe}f{\spe}s{\spe}c{\spe}h{\spe}r{\spe}i{\spe}f{\spe}t}

\vspace{\baselineskip}

\newcommand{\spa}{\hspace{0pt}}
\newcommand{\spb}{\hspace{0pt}}
\newcommand{\spc}{\hspace{0.5pt}}
\parbox{0.44\linewidth}
{ter \spa verkrijging \spa van \spa de \spa graad \spa van \spa doctor
     aan \spb de \spb Katholieke \spb Universiteit \spb Nijmegen,
     volgens \spc besluit \spc van \spc het \spc College \spc van \spc Decanen}

\vspace{4pt}

     in \spc het \spc openbaar \spc te \spc verdedigen \spc op

\vspace{\baselineskip}

     dinsdag 9 januari 2001\\
     des namiddags om 1.30 uur precies
     
\vspace{\baselineskip} 

\vspace{\baselineskip}

{door}

\vspace{\baselineskip} 

\vspace{\baselineskip}

{\large Andreas Ferdinand Willem van Hameren}

\vspace{\baselineskip} 

{geboren op 4 december 1973 te Horst}

\end{center}

\newpage
\thispagestyle{empty}
\noindent
\begin{tabular}{ll}
Promotor: &Prof. Dr. R.H.P. Kleiss\\
 & \\
Manuscriptcommissie: &Prof. Dr. W.J. Stirling (University of Durham)\\
		     &Prof. Dr. S.J. de Jong\\
                     &Dr. J.D.M. Maassen
\end{tabular}

\vspace{\fill}

\noindent\epsfig{figure=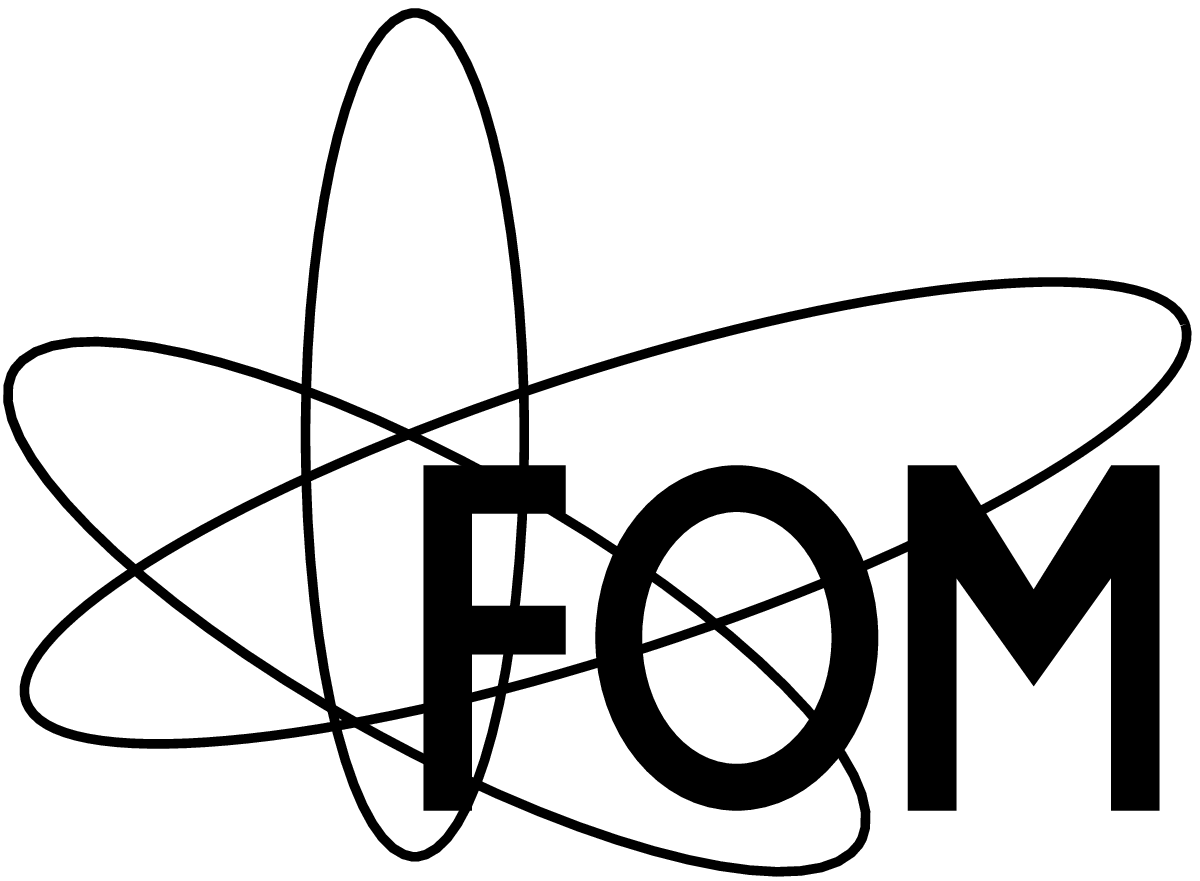,width=0.18\linewidth}

\noindent
Het werk beschreven in dit proefschrift maakt deel uit van het
onderzoeksprogramma van de Stichting voor Fundamenteel Onderzoek der Materie
(FOM), die financieel wordt gesteund door de Nederlandse Organisatie voor
Wetenschappelijk Onderzoek (NWO).\\

\noindent ISBN 90-9014235-5

\selectlanguage{english}
\clearemptydoublepage
\pagenumbering{roman}

\lhead[\fancyplain{}{\small\bfseries\thepage}]{}
\rhead[\fancyplain{}{\small\bfseries Contents}]
{\fancyplain{}{\small\bfseries\thepage}}
\cfoot[]{\fancyplain{}{}}
\dominitoc
\tableofcontents

\clearemptydoublepage
\pagenumbering{arabic}

\renewcommand{\chaptermark}[1]{\markboth{#1}{}}
\renewcommand{\sectionmark}[1]{\markright{\thesection\ #1}}
\lhead[\fancyplain{}{\small\bfseries\thepage}]
{\fancyplain{}{\small\bfseries\rightmark}}
\rhead[\fancyplain{}{\small\bfseries\leftmark}]
{\fancyplain{}{\small\bfseries\thepage}}
\cfoot[]{\fancyplain{}{}}

\chapter{Introduction}

The cement for the subjects this manuscript deals with is the Monte 
Carlo method of numerical integration. Therefore, the first section is 
endowed with an introduction to its aspects relevant for the second section, 
which digresses on the main contents of this thesis. 

\vspace{\baselineskip}

\minitoc

\section{Numerical integration}
Numerical integration is an approximation of the solution to an integration
problem. An integration problem consists of the task to integrate a function
$f$, the integrand. Sometimes,
the integral can be calculated analytically, but in most cases, this is not
possible. 
Let us, for simplicity, assume that the problem can be reduced to that of the
calculation of an integral on the $s$-dimensional hypercube $\Kube\df[0,1]^s$.
We denote the Lebesgue integral of the integrand $f$ by 
\begin{equation}
   \leb{f} \;\df\; \intk f(x)\,dx \;\;.
\label{IntEq001}   
\end{equation}
With numerical integration, this integral is estimated by a weighted average
of $f$ over a finite sample of $N$ points $x_k\in\Kube$, that is, by 
\begin{equation}
   \sum_{k=1}^Nw(x_k)f(x_k) \;\overset{?}{\approx}\; \leb{f} \;\;,
\end{equation}
where the numbers $w(x_k)$ are the weights coming with the particular method. 
Such a {\em method} is determined by the choice of the sample and the weights. 

In principle, the only restriction on a method to be acceptable is that, with
the estimate of the integral, it should give an estimate of the expected error
on the result. And of course, this expected error should not be too large. If a
certain method cannot give an error estimate, it is useless. In practice, there
is another restriction on a method to be acceptable, namely that the
computational complexity it introduces should not be too large. It should be
possible to do the computation within reasonable time. The computational
complexity is due to the generation of the sample, evaluating the weights and
evaluating the function values. Naturally, one expects that a result will
become more accurate if larger samples are used, because then more information
about the integrand is used. But if the evaluation of the function values is
very expensive (time consuming), then one would like to use small samples, and
indeed, there are methods that need smaller samples than other methods with the
same accuracy. For these methods, however, the generation of the samples is
more expensive. 

In the case of $s=1$, there are many acceptable and efficient methods. In most
of them, the sample is chosen to be distributed evenly over $[0,1]$, i.e., all
the distances between neighbors are the same and the whole of $[0,1]$ is
covered. Different weights can be chosen, depending on the smoothness of the
integrand. These methods give an expected error that decreases with the number
of points as $1/N^\al$, where $\al>0$, with the general rule that $\al$ is
larger for methods that can be applied to smoother integrands 
(cf.~\cite{PFTV}). 

Conceptually, it is a small step to extrapolate these one-dimensional methods
to more dimensions: the sample is taken to be the Cartesian product in the
coordinates of the one-dimensional samples, and the weights are the products
over the coordinates of the one-dimensional weights. Computationally, however,
is it a large step, for the expected error decreases with $N$ as $1/N^{\al/s}$.
So to get an expected error that is of the ``one-dimensional order'' with $N$
points, you need $N^s$ points. This small disaster is often called the ``curse
of dimensionality''.

A closer look at the choice of the samples reveals the cause of the curse. In
one dimension, the even distribution of the points is the most uniform
distribution possible. This makes the methods applicable to large classes of
functions, because in the choice of the sample no knowledge about the integrand
is assumed. As a result of this, the behavior with $N$ of the expected error
factorizes. In more dimensions, however, a Cartesian product of these
one-dimensional distributions is not at all `uniform'. The distances between
neighbors in different directions are not the same anymore. Therefore, these
methods can only be efficient for integrands that have the same kind of
Cartesian symmetry.  The error estimate, however, includes no knowledge about
the integrand, and as a result of this, increases rapidly with the number of
dimensions.  

\subsection{Monte Carlo integration\label{IntroMC}}
A popular remedy to the curse of dimensionality is the so called Monte Carlo
(MC) method of numerical integration \cite{James}. It is based on the
belief that the points of the sample will be distributed
fairly over $\Kube$ if they are chosen at random. To be
more precise, the points are chosen at random, independently and uniformly
distributed, and the estimate of the integral of a function $f$ is given by the
unweighted average 
\begin{equation}
   \lebM{f}{N}{} \;\df\; \frac{1}{N}\sum_{k=1}^Nf(x_k) \;\;.
\end{equation}
With this choice of the samples, the estimate of the integral becomes a random
variable, and probability theory can be applied to do statements about
it (some relevant topics are reviewed in \Sec{SecProb}).  For example, the
expectation value of $\lebM{f}{N}{}$ is given by 
\begin{equation}
   \Exp(\lebM{f}{N}{}) 
   \;=\; \frac{1}{N}\sum_{k=1}^N\leb{f}
   \;=\; \leb{f} \;\;.
\end{equation}
So the expectation value of the estimate of the integral is equal to the
integral itself. The variance of $\lebM{f}{N}{}$ is equal to 
\begin{equation}
   \Var(\lebM{f}{N}{}) 
   \;=\; \Exp(\lebM{f}{N}{2}) - \Exp(\lebM{f}{N}{})^2 
   \;=\; \frac{\leb{f^2} - \leb{f}^2}{N} \;\;,
\label{IntEq003}   
\end{equation}
where $f^2$ just denotes pointwise multiplication of $f$ with itself. This
means that, if $f$ is square integrable so that $\leb{f^2}$ and
$\Var(\lebM{f}{N}{})$ exist, then we can apply the Chebyshev inequality, with
the result that for large 
$N$, the estimate 
$\lebM{f}{N}{}$ converges to $\leb{f}$ with an expected error given by 
$\sqrt{\Var(\lebM{f}{N}{})}$\,. 
This is a very important result, for it states that the Monte Carlo method
works in any dimension with the same rate of convergence, given by the
$1/\sqrt{N}$-rule. The only restriction is that $f$ has to be square
integrable. If this is not the case, the Monte Carlo estimate of an integral
cannot be trusted. 

\subsubsection{Error estimation}
In practice, one of course does not know $\leb{f^2} - \leb{f}^2$, so that it
has to be estimated. This makes Monte Carlo integration a matter of statistics.
A good estimator for the squared error is given by 
\begin{equation}
   \lebM{f}{N}{[2]} 
   \;\df\;\frac{\lebM{f^2}{N}{}-\lebM{f}{N}{2}}{N-1} \;\;,
\end{equation}
which satisfies $\Exp(\lebM{f}{N}{[2]})=\Var(\lebM{f}{N}{})$. To get more
confidence in the result, an estimate of the squared error on the estimated
squared error can be calculated with 
\begin{equation}
   \lebM{f}{N}{[4]}
  \;\df\;  \frac{\lebM{f^4}{N}{} - 4\lebM{f^3}{N}{}\lebM{f}{N}{} 
                                 + 3\lebM{f^2}{N}{2}}
                {N(N-2)(N-3)}
	  -\frac{4N-6}{(N-2)(N-3)}\,\left(\lebM{f}{N}{[2]}\right)^2
\end{equation}
which satisfies $\Exp(\lebM{f}{N}{[4]})=\Var(\lebM{f}{N}{[2]})$. Notice that
$\leb{f^4}$ has to exist in order to credit any value to $\lebM{f}{N}{[4]}$.
One could, in principle, go on calculating higher errors on errors, but their
significance becomes less and less, {\em if} they converge at all.
 
\subsubsection{Sample generation}
Another question is how to obtain the random points. Monte Carlo integration
is preferably done with the help of a computer, and there exist algorithms
that produce sequences of numbers between $0$ and $1$ that are `as good as
random'.  They are called {\em pseudo-random number generators} (c.f.
\cite{Knuth}).  Since they are implemented on a computer, the algorithms are
deterministic, and the numbers they produce cannot be truly random. The
sequences, however, `look random' and are certainly suitable for the use in
Monte Carlo integration.  Another drawback is that, because computers
represent real numbers by a finite number of bits, the algorithms necessarily
have a period, that is, they can produce only a finite number of numbers, and
if they are recycled, they cannot be considered random anymore. Fortunately,
modern random number generators such as $\texttt{RANLUX}$ have very large
periods, up to $10^{165}$.

Finaly, the finiteness of a computer can cause problems when calculating a 
Lebesgue integral.
In the foregoing, we stated that the Monte
Carlo method is alway applicable if the integrand $f$ is square integrable.
For a computer, however, this is not enough. Consider the function
\begin{equation}
   f(x) \;\df\; \begin{cases}
                1 &\textrm{if $x\in\Qrat$} \\
		0 &\textrm{if $x\not\in\Qrat$} \;,
             \end{cases}
\end{equation}
which has Lebesgue integrals $\leb{f}=\leb{f^2}=0$. A computer represents
numbers with finite strings of bits, i.e., the numbers are always rational,
so that a Monte Carlo estimate will always give $\lebM{f}{N}{}=1$.
Fortunately, this kind of pathological cases do not appear often in physical
applications.

\subsection{Importance sampling\label{IntroIS}}
The original problem is usually not that of the integration of a function on a
hypercube. In general, it is the problem of integrating a function $F$ on a
more complicated manifold $\Mset$. As we have seen before, the problem {\em
has} to be reduced to that of integrating a function $f$ on a hypercube $\Kube$
in order to apply the MC method. This is done with a map
$\vhi:\Kube\mapsto\Mset$, and sometimes, an invertible map can be found in
which cases we simply have
\begin{equation}
   \int_{\Mset}F(y)\,dy 
   \;=\; \int_{\Kube}(F\circ\vhi)(x)|J_\vhi(x)|\,dx \;\;,
\end{equation}
where $J_\vhi$ is the determinant of the Jacobian matrix of $\vhi$, 
so that $f(x)=(F\circ\vhi)(x)|J_\vhi(x)|$.
In general however, this is not the case, and a suitable mapping
$\vhi:\Kube\mapsto\Mset$ and a function $g_\vhi:\Kube\mapsto\Real$ have to be
determined such that 
\begin{equation}
   \int_{\Kube}g_\vhi(x)\de(\vhi(x)-y)\,dx = 1  \;\;,
\end{equation}
where $\de$ is the Dirac delta-distribution on $\Mset$ (cf.~\cite{Choquet}).
The integral of $F$ over $\Mset$ is then given by 
\begin{equation}
   \int_{\Mset}F(y)\,dy = \leb{f_\vhi} \quad\textrm{with}\quad
   f_\vhi(x)=(F\circ\vhi)(x)g_\vhi(x)\;\;.
\end{equation}
If $\vhi$ is invertible, then $g_\vhi(x)=|J_{\vhi}(x)|$. We just used the word
``suitable'' in connection with the determination of $\vhi$, and therefore in
connection with the determination with the function $f_\vhi$, which is not
unique. Importance sampling is the effort to choose $f_\vhi$ such that
$\leb{f_\vhi^2}-\leb{f_\vhi}^2$ is as small as possible, so that the expected
error is as small as possible.  The optimal choice would be
such that it is zero, but this would mean that $f_\vhi(x)=1$ for all
$x\in\Kube$ and that the integration problem is solved analytically. 
In practice, $f_\vhi$ should be chosen as flat as possible.

\subsection{Quasi Monte Carlo integration\label{IntroQMC}}
The Monte Carlo method is very robust, but the $1/\sqrt{N}$ rate of convergence
can be considered rather slow: to get one more significant digit in the result,
100 times more sample points are needed. The Quasi Monte Carlo (QMC) method
tries to improve this behavior, by using samples the points of which are
distributed more uniformly {\em over} the integration region than independent
random points that are distributed uniformly {\em in} the integration region
(cf.~\cite{Nieder1}). 

The previous sentence seems a bit paradoxical, but notice the difference between
`uniformly over' and `uniformly in'. The latter is meant in the probabilistic
sense: a random point is distributed following a distribution {\em in} an
integration region, which can be the uniform distribution. The former is
meant for a set of points: the points can be distributed uniformly {\em over}
the integration region. In this case, the word `uniformly' does not really have 
a meaning yet, and has to be defined, which is done by introducing measures of 
rates of uniformity. They are called {\em discrepancies}, and return a number 
$D_N(\XN)$ for a sample, or {\em point set}, $\XN=(x_1,x_2,\ldots,x_N)$. The 
idea is then that, the higher the number $D_N(\XN)$, the less uniformly the 
points are distributed. 

The task in QMC integration is to find low-discrepancy point sets. The integral
of a function $f$ is then estimated again by the unweighted average
$\lebM{f}{N}{}\df N^{-1}\sum_{k=1}^Nf(x_k)$ over the point set. That this
approach can indeed improve the convergence of the error is, for example, shown
by the Koksma-Hlawka inequality, which states that 
\begin{equation}
   |\lebM{f}{N}{}-\leb{f}| 
   \;\leq\; \textrm{V}_{\scriptscriptstyle\!\textrm{HK}}[f]\,D_N^*(\XN)\;\;, 
\label{KHIneq}   
\end{equation}
where $D_N^*(\XN)$ is the so called {\em star discrepancy} of $\XN$, and
$\textrm{V}_{\scriptscriptstyle\!\textrm{HK}}[f]$ is the variation of $f$ in
the sense of Hardy and Krause. It is a complicated function of $f$ that is,
however, independent of the point set. This inequality states that the error,
made by estimating the integral by an unweighted average over the function
values 
at the points of the point set, decreases with the number of points $N$ at 
least as quickly as the star discrepancy of the point set.

\section{Contents of this thesis}
The main contents start in \Chp{ChapForm}, and can be divided into two
subjects: the calculation of discrepancy distributions, and phase space
integration with the emphasize on a special case of importance sampling.
\Chp{ChapProb} reviews some topics from probability theory and formalism of
Feynman diagrams.

\subsection{Calculation of discrepancy distributions}
As discussed before, the relatively slow convergence of the MC method
has inspired a search for other point sets whose discrepancy is lower than that
expected for truly random points.  The low-discrepancy point sets and {\em
low-discrepancy sequences\/} have developed into a veritable industry, and
sequences with, asymptotically, very low discrepancy are now available,
especially for problems for which the dimension of the integration region is
very large \cite{Tezuka}. For point sets that are extracted as the first $N$
elements of such a sequence, though, one is usually still compelled to compute
the discrepancy numerically, and compare it to the expectation for random
points in order to show that the point set is indeed `better than random'. This
implies, however, that one has to know, for a given discrepancy, its
expectation value for truly random points, or preferably even its probability
density (cf.~\Sec{HQDSec}). 

In \Chp{ChapForm}, we introduce the formalism of the so-called {\em quadratic 
discrepancies}, and derive a formula for the generating function of their 
probability distribution. Furthermore, we give Feynman rules to calculate the 
generating function perturbatively using Feynman diagrams, with $1/N$ as 
expansion parameter. \Chp{ChapInst} digresses on the question whether the 
asymptotic series obtained is correct, and concludes affirmative for two 
examples of discrepancies, with great confidence in the general case.

In \cite{hk1,hk2,hk3} the problem of calculating the probability distribution
of quadratic discrepancies under truly random point sets has been solved for
large classes of discrepancies. Although computable, the resulting
distributions are typically not very illuminating. The exception is usually the
case where the number of dimensions of the integration problem becomes very
large, in which case a normal distribution often arises \cite{jhk,Leeb}. In
\Chp{GausChap}, we investigate this phenomenon in more detail, and we shall
describe the conditions under which this `law of large dimensions' applies.

Throughout the discussion of \Chp{GausChap}, only the asymptotic limit of very
large $N$ is considered, which implies that no statements can be done on how
the number of points has to approach infinity with respect to the number of
dimensions, as was for instance done in \cite{Leeb}. This problem is tackled 
in \Chp{ChapCorr}, in which the diagrammatic expansions of the generating 
function is given and calculated to low order for a few examples. For the 
{\em Lego discrepancy}, which is equivalent with a $\chi^2$-statistic for $N$ 
data points distributed over a number of $M$ bins, cases in which $N$ as well 
as $M$ become large are considered, leading to surprising results. 
Also the {\em Fourier diaphony}, for which a limit is derived in 
\cite{Leeb}, is handled, leading to a stronger limit.

\subsection{Phase space integration\label{IntroPSI}}
A typical example in which the MC method is the only option is in the problem
of {\em phase space integration}. It occurs in particle physics, where the
connection between the model of the particles and the experiments with the
particles is made with the help of transition probabilities (cf.
\cite{Weinberg}). These give the probability to get, under certain conditions,
a transition from one certain state of particles (the {\em initial state}) to
another certain state of particles (the {\em final state}). On one side, these
probabilities can be determined statistically, by performing an experiment
several times, starting with the same initial state every time, and by counting
the number of times certain final states occur. The probabilities can also be
calculated from the model, and then two outcomes can be compared to evaluate
the model. 

Phase space is the space of all possible momentum configurations of the
final-state particles, and particle models predict probability densities on it.
Because of the need of very high statistics for acceptable precision, it is
usually difficult to determine them experimentally. A solution to this
experimental problem is the creation of a mathematical problem: averaging
transition probabilities over phase space.  In the analysis of the experimental
data this just means that final states, that differ only in momentum
configuration, are considered equivalent. In the analysis of the model, this
means that an integration of the probability density over phase space has to be
performed.

The actual quantity that physicists deal with is not the transition
probability, but the {\em cross section}. If the number of initial particles is
two, then it is the transition rate per unit of time, normalized with respect
to the flux of the initial particles, i.e., the density of the initial
particles times their relative velocity. The {\em differential} cross section
$d\si$ of a proces from a two particle initial state to a certain final state
is given by 
\begin{equation}
   d\si(\mathrm{i}\ra\mathrm{f}\,) 
   \;=\; \frac{(2\pi)^4}{v_{\mathrm{i}}}\,
         |M_{\mathrm{f},\mathrm{i}}|^2\, 
	 \de(p_{\mathrm{f}}-p_{\mathrm{i}})\,d\mathrm{f} \;\;.
\end{equation}
In this expression, $d\mathrm{f}$ represents the final state degrees of freedom
that have to be integrated or summed in order to get the desired cross section
$\si$. This includes the final-state momenta. The delta-distribution
represents momentum conservation between the initial and the final states, and
$v_{\mathrm{i}}$ is the relative velocity of the initial particles.  The
characteristics of the particular proces are contained in
$M_{\mathrm{f},\mathrm{i}}$\,, the transition {\em amplitude} or {\em matrix
element}, and has to be calculated using the particle model in the formalism of
quantum field theory. It determines the function that has to be intergrated 
over phase space.

Besides momentum conservation, there are other restrictions the momenta of the
particles have to satisfy, independent of the amplitude. Algorithms that
generate random momenta, satisfying these restrictions, are called {\em phase
space generators}, and in \Chp{ChapPhSp} \rambo\ is described, which generates 
momenta distibuted uniformly in phase space. This chapter also deals with some
techniques that are useful for MC integration in general.

For certain particle processes, the squared amplitude can have complicated peak
structures, that make it hard to be integrated if the momenta are generated
such that they are distributed uniformly in phase space. This is in particular
true if it concernes processes in which the strong interaction is involved, for
which the integrand contains peak structures that are governed by the so-called
{\em antenna pole structure}. In \Chp{ChapSar}, the algorithm \sarge\ is
introduced that generates random momenta, satisfying the restrictions that are
independent of the amplitude, and such that they are distributed following a
density that containes the antenna pole structure. It improves the MC
integration process through importance sampling.

\clearemptydoublepage

\chapter{Probability, measures and diagrams\label{ChapProb}}

Since this thesis is meant to be read by both theoretical physicists and
mathematicians, this chapter elaborates on some subjects that are probably not
everyday routine to the one or the other. This concerns probability theory, 
including a (very) short introduction to martingales, and 
Feynman diagrams. The hasty reader is advised to read at least \Sec{HQDSec} and 
\Sec{CPDSec}.

\vspace{\baselineskip}

\minitoc

\section{Some probability theory\label{SecProb}}
We start this section `at level zero' with respect to the probability theory,
but expect the reader to be familiar with  a bit of set theory, logic, measure
theory, complex analysis and so on. For more details, we refer to
\cite{Laha}, \cite{Bill2} and \cite{Hall}.

\subsection{Probability space}
A probability space consists of a triple $(\Omega,\Ffam,\Pro)$, where
$\Omega$ is a set, $\Ffam$ a $\si$-field of subsets of $\Omega$, and $\Pro$ a
probability measure defined on $\Ffam$. The collection $\Ffam$ of subsets is
called a $\si$-field if 
\begin{enumerate}
   \item $\emptyset\in\Ffam$ \;;
   \item $\Fset\in\Ffam\Longrightarrow \Omega\backslash\Fset\in\Ffam$\;; 
   \item $\Fset_1,\Fset_2,\ldots\in\Ffam\Longrightarrow
          \bigcup_n\Fset_n\in\Ffam$ \;,
\end{enumerate}
where, in the last property, the number of sets $\Fset_n$ has to be countable.
The probability measure $\Pro$ is a function $\Ffam\mapsto[0,1]$ with
$\Pro(\Omega)=1$.
We will only consider probability spaces for which $\Omega$ is a Lebesgue
measurable subset of $\Real^n$, $n=1,2,\ldots$ and for which $\Pro$ is given
by
\begin{equation}
   \Pro(\Fset) \;=\; \int_{\Fset}P(\om)\,d\om \;\;,
\end{equation}
where $d\om$ stands for the Lebesgue measure on $\Omega\subset\Real^n$ and $P$
is a function $\Omega\mapsto[0,\infty)$ with $\int_\Omega P(\om)\,d\om=1$.  $P$
is called the {\em probability density\/} or {\em probability distribution},
although the latter name is more appropriate for the set of doubles
$\{(\Fset,\Pro(\Fset))\mid\Fset\in\Ffam\}$. 

A simple example of a probability distribution is the {\em uniform\/}
distribution in $[0,1]$, for which $P(\om)=1$. This is often extended to more
dimensions, say $n$, by taking the Cartesian product of independent
one-dimensional variables, that is, $P_n(\om)=\prod_{k=1}^nP(\om^{(k)})$, where
$\om=(\om^{(1)},\om^{(2)},\ldots,\om^{(n)})\in[0,1]^n$ and $P(\om^{(k)})=1$
for all $k$. We say that $\om$ is distributed uniformly in $[0,1]^n$.

\subsection{Random variables\label{SecProbRV}}
A random variable $X$ is a function on $\Omega$. It is an object about which
statements $\Pi$ can be made. These statements are then `valued' with a number
between $0$ and $1$ by the probability measure. Probability theory concerns
itself with the calculation of these numbers; their interpretation depends on
the user. It can be a ``rate of belief'' (the Bayesian interpretation) or a
ratio of outcomes in the limit of an infinite number of repetitions of
experiments (the frequentist interpretation). In Monte Carlo integration, for
example, the latter applies.

Let $\Pi(X)$ denote a statement $\Pi$ about $X$, and let
$\Fset_{\Pi(X)}\df\{\omega\in\Omega\mid\Pi(X(\omega))\,\textrm{is true}\}$ be
the subspace of $\Omega$ for which $\Pi(X(\omega))$ is true, then we denote
\begin{equation}
   \Pro(\Pi(X))\df\Pro(\Fset_{\Pi(X)}) \;\;.
\end{equation}
An important operator in the theory of probability is the expectation value 
$\Exp$. It is the average of $X$ over $\Omega$, weighted with $P$:
\begin{equation}
   \Exp(X) \;\df\; \int_\Omega X(\om)P(\om)\,d\om \;\;.
\end{equation}
Especially expectation values of powers of $X$ are often considered, and they 
are called the moments of the probability distribution of $X$. This name 
anticipates the fact that a random variable has its own probability 
distribution, which is simply defined through
\begin{equation}
   \Pro_{\!X}(\Pi(Z)) \df \Pro(\Fset_{\Pi(Z(X))}) \;\;,
\end{equation}
where $\Fset_{\Pi(Z(X))}\df\{\omega\in\Omega\mid\Pi(Z(X(\omega)))\, \textrm{is
true}\}$. From now on, we will assume that $X$ is real, and introduce the 
{\em cumulative probability distribution\/} or {\em distribution function\/}
\begin{equation}
   F_X(x) 
   \;\df\; \Pro(X\leq x) \;\;.
\end{equation}
$F_X$ is a monotonously increasing function $\Real\mapsto[0,1]$. Its
derivative is the probability density $P_X$, and we have 
\begin{equation}
   F_X(x) 
   \;=\; \int_{-\infty}^xP_X(t)\,dt \;\;.
\end{equation}
Discontinuities in $F_X$ are represented by Dirac delta-distributions in $P_X$.
An interesting observation is, furthermore, that if $X$ is distributed
following $F_X$, then the random variable $Y\df F_X(X)$ is distributed
uniformly in $[0,1]$, since $\Pro(Y\leq y)=\Pro(X\leq F_X^{-1}(y))=y$.

We proceed with a translation of confidence 
levels, given by $\Pro$, into expectation values. This is done by the 
{\em Chebyshev inequality}, which states that, for a given number $a>0$,
\begin{equation}
   \Pro(|X|>a) \;\leq\; \frac{\Exp(|X|^{s})}{a^{s}} 
   \quad\textrm{for any $s\geq1$}\;\;.
\label{Cheby}	      
\end{equation}
Its proof is simple. We have  
\begin{align}
   \Pro(|X|>a)
   \;=\; \int_{a}^{\infty}P_X(t)\,dt + \int_{-\infty}^{-a}P_X(t)\,dt 
   \;\leq\;   \int_{a}^{\infty}\frac{|t|^{s}}{a^{s}}\,P_X(t)\,dt 
              +\int_{-\infty}^{-a}\frac{|t|^{s}}{a^{s}}\,P_X(t)\,dt\;\;,\notag
\end{align}
where the inequality holds because $|t|/a\geq1$ under the integrals. The final
expression is smaller than the integral over the whole of $\Real$, which is
equal to the r.h.s.~of \eqn{Cheby}. An example of its use is an estimate of the
probability that a variable $X$ will differ an amount $a$ from its expectation
value $\Exp(X)$. The Chebyshev inequality tells us that 
\begin{equation}
   \Pro(|X-\Exp(X)|>a)
   \;\leq\; \frac{\Exp(|X-\Exp(X)|^2)}{a^2} 
   \;=\; \frac{\Var(X)}{a^2} \;\;,
\end{equation}
where 
\begin{equation}
   \Var(X) \;\df\; \Exp(X^2) - \Exp(X)^2 
\end{equation}
is called the {\em variance\/} of $X$, and its square root
$\si(X)\df\sqrt{\Var(X)}$ is called the {\em standard deviation}. So if we take
$a=c\cdot\si(X)$, then we see that the probability of $|X-\Exp(X)|$ to be
larger than $c\cdot\si(X)$ is smaller than $1/c^2$. 

It is common not to consider the random variable itself, but {\em standardized 
variable\/} which is given by 
\begin{equation}
    \frac{X-\Exp(X)}{\si(X)} \;\;.
\end{equation}
It has its expectation value equal to zero and its variance equal to one.

\subsection{Generating functions\label{ProSecGen}}
If $X$ is real, its probability density can be calculated as follows.
Let $\theta$ denote the Heaviside step-function. It can, for example,
be represented by the integral in the complex plane 
\begin{equation}
   \theta(t) 
   \;=\; \frac{-1}{2\pi i}\int_{\Ga}
         \frac{e^{-zt}}{z}\,dz \;\;, 
\end{equation}
where the contour $\Ga$ is along the line $\RE\,z=-\ve$, and $\ve$ is positive
and small. If $t>0$, then the integration contour can be closed to the right
and the pole in $z=0$ contributes with a residue that is equal to $-1$. An extra
minus sign comes from the orientation of the contour. If $t<0$, then the
contour can be closed to the left, giving zero. The probability distribution 
function $F_X$ is then given by 
\begin{equation}
   F_X(t) 
   \;=\; \int_\Omega\theta(\,t-X(\om)\,)P(\om)\,d\om
   \;=\; \frac{-1}{2\pi i}\int_{\Ga}\frac{e^{-zt}}{z}
         \int_\Omega e^{zX(\om)}P(\om)\,d\om\,dz \;\;.
\end{equation}
The integral over $\Omega$ just gives the expectation value of $e^{zX}$, which 
is called the {\em moment generating function}
\begin{equation}
   G_X(z) \df \Exp(\,e^{zX}\,) \;\;.
\end{equation}
It carries this name, because its derivatives in $z=0$ give the moments
$\Exp(X^n)$ of $X$. In literature, the {\em characteristic function\/} is often 
used, which is just given by $G_X(iz)$.
The final result is that $F_X$ is
given by
\begin{equation}
   F_X(t) \;=\; \frac{-1}{2\pi i}\int_{\Ga}\frac{e^{-zt}}{z}\,G_X(z)\,dz  \;\;.
\label{ProEq001}   
\end{equation}
We can translate this into a formula for the probability density $P_X$ by 
differentiation with respect to $t$. The result is that
\begin{equation}
   P_X(t) \;=\; \frac{1}{2\pi i}\int_{\Ga}e^{-zt}\,G_X(z)\,dz \;\;,
\label{ProEq002}   
\end{equation}
i.e., it is the inverse Laplace transform of the moment generating function of
$X$. Notice that the generating function satisfies $G(0)=1$, because the 
probability density $P_X$ is properly normalized: 
$\int_{\Real}P_X(t)\,dt=1$.

Another generating function that is often used, the cumulant generating
function $W_X$, is simply given by $W_X(z)=\log(G_X(z))$. The first
cumulant is equal to $\Exp(X)$ itself, and the second is the variance
$\Var(X)$.

The generating function of the standardized variable can be expressed in terms
of the original generating function $G_X$ through
\begin{equation}
   \Exp(\,e^{z(X-\Exp(X))\si(X)^{-1}}\,)
   \;=\; e^{-\Exp(X)\si(X)^{-1}}G_X(z\si(X)^{-1}) \;\;.
\end{equation}

\subsection{Convergence of random variables and distributions\label{ProSecCon}}
Sequences $\{X_n\mid n=1,2,\ldots\}$ of random variables are often considered in
probabilistic analyses, and in particular their limiting behavior. Therefore, 
notions of convergence are needed, and we distinguish various types. 
First there is {\em convergence in probability}, and we write
\begin{equation}
   X_n \pcon X  \quad\quad\textrm{if}\quad\quad 
   \Pro_{\!n}(|X_n-X|\geq\ve)\ra0 \quad \forall\,\ve>0 \;\;.
\end{equation}
With the Chebyshev inequality, we see that the requirement for convergence in
probability is satisfied if there is a $p\geq1$ such that
$\Exp(|X_n-X|^p)/\ve^p\ra0$ for all $\ve>0$. This observation suggests to 
introduce {\em convergence in $p^{\textit{th}}$ mean}, and we write
\begin{equation}
   X_n \Lcon X \quad\quad\textrm{if}\quad\quad
   \Exp(|X_n-X|^p)\ra0\;\;.
\end{equation}
The case of $p\ra\infty$ can be considered special, and leads to {\em almost
sure convergence}:  
\begin{equation}
   X_n \acon X \quad\quad\textrm{if}\quad\quad
   X_n(\om)\ra X(\om)\;\;\textrm{for all}\;\om\in\Omega\backslash\Fset\;\;,
\end{equation}
where $\Fset\in\Ffam$ with $\Pro(\Fset)=0$. 
To compare these notions of convergence, we note that (cf.~\cite{Laha})
\begin{align}
   X_n \acon X \quad&\Longrightarrow\quad X_n \pcon X \;\;,\\
   X_n \Lcon X\;\;\textrm{for some $p>0$}
               \quad&\Longrightarrow\quad X_n \pcon X \;\;.
\end{align}
Finally, there is {\em convergence in distribution\/} or {\em convergence in
law}, and we write
\begin{equation}
   X_n \dcon X \quad\quad\textrm{if}\quad\quad P_n \Wcon P \;\;,
\end{equation}
where the latter denotes {\em weak convergence\/} of the distributions $P_n$ of
the variables $X_n$:
\begin{equation}
   P_n \Wcon P \quad\quad\textrm{if}\quad\quad
   \Exp(f(X_n)) \ra \Exp(f(X)) \quad\textrm{for any bounded function $f$.}
\end{equation}
Notice that, in general, the moments of the variables $X_n$ are not bounded
functions.  The generating functions $G_n(z)$, however, are bounded for
imaginary $z$.  We actually have, (cf.~\cite{Bill2}) 
\begin{equation}
   P_n \Wcon P \quad\Longleftrightarrow\quad 
   G_n(z) \ra G(z) \quad\textrm{for each imaginary $z$.}
\end{equation}
The notion of weak convergence is also used in connection with distribution 
functions, and we write
\begin{equation}
   F_n \wcon F \quad\quad\textrm{if}\quad\quad 
   F_n(x)\ra F(x) \quad\textrm{at all continuity points $x$ of $F$.}
\end{equation}
Distribution functions 
are right-continuous and satisfy $F_n(-\infty)=0$ and $F_n(\infty)=1$. Because
$F$ does not have to be a distribution function in case of weak convergence, 
it is useful to define {\em complete convergence}, and we write
\begin{equation}
   F_n \ccon F \quad\quad\textrm{if}\quad\quad 
   F_n\wcon F  \;\;\textrm{and}\;\; F_n(\pm\infty)\ra F(\pm\infty)\;\;.
\end{equation}
We note that weak convergence of a distribution is {\em not\/} necessarily
equivalent with weak convergence of the density, but that (cf.~\cite{Laha})
\begin{equation}
   P_n \Wcon P \quad\Longleftrightarrow\quad F_n \ccon F \;\;,
   \quad\quad\textrm{so that}\quad\quad
   X_n \dcon X \quad\textrm{if}\quad F_n \ccon F \;\;.
\end{equation}
We end this section with the remark that $X_n \pcon X$ implies $X_n \dcon X$
(cf.~\cite{Laha}), so that
\begin{equation}
   X_n \acon X \quad\Longrightarrow\quad 
   X_n \pcon X \quad\Longrightarrow\quad X_n \dcon X \;\;.
\end{equation}

\subsection{Martingales}
With a sequence of random variables $X_n$ should come a sequence of
$\si$-fields $\Ffam_n$. A sequence $\{Z_n,\Ffam_n\mid n=1,2,\ldots\}$ is
called a {\em martingale\/} if 
\begin{enumerate}
  \item $Z_n$ is measurable with respect to $\Ffam_n$ \;;
  \item $\Exp(|Z_n|)<\infty$ \;;
  \item $\Exp(Z_n|\Ffam_m)=Z_m$ with probability one for all $m<n$\;.
\end{enumerate}
The idea is that $Z_n$ depends on a number of $k_n$ variables $\om_i$ that take
their values in $\Omega$.
In $\Exp(Z_n|\Ffam_m)$, the first $k_m<k_n$ variables have to be taken fixed, 
and only the average over the remaining $k_n-k_m$ variables has to be taken.
This average can then be considered to depend on the first $k_m$ variables 
again, and this dependence should be the same as the one of $Z_m$. 

A martingale is called zero-mean if $\Exp(Z_n)=0$ for all $n$. Furthermore, it
is called square-integrable if $\Exp(Z_n^2)$ exists for all $n$. A double
sequence $\{Z_{n,i},\Ffam_{n,i}\mid 1\leq i\leq k_n,n=1,2,\ldots\}$ is called a
martingale array, if $\{Z_{n,i},\Ffam_{n,i}\mid 1\leq i\leq k_n\}$ is a
martingale for each $n\geq1$. The variables $X_{n,i}\df Z_{n,i}-Z_{n,i-1}$ are
called the martingale differences. These are the ingredients needed for the
powerful (c.f. \cite{Hall})

\subsubsection{Central Limit Theorem:\label{CeLiTh}} 
{\sl Let $\{Z_{n,i},\Ffam_{n,i}\mid 1\leq i\leq k_n,n=1,2,\ldots\}$ be a
zero-mean, square-integrable martingale array with differences $X_{n,i}$, and
suppose that} 
\begin{gather}
   \max_i|X_{n,i}| \pcon 0 \;,\label{CLT1}\\
   \sum_i X_{n,i}^2 \pcon 1 \;,\label{CLT2}\\
   \Exp(\,\max_i X_{n,i}^2\,) \quad\textrm{is bounded in $n$}\;,\label{CLT3} \\
   \Ffam_{n,i}\subseteq\Ffam_{n+1,i}\quad\textrm{for}\quad
                                  1\leq i\leq k_n\;,\quad n\geq1 \;\;.
   \label{CLT4}				  
\end{gather}
{\sl Then $Z_{n,k_n}\dcon Z$, where $Z$ is a normal variable.}

A normal variable has a Gaussian distribution with zero mean and unit variance,
given by a density $P(t)=(2\pi)^{-1/2}\exp(-\half t^2)$ and generating function
$G(z)=\exp(\half z^2)$.

\newcommand{\GG}{g}
\newcommand{\FF}{f}
\newcommand{\N}{n}
\subsubsection{Adaptive Monte Carlo integration\label{ProSecAMC}} 
We apply this theorem to adaptive Monte Carlo integration, as a small exercise.
It concerns the problem of calculating the Lebesgue integral
$\lebM{\FF}{\Omega}{}$ of a function $\FF$ on an integration region $\Omega$.
Let $\GG_1,\GG_2,\ldots$ be a sequence of positive functions, where $\GG_k$
depends on $k$ variables $x_i$ that take their values in $\Omega$. We denote
such a set of $k$ variables by $\{x\}_k\df\{x_1,\ldots,x_k\}$. Assume that
$\int_{\Omega}g_k(\{x\}_k)\,dx_k=1$ for all values of $\{x\}_{k-1}$, so that
\begin{equation}
   \GG_{\{x\}_{k-1}} : x_k\mapsto g_k(\{x\}_k)
\end{equation}
is a probability density in $x_k$. Let us also introduce the functions
\begin{equation}
   \bar{\GG}^{-r}_k: x_k\mapsto 
   \int_{\Omega^{k-1}}\frac{\GG_1(x_1)\cdots\GG_{k-1}(\{x\}_{k-1})}
                     {\GG_k(\{x\}_k)^r}\,
   dx_1\cdots dx_{k-1} \;\;.
\end{equation}
In adaptive Monte Carlo integration, one generates a random point $x_1$ in
$\Omega$ following a density $\GG_1$, and with this point a density
$\GG_{\{x\}_{1}}$ is constructed to generate $x_2$, so that $\GG_{\{x\}_{2}}$
can be constructed to generate $x_3$ and so on. Then, one tries to estimate the
integral $\lebM{\FF}{\Omega}{}$ with 
\begin{equation}
   \lebM{\FF}{\N}{} 
   \;\df\; \frac{1}{\N}\sum_{k=1}^\N\frac{\FF(x_k)}{\GG_{\{x\}_{k-1}}(x_k)} 
   \;\;.
\end{equation}
The expectation value and the variance of $\lebM{\FF}{\N}{}$ can easily be
calculated, with the result that $\Exp(\lebM{\FF}{\N}{})=\lebM{\FF}{\Omega}{}$
and $\Var(\lebM{\FF}{\N}{})=\VAR_{\!\N}[\FF]/\N$, where
\begin{equation}
   \VAR_{\!\N}[\FF] 
   \;\df\; \frac{1}{\N}\sum_{k=1}^\N
            \lebM{\bar{\GG}^{-1}_k\FF^2}{\Omega}{}-\lebM{\FF}{\Omega}{2} \;\;.
\end{equation}
Monte Carlo integration is based on the observation that if
$\lebM{\bar{\GG}^{-1}_k\FF^2}{\Omega}{}$ exists for every $k$, so that
$\VAR_{\!\N}[\FF]$ is a finite number, the Chebyshev inequality gives 
\begin{equation}
   \Pro(\,|\lebM{\FF}{\N}{}-\lebM{\FF}{\Omega}{}|>\ve\,)
   \;\leq\; \frac{\VAR_{\!\N}[\FF]}{\ve^2\N}
   \qquad\Longrightarrow\qquad
   \lebM{\FF}{\N}{} \;\pcon\; \lebM{\FF}{\Omega}{} \;\;,
\end{equation}
which suggests to use $\lebM{\FF}{\N}{}$ as an estimator of
$\lebM{\FF}{\Omega}{}$, and to interpret $\VAR_{\!\N}[\FF]/\N$ as the square
of the expected integration error.

We shall prove\footnote{This is a correction of the erroneous proof in the 
original thesis.}
now, that $\lebM{\FF}{\N}{}$ converges to $\lebM{\FF}{\Omega}{}$
with Gaussian confidence levels. Except of the existence of
$\lebM{\bar{\GG}^{-1}_k\FF^2}{\Omega}{}$, we shall need some more requirements,
but first let us introduce the variables
\begin{equation}
   Z_{\N,i} \df \sum_{k=1}^i\bar{X}_{n,k} 
   \quad,\qquad
   \bar{X}_{\N,k} \df \frac{X_{k}}{\sqrt{\N\VAR_{\!\N}[\FF]}}
   \quad,\qquad
   X_{k} \df
   \frac{\FF(x_k)}{\GG_{\{x\}_{k-1}}(x_k)} 
         - \lebM{\FF}{\Omega}{} \;\;. 
\notag	 
\end{equation}
Because we define the variables $Z_{\N,i}$ explicitly as the sum of the
differences $\bar{X}_{\N,k}$, we are clearly dealing with a martingale array
(with $k_{\N}=\N$) satisfying (\ref{CLT4}). It obviously is zero mean, and it
is square integrable by the requirement that
$\lebM{\bar{\GG}^{-1}_k\FF^2}{\Omega}{}$ exists for all $k$. 
For the proof, we shall furthermore need the requirements that  
\begin{equation}
   \lim_{\N\to\infty}\frac{1}{\N^2}\sum_{i=1}^\N\Exp(X_{i}^4) = 0  
   \qquad\textrm{and}\qquad
   \lim_{\N\to\infty}\frac{1}{\N^2}\sum_{i\neq j}^\N
    \left|\Exp(X_i^2X_j^2)-\Exp(X_i^2)\Exp(X_j^2)\right| = 0  \;\;.  
\label{AMCREQ}
\end{equation}
The first one is satisfied if $\Exp(X_{i}^4)$ exists for all $i$, which can be translated
in the demand that $\lebM{\bar{\GG}^{-2}_k\FF^3}{\Omega}{}$ and
$\lebM{\bar{\GG}^{-3}_k\FF^4}{\Omega}{}$ exist for all $k$. The second one
puts a restriction on how strong the dependencies between the variables may be.
This demand is, for example, satisfied if for every $\N$ there are numbers 
$K_\N(i,j)$ such that
\begin{equation}
   \int_{\Omega^{j-i}}\left|  
     \int_{\Omega^{i}}\frac{\GG_1\GG_2\cdots\GG_{j-1}\FF_j^2}
                           {\GG_j}\,dx_1\cdots dx_i
     - \frac{\GG_{i+1}\cdots\GG_{j-1}\FF_j^2}{\GG_j}
			     \right|dx_{i+1}\cdots dx_j \;<\; K_\N(i,j)
\end{equation}
and that satisfy 
$\lim_{\N\to\infty}\frac{1}{\N}\sum_{j=i+1}^\N K_\N(i,j) = 0$.
This is, for example, the case if $K_\N(i,j)\sim 1/|i-j|$.
We prove along
the line of argument as presented in \cite{Leeb}, that the first three
requirements of the theorem are satisfied. 
First we observe that the martingale is constructed such that
\begin{equation}
   \sum_{i=1}^\N\Exp(\bar{X}_{\N,i}^2) 
   \;=\; \frac{1}{\N\VAR_{\!\N}[\FF]}\sum_{i=1}^\N\Exp(X_{i}^2)
   \;=\; \frac{1}{\N\VAR_{\!\N}[\FF]}
         \sum_{i=1}^\N(\,\lebM{\bar{\GG}^{-1}_i\FF^2}{\Omega}{} 
	              - \lebM{f}{\Omega}{2}\,)
   \;=\;1 \;\;,
\label{AMCEQ1}   
\end{equation}
so that, for requirement (\ref{CLT3}), we have 
$\Exp(\,\max_i\bar{X}_{\N,i}^2\,)\leq\sum_{i=1}^\N\Exp(\bar{X}_{\N,i}^2)=1$.
For (\ref{CLT1}) we use the Chebyshev inequality to find that
\begin{align}
   \Pro(\max_i|\bar{X}_{\N,i}|>\ve)
   \;\leq\; \sum_{i=1}^\N\Pro(|\bar{X}_{\N,i}|>\ve) 
   \;\leq\; \frac{1}{\ve^4}\sum_{i=1}^\N\Exp(\bar{X}_{\N,i}^4)
   \;=\;    \frac{1}{\ve^4\VAR_{\!\N}[\FF]^2\N^2}\,
            \sum_{i=1}^\N\Exp(X_{i}^4)
   \;\;,\notag
\end{align}
which goes to zero for all $\ve>0$ by (\ref{AMCREQ}).
Requirement (\ref{CLT2}) goes the same way:
\begin{align}
   \Pro\Big(\Big|\sum_{i=1}^n\bar{X}_{\N,i}^2-1\Big|>\ve\Big)
   &\leq \frac{1}{\ve^2}\Big(
                \sum_{i,j=1}^\N\Exp(\bar{X}_{\N,i}^2\bar{X}_{\N,j}^2)
	     - 2\sum_{i=1}^\N\Exp(\bar{X}_{\N,i}^2) + 1\Big)  \notag\\
   &= \frac{1}{\ve^2\VAR_{\!\N}[\FF]^2\N^2}\Big( 
          \sum_{i=1}^\N\left(\Exp(X_{i}^4)-\Exp(X_{i}^2)^2\right) \notag\\
   &\qquad\qquad\qquad\quad
      + \sum_{i\neq j}^\N\left(\Exp(X_i^2X_j^2)-\Exp(X_i^2)\Exp(X_j^2)\right)
	  \Big)  \;\;,
\notag
\end{align}
where we used (\ref{AMCEQ1}) again. The final expression 
goes to zero for all $\ve>0$ by (\ref{AMCREQ}). The
result is that, because the variables $Z_{n,n}$ converge to a Gaussian variable
with zero-mean and variance one, the random variables 
\begin{equation}
   \lebM{\FF}{\N}{} 
   \;=\; \sqrt{\frac{\VAR_{\!\N}[\FF]}{\N}}\,Z_{n,n} + \lebM{\FF}{\Omega}{} 
\end{equation}
converge to a Gaussian variable with mean $\lebM{\FF}{\Omega}{}$ 
and variance $\VAR_{\!\N}[\FF]/\N$. 
Note that for non-adaptive Monte Carlo integration, for which the densities
$\GG_k$ are equal to a fixed density $g$ for all $k$, the L\'evy Central Limit
Theorem applies (cf.~\cite{Laha}) and only the existence of
$\lebM{\FF^2/\GG}{\Omega}{}$ is needed.

\subsection{Hypothesis testing, qualification and discrepancies\label{HQDSec}}
Probability theory is extensively used in the field of statistics.
Statisticians try to derive probability distributions from empirical data,
which are believed to be distributed following an existing, but unknown,
distribution. In this section, some statistical procedures and their relevance
to the main subjects of this thesis are discussed.

\subsubsection{Hypothesis testing}
One way to test a model of a physical system is by deriving from this model the
probability distribution according to which certain data from the system are
supposed to be distributed. Then a test has to be developed, which measures the
deviation between the probability distribution from the model, and the
empirical distribution of the data. Because the actual probability distribution
that the data seem to be drawn from is not known, this procedure belongs to the
field of statistics, and it goes under the name of {\em hypothesis testing}. 

Let $\XN=\{x_1,\ldots,x_N\}$ be a sample of physical data, $P_N$ the 
probability density derived from the model (the hypothesis), and $T_N$ the
statistical test. In order for the test to be suitable, it should be developed
such that, if $\XN$ is distributed following $P_N$, then 
\begin{equation}
   \lim_{N\ra\infty}T_N(\XN) = 0 \;\;.
\end{equation}
The idea is then that, for a finite number of data, $T_N(\XN)$ also has to be
small if $\XN$ is distributed following $P_N$. If $T_N(\XN)$ happens to be to
large, the hypothesis has to be rejected. The question is now: what is {\em
small\/} or {\em large}? In order to answer this question, the probability
distribution of $T_N$ under $P_N$ has to be calculated. The probability 
distribution function $F_{T,N}$ is given by 
\begin{equation}
   F_{T,N}(t) \;\df\; \int_{\Omega_N}\theta(\,t-T_N(\om)\,)P_N(\om)\,d\om\;\;,
\end{equation}
where $\Omega_N$ is the space the data $\XN$ can take their values in.
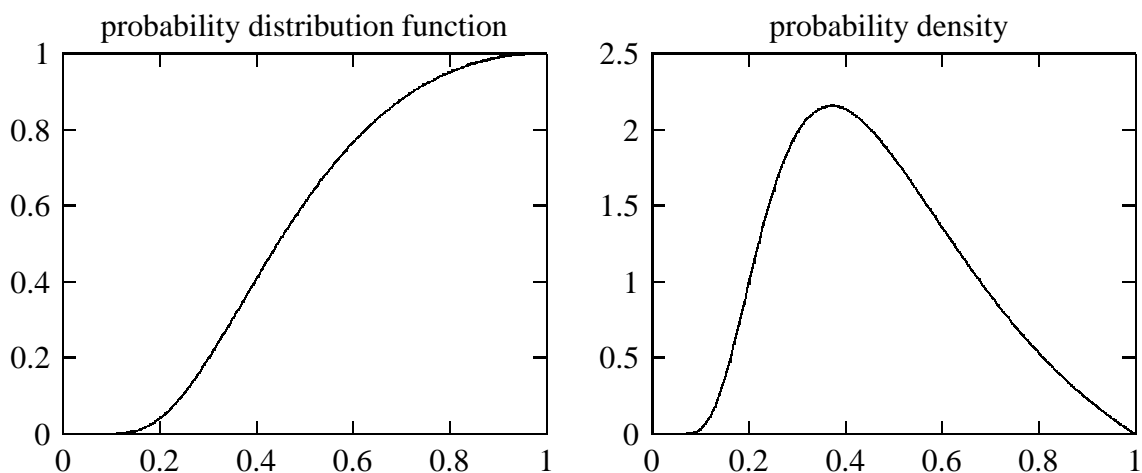
\begin{figure}
\begin{center}
\hspace{-20pt}
%
\setlength{\unitlength}{0.240900pt}
\ifx\plotpoint\undefined\newsavebox{\plotpoint}\fi
\begin{picture}(900,720)(0,0)
\sbox{\plotpoint}{\rule[-0.200pt]{0.400pt}{0.400pt}}%
\put(120.0,82.0){\rule[-0.200pt]{4.818pt}{0.400pt}}
\put(100,82){\makebox(0,0)[r]{0}}
\put(859.0,82.0){\rule[-0.200pt]{4.818pt}{0.400pt}}
\put(120.0,202.0){\rule[-0.200pt]{4.818pt}{0.400pt}}
\put(100,202){\makebox(0,0)[r]{0.2}}
\put(859.0,202.0){\rule[-0.200pt]{4.818pt}{0.400pt}}
\put(120.0,321.0){\rule[-0.200pt]{4.818pt}{0.400pt}}
\put(100,321){\makebox(0,0)[r]{0.4}}
\put(859.0,321.0){\rule[-0.200pt]{4.818pt}{0.400pt}}
\put(120.0,441.0){\rule[-0.200pt]{4.818pt}{0.400pt}}
\put(100,441){\makebox(0,0)[r]{0.6}}
\put(859.0,441.0){\rule[-0.200pt]{4.818pt}{0.400pt}}
\put(120.0,560.0){\rule[-0.200pt]{4.818pt}{0.400pt}}
\put(100,560){\makebox(0,0)[r]{0.8}}
\put(859.0,560.0){\rule[-0.200pt]{4.818pt}{0.400pt}}
\put(120.0,680.0){\rule[-0.200pt]{4.818pt}{0.400pt}}
\put(100,680){\makebox(0,0)[r]{1}}
\put(859.0,680.0){\rule[-0.200pt]{4.818pt}{0.400pt}}
\put(120.0,82.0){\rule[-0.200pt]{0.400pt}{4.818pt}}
\put(120,41){\makebox(0,0){0}}
\put(120.0,660.0){\rule[-0.200pt]{0.400pt}{4.818pt}}
\put(272.0,82.0){\rule[-0.200pt]{0.400pt}{4.818pt}}
\put(272,41){\makebox(0,0){0.2}}
\put(272.0,660.0){\rule[-0.200pt]{0.400pt}{4.818pt}}
\put(424.0,82.0){\rule[-0.200pt]{0.400pt}{4.818pt}}
\put(424,41){\makebox(0,0){0.4}}
\put(424.0,660.0){\rule[-0.200pt]{0.400pt}{4.818pt}}
\put(575.0,82.0){\rule[-0.200pt]{0.400pt}{4.818pt}}
\put(575,41){\makebox(0,0){0.6}}
\put(575.0,660.0){\rule[-0.200pt]{0.400pt}{4.818pt}}
\put(727.0,82.0){\rule[-0.200pt]{0.400pt}{4.818pt}}
\put(727,41){\makebox(0,0){0.8}}
\put(727.0,660.0){\rule[-0.200pt]{0.400pt}{4.818pt}}
\put(879.0,82.0){\rule[-0.200pt]{0.400pt}{4.818pt}}
\put(879,41){\makebox(0,0){1}}
\put(879.0,660.0){\rule[-0.200pt]{0.400pt}{4.818pt}}
\put(120.0,82.0){\rule[-0.200pt]{182.843pt}{0.400pt}}
\put(879.0,82.0){\rule[-0.200pt]{0.400pt}{144.058pt}}
\put(120.0,680.0){\rule[-0.200pt]{182.843pt}{0.400pt}}
\put(180,720){\makebox(0,0)[l]{probability distribution function}}
\put(120.0,82.0){\rule[-0.200pt]{0.400pt}{144.058pt}}
\put(128,82){\usebox{\plotpoint}}
\put(204,81.67){\rule{1.927pt}{0.400pt}}
\multiput(204.00,81.17)(4.000,1.000){2}{\rule{0.964pt}{0.400pt}}
\put(212,82.67){\rule{1.927pt}{0.400pt}}
\multiput(212.00,82.17)(4.000,1.000){2}{\rule{0.964pt}{0.400pt}}
\put(220,83.67){\rule{1.686pt}{0.400pt}}
\multiput(220.00,83.17)(3.500,1.000){2}{\rule{0.843pt}{0.400pt}}
\put(227,85.17){\rule{1.700pt}{0.400pt}}
\multiput(227.00,84.17)(4.472,2.000){2}{\rule{0.850pt}{0.400pt}}
\multiput(235.00,87.61)(1.579,0.447){3}{\rule{1.167pt}{0.108pt}}
\multiput(235.00,86.17)(5.579,3.000){2}{\rule{0.583pt}{0.400pt}}
\multiput(243.00,90.61)(1.355,0.447){3}{\rule{1.033pt}{0.108pt}}
\multiput(243.00,89.17)(4.855,3.000){2}{\rule{0.517pt}{0.400pt}}
\multiput(250.00,93.60)(1.066,0.468){5}{\rule{0.900pt}{0.113pt}}
\multiput(250.00,92.17)(6.132,4.000){2}{\rule{0.450pt}{0.400pt}}
\multiput(258.00,97.59)(0.821,0.477){7}{\rule{0.740pt}{0.115pt}}
\multiput(258.00,96.17)(6.464,5.000){2}{\rule{0.370pt}{0.400pt}}
\multiput(266.00,102.59)(0.581,0.482){9}{\rule{0.567pt}{0.116pt}}
\multiput(266.00,101.17)(5.824,6.000){2}{\rule{0.283pt}{0.400pt}}
\multiput(273.00,108.59)(0.671,0.482){9}{\rule{0.633pt}{0.116pt}}
\multiput(273.00,107.17)(6.685,6.000){2}{\rule{0.317pt}{0.400pt}}
\multiput(281.00,114.59)(0.569,0.485){11}{\rule{0.557pt}{0.117pt}}
\multiput(281.00,113.17)(6.844,7.000){2}{\rule{0.279pt}{0.400pt}}
\multiput(289.59,121.00)(0.485,0.569){11}{\rule{0.117pt}{0.557pt}}
\multiput(288.17,121.00)(7.000,6.844){2}{\rule{0.400pt}{0.279pt}}
\multiput(296.59,129.00)(0.488,0.560){13}{\rule{0.117pt}{0.550pt}}
\multiput(295.17,129.00)(8.000,7.858){2}{\rule{0.400pt}{0.275pt}}
\multiput(304.59,138.00)(0.488,0.560){13}{\rule{0.117pt}{0.550pt}}
\multiput(303.17,138.00)(8.000,7.858){2}{\rule{0.400pt}{0.275pt}}
\multiput(312.59,147.00)(0.485,0.721){11}{\rule{0.117pt}{0.671pt}}
\multiput(311.17,147.00)(7.000,8.606){2}{\rule{0.400pt}{0.336pt}}
\multiput(319.59,157.00)(0.488,0.692){13}{\rule{0.117pt}{0.650pt}}
\multiput(318.17,157.00)(8.000,9.651){2}{\rule{0.400pt}{0.325pt}}
\multiput(327.59,168.00)(0.488,0.692){13}{\rule{0.117pt}{0.650pt}}
\multiput(326.17,168.00)(8.000,9.651){2}{\rule{0.400pt}{0.325pt}}
\multiput(335.59,179.00)(0.485,0.874){11}{\rule{0.117pt}{0.786pt}}
\multiput(334.17,179.00)(7.000,10.369){2}{\rule{0.400pt}{0.393pt}}
\multiput(342.59,191.00)(0.488,0.692){13}{\rule{0.117pt}{0.650pt}}
\multiput(341.17,191.00)(8.000,9.651){2}{\rule{0.400pt}{0.325pt}}
\multiput(350.59,202.00)(0.488,0.824){13}{\rule{0.117pt}{0.750pt}}
\multiput(349.17,202.00)(8.000,11.443){2}{\rule{0.400pt}{0.375pt}}
\multiput(358.59,215.00)(0.485,0.874){11}{\rule{0.117pt}{0.786pt}}
\multiput(357.17,215.00)(7.000,10.369){2}{\rule{0.400pt}{0.393pt}}
\multiput(365.59,227.00)(0.488,0.824){13}{\rule{0.117pt}{0.750pt}}
\multiput(364.17,227.00)(8.000,11.443){2}{\rule{0.400pt}{0.375pt}}
\multiput(373.59,240.00)(0.488,0.758){13}{\rule{0.117pt}{0.700pt}}
\multiput(372.17,240.00)(8.000,10.547){2}{\rule{0.400pt}{0.350pt}}
\multiput(381.59,252.00)(0.485,0.950){11}{\rule{0.117pt}{0.843pt}}
\multiput(380.17,252.00)(7.000,11.251){2}{\rule{0.400pt}{0.421pt}}
\multiput(388.59,265.00)(0.488,0.824){13}{\rule{0.117pt}{0.750pt}}
\multiput(387.17,265.00)(8.000,11.443){2}{\rule{0.400pt}{0.375pt}}
\multiput(396.59,278.00)(0.488,0.824){13}{\rule{0.117pt}{0.750pt}}
\multiput(395.17,278.00)(8.000,11.443){2}{\rule{0.400pt}{0.375pt}}
\multiput(404.59,291.00)(0.485,0.950){11}{\rule{0.117pt}{0.843pt}}
\multiput(403.17,291.00)(7.000,11.251){2}{\rule{0.400pt}{0.421pt}}
\multiput(411.59,304.00)(0.488,0.824){13}{\rule{0.117pt}{0.750pt}}
\multiput(410.17,304.00)(8.000,11.443){2}{\rule{0.400pt}{0.375pt}}
\multiput(419.59,317.00)(0.488,0.824){13}{\rule{0.117pt}{0.750pt}}
\multiput(418.17,317.00)(8.000,11.443){2}{\rule{0.400pt}{0.375pt}}
\multiput(427.59,330.00)(0.485,0.950){11}{\rule{0.117pt}{0.843pt}}
\multiput(426.17,330.00)(7.000,11.251){2}{\rule{0.400pt}{0.421pt}}
\multiput(434.59,343.00)(0.488,0.824){13}{\rule{0.117pt}{0.750pt}}
\multiput(433.17,343.00)(8.000,11.443){2}{\rule{0.400pt}{0.375pt}}
\multiput(442.59,356.00)(0.488,0.758){13}{\rule{0.117pt}{0.700pt}}
\multiput(441.17,356.00)(8.000,10.547){2}{\rule{0.400pt}{0.350pt}}
\multiput(450.59,368.00)(0.485,0.950){11}{\rule{0.117pt}{0.843pt}}
\multiput(449.17,368.00)(7.000,11.251){2}{\rule{0.400pt}{0.421pt}}
\multiput(457.59,381.00)(0.488,0.758){13}{\rule{0.117pt}{0.700pt}}
\multiput(456.17,381.00)(8.000,10.547){2}{\rule{0.400pt}{0.350pt}}
\multiput(465.59,393.00)(0.488,0.758){13}{\rule{0.117pt}{0.700pt}}
\multiput(464.17,393.00)(8.000,10.547){2}{\rule{0.400pt}{0.350pt}}
\multiput(473.59,405.00)(0.485,0.798){11}{\rule{0.117pt}{0.729pt}}
\multiput(472.17,405.00)(7.000,9.488){2}{\rule{0.400pt}{0.364pt}}
\multiput(480.59,416.00)(0.488,0.758){13}{\rule{0.117pt}{0.700pt}}
\multiput(479.17,416.00)(8.000,10.547){2}{\rule{0.400pt}{0.350pt}}
\multiput(488.59,428.00)(0.488,0.692){13}{\rule{0.117pt}{0.650pt}}
\multiput(487.17,428.00)(8.000,9.651){2}{\rule{0.400pt}{0.325pt}}
\multiput(496.59,439.00)(0.485,0.798){11}{\rule{0.117pt}{0.729pt}}
\multiput(495.17,439.00)(7.000,9.488){2}{\rule{0.400pt}{0.364pt}}
\multiput(503.59,450.00)(0.488,0.692){13}{\rule{0.117pt}{0.650pt}}
\multiput(502.17,450.00)(8.000,9.651){2}{\rule{0.400pt}{0.325pt}}
\multiput(511.59,461.00)(0.488,0.626){13}{\rule{0.117pt}{0.600pt}}
\multiput(510.17,461.00)(8.000,8.755){2}{\rule{0.400pt}{0.300pt}}
\multiput(519.59,471.00)(0.485,0.798){11}{\rule{0.117pt}{0.729pt}}
\multiput(518.17,471.00)(7.000,9.488){2}{\rule{0.400pt}{0.364pt}}
\multiput(526.59,482.00)(0.488,0.560){13}{\rule{0.117pt}{0.550pt}}
\multiput(525.17,482.00)(8.000,7.858){2}{\rule{0.400pt}{0.275pt}}
\multiput(534.59,491.00)(0.488,0.626){13}{\rule{0.117pt}{0.600pt}}
\multiput(533.17,491.00)(8.000,8.755){2}{\rule{0.400pt}{0.300pt}}
\multiput(542.59,501.00)(0.485,0.645){11}{\rule{0.117pt}{0.614pt}}
\multiput(541.17,501.00)(7.000,7.725){2}{\rule{0.400pt}{0.307pt}}
\multiput(549.59,510.00)(0.488,0.560){13}{\rule{0.117pt}{0.550pt}}
\multiput(548.17,510.00)(8.000,7.858){2}{\rule{0.400pt}{0.275pt}}
\multiput(557.59,519.00)(0.488,0.560){13}{\rule{0.117pt}{0.550pt}}
\multiput(556.17,519.00)(8.000,7.858){2}{\rule{0.400pt}{0.275pt}}
\multiput(565.59,528.00)(0.485,0.645){11}{\rule{0.117pt}{0.614pt}}
\multiput(564.17,528.00)(7.000,7.725){2}{\rule{0.400pt}{0.307pt}}
\multiput(572.00,537.59)(0.494,0.488){13}{\rule{0.500pt}{0.117pt}}
\multiput(572.00,536.17)(6.962,8.000){2}{\rule{0.250pt}{0.400pt}}
\multiput(580.00,545.59)(0.494,0.488){13}{\rule{0.500pt}{0.117pt}}
\multiput(580.00,544.17)(6.962,8.000){2}{\rule{0.250pt}{0.400pt}}
\multiput(588.00,553.59)(0.492,0.485){11}{\rule{0.500pt}{0.117pt}}
\multiput(588.00,552.17)(5.962,7.000){2}{\rule{0.250pt}{0.400pt}}
\multiput(595.00,560.59)(0.494,0.488){13}{\rule{0.500pt}{0.117pt}}
\multiput(595.00,559.17)(6.962,8.000){2}{\rule{0.250pt}{0.400pt}}
\multiput(603.00,568.59)(0.569,0.485){11}{\rule{0.557pt}{0.117pt}}
\multiput(603.00,567.17)(6.844,7.000){2}{\rule{0.279pt}{0.400pt}}
\multiput(611.00,575.59)(0.492,0.485){11}{\rule{0.500pt}{0.117pt}}
\multiput(611.00,574.17)(5.962,7.000){2}{\rule{0.250pt}{0.400pt}}
\multiput(618.00,582.59)(0.671,0.482){9}{\rule{0.633pt}{0.116pt}}
\multiput(618.00,581.17)(6.685,6.000){2}{\rule{0.317pt}{0.400pt}}
\multiput(626.00,588.59)(0.671,0.482){9}{\rule{0.633pt}{0.116pt}}
\multiput(626.00,587.17)(6.685,6.000){2}{\rule{0.317pt}{0.400pt}}
\multiput(634.00,594.59)(0.581,0.482){9}{\rule{0.567pt}{0.116pt}}
\multiput(634.00,593.17)(5.824,6.000){2}{\rule{0.283pt}{0.400pt}}
\multiput(641.00,600.59)(0.671,0.482){9}{\rule{0.633pt}{0.116pt}}
\multiput(641.00,599.17)(6.685,6.000){2}{\rule{0.317pt}{0.400pt}}
\multiput(649.00,606.59)(0.671,0.482){9}{\rule{0.633pt}{0.116pt}}
\multiput(649.00,605.17)(6.685,6.000){2}{\rule{0.317pt}{0.400pt}}
\multiput(657.00,612.59)(0.710,0.477){7}{\rule{0.660pt}{0.115pt}}
\multiput(657.00,611.17)(5.630,5.000){2}{\rule{0.330pt}{0.400pt}}
\multiput(664.00,617.59)(0.821,0.477){7}{\rule{0.740pt}{0.115pt}}
\multiput(664.00,616.17)(6.464,5.000){2}{\rule{0.370pt}{0.400pt}}
\multiput(672.00,622.59)(0.821,0.477){7}{\rule{0.740pt}{0.115pt}}
\multiput(672.00,621.17)(6.464,5.000){2}{\rule{0.370pt}{0.400pt}}
\multiput(680.00,627.60)(0.920,0.468){5}{\rule{0.800pt}{0.113pt}}
\multiput(680.00,626.17)(5.340,4.000){2}{\rule{0.400pt}{0.400pt}}
\multiput(687.00,631.60)(1.066,0.468){5}{\rule{0.900pt}{0.113pt}}
\multiput(687.00,630.17)(6.132,4.000){2}{\rule{0.450pt}{0.400pt}}
\multiput(695.00,635.60)(1.066,0.468){5}{\rule{0.900pt}{0.113pt}}
\multiput(695.00,634.17)(6.132,4.000){2}{\rule{0.450pt}{0.400pt}}
\multiput(703.00,639.60)(0.920,0.468){5}{\rule{0.800pt}{0.113pt}}
\multiput(703.00,638.17)(5.340,4.000){2}{\rule{0.400pt}{0.400pt}}
\multiput(710.00,643.60)(1.066,0.468){5}{\rule{0.900pt}{0.113pt}}
\multiput(710.00,642.17)(6.132,4.000){2}{\rule{0.450pt}{0.400pt}}
\multiput(718.00,647.61)(1.579,0.447){3}{\rule{1.167pt}{0.108pt}}
\multiput(718.00,646.17)(5.579,3.000){2}{\rule{0.583pt}{0.400pt}}
\multiput(726.00,650.61)(1.355,0.447){3}{\rule{1.033pt}{0.108pt}}
\multiput(726.00,649.17)(4.855,3.000){2}{\rule{0.517pt}{0.400pt}}
\multiput(733.00,653.61)(1.579,0.447){3}{\rule{1.167pt}{0.108pt}}
\multiput(733.00,652.17)(5.579,3.000){2}{\rule{0.583pt}{0.400pt}}
\multiput(741.00,656.61)(1.579,0.447){3}{\rule{1.167pt}{0.108pt}}
\multiput(741.00,655.17)(5.579,3.000){2}{\rule{0.583pt}{0.400pt}}
\multiput(749.00,659.61)(1.355,0.447){3}{\rule{1.033pt}{0.108pt}}
\multiput(749.00,658.17)(4.855,3.000){2}{\rule{0.517pt}{0.400pt}}
\put(756,662.17){\rule{1.700pt}{0.400pt}}
\multiput(756.00,661.17)(4.472,2.000){2}{\rule{0.850pt}{0.400pt}}
\put(764,664.17){\rule{1.700pt}{0.400pt}}
\multiput(764.00,663.17)(4.472,2.000){2}{\rule{0.850pt}{0.400pt}}
\put(772,666.17){\rule{1.500pt}{0.400pt}}
\multiput(772.00,665.17)(3.887,2.000){2}{\rule{0.750pt}{0.400pt}}
\put(779,668.17){\rule{1.700pt}{0.400pt}}
\multiput(779.00,667.17)(4.472,2.000){2}{\rule{0.850pt}{0.400pt}}
\put(787,670.17){\rule{1.700pt}{0.400pt}}
\multiput(787.00,669.17)(4.472,2.000){2}{\rule{0.850pt}{0.400pt}}
\put(795,671.67){\rule{1.686pt}{0.400pt}}
\multiput(795.00,671.17)(3.500,1.000){2}{\rule{0.843pt}{0.400pt}}
\put(802,673.17){\rule{1.700pt}{0.400pt}}
\multiput(802.00,672.17)(4.472,2.000){2}{\rule{0.850pt}{0.400pt}}
\put(810,674.67){\rule{1.927pt}{0.400pt}}
\multiput(810.00,674.17)(4.000,1.000){2}{\rule{0.964pt}{0.400pt}}
\put(818,675.67){\rule{1.686pt}{0.400pt}}
\multiput(818.00,675.17)(3.500,1.000){2}{\rule{0.843pt}{0.400pt}}
\put(825,676.67){\rule{1.927pt}{0.400pt}}
\multiput(825.00,676.17)(4.000,1.000){2}{\rule{0.964pt}{0.400pt}}
\put(128.0,82.0){\rule[-0.200pt]{18.308pt}{0.400pt}}
\put(841,677.67){\rule{1.686pt}{0.400pt}}
\multiput(841.00,677.17)(3.500,1.000){2}{\rule{0.843pt}{0.400pt}}
\put(833.0,678.0){\rule[-0.200pt]{1.927pt}{0.400pt}}
\put(856,678.67){\rule{1.927pt}{0.400pt}}
\multiput(856.00,678.17)(4.000,1.000){2}{\rule{0.964pt}{0.400pt}}
\put(848.0,679.0){\rule[-0.200pt]{1.927pt}{0.400pt}}
\put(864.0,680.0){\rule[-0.200pt]{3.613pt}{0.400pt}}
\end{picture}
%
%
\setlength{\unitlength}{0.240900pt}
\ifx\plotpoint\undefined\newsavebox{\plotpoint}\fi
\begin{picture}(900,720)(0,0)
\sbox{\plotpoint}{\rule[-0.200pt]{0.400pt}{0.400pt}}%
\put(120.0,82.0){\rule[-0.200pt]{4.818pt}{0.400pt}}
\put(100,82){\makebox(0,0)[r]{0}}
\put(859.0,82.0){\rule[-0.200pt]{4.818pt}{0.400pt}}
\put(120.0,202.0){\rule[-0.200pt]{4.818pt}{0.400pt}}
\put(100,202){\makebox(0,0)[r]{0.5}}
\put(859.0,202.0){\rule[-0.200pt]{4.818pt}{0.400pt}}
\put(120.0,321.0){\rule[-0.200pt]{4.818pt}{0.400pt}}
\put(100,321){\makebox(0,0)[r]{1}}
\put(859.0,321.0){\rule[-0.200pt]{4.818pt}{0.400pt}}
\put(120.0,441.0){\rule[-0.200pt]{4.818pt}{0.400pt}}
\put(100,441){\makebox(0,0)[r]{1.5}}
\put(859.0,441.0){\rule[-0.200pt]{4.818pt}{0.400pt}}
\put(120.0,560.0){\rule[-0.200pt]{4.818pt}{0.400pt}}
\put(100,560){\makebox(0,0)[r]{2}}
\put(859.0,560.0){\rule[-0.200pt]{4.818pt}{0.400pt}}
\put(120.0,680.0){\rule[-0.200pt]{4.818pt}{0.400pt}}
\put(100,680){\makebox(0,0)[r]{2.5}}
\put(859.0,680.0){\rule[-0.200pt]{4.818pt}{0.400pt}}
\put(120.0,82.0){\rule[-0.200pt]{0.400pt}{4.818pt}}
\put(120,41){\makebox(0,0){0}}
\put(120.0,660.0){\rule[-0.200pt]{0.400pt}{4.818pt}}
\put(272.0,82.0){\rule[-0.200pt]{0.400pt}{4.818pt}}
\put(272,41){\makebox(0,0){0.2}}
\put(272.0,660.0){\rule[-0.200pt]{0.400pt}{4.818pt}}
\put(424.0,82.0){\rule[-0.200pt]{0.400pt}{4.818pt}}
\put(424,41){\makebox(0,0){0.4}}
\put(424.0,660.0){\rule[-0.200pt]{0.400pt}{4.818pt}}
\put(575.0,82.0){\rule[-0.200pt]{0.400pt}{4.818pt}}
\put(575,41){\makebox(0,0){0.6}}
\put(575.0,660.0){\rule[-0.200pt]{0.400pt}{4.818pt}}
\put(727.0,82.0){\rule[-0.200pt]{0.400pt}{4.818pt}}
\put(727,41){\makebox(0,0){0.8}}
\put(727.0,660.0){\rule[-0.200pt]{0.400pt}{4.818pt}}
\put(879.0,82.0){\rule[-0.200pt]{0.400pt}{4.818pt}}
\put(879,41){\makebox(0,0){1}}
\put(879.0,660.0){\rule[-0.200pt]{0.400pt}{4.818pt}}
\put(120.0,82.0){\rule[-0.200pt]{182.843pt}{0.400pt}}
\put(879.0,82.0){\rule[-0.200pt]{0.400pt}{144.058pt}}
\put(120.0,680.0){\rule[-0.200pt]{182.843pt}{0.400pt}}
\put(305,720){\makebox(0,0)[l]{probability density}}
\put(120.0,82.0){\rule[-0.200pt]{0.400pt}{144.058pt}}
\put(128,82){\usebox{\plotpoint}}
\put(174,81.67){\rule{1.686pt}{0.400pt}}
\multiput(174.00,81.17)(3.500,1.000){2}{\rule{0.843pt}{0.400pt}}
\put(181,83.17){\rule{1.700pt}{0.400pt}}
\multiput(181.00,82.17)(4.472,2.000){2}{\rule{0.850pt}{0.400pt}}
\multiput(189.00,85.59)(0.821,0.477){7}{\rule{0.740pt}{0.115pt}}
\multiput(189.00,84.17)(6.464,5.000){2}{\rule{0.370pt}{0.400pt}}
\multiput(197.59,90.00)(0.485,0.569){11}{\rule{0.117pt}{0.557pt}}
\multiput(196.17,90.00)(7.000,6.844){2}{\rule{0.400pt}{0.279pt}}
\multiput(204.59,98.00)(0.488,0.692){13}{\rule{0.117pt}{0.650pt}}
\multiput(203.17,98.00)(8.000,9.651){2}{\rule{0.400pt}{0.325pt}}
\multiput(212.59,109.00)(0.488,1.088){13}{\rule{0.117pt}{0.950pt}}
\multiput(211.17,109.00)(8.000,15.028){2}{\rule{0.400pt}{0.475pt}}
\multiput(220.59,126.00)(0.485,1.484){11}{\rule{0.117pt}{1.243pt}}
\multiput(219.17,126.00)(7.000,17.420){2}{\rule{0.400pt}{0.621pt}}
\multiput(227.59,146.00)(0.488,1.550){13}{\rule{0.117pt}{1.300pt}}
\multiput(226.17,146.00)(8.000,21.302){2}{\rule{0.400pt}{0.650pt}}
\multiput(235.59,170.00)(0.488,1.748){13}{\rule{0.117pt}{1.450pt}}
\multiput(234.17,170.00)(8.000,23.990){2}{\rule{0.400pt}{0.725pt}}
\multiput(243.59,197.00)(0.485,2.247){11}{\rule{0.117pt}{1.814pt}}
\multiput(242.17,197.00)(7.000,26.234){2}{\rule{0.400pt}{0.907pt}}
\multiput(250.59,227.00)(0.488,2.013){13}{\rule{0.117pt}{1.650pt}}
\multiput(249.17,227.00)(8.000,27.575){2}{\rule{0.400pt}{0.825pt}}
\multiput(258.59,258.00)(0.488,2.079){13}{\rule{0.117pt}{1.700pt}}
\multiput(257.17,258.00)(8.000,28.472){2}{\rule{0.400pt}{0.850pt}}
\multiput(266.59,290.00)(0.485,2.399){11}{\rule{0.117pt}{1.929pt}}
\multiput(265.17,290.00)(7.000,27.997){2}{\rule{0.400pt}{0.964pt}}
\multiput(273.59,322.00)(0.488,2.079){13}{\rule{0.117pt}{1.700pt}}
\multiput(272.17,322.00)(8.000,28.472){2}{\rule{0.400pt}{0.850pt}}
\multiput(281.59,354.00)(0.488,2.013){13}{\rule{0.117pt}{1.650pt}}
\multiput(280.17,354.00)(8.000,27.575){2}{\rule{0.400pt}{0.825pt}}
\multiput(289.59,385.00)(0.485,2.171){11}{\rule{0.117pt}{1.757pt}}
\multiput(288.17,385.00)(7.000,25.353){2}{\rule{0.400pt}{0.879pt}}
\multiput(296.59,414.00)(0.488,1.748){13}{\rule{0.117pt}{1.450pt}}
\multiput(295.17,414.00)(8.000,23.990){2}{\rule{0.400pt}{0.725pt}}
\multiput(304.59,441.00)(0.488,1.616){13}{\rule{0.117pt}{1.350pt}}
\multiput(303.17,441.00)(8.000,22.198){2}{\rule{0.400pt}{0.675pt}}
\multiput(312.59,466.00)(0.485,1.713){11}{\rule{0.117pt}{1.414pt}}
\multiput(311.17,466.00)(7.000,20.065){2}{\rule{0.400pt}{0.707pt}}
\multiput(319.59,489.00)(0.488,1.352){13}{\rule{0.117pt}{1.150pt}}
\multiput(318.17,489.00)(8.000,18.613){2}{\rule{0.400pt}{0.575pt}}
\multiput(327.59,510.00)(0.488,1.154){13}{\rule{0.117pt}{1.000pt}}
\multiput(326.17,510.00)(8.000,15.924){2}{\rule{0.400pt}{0.500pt}}
\multiput(335.59,528.00)(0.485,1.179){11}{\rule{0.117pt}{1.014pt}}
\multiput(334.17,528.00)(7.000,13.895){2}{\rule{0.400pt}{0.507pt}}
\multiput(342.59,544.00)(0.488,0.890){13}{\rule{0.117pt}{0.800pt}}
\multiput(341.17,544.00)(8.000,12.340){2}{\rule{0.400pt}{0.400pt}}
\multiput(350.59,558.00)(0.488,0.758){13}{\rule{0.117pt}{0.700pt}}
\multiput(349.17,558.00)(8.000,10.547){2}{\rule{0.400pt}{0.350pt}}
\multiput(358.59,570.00)(0.485,0.645){11}{\rule{0.117pt}{0.614pt}}
\multiput(357.17,570.00)(7.000,7.725){2}{\rule{0.400pt}{0.307pt}}
\multiput(365.00,579.59)(0.569,0.485){11}{\rule{0.557pt}{0.117pt}}
\multiput(365.00,578.17)(6.844,7.000){2}{\rule{0.279pt}{0.400pt}}
\multiput(373.00,586.59)(0.671,0.482){9}{\rule{0.633pt}{0.116pt}}
\multiput(373.00,585.17)(6.685,6.000){2}{\rule{0.317pt}{0.400pt}}
\multiput(381.00,592.61)(1.355,0.447){3}{\rule{1.033pt}{0.108pt}}
\multiput(381.00,591.17)(4.855,3.000){2}{\rule{0.517pt}{0.400pt}}
\put(388,595.17){\rule{1.700pt}{0.400pt}}
\multiput(388.00,594.17)(4.472,2.000){2}{\rule{0.850pt}{0.400pt}}
\put(396,596.67){\rule{1.927pt}{0.400pt}}
\multiput(396.00,596.17)(4.000,1.000){2}{\rule{0.964pt}{0.400pt}}
\put(404,596.67){\rule{1.686pt}{0.400pt}}
\multiput(404.00,597.17)(3.500,-1.000){2}{\rule{0.843pt}{0.400pt}}
\put(411,595.17){\rule{1.700pt}{0.400pt}}
\multiput(411.00,596.17)(4.472,-2.000){2}{\rule{0.850pt}{0.400pt}}
\multiput(419.00,593.94)(1.066,-0.468){5}{\rule{0.900pt}{0.113pt}}
\multiput(419.00,594.17)(6.132,-4.000){2}{\rule{0.450pt}{0.400pt}}
\multiput(427.00,589.93)(0.710,-0.477){7}{\rule{0.660pt}{0.115pt}}
\multiput(427.00,590.17)(5.630,-5.000){2}{\rule{0.330pt}{0.400pt}}
\multiput(434.00,584.93)(0.821,-0.477){7}{\rule{0.740pt}{0.115pt}}
\multiput(434.00,585.17)(6.464,-5.000){2}{\rule{0.370pt}{0.400pt}}
\multiput(442.00,579.93)(0.569,-0.485){11}{\rule{0.557pt}{0.117pt}}
\multiput(442.00,580.17)(6.844,-7.000){2}{\rule{0.279pt}{0.400pt}}
\multiput(450.00,572.93)(0.492,-0.485){11}{\rule{0.500pt}{0.117pt}}
\multiput(450.00,573.17)(5.962,-7.000){2}{\rule{0.250pt}{0.400pt}}
\multiput(457.00,565.93)(0.494,-0.488){13}{\rule{0.500pt}{0.117pt}}
\multiput(457.00,566.17)(6.962,-8.000){2}{\rule{0.250pt}{0.400pt}}
\multiput(465.00,557.93)(0.494,-0.488){13}{\rule{0.500pt}{0.117pt}}
\multiput(465.00,558.17)(6.962,-8.000){2}{\rule{0.250pt}{0.400pt}}
\multiput(473.59,548.45)(0.485,-0.645){11}{\rule{0.117pt}{0.614pt}}
\multiput(472.17,549.73)(7.000,-7.725){2}{\rule{0.400pt}{0.307pt}}
\multiput(480.59,539.51)(0.488,-0.626){13}{\rule{0.117pt}{0.600pt}}
\multiput(479.17,540.75)(8.000,-8.755){2}{\rule{0.400pt}{0.300pt}}
\multiput(488.59,529.51)(0.488,-0.626){13}{\rule{0.117pt}{0.600pt}}
\multiput(487.17,530.75)(8.000,-8.755){2}{\rule{0.400pt}{0.300pt}}
\multiput(496.59,519.21)(0.485,-0.721){11}{\rule{0.117pt}{0.671pt}}
\multiput(495.17,520.61)(7.000,-8.606){2}{\rule{0.400pt}{0.336pt}}
\multiput(503.59,509.51)(0.488,-0.626){13}{\rule{0.117pt}{0.600pt}}
\multiput(502.17,510.75)(8.000,-8.755){2}{\rule{0.400pt}{0.300pt}}
\multiput(511.59,499.30)(0.488,-0.692){13}{\rule{0.117pt}{0.650pt}}
\multiput(510.17,500.65)(8.000,-9.651){2}{\rule{0.400pt}{0.325pt}}
\multiput(519.59,487.98)(0.485,-0.798){11}{\rule{0.117pt}{0.729pt}}
\multiput(518.17,489.49)(7.000,-9.488){2}{\rule{0.400pt}{0.364pt}}
\multiput(526.59,477.30)(0.488,-0.692){13}{\rule{0.117pt}{0.650pt}}
\multiput(525.17,478.65)(8.000,-9.651){2}{\rule{0.400pt}{0.325pt}}
\multiput(534.59,466.09)(0.488,-0.758){13}{\rule{0.117pt}{0.700pt}}
\multiput(533.17,467.55)(8.000,-10.547){2}{\rule{0.400pt}{0.350pt}}
\multiput(542.59,453.98)(0.485,-0.798){11}{\rule{0.117pt}{0.729pt}}
\multiput(541.17,455.49)(7.000,-9.488){2}{\rule{0.400pt}{0.364pt}}
\multiput(549.59,443.30)(0.488,-0.692){13}{\rule{0.117pt}{0.650pt}}
\multiput(548.17,444.65)(8.000,-9.651){2}{\rule{0.400pt}{0.325pt}}
\multiput(557.59,432.30)(0.488,-0.692){13}{\rule{0.117pt}{0.650pt}}
\multiput(556.17,433.65)(8.000,-9.651){2}{\rule{0.400pt}{0.325pt}}
\multiput(565.59,420.74)(0.485,-0.874){11}{\rule{0.117pt}{0.786pt}}
\multiput(564.17,422.37)(7.000,-10.369){2}{\rule{0.400pt}{0.393pt}}
\multiput(572.59,409.30)(0.488,-0.692){13}{\rule{0.117pt}{0.650pt}}
\multiput(571.17,410.65)(8.000,-9.651){2}{\rule{0.400pt}{0.325pt}}
\multiput(580.59,398.30)(0.488,-0.692){13}{\rule{0.117pt}{0.650pt}}
\multiput(579.17,399.65)(8.000,-9.651){2}{\rule{0.400pt}{0.325pt}}
\multiput(588.59,386.98)(0.485,-0.798){11}{\rule{0.117pt}{0.729pt}}
\multiput(587.17,388.49)(7.000,-9.488){2}{\rule{0.400pt}{0.364pt}}
\multiput(595.59,376.30)(0.488,-0.692){13}{\rule{0.117pt}{0.650pt}}
\multiput(594.17,377.65)(8.000,-9.651){2}{\rule{0.400pt}{0.325pt}}
\multiput(603.59,365.30)(0.488,-0.692){13}{\rule{0.117pt}{0.650pt}}
\multiput(602.17,366.65)(8.000,-9.651){2}{\rule{0.400pt}{0.325pt}}
\multiput(611.59,353.98)(0.485,-0.798){11}{\rule{0.117pt}{0.729pt}}
\multiput(610.17,355.49)(7.000,-9.488){2}{\rule{0.400pt}{0.364pt}}
\multiput(618.59,343.30)(0.488,-0.692){13}{\rule{0.117pt}{0.650pt}}
\multiput(617.17,344.65)(8.000,-9.651){2}{\rule{0.400pt}{0.325pt}}
\multiput(626.59,332.51)(0.488,-0.626){13}{\rule{0.117pt}{0.600pt}}
\multiput(625.17,333.75)(8.000,-8.755){2}{\rule{0.400pt}{0.300pt}}
\multiput(634.59,321.98)(0.485,-0.798){11}{\rule{0.117pt}{0.729pt}}
\multiput(633.17,323.49)(7.000,-9.488){2}{\rule{0.400pt}{0.364pt}}
\multiput(641.59,311.51)(0.488,-0.626){13}{\rule{0.117pt}{0.600pt}}
\multiput(640.17,312.75)(8.000,-8.755){2}{\rule{0.400pt}{0.300pt}}
\multiput(649.59,301.51)(0.488,-0.626){13}{\rule{0.117pt}{0.600pt}}
\multiput(648.17,302.75)(8.000,-8.755){2}{\rule{0.400pt}{0.300pt}}
\multiput(657.59,291.21)(0.485,-0.721){11}{\rule{0.117pt}{0.671pt}}
\multiput(656.17,292.61)(7.000,-8.606){2}{\rule{0.400pt}{0.336pt}}
\multiput(664.59,281.51)(0.488,-0.626){13}{\rule{0.117pt}{0.600pt}}
\multiput(663.17,282.75)(8.000,-8.755){2}{\rule{0.400pt}{0.300pt}}
\multiput(672.59,271.72)(0.488,-0.560){13}{\rule{0.117pt}{0.550pt}}
\multiput(671.17,272.86)(8.000,-7.858){2}{\rule{0.400pt}{0.275pt}}
\multiput(680.59,262.21)(0.485,-0.721){11}{\rule{0.117pt}{0.671pt}}
\multiput(679.17,263.61)(7.000,-8.606){2}{\rule{0.400pt}{0.336pt}}
\multiput(687.59,252.72)(0.488,-0.560){13}{\rule{0.117pt}{0.550pt}}
\multiput(686.17,253.86)(8.000,-7.858){2}{\rule{0.400pt}{0.275pt}}
\multiput(695.59,243.72)(0.488,-0.560){13}{\rule{0.117pt}{0.550pt}}
\multiput(694.17,244.86)(8.000,-7.858){2}{\rule{0.400pt}{0.275pt}}
\multiput(703.59,234.69)(0.485,-0.569){11}{\rule{0.117pt}{0.557pt}}
\multiput(702.17,235.84)(7.000,-6.844){2}{\rule{0.400pt}{0.279pt}}
\multiput(710.59,226.72)(0.488,-0.560){13}{\rule{0.117pt}{0.550pt}}
\multiput(709.17,227.86)(8.000,-7.858){2}{\rule{0.400pt}{0.275pt}}
\multiput(718.00,218.93)(0.494,-0.488){13}{\rule{0.500pt}{0.117pt}}
\multiput(718.00,219.17)(6.962,-8.000){2}{\rule{0.250pt}{0.400pt}}
\multiput(726.59,209.45)(0.485,-0.645){11}{\rule{0.117pt}{0.614pt}}
\multiput(725.17,210.73)(7.000,-7.725){2}{\rule{0.400pt}{0.307pt}}
\multiput(733.00,201.93)(0.494,-0.488){13}{\rule{0.500pt}{0.117pt}}
\multiput(733.00,202.17)(6.962,-8.000){2}{\rule{0.250pt}{0.400pt}}
\multiput(741.00,193.93)(0.569,-0.485){11}{\rule{0.557pt}{0.117pt}}
\multiput(741.00,194.17)(6.844,-7.000){2}{\rule{0.279pt}{0.400pt}}
\multiput(749.59,185.69)(0.485,-0.569){11}{\rule{0.117pt}{0.557pt}}
\multiput(748.17,186.84)(7.000,-6.844){2}{\rule{0.400pt}{0.279pt}}
\multiput(756.00,178.93)(0.569,-0.485){11}{\rule{0.557pt}{0.117pt}}
\multiput(756.00,179.17)(6.844,-7.000){2}{\rule{0.279pt}{0.400pt}}
\multiput(764.00,171.93)(0.494,-0.488){13}{\rule{0.500pt}{0.117pt}}
\multiput(764.00,172.17)(6.962,-8.000){2}{\rule{0.250pt}{0.400pt}}
\multiput(772.00,163.93)(0.492,-0.485){11}{\rule{0.500pt}{0.117pt}}
\multiput(772.00,164.17)(5.962,-7.000){2}{\rule{0.250pt}{0.400pt}}
\multiput(779.00,156.93)(0.569,-0.485){11}{\rule{0.557pt}{0.117pt}}
\multiput(779.00,157.17)(6.844,-7.000){2}{\rule{0.279pt}{0.400pt}}
\multiput(787.00,149.93)(0.671,-0.482){9}{\rule{0.633pt}{0.116pt}}
\multiput(787.00,150.17)(6.685,-6.000){2}{\rule{0.317pt}{0.400pt}}
\multiput(795.00,143.93)(0.492,-0.485){11}{\rule{0.500pt}{0.117pt}}
\multiput(795.00,144.17)(5.962,-7.000){2}{\rule{0.250pt}{0.400pt}}
\multiput(802.00,136.93)(0.671,-0.482){9}{\rule{0.633pt}{0.116pt}}
\multiput(802.00,137.17)(6.685,-6.000){2}{\rule{0.317pt}{0.400pt}}
\multiput(810.00,130.93)(0.671,-0.482){9}{\rule{0.633pt}{0.116pt}}
\multiput(810.00,131.17)(6.685,-6.000){2}{\rule{0.317pt}{0.400pt}}
\multiput(818.00,124.93)(0.581,-0.482){9}{\rule{0.567pt}{0.116pt}}
\multiput(818.00,125.17)(5.824,-6.000){2}{\rule{0.283pt}{0.400pt}}
\multiput(825.00,118.93)(0.671,-0.482){9}{\rule{0.633pt}{0.116pt}}
\multiput(825.00,119.17)(6.685,-6.000){2}{\rule{0.317pt}{0.400pt}}
\multiput(833.00,112.93)(0.671,-0.482){9}{\rule{0.633pt}{0.116pt}}
\multiput(833.00,113.17)(6.685,-6.000){2}{\rule{0.317pt}{0.400pt}}
\multiput(841.00,106.93)(0.710,-0.477){7}{\rule{0.660pt}{0.115pt}}
\multiput(841.00,107.17)(5.630,-5.000){2}{\rule{0.330pt}{0.400pt}}
\multiput(848.00,101.93)(0.671,-0.482){9}{\rule{0.633pt}{0.116pt}}
\multiput(848.00,102.17)(6.685,-6.000){2}{\rule{0.317pt}{0.400pt}}
\multiput(856.00,95.93)(0.821,-0.477){7}{\rule{0.740pt}{0.115pt}}
\multiput(856.00,96.17)(6.464,-5.000){2}{\rule{0.370pt}{0.400pt}}
\multiput(864.00,90.93)(0.710,-0.477){7}{\rule{0.660pt}{0.115pt}}
\multiput(864.00,91.17)(5.630,-5.000){2}{\rule{0.330pt}{0.400pt}}
\multiput(871.00,85.93)(0.821,-0.477){7}{\rule{0.740pt}{0.115pt}}
\multiput(871.00,86.17)(6.464,-5.000){2}{\rule{0.370pt}{0.400pt}}
\put(128.0,82.0){\rule[-0.200pt]{11.081pt}{0.400pt}}
\end{picture}
\end{center}
\vspace{-20pt}
\caption{A probability distribution function 
         and the probability density.}
\label{ProbFig1}
\end{figure}
The generic shape of $F_{T,N}$ and its derivative, the probability density, are
depicted in \fig{ProbFig1}. They tell us what the probability would be to find
certain values for $T_N(\XN)$ if $\XN$ would be distributed following the
hypothesis, i.e., they give the {\em confidence levels}. For example, we can
read off the first graph that the probability for $T_N(\XN)$ to be larger than
$0.80$ is about $1-0.95=0.05$. This means that it is not very probable to find
a value for $T_N(\XN)$ this large, so that this number {\em can} be considered
large. 

\subsubsection{Qualification of samples}
Instead of for hypothesis testing, a test $T_N$ can also be used to {\em
qualify} a sample of data $\XN$. Suppose that there is a notion of {\em good}
and {\em bad} samples, and that this notion is translated into the test $T_N$:
if $T_N(\XN)$ is small, then $\XN$ is {\em good}, and if $T_N(\XN)$ is large,
then $\XN$ is {\em bad}.  A first question can be whether the test makes sense,
and an answer can again be given by the probability distribution. Suppose that
the information available about the source of the data leads to a probability
density $P_N$ according to which the data seem to be randomly distributed. If
the probability density of $T_N$ looks like the one in \fig{ProbFig1}, i.e., if
it goes to zero for small values of $T_N(\XN)$, then the test makes sense. It
means that the test is capable of distinguishing between {\em good} samples and
the kind of samples that occur most often. The next question is then: what {\em
are} small values of $T_N(\XN)$? The answer is that values are small if it is
improbable to find them. 

\subsubsection{Qualification of algorithms and discrepancy}
It can also be the case that the data come from an (expensive)
algorithm that was specially designed to produce {\em good} samples (for
example integration points for numerical integration), and the question is
whether the algorithm makes sense. Suppose there is another (cheap) algorithm
that produces data distributed with density $P_N$. The
probability distribution of $T_N$ determines the notion of {\em smallness} for
the values of $T_N(\XN)$ again, and the expensive algorithm only makes sense
if it produces samples with low values of $T_N(\XN)$ that are improbable to
find. In the mentioned case of numerical integration, {\em good} samples are 
the point sets that are distributed uniformly over the integration space, 
and the tests are called {\em discrepancies} (\Sec{DefDisSec}). 

Discrepancies have the structure of tests that measure the deviation between
the empirical distribution of the point set, and the uniform distribution in
the integration space. Algorithms to generate point sets following the uniform
distribution (cf.~\cite{Knuth}) can be considered `cheap' compared to the
special algorithms developed for numerical integration (cf.~\cite{Tezuka}).
This seems paradoxical, since numerical integration asks for point sets that
are distributed over the integration space as uniformly as possible 
(\Sec{IntroQMC}). The clue is that (random) point sets, generated following the
uniform distribution, are not necessarily those that are distributed over the
integration space as uniformly as possible. A simple example is one-dimensional
space. For a given number of points, the most uniform distribution possible
clearly is the one for which all distances between the points are the same.
However, if the points are distributed randomly following the uniform
distribution, this situation will never occur. 

The example above gives a simple algorithm to generate {\em good} samples in
one-dimensional space. Algorithms become `expensive' if they have to be 
generated in more-dimensional spaces.

\subsection{Calculation of probability distributions\label{CPDSec}}
In statistics, probability distributions are used, and in probability theory,
they are calculated. Part of this thesis deals with their calculation for
discrepancies. The way this will be done is by calculating the generating
function. The probability density can then be found using the inverse Laplace
transform (\eqn{ProEq002}), which can, if necessary, be calculated through a
numerical integral over a contour in the complex plane. 

The distribution is often calculated in certain limits, such as an infinite
number of random variables or degrees of freedom. This is, in most cases, done
because it simplifies the calculation. These limits can, however, often be
considered as the limiting cases in certain stochastic processes: in Monte
Carlo integration, for example, the limit of an infinite number of integration
points can be interpreted as the limit of an infinite run-time for a computer.

If the generating function is considered, these limits correspond with 
weak convergence. However, if the generating function is calculated
through all moments of the distribution, this corresponds with a stronger
convergence: if $z\mapsto\sum_{p=0}^\infty a_pz^p/p!$ is a generating
function and 
\begin{equation}
   \Exp(X_n^p) \ra a_p\;,\;\;p=0,1,2,\ldots
   \qquad\textrm{then}\qquad
   G_{n}(z) \ra \sum_{p=0}^\infty\frac{a_pz^p}{p!} \;\;,
\end{equation}
but the opposite does not have to be true. The moments might even go to
infinity, while the generating function converges to an analytic function, and 
we will encounter an explicit example in which this happens.

\section{Feynman diagrams and Gaussian measures}
Part of this thesis deals with the calculation of probability distributions of
measures of non-uniformity of point sets $\{x_1,\ldots,x_N\}$ in an integration
space. The measures that are considered can be written in terms of two-point
functions as $N^{-1}\sum_{k,l}^N\twoB(x_k,x_l)$.  Consequently, the
calculation of the moments of the distributions involves the calculation of
multiple convolutions of these two-point functions, and Feynman diagrams can be
of help.

\subsection{Feynman diagrams\label{GmFdSec1}}
Feynman diagrams are drawings obtained by connecting vertices following some
rules. Let us illustrate this with an example, in which three vertices are
connected to a diagram:
\begin{equation}
   \diagram{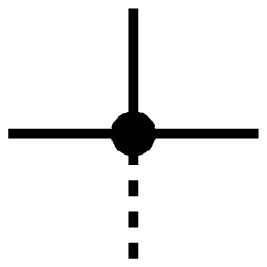}{22}{10}\quad
   \diagram{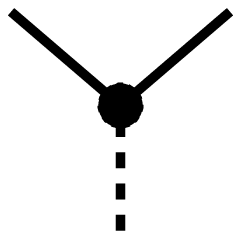}{19}{10}\quad
   \diagram{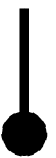}{4}{4}\quad\quad\longrightarrow\quad\quad
   \diagram{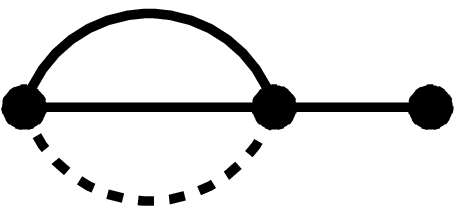}{38}{7} \quad.
\end{equation}
The vertices have a number of legs and, in this case, there are two kinds of
legs. The rule to get from these vertices to the particular diagram could be
that legs of the same kind have to be connected. One rule that will always
apply to cases we consider is that {\em all legs have to be connected to other
legs}. Notice that, with this rule and the one that connected legs have to be
of the same kind, the diagram drawn above is not the only possible one. Also
\begin{equation}
   \diagram{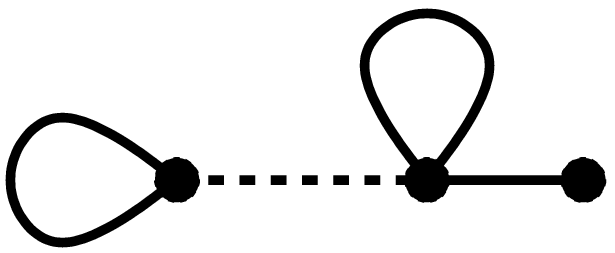}{53}{8}
\end{equation}
is a permitted diagram. The vertices in the diagrams are connected by 
{\em lines}. The previous two diagrams we also call {\em connected} as a whole, 
because one can walk from any vertex to any other vertex over lines. 
An example of a {\em disconnected} diagram can be obtained with twice as many 
vertices:
\begin{equation}
   \diagram{V31}{22}{10}\quad
   \diagram{V31}{22}{10}\quad
   \diagram{V21}{19}{10}\quad
   \diagram{V21}{19}{10}\quad
   \diagram{V1}{4}{4}\quad
   \diagram{V1}{4}{4}\quad\quad\longrightarrow\quad\quad
   \diagram{D31211}{38}{7} \; \diagram{D31211_2}{53}{8} \quad.
\end{equation}
This is a possible diagram if the previous rules are applied. 

If there are also rules how to assign a number to a diagram, these, together
with the rules how to construct the diagrams, are called the {\em Feynman
rules}. The Feynman rules make the diagrams of practical use.  Certain
calculations can be reduced to the assignment of numbers to a set of diagrams,
which then have to be added to finish the whole calculation. We call such a
number the {\em contribution} of the diagram, but this word shall often be
omitted, and we will refer to `sums of diagrams' instead of `sums of
contributions of diagrams'. In the following sections, we will give some
examples, but first we derive 

\subsubsection{A few general relations}
Let $\eta_1,\eta_2,\eta_3,\ldots$ be elements of a commutative algebra over 
$\Comp$, and assume that there is an operation 
$\leb{\!\leb{\cdot}\!}:
 \prod_{i}\eta_{k_i}\mapsto\leb{\!\leb{\prod_{i}\eta_{k_i}}\!}\in\Comp$ 
which is linear over $\Comp$. Suppose that every $\eta_k$ represents a type of 
vertex, and that 
\begin{equation}
   \frac{\leb{\!\leb{\,\eta_1^{p_1}\,
                       \eta_2^{p_2}\cdots
	               \eta_n^{p_n}\,}\!}}
        {p_1!\cdot p_2!\cdots p_n!}	
\end{equation}
can be interpreted as the sum of all possible diagrams with $p_1$ vertices of
type $\eta_1$, $p_2$ vertices of type $\eta_2$ and so on.  Then, the sum of all
possible diagrams is given by 
\begin{equation}
   \sum_{n}\sum_{p(n)}
         \frac{\leb{\!\leb{\,\eta_1^{p_1(n)}\,
                       \eta_2^{p_2(n)}\cdots
	               \eta_n^{p_n(n)}\,}\!}}
        {p_1(n)!\cdot p_2(n)!\cdots p_n(n)!}\;\;,
\end{equation}
where the second sum is over all partitions $p(n)$ of $n$.  Using the
combinatorial rule of \eqn{GmFdEq008}, which is derived in \App{GmFdSec2}, 
and the linearity of $\leb{\!\leb{\cdot}\!}$, we find that this is equal to 
\begin{align}
   \sum_{m=1}^\infty\frac{1}{m!}\sum_{k_1,\ldots,k_m}
             \leb{\!\leb{\eta_{k_1}\cdots\eta_{k_m}}\!}
   \;=\; \sum_{m=1}^\infty\frac{1}{m!}
             \leb{\!\leb{\,\Big(\sum_{k}\eta_k\Big)^m\,}\!} 
   \;=\; \leb{\!\leb{\,\exp\Big(\sum_k\eta_k\Big)\,}\!} - 1 \;\;,
\end{align}
so that 
\begin{equation}
   G(\{\eta\})
   \;\df\;  1 + \textsl{the sum of all possible diagrams}
   \;=\; \leb{\!\leb{\,\exp\Big(\sum_k\eta_k\Big)\,}\!} \;\;.
\end{equation}
Now, we show that the sum of all 
possible connected diagrams is given by $\log G(\{\eta\})$.
Define 
\begin{equation}
    G_n[\eta_m]
   \;\df\; \frac{1}{n!}
           \leb{\!\leb{\,\eta_m^n\exp\Big(\sum_{k\neq m}\eta_k\Big)\,}\!}\;\;,
\end{equation}
so that
\begin{equation}
   G(\{\eta\}) \;=\; \sum_{n=0}^\infty G_n[\eta_m] \;\;.
\end{equation}
$ G_n[\eta_m]$ contains all diagrams with $n$
vertices of type $\eta_m$. The sum of all diagrams for which these
$n$ vertices are contained in the same connected piece is denoted
$ C_n[\eta_m]$, so that
\begin{equation}
    G_n[\eta_m] 
    \;=\; \sum_{p(n)} \prod_{i=1}^n\frac{ C_i[\eta_m]^{p_i(n)}}{p_i(n)!} \;\;,
\end{equation}
where the sum in the r.h.s.~is over all partitions $p(n)$ of $n$. Using 
\eqn{GmFdEq008} again, we find that
\begin{equation}
   G(\{\eta\}) 
   \;=\; \sum_{n=0}^\infty\frac{1}{n!}\sum_{i_1,\ldots,i_n=1}^\infty
                           C_{i_1}[\eta_m]\cdots C_{i_n}[\eta_m]
   \;=\; \exp(W[\eta_m]) \;\;,
\end{equation}
where
\begin{equation}
   W[\eta_m] \;\df\; \sum_{n=1}^\infty C_n[\eta_m]
\end{equation}
is the sum of all diagrams for which all vertices of the kind
$\eta_m$ are contained in the same connected piece. Because we can take any
kind of vertex for $\eta_m$, the sum of all diagrams has to be given
by the exponential of the sum of all connected diagrams, and we find 
that
\begin{equation}
   \log G(\{\eta\}) \;=\; \textsl{the sum of all possible connected diagrams.} 
\end{equation}

\subsection{Gaussian measures\label{GmFdSec3}}
An example of the use of Feynman diagrams is in calculations with Gaussian 
measures. We refer to \cite{Choquet} for more details about the formalism used. 

We are going to look at measures on spaces $\Cfam$ of real bounded functions
on a subset $\Kube$ of $\Real^s$, where $s=1,2,\ldots$. The Lebesgue measure on
$\Kube$ we just denote $dx$. $\Kube$ can also be a finite set, in which case 
the Lebesgue integral becomes a finite sum. 
A measure on $\Cfam$ will be denoted $\mu$, and for, not necessarily linear, 
functionals 
$\eta_1,\eta_2,\ldots,\eta_n$ on $\Cfam$, we denote
\begin{equation}
   \leb{\eta_1\,\eta_2\,\cdots\,\eta_n}_\mu 
   \;\df\; \int_{\Cfam}\eta_1[\phi]\eta_2[\phi]\cdots\eta_n[\phi]\,d\mu[\phi] 
   \;\;.
\end{equation}
The space of continuous linear functionals on $\Cfam$ is denoted $\Cfam'$, and
a typical member is the Dirac measure $\de_x$, which is for every $x\in\Kube$
defined by 
\begin{equation}
   \de_x[\phi] \df \phi(x) \;\;.
\end{equation}
Furthermore, we introduce the so called 
$n$-point functions, which are given by 
\begin{equation}
   \prop_n(x_1,x_2,\ldots,x_n) 
   \;\df\; \leb{\de_{x_1}\,\de_{x_2}\,\cdots\,\de_{x_n}}_\mu \;\;.
\label{GmFdEq002}   
\end{equation}
Notice that they are symmetric in their arguments. We will always assume that 
$\mu$ is normalized, so that $\prop_0\df\int_{\Cfam}d\mu[\phi]=1$.
For linear functionals, we will use the notation
\begin{equation}
   \eta[\phi] \;=\; \intk\eta(x)\phi(x)\,dx \;\;,
\end{equation}
although `$\eta(x)$' cannot always be seen as a function value. For example, 
$\de_x(x)$ does not exist.
If we combine this notation 
with the notation of \eqn{GmFdEq002}, we can write
\begin{equation}
   \leb{\eta_1\,\eta_2\,\cdots\,\eta_n}_\mu
   \;=\; \int_{\Kube^n}\prop_n(x_1,x_2,\ldots,x_n)
         \eta_1(x_1)\eta_2(x_2)\cdots\eta_n(x_n)\,dx_1dx_2\cdots dx_n \;.
\end{equation}
The Fourier transform of a measure $\mu$ on $\Cfam$ is the function on $\Cfam'$
given by 
\begin{equation}
   \eta \mapsto \leb{\,\exp(i\eta)\,}_\mu \;\;,
\end{equation}
and $\mu$ is {\em Gaussian} if there is a quadratic form
$\Quad$ on $\Cfam'$ such that the Fourier transform is given by
\begin{equation}
   \leb{\,\exp(i\eta)\,}_\mu
   \;=\; \exp(-\half\Quad[\eta]) \;\;.
\label{GmFdEq001}   
\end{equation}
$\Quad$ can be written in terms of the two-point function, for take 
$\eta\df\la\zeta$, where $\la$ is a real 
variable, and differentiate \eqn{GmFdEq001} twice with respect to $\la$ 
before putting it to zero. Then it is easy to see that
\begin{equation}
   \Quad[\zeta] 
   \;=\; \leb{\zeta\,\zeta}_\mu 
   \;=\; \intk\zeta(x_1)\prop_2(x_1,x_2)\zeta(x_2) \,dx  \;\;.
\end{equation}
With this result, we can express the $n$-point functions in terms of the
two-point function.  If we take $\eta\df\sum_{i=1}^n\la_i\de_{x_i}$, where
$\la_i$, $i=1,\ldots,n$ are real variables, and differentiate \eqn{GmFdEq001}
once with respect of each of these variables  before putting them to zero, we
find that, for odd $n$, $\prop_n=0$, and for even $n$ that
\begin{equation}
   \prop_n(x_1,x_2,\ldots,x_n)
   \;=\; \sum_{\textrm{pairs $(i<j)$}}\prop_2(x_i,x_j) \;\;,
\label{GmFdEq003}   
\end{equation}
where the sum is over all pairs $(i,j)$ for which $i<j$. 

\subsubsection{Diagrams}
The previous formula suggests to interpret $\prop_2(x_i,x_j)$ as a {\em line}
that connects the arguments $x_i$ and $x_j$, so that the r.h.s.~consists of all
possible ways to connect the arguments $x_1,x_2,\ldots,x_n$ in pairs with
lines. If there is a prescription to identify a number of $m\leq n$ arguments,
then they can represent a vertex.  The number of arguments in a vertex, the
number of legs, we call the {\em order} of the vertex. 

A typical case in which arguments are identified is when integrals of the 
following type are calculated. Let $\eta_1,\eta_2,\eta_3,\ldots$ be a sequence 
of functionals acting on $\Cfam$ as
\begin{equation}
   \eta_k[\phi] 
   \;\df\; \int_{\Kube^k}\eta_k(x_1,\ldots,x_k)\,\phi(x_1)\cdots\phi(x_k)\,
           dx_1\cdots dx_k  \;\;.
\end{equation}
Let, furthermore, $k_1,\ldots,k_m$ be a set of integers larger than zero, and
denote $k_{(i)}=\sum_{j=1}^ik_j$.
The integrals we want to consider are given by 
\begin{align}
    \leb{\eta_{k_1}\cdots\eta_{k_m}}_\mu
    \;=\; \int_{\Kube^{k_{(m)}}}
         \prop_{k_{(m)}}(\{x\}^{k_{(m)}}_0)\;
	 \eta_{k_1}(\{x\}^{k_{(1)}}_0)\cdots
	 \eta_{k_m}(\{x\}^{k_{(m)}}_{k_{(m-1)}})\;
	 dx_1\cdots dx_{k_{(m)}}  \;\;,
\end{align}
where we use the notation $\{x\}^i_j=\{x_{j+1},x_{j+2},\ldots,x_{i}\}$. The
set of arguments that are the integration variables of the same $\eta_i$ can
be considered identical.  As a result, the whole integral is given by
the sum of all possible diagrams with the $m$
vertices of the orders $k_1,\ldots,k_m$. The contribution of a diagram is
obtained by convoluting the two-point functions with the functionals
$\eta_{k_i}$ in the vertices.

The question we want to answer now is, given the Gaussian measure $\mu$ and the
functionals $\eta_k$, what the sum of all possible diagrams
is. Because $\prop$ is symmetric, it suffices to consider the integrals
\begin{equation}
   \leb{\,\eta_1^{p_1}\,
                  \eta_2^{p_2}\,\cdots\,
	 	  \eta_n^{p_n}\,}_\mu \;\;.
\label{GmFdEq011}		  
\end{equation}
The diagrams that contribute have $p_1$
vertices of order $1$, $p_2$ vertices of order $2$ and so on. In the set of
diagrams that contribute to this integral, there are many diagrams that look
exactly the same because they only differ in the exchange of integration
variables of the same vertex, or in the exchange of vertices of the same order.
We do not want to count them separately, and therefor include in the
contribution of a diagram the number of ways it can be obtained. We turn this
number into a {\em symmetry factor}, by considering
\begin{equation}
   \frac{\leb{\,  \bar{\eta}_1^{p_1}\,
                  \bar{\eta}_2^{p_2}\,\cdots\,
	 	  \bar{\eta}_n^{p_n}\,}_\mu}
        {p_1!\cdot p_2!\cdots p_n!}\;\;,\quad\quad
   \bar{\eta}_k \df \frac{\eta_k}{k!}\;\;,
\end{equation}
instead of (\ref{GmFdEq011}). As a consequence, 
every vertex of order $k$ accounts for a factor $1/k!$, and the set of
vertices of order $k$ accounts for a factor $1/p_k!$. The contribution of a
diagram is then given by the number obtained calculating the convolutions of
the $\eta_k$'s, represented by the vertices, with the $\prop_2$'s, represented
by the lines, multiplied with the symmetry factor. This factor is the number of
ways the diagram can be obtained, considering all vertices and all legs of
vertices distinct, divided by $\prod_{i}p_i!(i!)^{p_i}$, where $p_i$ is the
number of vertices of order $i$.  We can use the results of 
\Sec{GmFdSec1} now and find that 
\begin{equation}
   \textsl{$1\,+$ the sum of all possible diagrams}
   \;=\; \leb{\,\exp\Big(\sum_{k=1}^\infty\frac{\eta_k}{k!}\Big)\,}_\mu \;\;,
\end{equation}
and that this equal to the exponential of the sum of all connected diagrams.

\subsection{Falling powers, diagrams and Grassmann variables\label{GmFdSec4}}
Another, small, example of the use of Feynman diagrams is in the representation 
of the numbers 
\begin{equation}
   N^{\underline{k}}
   \;\df\; N(N-1)(N-2)\cdots(N-k+1) \;\;,\quad N,k\in\Natu \;\;.
\end{equation}
It can be derived from the relation 
\begin{equation}
   N^{\underline{k}}
   \;=\; \sum_{i_1,\ldots,i_k=1}^N\sum_{\pi\in S_k}\sgn(\pi)
         \de_{i_1,i_{\pi(1)}}\de_{i_2,i_{\pi(2)}}\cdots
	 \de_{i_k,i_{\pi(k)}}\;\;,
\label{GmFdEq007}	 
\end{equation}
where the second sum on the r.h.s.~is over all permutations of
$(1,2,\ldots,k)$. This relation is derived in \App{GmFdSec2}. 
It allows for following diagrammatic interpretation.

Consider `arrowed' vertices of order two, that is, vertices of order two with
distinct legs: one incoming and one outgoing. They can be connected with the
rule that outgoing legs may only be connected to incoming legs and vice versa.
The legs are connected with an `arrowed' line, representing a $\de_{i_1,i_2}$,
and the vertices represent the convolution
$\sum_{i_2=1}^N\de_{i_1,i_2}\de_{i_2,i_3}$  of the two lines arriving at and
starting from that vertex. Up to an overall minus sign, the r.h.s.~of
\eqn{GmFdEq007} is equal to the sum of all possible
diagrams with $k$ distinct `arrowed' vertices, with the extra rule that every
closed loop gives a factor $-1$. The overall minus sign is equal to $(-1)^k$.
For example, 
\begin{align}
   (-1)^3N^{\underline{3}} 
   &\;=\; \diagram{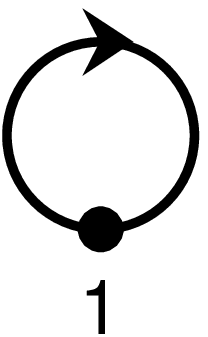}{18}{15}\,\diagram{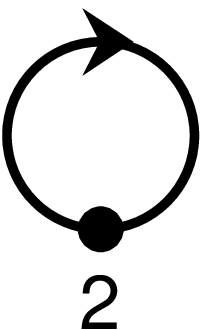}{18}{15}\,
          \diagram{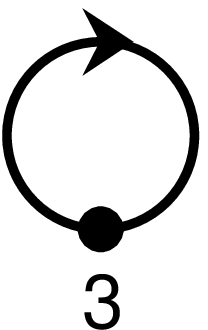}{18}{15}
          \;+\; \diagram{arlo1_1}{18}{15}\;\diagram{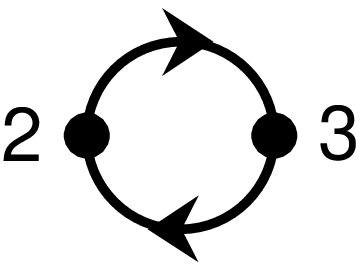}{32}{9}
	  \;+\; \diagram{arlo1_2}{18}{15}\;\diagram{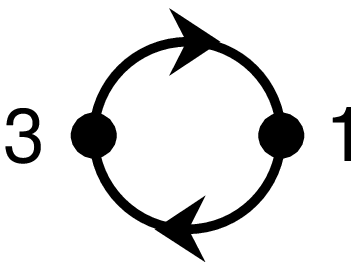}{32}{9}
	  \;+\; \diagram{arlo1_3}{18}{15}\;\diagram{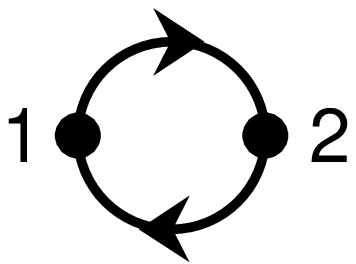}{32}{9}
	  \;+\; \diagram{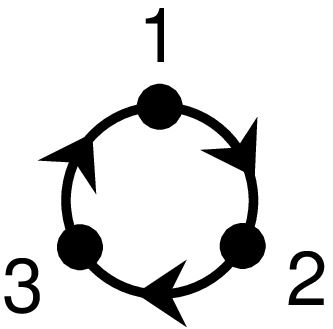}{30}{10} 
	  \;+\; \diagram{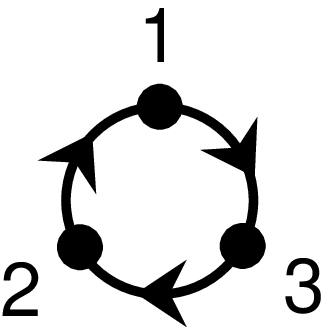}{30}{10}  \notag\\
   &\;=\; -N^3 + 3N^2 - 2N  \;\;.
\end{align}
It is useful to consider diagrams that look exactly the same as one diagram
again. In the example above, this applies to the last two diagrams, and to the
second, the third and the fourth diagram. The extra number the contribution of
a diagram has to be multiplied with is turned into a symmetry factor by
considering $(-1)^kN^{\underline{k}}/k!$ instead of
$(-1)^kN^{\underline{k}}$.

\subsubsection{Grassmann variables}
The numbers $(-1)^kN^{\underline{k}}$ can also be written in another way. We 
introduce $2N$ {\em Grassmann variables} $\psi_i$ and $\psib_i$, $i=1,\ldots,N$.
They all anti-commute with each other
and commute with complex numbers:
\begin{align}
     \psib_i\psib_j + \psib_j\psib_i = 0 \;&,\;\; 
     \psib_i\psi_j + \psi_j\psib_i = 0 \;,\;\; 
     \psi_i\psi_j + \psi_j\psi_i = 0  \quad\quad i,j=1,2,\ldots,N \\
     c\psib_i - \psib_i c = 0 \;&,\;\; 
     c\psi_i - \psi_i c = 0 \quad\quad 
     i=1,2,\ldots,N \;\;,\quad c\in\Comp \;\;.
\end{align}
These variables are nilpotent, i.e., $\psi_i\psi_i=\psib_i\psib_i=0$ for all
$i$. Products of even numbers of these variables commute with all other combinations. Furthermore, we introduce the `integral' of these variables, which maps 
sums of products of them onto $\Comp$. It is linear over $\Comp$ and defined by the relations 
\begin{align}
   \int[d\psib d\psi]\,\psi_{i_1}\psi_{i_2}\cdots\psi_{i_k}\,
                       \psib_{j_1}\psib_{j_2}\cdots\psib_{j_l}
   &\;\df\; 0 \quad\textrm{if $k,l<N$} \;\;,\\
   \int[d\psib d\psi]\,\psi_1\psib_1\,\psi_2\psib_2\cdots\psi_N\psib_N 
   &\;\df\; 1 \;\;.
\end{align}
Notice that the first integral is also zero if $k>N$, because then there has to
be a pair $(\psi_{i_r},\psi_{i_s})$ with $i_r=i_s$ in the product of
$\psi$'s, so that $\psi_{i_r}\psi_{i_s}=0$. The same holds if $l>N$. 
A useful calculation of such an integral is
\begin{align}
   \int[d\psib d\psi]\Big(\sum_{i=1}^N\psib_i\psi_i\Big)^N 
  &\;=\; \sum_{\pi\in S_N}\int[d\psib d\psi]\,
         \psib_{\pi(1)}\psi_{\pi(1)}\,\psib_{\pi(2)}\psi_{\pi(2)}\cdots
	 \psib_{\pi(N)}\psi_{\pi(N)}\notag\\
  &\;=\; \sum_{\pi\in S_N}(-1)^N\sgn(\pi)^2 \;=\; (-1)^NN! \;\;.	 
\end{align}
Another useful relation is the following. If $A$ is a complex number, then
\begin{equation}
   \int[d\psib d\psi]\,\exp\Big(-\sum_{j=1}^N\psib_j\psi_jA\Big)
   \;=\; \sum_{k=0}^N\frac{(-)^kA^k}{k!}
         \int[d\psib d\psi]\Big(\sum_{i=1}^N\psib_i\psi_i\Big)^k
   \;=\; A^N \;,	 
\end{equation}
since only the term with $k=N$ is non-zero.
Let us now introduce the `measure' $\leb{\cdot}_\psi$, defined by 
\begin{equation}
   \leb{f}_\psi
   \;\df\; \int[d\psib d\psi]\,f(\psi_1,\ldots,\psi_N,\psib_1,\ldots,\psib_N)\,
           \exp\Big(-\sum_{j=1}^N\psib_j\psi_j\Big) \;\;.
\end{equation}
Using the result of the previous calculations, we find 
\begin{align}
   \leb{\,\Big(\sum_{i=1}^N\psib_i\psi_i\Big)^k\,}_\psi
   \;=\; \sum_{m=0}^N\frac{(-1)^m}{m!}\int[d\psib d\psi]
                           \Big(\sum_{i=1}^N\psib_i\psi_i\Big)^{k+m}
   \;=\; (-)^kN^{\underline{k}}  \;\;,	  
\end{align}
since only the term with $m=N-k$ contributes.
If we combine the two representations of the numbers $(-)^kN^{\underline{k}}$,
we can draw the conclusion that 
\begin{equation}
   \frac{1}{k!}\leb{\,\Big(\sum_{i=1}^N\psib_i\psi_i\Big)^k\,}_\psi
   \;=\; \textsl{the sum of all possible diagrams with $k$ 
                 `arrowed' two-point vertices.}
\end{equation}

\subsection{Gaussian measures and Grassmann variables}
As a final application, we are going to combine the previous two examples. 
Let $\mu$ be a Gaussian measure on a space of real bounded functions
on a subset $\Kube$ of $\Real^s$, let $\psi_i$ and $\psib_i$, $i=1,\ldots,N$ 
be a set of Grassmann variables, and let us denote
\begin{equation}
   \leb{\!\leb{\,\chi\,}\!}
   \;\df\; \leb{\leb{\,\chi\,}_\psi}_\mu 
   \;=\;  \leb{\leb{\,\chi\,}_\mu}_\psi  \;\;.
\end{equation}
If $\eta_1,\eta_2,\eta_3,\ldots$ is a sequence 
of functionals acting on $\Cfam$ as
\begin{equation}
   \eta_k[\phi] 
   \;\df\; \int_{\Kube^k}\eta_k(x_1,\ldots,x_k)\,\phi(x_1)\cdots\phi(x_k)\,
           dx_1\cdots dx_k  \;\;, 
\label{GmFdEq010}   
\end{equation}
then 
\begin{equation}
   \frac{\leb{\,\eta_1^{p_1}\,\eta_2^{p_2}\,\cdots\,\eta_n^{p_n}\,}_\mu}
        {p_1!\cdot p_2!\cdots p_n!} 
\end{equation}
can be calculated using diagrams with $p_1$ vertices of type $\eta_1$, $p_2$
vertices of type $\eta_2$ and so on, as described in \Sec{GmFdSec3}.
If we apply the results of \Sec{GmFdSec4}, we see that
\begin{equation}
   \frac{(-1)^{p_{(n)}}N^{\underline{p_{(n)}}}\,
         \leb{\,\eta_1^{p_1}\,\eta_2^{p_2}\,\cdots\,\eta_n^{p_n}\,}_\mu}
        {p_1!\cdot p_2!\cdots p_n!}  \;\;,\quad\quad 
   p_{(n)}\df p_1+p_2+\cdots+p_n \;\;,
\end{equation}
can be calculated by attaching an incoming and an outgoing `arrowed' leg to
each type of vertex, and using the Feynman rules of \Sec{GmFdSec4}. 
Furthermore, we see that 
\begin{equation}
   (-1)^{p_{(n)}}N^{\underline{p_{(n)}}}\,
   \leb{\,\eta_1^{p_1}\,\eta_2^{p_2}\,\cdots\,\eta_n^{p_n}\,}_\mu
   \;=\; \leb{\!\leb{\,\chi_1^{p_1}\,\chi_2^{p_2}\,\cdots\,\chi_n^{p_n}\,}\!}
   \;\;,\quad \chi_k\df\eta_k\sum_{i=1}^N\psib_i\psi_i\;\;.
\end{equation}
Each $\chi_k$ represents a vertex of the kind $\eta_k$, with attached to it 
an incoming and an outgoing `arrowed' leg.
Now, we can apply the relations of \Sec{GmFdSec1} to arrive at the 
result that
\begin{align}
   G(\{\chi\}) \;\df&\;
   \textsl{$1\,+$ the sum of all possible diagrams with $\chi_k$-vertices}  
   \notag\\
   \;=&\; \leb{\!\leb{\,\exp\Big(\sum_k\frac{\chi_k}{k!}\Big)\,}\!}   \;\;,
\label{GmFdEq009}	 
\end{align}
and that the sum of the connected diagrams is equal to  $\log G(\{\chi\})$.

\Appendix{:\; Some combinatorial relations \label{GmFdSec2}}
\addcontentsline{toc}{subsection}{Appendix \theappendix}%
Consider a sequence $f_1,f_2,f_3,\ldots$ of
functions of integer arguments that are completely symmetric in those
arguments. We want to establish a relation of the kind 
\begin{equation}
   \sum_{m=1}^\infty\sum_{k_1,\ldots,k_m=1}^\infty
  f_m(k_1,\ldots,k_m)
   \;\overset{?}{=}\; 
   \sum_{n=1}^\infty\sum_{p(n)}
   f_n(\overbrace{1,\ldots,1}^{p_1(n)},
             \overbrace{2,\ldots,2}^{p_2(n)},\ldots,
	     \overbrace{\overset{}{n}}^{p_n(n)})  \;\;,
\end{equation}
where the second sum on the r.h.s.~is over all partitions $p(n)$ of $n$.  Put
like this, the relation is obviously incorrect, since on the l.h.s.~all
permutations $f_m(\pi(1),\pi(2),\ldots,\pi(m))$ are counted separately, whereas
on the r.h.s., only $f_m(1,2,\ldots,m)$ is counted. At first instance, it seems
natural to correct for this by including a factor $1/m!$ on the l.h.s.. This
is, however, too crude, because permutations of equal $k_i$'s are not counted
separately. This can again be cured by including a factor $p_1(n)!\cdot
p_2(n)!\cdots p_n(n)!$ on the r.h.s., and we arrive at
\begin{equation}
   \sum_{m=1}^\infty\sum_{k_1,\ldots,k_m=1}^\infty
   \frac{f_m(k_1,\ldots,k_m)}{m!}
   \;=\; 
   \sum_{n=1}^\infty\sum_{p(n)}
   \frac{f_n(\overbrace{1,\ldots,1}^{p_1(n)},
             \overbrace{2,\ldots,2}^{p_2(n)},\ldots,
	     \overbrace{\overset{}{n}}^{p_n(n)})}
        {\;\;p_1(n)!\;\cdot\; p_2(n)!\;\cdots\; p_n(n)!} \;\;.
\label{GmFdEq008}	
\end{equation}
Note that $p_n(n)$ is equal to $0$ or $1$. 

Consider the $k\times k$ matrix $A(\{i\}^k)$, depending on $k$ integer
variables $\{i\}^k\df\{i_1,i_2,\ldots,i_k\}$ that run from $1$ to $N$. The
matrix is defined by 
\begin{equation}
   A_{r,s}(\{i\}^k) 
   \;\df\; \de_{i_r,i_s} 
   \;\df\; \begin{cases}
              1 &\textrm{if $i_r=i_s$}\;, \\
              0 &\textrm{if $i_r\neq i_s$}\;.
           \end{cases}
\end{equation}
Every diagonal element of this matrix is equal to $1$ for every configuration 
$\{i\}$, because $i_r=i_s$ if $r=s$. Now consider a configuration $\{i\}$ 
for which all $i$'s are not equal, except of one pair $i_r=i_s$ with $r\neq s$.
Then $A_{r,s}(\{i\}^k)=A_{s,r}(\{i\}^k)=1$, and we see that row $r$ and row $s$
are the same, so that $\det A(\{i\}^k)=0$. It is easy to see that this will 
always be the case if there are pairs $i_r=i_s$ with $r\neq s$. The number of 
configurations $\{i\}$ for which all $i$'s are not equal is precisely 
$N^{\underline{k}}$,
so that we can write down the following identity
\begin{equation}
   N^{\underline{k}}
   \;=\; \sum_{i_1,\ldots,i_k=1}^N\det A(\{i\}^k)
   \;=\; \sum_{i_1,\ldots,i_k=1}^N\sum_{\pi\in S_k}\sgn(\pi)
         \de_{i_1,i_{\pi(1)}}\de_{i_2,i_{\pi(2)}}\cdots
	 \de_{i_k,i_{\pi(k)}}\;\;,
\end{equation}
where the second sum on the r.h.s.~is over all permutations of
$(1,2,\ldots,k)$.  We just used the formula $\det A=\sum_{\pi\in
S_k}\sgn(\pi)\prod_{r=1}^kA_{r,\pi(r)}$ to arrive at this result.

\clearemptydoublepage

\chapter{The formalism of quadratic discrepancies\label{ChapForm}}

Discrepancies are measures of non-uniformity of point sets in subsets of
$s$-dimen\-sional Euclidean space. They are interesting in connection with
numerical integration, because the integration error can be estimated in terms
of the discrepancy of the point set used (\Sec{IntroQMC}). 
Their definition will be given in the first section of this chapter.
An interesting feature of discrepancies is their probability distribution
(\Sec{HQDSec}), and large part of this chapter
concerns with techniques, borrowed from quantum field theory, to
calculate them for the so called quadratic discrepancies. In the last section,
some examples of quadratic discrepancies are given.

\vspace{\baselineskip}

\minitoc

\section{Definition of discrepancy\label{DefDisSec}}
The only subspace of $\Real^s$, $s=1,2,\ldots$ that will be considered 
is the $s$-dimensional unit hypercube $\Kube\df[0,1]^s$ since, in practice, an
integration problem can always be reduced to one on $\Kube$. 
A point set $\XN$ consists of $N$ points
$x_k\in\Kube$, $1\leq k\leq N$. The coordinates of the points will be labeled
with an upper index as $x_k^\nu$, $1\leq\nu\leq s$. For an arbitrary subset
$\Aset$ of $\Kube$, we define the characteristic function $\vt_\Aset$ such that 
\begin{equation}
   \vt_\Aset(x) \;\df\; \begin{cases}
                         1 \;\;&\textrm{if}\;\; x\in\Aset \\
                         0 \;\;&\textrm{if}\;\; x\not\in\Aset \;\;.
                      \end{cases}
\end{equation}
The integral of a function $f$, Lebesgue integrable on $\Kube$, we denote by
\begin{equation}
   \leb{f} \;\df\; \intk f(x)\,dx  \;\;,
\end{equation}
so that the Lebesgue measure of a region $\Aset\subset\Kube$ is given by
$\leb{\vt_\Aset}$. For every point set $\XN$, we introduce the
estimate $\lebM{f}{N}{}$ of $\leb{f}$ using $\XN$ by  
\begin{equation}
   \lebM{f}{N}{} \;\df\; \frac{1}{N}\sum_{i=1}^Nf(x_i) \;\;.
\end{equation}

\subsection{The original definition}
Naturally, a discrepancy of the point set is defined with respect to a certain
family $\Afam$ of measurable subsets of $\Kube$ as follows
\begin{equation}
   D_N^{[\Afam]}(\XN) 
   \;\df\; \sup_{\Aset\in\Afam}
         \big|\lebM{\vt_\Aset}{N}{}-\leb{\vt_\Aset}\big| \;\;.
\end{equation}
It is the largest absolute error one makes if one tries to estimate the measure
of every subset $\Aset\in\Afam$ by counting the number of points from $\XN$ in
the subset. The idea is that, if a point set is suitable for estimating the
measures of all subsets well, so that $D_N^{[\Afam]}(\XN)$ is small, then the
point set must be distributed very ``uniformly'' over $\Kube$.   In order to
arrive at a natural notion of uniformity, the family $\Afam$ of subsets has to
be chosen sensibly. In principle, for every finite point set a subset of
$\Kube$ can be found, such that the discrepancy takes its maximum value, which
is $1$. A first restriction on the subsets one can for example take is that
they have to be convex, i.e., for every $x_1,x_2$ in $\Aset$ and every
$t\in[0,1]$ also $tx_1+(1-t)x_2$ is in $\Aset$.  

This restriction still leaves many possible choices for the subsets, leading to
different discrepancies (cf.~\cite{Drmota}). An important example is the so
called {\em star discrepancy}, denoted by $D_N^*$, for which the family 
$\Afam^*$ consists of all subsets 
\begin{equation}
   \Aset_y \;\df\; [0,y^1)\times[0,y^2)\times\cdots\times[0,y^s)\;\;,\quad
   y\in\Kube \;\;.
\end{equation}
It consists of all hyper-rectangles spanned by the origin and points
$y\in\Kube$. For this discrepancy, various theorems are derived 
(cf.~\cite{Drmota}), such as Koksma-Hlawka's inequality (\eqn{KHIneq}). In one 
dimension, $D_N^*$ is equal to the statistic of the 
Kolmogorov-Smirnov test for the hypothesis that the points are 
distributed randomly following the uniform distribution (cf.~\cite{Knuth}).

In order to proceed in a direction that leads to a definition of discrepancy
that we will use, we introduce the $L_p^*$-{\em discrepancy}. If we denote
$\vt_y\df\vt_{\Aset_y}$, then 
\begin{equation}
   D_N^{[p]}(\XN)
   \;\df\; \left(\intk\big|\lebM{\vt_y}{N}{}
                           -\leb{\vt_y}\big|^pdy\right)^{1/p} \;\;.
\label{defEq005} 
\end{equation}
It is the average over $\Afam^*$ of the $p^{\textrm{th}}$ power of the error
made by estimating the measures of the subsets using $\XN$. This definition
assures the limit $D_N^*(\XN)=\lim_{p\ra\infty}D_N^{[p]}(\XN)$. Furthermore,
it satisfies the bounds
\begin{equation}
   D_N^{[p]}(\XN)
   \leq D_N^*(\XN)\leq c(s,p)D_N^{[p]}(\XN)^{\frac{p}{p+1}} \;\;,
\end{equation}
where $c(s,p)$ is independent of the point set \cite{Drmota}. For us, the case
of $p=2$ is in particular interesting. The expression for $D_N^{[2]}(\XN)$ can
be evaluated further, with the result that
\begin{equation}
   D_N^{[2]}(\XN) 
   \;=\; \left(\frac{1}{N^2}\sum_{k,l=1}^N\twoB(x_k,x_l)\right)^{1/2} \;\;,
\label{defEq006}
\end{equation}
where
\begin{equation}
   \twoB(x_k,x_l)
   \;=\; \twoC(x_k,x_l) - \intk\twoC(x_k,y)\,dy - \intk\twoC(y,x_l)\,dy 
                        + \int_{\Kube^2}\twoC(y_1,y_2)\,dy_1dy_2 \;\;,
\label{defEq003}	 
\end{equation}
and
\begin{equation}
   \twoC(x_k,x_l) 
   \;=\; \intk\vt_y(x_k)\vt_y(x_l)\,dy
   \;=\; \prod_{\nu=1}^s\min(1-x_k^\nu,1-x_l^\nu) \;\;.
\end{equation}
In this case of $p=2$, the discrepancy is called {\em quadratic\/}, and is
completely determined by the two-point function $\twoC$.

\subsection{Quadratic discrepancies}
The quadratic discrepancy invites generalizations. The number
$\twoB(x_1,x_2)$ can be interpreted as a correlation between the points $x_1$
and $x_2$, and the discrepancy is a function of the sum over all correlations
in the point set. Various quadratic discrepancies can be defined by choosing
different, and sensible, two-point correlation functions.  The two-point
functions can, however, also be interpreted differently, leading to another
approach to quadratic discrepancies. This approach is based on the insight by 
H.~Wo\'{z}niakowski \cite{Woz}, that $D_N^{[2]}$ can be written as an 
{\em average case complexity}. 
We will demonstrate this here by constructing the probability measure with
respect to which the discrepancy can be written as an average. For more details
about the formalism, we refer to \cite{Choquet}.

\newcommand{\CfamW}{\Cfam_{\textrm{W}}}
\newcommand{\muW}{\mu^{}_{\textrm{W}}}
Consider the Hilbert space $\Hilb\df L_2(\Kube)$ of (equivalence classes of
almost everywhere equal) real quadratically integrable functions on $\Kube$. We
denote the inner product
and the norm
on $\Hilb$ by 
\begin{equation}
   \inp{f}{h} \;\df\; \intk f(x)h(x)\,dx  
   \quad,\quad\quad
   \norm{f} \df \sqrt{\inp{f}{f}}\quad\textrm{for real $f$} \;\;.
\end{equation}
Let us, as before, denote $\vt_y(x)=\prod_{\nu=1}^s\theta(y^\nu-x^\nu)$ and let
$\Prim$ be the ``primitivation'' operator, defined by 
\begin{equation}
   (\Prim\Pspc f)(x) \;\df\; \intk\vt_y(x)f(y)\,dy \;\;.
\end{equation}
$\Prim$ is a continuous linear map from $\Hilb$ to the space $\CfamW$ of
continuous functions that vanish if any coordinate $x^\nu=1$. It even is a
Hilbert-Schmidt operator: if $\{u_n\}$ is a basis of $\Hilb$, then
$\sum_n\norm{\Prim\Pspc u_n}^2<\infty$.

The dual space $\CfamW'$, i.e.~the space of all continuous linear functionals
on $\CfamW$, consists of all bounded measures on $[0,1)^s$. For such a measure
$\eta$, we will use the notation
\begin{equation}
   \eta[f] \;\df\; \intk f(x)\eta(x)\,dx \;\;,
\end{equation}
although ``$\eta(x)$'' cannot always be seen as a function value.  The
transposed $\Prit$, which acts on $\CfamW'$ through the definition
$(\Prit\Pspc\eta)[f]\df\eta[\Prim\Pspc f]$, is then simply given by
\begin{equation}
   (\Prit\Pspc\eta)(x) \;=\; \intk\vt_x(y)\eta(y)\,dy \;\;,
\end{equation}
and $\Prit\Pspc\eta$ {\it is} a bounded function. 
Notice that, because $\Hilb$ is isomorphic to its dual, we can make the 
straightforward identification $\Prit=\Prid$, where $\Prid$ is the adjoint of 
$\Prim$.
There is a unique Gaussian
probability measure $\muW$ on $\CfamW$ which has Fourier transform
\begin{equation}
   \int_{\CfamW}\exp(i\eta[\phi])\,d\muW[\phi] 
   \;=\; \exp(-\sfrac{1}{2}\qnm{\Prit\Pspc\eta}) \;\;.
\end{equation}
It is going to serve as the probability measure mentioned before. In
literature, it is known as the {\em Wiener sheet measure}. By taking
$\eta\df\la(\la_1\de_{x_1}+\la_1\de_{x_1})$, where $\la,\la_1,\la_2$ are real
variables and $\de_x$ denotes the Dirac measure $\de_x[\phi]\df\phi(x)$, and
differentiating the above equation twice with respect to $\la$, it is easy to
see that $\muW$ has two point function 
\begin{equation}
   \int_{\CfamW}\phi(x_1)\phi(x_2)\,d\muW[\phi]
   \;=\; \inp{\Prit\de_{x_1}}{\Prit\de_{x_2}}
   \;=\; \prod_{\nu=1}^s\min(1-x_1^\nu,1-x_2^\nu) \;\;.
\end{equation}
So the two-point function with which the discrepancy is defined, is the
two-point function of the Gaussian measure $\muW$ on $\CfamW$. Using
this equation and \eqn{defEq006} and \eqn{defEq003}, it is easy to see that
\begin{equation}
   \big(\,D_N^{[2]}(\XN)\,\big)^2 
   \;=\; \int_{\CfamW}\big|\lebM{\phi}{N}{}
                                - \leb{\phi}\big|^2\,d\muW[\phi]\;\;.
\label{defEq001}		   
\end{equation}
So we can identify the square of the discrepancy as the average case
complexity, defined as the squared integration error averaged over $\CfamW$.

The particular choice of $\Prim$ led to the Wiener sheet measure. In principle,
any Hilbert-Schmidt operator can be used, leading to another Gaussian measure
$\mu$. We want to apply this generalization. For further analysis, it will
therefore appear to be convenient to use the square, as it stands on the
l.h.s., as definition for quadratic discrepancy.  Furthermore, the definition
of discrepancy is such, that it goes to zero, if the number of points in a
uniformly distributed point set goes to infinity. This is immediately clear
from inequality (\ref{KHIneq}), the l.h.s.~of which goes to zero if $\XN$
consists of independent random points and $N$ goes to infinity. In fact, Monte
Carlo integration tells us that it goes to zero as $1/\sqrt{N}$. Therefore, it
seems natural to use $N$ times the square of the original definition of the
quadratic discrepancy, especially since we want to calculate probability
distributions of discrepancies for large $N$. This multiplication with the
factor $N$ is equivalent with considering $\sqrt{N}$ times the average of $N$
random variables (with zero mean) when applying the central limit theorem in
probabilistic analyses. 

\subsubsection{Definition quadratic discrepancy}
We conclude this section with the definition of discrepancy we will further
use. Given a Hilbert-Schmidt operator $\Prim$ on $\Hilb\df L_2(\Kube)$,
there is a Gaussian measure $\mu^{}_{\Prim}$ on $\Hilb$ with Fourier
transform
\begin{equation}
  \int_{\Hilb}\exp(i\eta[\phi])\,d\mu^{}_{\Prim}[\phi]
  \;=\; \exp(-\half\qnm{\Prit\Pspc\eta})  \;\;.
\label{defEq007}  
\end{equation}
In the case of the Wiener sheet measure, it even is a measure on a space of
continuous functions, but for the general case this is not necessary. The
operator $\Prim$ should only be such, that it maps $\Hilb$ continuously on a
space $\Cfam$ of continuous functions, so that there is a number $p$ such that
$\sup_{x\in\Kube}|(\Prim\Pspc f)(x)|\leq p\norm{f}$ for any $f\in\Hilb$. In
\App{App3A}, we show that in that case the Dirac measure can be properly
defined under the measure $\mu^{}_{\Prim}$, which we will need. We shall omit
the label $\Hilb$ at the integral symbol from now on.

We define the discrepancy of a point sets $\XN$ in $\Kube$ as the quadratic
integration error, made by using $\XN$, averaged under $\mu^{}_{\Prim}$ over
$\Hilb$: 
\begin{equation}
   D_N(\XN)
   \;\df\; N\int\big|\lebM{\phi}{N}{}
                              - \leb{\phi}\big|^2\,d\mu^{}_{\Prim}[\phi] \;\;.
\end{equation}
From now on, we will omit the argument $\XN$ when we denote the discrepancy.
Using this definition, the discrepancy can again be written in terms of
two-point functions.  With the Hilbert-Schmidt operator comes a two-point
function
\begin{equation}
   \twoC_{\Prim}(x_1,x_2) 
   \;\df\; \int\phi(x_1)\phi(x_2)\,d\mu^{}_{\Prim}[\phi] 
   \;=\; \inp{\Prit\de_{x_1}}{\Prit\de_{x_2}} \;\;.
\end{equation}
Because the combination $\phi-\leb{\phi}$ appears in the average, it is useful
to introduce the notation
\begin{equation}
   \hphi(x) \;\df\; \phi(x) - \leb{\phi} \;\;,
\end{equation}
and the {\em reduced\/} two-point function 
\begin{align}
   \twoB_{\Prim}(x_1,x_2) 
   \;\df\; \int\hphi(x_1)\hphi(x_2)\,d\mu^{}_{\Prim}[\phi] \;\;,
\end{align}
which can be written in terms of $\twoC_{\Prim}$ like in \eqn{defEq003}. It has
the important feature that it integrates to zero with respect to each of its
arguments. The discrepancy is given by
\begin{equation}
   D_N \;=\; \frac{1}{N}\sum_{k,l=1}^N\twoB_{\Prim}(x_k,x_l)  \;\;,
\end{equation}
i.e., as a sum over two-point correlations between the points of $\XN$. The
correlation function $\twoB_{\Prim}$ is determined by the operator $\Prim$ in
this formulation.

\section{The generating function}
When $\XN$ consists of uniformly distributed random points, then the
discrepancy $D_N$ is a random variable with a certain probability density.
In \cite{jhk,hk1,hk2}, the generating function
\begin{equation} 
    G(z) \df \Exp(\,e^{zD_N}\,) 
\end{equation}
has been used to calculate it. 
We will also concentrate on the calculation of $G(z)$. Given $G$, the
probability density can then be calculated by the inverse Laplace
transform (\Sec{ProSecGen}). 

It will turn out that it is far to complicated to calculate $G(z)$
analytically.  For large number $N$ of points, however, a series expansion in
$1/N$ can be made which can be calculated term by term.  
We intend to calculate the generating function from an explicit
expression in terms of $\mu^{}_{\Prim}$, which we will now derive. 

\subsection{The generating function as an average over functions}
First, we introduce the following bounded measure on $\Kube$, which consists of
a sum of Dirac measures, centered around the points of the point set, minus
one:
\begin{equation}
   \eta_N(x) 
   \;\df\; \frac{i}{N}\sum_{k=1}^N\left[\de_{x_k}(x) - 1\right] \;\;.
\label{FieEq003}   
\end{equation}
The integration error of a function $\phi$ and the discrepancy can be
written in terms of $\eta_N$: 
\begin{equation}
   \lebM{\phi}{N}{} - \leb{\phi} = -i\eta_N[\phi] 
   \quad\quad\textrm{and}\quad\quad
   D_N = -N\qnm{\Prit\Pspc\eta_N} \;\;.
\end{equation}
Using this expression for the discrepancy and the relation of \eqn{defEq007},
we can write 
\begin{equation}
   \exp(zD_N) \;=\; \exp(-zN\qnm{\Prit\Pspc\eta_N})
   \;=\; \int\exp\left(\sqrt{2zN}\,\eta_N[\phi]\right)
         d\mu^{}_{\Prim}[\phi] \;\;.
\end{equation}
If now the definition of $\eta_N$ is used, and the integrals over
$x_1,\ldots,x_N$ are performed on the l.h.s.~and the r.h.s., we arrive at
\begin{equation}
   G(z) 
   \;=\; \int\leb{\,e^{g\hphi}\,}^Nd\mu^{}_{\Prim}[\phi] 
         \quad,\quad\quad  g \;=\; \sqrt{\frac{2z}{N}} \;\;,
\label{FieEq005}
\end{equation}
where we denote
\begin{equation}
   \leb{\,e^{g\hphi}\,}\df\intk\exp(\,g\hphi(x)\,)\,dx \;\;.
\end{equation}

\subsection{Gauge freedom}
For the calculation of the generating function of the probability density of
the discrepancy, there exists a freedom in the choice of the operator $\Prim$
with which the measure is defined, as we will show now. Let $\Tran$ act on
$\Hilb\df L_2(\Kube)$ such that $\Tran\Prim$ is a Hilbert-Schmidt operator on
$\Hilb$ that maps $\Hilb$ continuously on a space of continuous
functions. For each functional $F$ on $\Hilb$ there is a functional
$F\circ\Tran$ which maps $\phi\in\Hilb$ onto $F[\Tran\phi]$. We use this to
define the measure $\mu^{}_{\Tran\Prim}$ by
\begin{equation}
   \int F[\phi]\,d\mu^{}_{\Tran\Prim}[\phi]
   \;\df\; \int(F\circ\Tran)[\phi]\,d\mu^{}_{\Prim}[\phi] \;\;,
\end{equation}
so that its Fourier transform is given by
\begin{equation}
   \int\exp(i\eta[\phi])\,d\mu^{}_{\Tran\Prim}[\phi]
   \;=\; \exp(-\half\qnm{\Prit\Pspc\tilde{\Tran}\eta}) \;\;,
\end{equation}
and its two-point function is given by 
\begin{equation}
   \twoC_{\Tran\Prim}(x_1,x_2) 
   \;=\; \inp{\Prit\Pspc\tilde{\Tran}\de_{x_1}}{\Prit\tilde{\Tran}\de_{x_2}} 
   \;\;.
\end{equation}
If $\Tran$ is such that $\widehat{\Tran\phi}=\hphi$ for all
$\phi$, then
\begin{equation}
   G(z) 
   \;=\; \int\leb{\,e^{g\hphi}\,}^Nd\mu^{}_{\Prim}[\phi]
   \;=\; \int\leb{\,e^{g\hphi}\,}^Nd\mu^{}_{\Tran\Prim}[\phi] \;\;,
\end{equation}
and we call this property the {\em gauge freedom}. It leads to a freedom in the
choice of the operator $\Prim$ with which the measure is defined, and we call
these choices the {\em gauges}. Most {\em gauge transformations\/} $\Tran$ we
consider are {\em global translations\/} that are characterized by a functional
$t:\Hilb\mapsto\Real$, and are given by
$(\Tran\phi)(x)\df\phi(x)+t[\phi]$ for all $x$. They trivially satisfy
$\widehat{\Tran\phi}=\hphi$.

An example of a gauge transformation that satisfies the criteria is simply
given by $\Tran\phi\df\hphi$. It results in the {\em Landau gauge}, for which
all $\phi$ satisfy $\leb{\phi}=0$. This is, actually, the natural gauge to
choose, because it restricts the analysis to functions that integrate to zero,
so that the integration error becomes equal to the average of the function over
the point set. The existence of the gauge freedom originates from the fact that
the integration error is the same for integrands that differ only by a
constant.  The two-point function is equal to the reduced two-point function in
the Landau gauge: $\twoC_{\Tran\Prim}=\twoB_{\Tran\Prim}$.

From now on, we will omit the label `$\Prim$' in the notation of the measures
and the two-point functions.

\subsection{The path integral, perturbation theory and instantons
            \label{SecPath}}
The connection of the foregoing with Euclidean quantum field theory is made via
the path integral formulation. The emphasis is put on the use of perturbation
theory.

\subsubsection{The path integral}
We want to express the generating function, as given by \eqn{FieEq005}, in
terms of a {\em Euclidean path integral\/} (cf.~\cite{Rivers}). We have to
introduce the {\em free action}, which is a quadratic functional 
\begin{equation}
   S_0[\phi] \;\df\; \half\qnm{\Prim^{-1}\phi} \;\;,
\end{equation}
and we arrive at the path integral formulation of the measure
$\mu$ by making the identification 
\begin{equation}
   d\mu[\phi] \;=\; \Dpath \exp(-S_0[\phi]) \;\;.
\end{equation}
It is a formal expression, where $\Dpath$ represents the product over the whole
of $\Kube$ of the ``infinitesimal volume elements'' $d\phi(x)$. This is, of
course, ill-defined, and it gets even worse since the set of
functions $\phi$, for which $S_0[\phi]$ is finite, in general has measure zero.

One thing we want to be more precise about is the fact that, in first instance,
$\Prim^{-1}$ is only well-defined on the image $\Prim\Hilb$ of $\Hilb\df
L_2(\Kube)$ under $\Prim$, while this set has measure zero. The members of the
subsets of $\Hilb$ that do not have measure zero, however, usually {\em do\/}
satisfy the boundary conditions imposed by $\Prim$. We assume that these 
boundary conditions can be expressed by a finite number of linear equations 
\begin{equation}
   \varUpsilon_i[\phi] = 0 \;\;,\quad i=1,2,\ldots \quad.
\end{equation}
Then, the action should be extended as follows: 
\begin{equation}
   S_0[\phi] \;=\;  \half\qnm{\Prim^{-1}\phi}
                   + \sum_i\half M_i\varUpsilon_i[\phi]^2 \;\;,
\end{equation}
where $M_i\ra\infty$ for all labels $i$. The ``infinitesimal volume element''
$\Dpath$ gets a factor $\sqrt{2\pi/M_i}$ for every $i$ in order for the
measure to stay normalized to one. The extra terms in the action assure that
the measure is zero if a function does not satisfy the boundary conditions.
Notice that the action is still quadratic in $\phi$. 

If we apply all this to the expression of \eqn{FieEq005} for the generating
function, we find that 
\begin{equation}
   G(z) \;=\; \int\Dpath\exp(-S[\phi])  \;\;,
\end{equation}
with an action $S$ given by 
\begin{equation}
   S[\phi] 
   \;=\; S_0[\phi] - N\log\leb{\,e^{g\hphi}\,} \;\;,\quad\quad
   g=\sqrt{\frac{2z}{N}} \;\;.
\label{FieEq006}		    
\end{equation}

\subsubsection{Perturbation theory and instantons}
In the action of \eqn{FieEq006}, $g$ appears to be a natural expansion
parameter if $N$ is large. An expansion around $g=0$ will automatically result
in an expansion of the generating function around $z=0$. Furthermore, it
corresponds to an expansion of the action around $\phi=0$. An expansion of the
action to evaluate the generating function, however, only makes sense when it
is an expansion around a minimum, so that it represents a saddle point
approximation of the path integral. Therefore, a straightforward expansion such
as just proposed, is only correct if it is an expansion around the minimum of
the action, that is, if the trivial solution $\phi=0$ gives the only minimum of
the action.  General extrema of the action are given by solutions of the {\em
field equation}
\begin{equation}
   (\vLa\phi)(x) 
   + Ng - Ng\frac{e^{g\phi(x)}}{\leb{e^{g\phi}}} \;=\; 0 \;\;,
\label{FieEq013}    
\end{equation}
where $\vLa$ represents the self-adjoint operator
$(\Prim^{-1})^\dagger\Prim^{-1}$ including the boundary conditions
$\varUpsilon_i$ {\em and\/} possible boundary conditions coming from the fact
that $\Prim^{-1}$ is not necessarily self-adjoint~\footnote{For example, if
$\Prim^{-1}\phi=\phi'\df\frac{d\phi}{dx}$ and $\varUpsilon[\phi]=\phi(0)$, then
$\qnm{\Prim^{-1}\phi}=\phi(1)\phi'(1)-\inp{\phi}{\phi''}$, so
that the extra boundary condition is $\phi'(1)=0$.}.  Depending on
the value of $z$, non-trivial solutions may also exist.  At this point it can
be said that, because $\vLa\phi$ is real, non-trivial solutions only exist if
$z$ is real and non-zero so that $g\in\Real$.  In the analysis of the solutions
we therefore can do a scaling $\phi(x)\mapsto \phi(x)/g$ so that the action for
these solutions is given by
\begin{equation}
   \vSi[\phi]\;\df\;
   \frac{1}{N}\,S[{\textstyle\frac{1}{g}}\,\phi] 
   \;=\; \frac{1}{2}\,\frac{\leb{\phi\,e^{\phi}}}{\leb{e^{\phi}}}
         + \frac{1}{2}\leb{\phi}
         - \log\leb{e^{\phi}}\;\;.
\label{FieEq014}	 
\end{equation}
These non-trivial solutions we call instantons (cf.~\cite{Coleman}), although
this may not be a rigorously correct nomenclature, in the field theoretical
sense, for all situations we will encounter.  Notice that instantons under
different gauges only differ by a constant: if two gauges are connected by a
global translation $\Tran$, and $\phi$ is an instanton in the $\Prim$-gauge,
then $\Tran\phi$ is an instanton in the $\Tran\Prim$-gauge.  The values of $z$
for which they appear and the value of the action are gauge invariant, as can
be concluded from \eqn{FieEq013} and \eqn{FieEq014}.

If $N$ becomes large, then the contribution of an instanton to the path
integral will behave as $e^{-N\vSi[\phi]}$, where $\vSi[\phi]$ does not depend on
$N$ (Notice that $\phi(x)$ does not depend on $N$ because the field equation
for these rescaled functions does not depend on $N$.).  The
$e^{-N\vSi[\phi]}$-like behavior of the instanton contribution makes it
invisible in the perturbative expansion around $1/N=0$. If $\vSi[\phi]$ is
larger than zero, this will not be a problem, because the contribution will be
very small. If, however, $\vSi[\phi]$ is equal to zero, then the contribution
will be more substantial, and it will even explode if $\vSi[\phi]$ is
negative. Notice that, to be able to do make a perturbation series
around $\phi=0$, the action has to be zero for this solution, for else the
terms would all become zero or would explode for large $N$.

The escape from this possible disaster is given by the fact that $z$ has to be
real and larger than zero for instantons to exist, and we want to integrate
$G(z)$ along the imaginary $z$-axis (\Sec{ProSecGen}). Also in the end,
when we want to close the integration contour in the complex $z$-plane to the
right, we will not meet the problem, because the function we want to integrate
is an expansion in $z$ around $z=0$ that can be integrated term by term.
Problems might only occur when instantons exist for values of $z$ that are
arbitrarily close to $0$. We will confirm for a few cases that this does not
happen.

\subsection[Feynman rules to calculate the $1/N$ corrections]
           {Feynman rules to calculate the $1/N$ corrections
	    \label{FormSec1}}
We just suggested a straight-forward expansion in $1/N$ of $\exp(-S)$
to calculate $G$ perturbatively. This way, however, the calculation of the
perturbation series becomes very cumbersome, and the reason for this is the
following. We want to use the fact that an expansion in $1/N$ corresponds to an
expansion around $\phi=0$ of the part of the action that is non-quadratic in
$\phi$. The subsequent terms in the expansions are therefore proportional to
moments of a Gaussian measure, and can be calculated using diagrams
(\Sec{GmFdSec3}). These diagrams, the {\em Feynman diagrams}, consist of lines
representing two-point functions and vertices representing convolutions of
two-point functions. Because the action is non-local, i.e.~it cannot be written
as a single integral over a Lagrangian density because of the logarithm in
\eqn{FieEq006}, the total path integral, thus the total sum of all diagrams,
cannot be seen as the exponential of all {\em connected\/} diagrams, and it is
this that makes the calculations difficult.

In order to circumvent this obstacle, we first of all use the Landau gauge, so
that $\hphi=\phi$ for all $\phi$. Secondly, we introduce $2N$ Grassmann
variables $\psib_i$ and $\psi_i$, $i=1,2,\ldots,N$, as in \Sec{GmFdSec4}, so
that we can write
\begin{equation}
   G(z) 
   \;=\; \int\leb{\,e^{g\phi}\,}^N\,d\mu[\phi]
   \;=\; \int\int[d\psib d\psi]
         \exp\Big(-\sum_{i=1}^N\psib_i\psi_i\,\leb{\,e^{g\phi}\,}\Big)\,
	 d\mu[\phi] \;\;.
\end{equation}
If we now define 
\begin{equation}
   \eta_k[\phi] \df -g^k\leb{\phi^{k}} \;\;,\quad
   \chi_k       \df \eta_k\sum_{i=1}^N\psib_i\psi_i\;\;,
\end{equation}
and
\begin{equation}
   \leb{\!\leb{\,f\,}\!} 
   \;\df\; \int\int[d\psib d\psi] \,f(\{\chi\})\,
           \exp\Big(-\sum_{i=1}^N\psib_i\psi_i\Big)\,d\mu[\phi] \;\;,
\end{equation}
we can write
\begin{equation}
   G(z)
  \;=\; \leb{\!\leb{\,
         \exp\Big(\sum_{k=2}^\infty\frac{\chi_k}{k!}\Big)\,}\!} \;\;,	 
\end{equation}
which has exactly the form of the r.h.s.~of \eqn{GmFdEq009}. Because the 
functionals $\eta_k$ are also of the kind of \eqn{GmFdEq010}, we can use the 
statement of \eqn{GmFdEq009}, that $G(z)$ is equal to the sum of the
contributions of all Feynman diagrams that can be constructed with the vertices
\begin{equation}
   \diagram{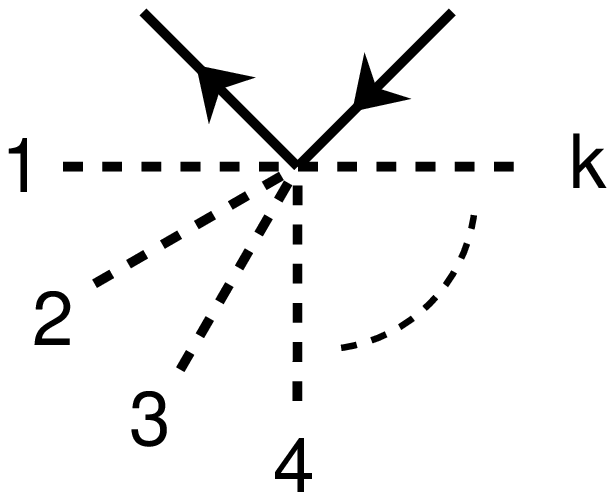}{50}{25}  \;\;,\quad k\geq 2 \;\;,
\label{CorEq007}
\end{equation}
and with the rules that all incoming legs have to be connected to outgoing legs
and vice versa, and all dashed legs have to be connected to dashed legs. The 
lines in the obtained diagrams stand for {\em propagators}:
\begin{align}
   &\textrm{boson propagator:}\hspace{6pt}\quad
    x\,\diagram{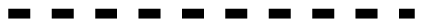}{30}{-1}\,y \;=\; \twoB(x,y) \;\;; \\
   &\textrm{fermion propagator:}\quad		  
    i\,\diagram{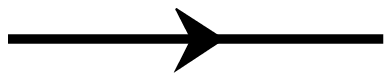}{30}{0}\,j \;=\; \de_{i,j}   \;\;,
\end{align}
and to calculate the contribution of a diagram, boson propagators have to be
convoluted in the vertices as
$\intk\twoB(y,x_1)\twoB(y,x_2)\cdots\twoB(y,x_k)\,dy$, fermion propagators as
$\sum_{j=1}^{\raisebox{-1pt}{${\scriptstyle N}$}}\de_{i_1,j}\de_{j,i_2}$, and
then these convolutions have to be multiplied. To get the final result for a diagram, a factor
$-g^k$ has to be included for every vertex of order $k$, and the symmetry
factor has to be included. 

The contribution of the fermionic part can easily be determined, for every
fermion loop only gives a factor $-N$. The main problem is to calculate the
bosonic part.  Furthermore, only the connected diagrams have to be calculated,
since the sum of their contributions is equal to 
\begin{equation}
   W(z) \df \log G(z) \;\;.
\end{equation}
Because every vertex carries a power of $g$ that is equal to its
order, the expansion in $g$ is an expansion in the complexity of the diagrams,
which can be systematically evaluated.

\subsection{Gaussian measures on a countable basis\label{GMOACB}}
Because $\Prim$ is a Hilbert-Schmidt operator on $\Hilb\df L_2(\Kube)$,
$\Prid\Prim$ is a self adjoint compact operator on $\Hilb$, and
there exists an orthonormal basis $\{u_n\}$ of $\Hilb$, consisting 
of eigenvectors of $\Prid\Prim$. If we denote the eigenvalues by 
$\si_n^2$, then the eigenvalue equation is given by 
\begin{equation}
   (\Prid\Prim\Pspc u_n)(x) 
   \;=\; \intk\twoC(x,y)u_n(y)\,dy \;=\; \si_n^2u_n(x) \;\;.
\end{equation}
As the notation suggests, they are positive since 
$0<\norm{\Prim\Pspc u_n}^2=\inp{u_n}{\Prid\Prim\Pspc u_n}=\si_n^2$.
Notice that, because $\Prim$ is Hilbert-Schmidt,
$\sum_n\si_n^2\norm{u_n}^2<\infty$, and this leads immediately to the spectral
decomposition of $\twoC$, which is simply given by 
\begin{equation}
   \twoC(x,y)
   \;=\; \sum_n\si_n^2u_n(x)u_n(y) \;\;.
\label{CorEq005}   
\end{equation}
In principle, the basis and the eigenvalues can be used as an alternative 
definition of a quadratic discrepancy. They naturally introduce the spectral 
decomposition of a two-point function and a reduced two-point function. 
The reasonable requirement of the existence of $\Exp(D_N)$ leads to 
\begin{equation}
   \Exp(D_N)
   \;=\; \sum_n\si_n^2 \left(\norm{u_n}^2-\leb{u_n}^2\right)
   \;<\; \infty\;\;, 
\end{equation}
which is satisfied if $\twoC$ comes from a Hilbert-Schmidt operator.
If we denote the expansion of a function $\phi\in\Hilb$ by
\begin{equation}
   \phi(x) 
   \;=\; \sum_n \phi_nu_n(x) \;\;,\quad \phi_n\in\Real \;\;,
\end{equation}
then the probability measure $\mu$ can be written as
\begin{equation}
  d\mu[\phi] 
  \;=\;
  \prod_{n}\frac{\exp(-\phi_n^2/2\si_n^2)}{\sqrt{2\pi\si_n^2}}\,d\phi_n
  \;\;,\quad \si_n\in\Real \;\;.
\label{CorEq004}  
\end{equation}
The basis functions will often be refered to as {\em modes}, originating from 
an example of a quadratic discrepancy (the Fourier diaphony), for which the 
basis is the Fourier basis without the constant mode.

With different gauges come different bases and strengths. We call
a gauge in which the basis is orthonormal a {\em Feynman\/} gauge. If the
Landau gauge is used, in which $\leb{\phi}=0$, then the basis functions
have to integrate to zero:
\begin{equation}
   \leb{\,u^{(\textrm{L})}_n\,} \;=\; 0 \quad \forall\,n \;\;,
\end{equation}
where the label $\textrm{L}$ indicates the Landau gauge. These functions are 
the solutions of the eigenvalue equation
\begin{equation}
   \intk\twoB(x,y)\,u^{(\textrm{L})}_n(y)\,dx 
   \;=\; \si_{\textrm{L},n}^2u^{(\textrm{L})}_n(x) \;\;.
\end{equation}
It will not always be possible to find the Landau basis.  In
terms of a basis that is not in the Landau gauge, $\twoB$ is given by 
\begin{equation}
   \twoB(x,y)
   \;=\; \sum_n\si_n^2(u_n(x)-\leb{u_n})(u_n(y)-\leb{u_n}) \;\;.
\label{defEq008}   
\end{equation}

\section{Examples}
Some explicit, and well known, examples of quadratic discrepancies are introduced, and cast in the formalism of this chapter.
\subsection[The $L_2^*$-discrepancy]
           {The $L_2^*$-discrepancy\label{SecDefL2}}
In our definition, the $L_2^*$-discrepancy is $N$ times the square of the case
of $p=2$ in \eqn{defEq005}. The operator $\Prim$ and the two-point function 
$\twoC$ are given by
\begin{equation}
   (\Prim\phi)(x)
   \df\intk\vt_y(x)\phi(y)\,dy 
   \;\;,\quad
   \twoC(x_1,x_2) \;=\; \prod_{\nu=1}^s\min(1-x_1^\nu,1-x_2^\nu) \;\;,
\end{equation}
where $\vt_y(x)\df\prod_{\nu=1}^s\theta(y^\nu-x^\nu)$. The boundary conditions
imposed by $\Prim$ are given by
$\phi(x)=0$ if at least one of the
coordinates $x^\nu=1$. The basis functions can now be found by solving the
eigenvalue equation $\intk\twoC(x,y)u(y)=\si^2u(x)$. The equation factorizes
for the different coordinates, and is most easily solved by differentiating
twice on the l.h.s.~and the r.h.s.. The one-dimensional solutions, that satisfy
the boundary conditions, are 
\begin{equation}
   u_n(x^\nu)=\sqrt{2}\,\cos\left((n+\half)\pi x^\nu\right) \;\;,\quad
   \si_n^{2}=\pi^{-2}(n+\sfrac{1}{2})^{-2} \;\;,\quad
   n=0,1,2,\ldots \;\;.
\label{L2def003}   
\end{equation}
The set $\{u_n\}$ clearly is orthonormal, and it is the basis 
in the one-dimensional case. For $s>1$, the basis and the strengths are 
given by all possible products
\begin{equation}
   u_{\vn}(x) 
   \;=\; 2^{s/2}\prod_{\nu=1}^s\cos\left((n_\nu+\half)\pi x^\nu\right) 
   \;\;,\quad
   \si_{\vn}^{2} \;=\; \pi^{-2s}\prod_{\nu=1}^s(n_\nu+\half)^{-2} \;\;,
\end{equation}
where $\vn\df(n_1,n_2,\ldots,n_s)$ and $n_\nu=0,1,2,\ldots$ for 
$\nu=1,\ldots,s$. The reduced two-point function is given by
\vspace*{-1pt}\begin{equation}
   \twoB(x_1,x_2)
   \;=\; \prod_{\nu=1}^s\min(1-x_1^\nu,1-x_2^\nu)
         - (\half)^s\prod_{\nu=1}^s(1-(x_1^\nu)^2)
         - (\half)^s\prod_{\nu=1}^s(1-(x_2^\nu)^2)
	 + (\sfrac{1}{3})^s \;\;.
\end{equation}
In one dimension, the eigenfunctions and eigenvalues are
\begin{equation}
   u^{(\textrm{L})}_n(x)=\sqrt{2}\,\cos(n\pi x) \;\;,\quad
   \si_{\textrm{L},n}^{2}=\pi^{-2}n^{-2} \;\;,\quad
   n=1,2,\ldots \;\;.
\end{equation}
For $s>1$, it is difficult to find all solutions to the eigenvalue equation,
and we will address this problem in \Sec{GauSec01}.

\subsection{The Cram\'er-von Mises goodness-of-fit test}
The $L_2^*$-discrepancy is equivalent with the statistic of the Cram\'er-von
Mises goodness-of-fit test, which tests the hypotheses that $N$ data $x_k$ are
distributed independently following a cumulative distribution function $F$
(cf.~\cite{Knuth,Anderson}).
Consider, for simplicity, the one-dimensional case, so that $x_k\in\Real$, and
denote $\vt_{x_1}(x_2)\df\theta(x_1-x_2)$ and $\lebM{\phi}{N}{}\df
N^{-1}\sum_{k=1}^N\phi(x_k)$. The statistic is given by 
\begin{equation}
   W_N^2 
   \;\df\; N\int_{\Real}\big|\lebM{\vt_x}{N}{}-F(x)\big|^2\,dF(x) \;\;,
\end{equation}
where we put the extra factor $N$ again, just as in the case of the
discrepancies. Because $F$ is a cumulative distribution function, its inverse
$F^{-1}:[0,1]\mapsto\Real$ is uniquely defined, and we can re-write the
statistic as
\begin{equation}
   W_N^2
   \;=\;   N\int_{\Kube}\big|\lebM{\vt_{F^{-1}(y)}}{N}{}-y\big|^2\,dy \;\;,
\end{equation}
where we denote $\Kube\df[0,1]$. But $\vt_{F^{-1}(y)}(x)=\vt_{y}(F(x))$, so
that $W_N^2$ is equal to the $L_2^*$-discrepancy of the points $F(x_k)$. The
interpretation of the statistic is slightly different from the
$L_2^*$-discrepancy, but the probability distribution is exactly the same.

\subsection{The Fourier diaphony\label{SecDefDia}}
For the Fourier diaphony, $\Prim$ should impose periodic boundary conditions.
In one dimension, a simple Hilbert-Schmidt operator that achieves this is given
by 
\begin{equation}
   (\Prim\phi)(x) \;\df\;\frac{\sqrt{3}}{\pi} \intk k_1(x-y)\phi(y)\,dy 
   \quad,\quad\quad k_1(x) \df 2\pi(\{x\}-\half) \;\;,
\end{equation}
where $\{x\}\df x\!\!\mod 1$. The term of a half in the integration kernel
assures that $\Prim\phi$ integrates to zero, so that the discrepancy is
formulated in the Landau gauge from the start. The choice of the factors seems
odd, but will appear to be the natural choice for the extension to more
dimensions.  The two-point function is given by 
\begin{equation}
   \twoB(x_1,x_2) \;=\; \inp{\Prit\de_{x_1}}{\Prit\de_{x_2}}
                  \;=\; 1-6\{x_1-x_2\}(1-\{x_1-x_2\}) \;\;.
\end{equation}
Notice that the two-point function only depends on $x_1-x_2$ and therefore is
translation invariant, i.e., 
$\twoB(x_1+a,x_2+a) \;=\; \twoB(x_1,x_2)$ for all $a\in[0,1]$.
As a result of this, all information
about $\twoB$ is contained in the function $\twoB_1:x\mapsto\twoB(x,0)$, and we 
have 
\begin{equation}
   \twoB(x_1,x_2) \;=\; \twoB_1(x_1-x_2) \;\;.
\end{equation}
The factor $\sqrt{3}/\pi$ was chosen such, that $\twoB_1(0)=1$.  The set
$\{u_n\}$ of solutions of the eigenvalue equation
$\intk\twoB(x,y)u(y)\,dy=\si^2u(x)$ is just the Fourier basis on $[0,1]$
without the constant mode: 
\begin{equation}
   u_{2n-1}(x)=\sqrt{2}\sin(2\pi nx) \;\;,\quad
   u_{2n}(x)  =\sqrt{2}\cos(2\pi nx) \;\;,\quad n=1,2,\ldots \;\;,
\end{equation}
with eigenvalues
\begin{equation}
   \si_{2n-1}^{2} = \si_{2n}^{2} \;=\; 3\,\pi^{-2}n^{-2}
   \;\;,\quad n=1,2,\ldots \;\;. 
\end{equation}
The function $u_0:x\mapsto 1$ is not a member of the basis because of the
Landau gauge. Only functions that integrate to zero are present. 

In $s>1$ dimensions, the operator $\Prim$ is extended as follows. Let $\ominus$ 
denote coordinate wise subtraction, then
\begin{equation}
   (\Prim\phi)(x) \;\df\; \frac{1}{\sqrt{(1+\pi^2/3)^s-1}}
                          \intk k_s(x\ominus y)\phi(y)\,dy \;\;,
\end{equation}
with 
\begin{equation}
   k_s(x) \;\df\; -1+\prod_{\nu=1}^s[1+k_1(x^\nu)] \;\;.
\end{equation}
The $s$-dimensional integration kernel is obtained from the one-dimensional one
by adding the constant mode and taking the product over the coordinates. The
extra term of $-1$ assures that the constant mode in $s$ dimensions disappears
again. The new factor assures that the $s$-dimensional two-point function is 
equal to one in the origin:  
\begin{equation}
  \twoB_s(x) 
  \;=\; \frac{1}{(1+\pi^2/3)^s-1}
        \left(-1+\prod_{\nu=1}^s\left[1+
	                 \frac{\pi^2}{3}\,\twoB_1(x^\nu)\right]\right) \;\;.
\end{equation}
The basis in $s$-dimensions consists of all products over coordinates of the
one-dimensional basis including the constant mode $u_0$. The only product that
does not appear is, of course, $\prod_{\nu=1}^su_{0}(x^\nu)$. The eigenvalue
coming with $u_0$ is determined by the choice of $k_s$, and equal to $1$. The
eigenvalues in $s$ dimensions are just the properly normalized products of the
one-dimensional ones. If we denote $\vn=(n_1,n_2,\ldots,n_s)$ and introduce
\begin{equation}
   k_\nu(\vn)\df\begin{cases}
           \half n_\nu    &\textrm{if $n_\nu$ is even,} \\
           \half(n_\nu+1) &\textrm{if $n_\nu$ is odd,} 
          \end{cases}
\end{equation}
then
\begin{equation}
   \si_{\vn}^{2} \;\df\; \si^{2}(\vec{k}(\vn)) 
   \;\df\; \frac{1}{(1+\pi^2/3)^s-1} \prod_{\nu=1}^s \frac{1}{r(k_\nu)^2} 
   \quad,\quad\quad
   r(k_\nu)=\begin{cases}
           k_\nu &\textrm{if $k_\nu\neq0$ ,}\\
           1 &\textrm{if $k_\nu=0$ .}
        \end{cases}
\label{defEq010}	
\end{equation}
The Fourier diaphony is often written in terms of the complex Fourier basis of 
$\Kube$. Then, it attains the form 
\begin{equation}
   D_N 
   \;=\; \frac{1}{N}\sum_{\vec{k}}\si^2(\vec{k}\,)
         \left|\sum_{l=1}^Ne^{2i\pi\vec{k}\cdot x_l}\right|^2  \;\;,
\end{equation}
where $\vec{k}\cdot x\df k_1x^1+k_2x^2+\cdots+k_sx^s$, and the first sum is
over all $\vec{k}\in\Zatu^s$ except the constant mode $\vec{k}=(0,0,\ldots,0)$.
Introduced as in this section, the diaphony is again $N$ times the square of 
the definition as given in, for example, \cite{Hellek98}. 

\subsection[The Lego discrepancy and the $\chi^2$-statistic]
{The Lego discrepancy and the $\chi^2$-statistic\label{SecDefLego}}
For the Lego discrepancy, the image $\Cfam\df\Prim\Hilb$ is a finite
dimensional vector space.  It is obtained by dividing $\Kube$ into $M$ disjoint
`bins' $\Aset_n$ with $\bigcup_{n=1}^M\Aset_n=\Kube$, and taking 
\begin{equation}
   (\Prim\phi)(x) 
   \;\df\; \sum_{n=1}^M\frac{\si_n}{\sqrt{w_n}}\,\vt_n(x)\inp{\vt_n}{\phi}\;\;,
\end{equation}
where
\begin{equation}
   \vt_n \df \vt_{\Aset_n} \quad,\quad\quad
   w_n \df \leb{\vt_n} \;\;,
\end{equation}
and, in first instance, the strengths $\si_n$ are not specified.
$\Prim$ maps $\Hilb$ onto the space of functions that are defined 
with a precision up to the size of the bins $\Aset_n$.
Notice that $\vt_n\vt_m=\de_{n,m}\vt_m$ where $\de_{n,m}$ is the Kronecker
delta symbol, and that $\sum_{n=1}^M\vt_n=1$, $\sum_{n=1}^Mw_n=1$. 
The two-point function is given by 
\begin{equation}
   \twoC(x_1,x_2)\;=\;\sum_{n=1}^M\si_n^2\vt_n(x_1)\vt_n(x_2) \;\;.
\end{equation}
Clearly, this model is dimension-independent, in the sense that the only
information on the dimension of $\Kube$ is that contained in the value of $M$:
if the dissection of $\Kube$ into bins is of the hyper-cubic type with $p$ bins
along each axis, then we shall have $M=p^s$. Also, a general area-preserving
mapping of $\Kube$ onto itself will leave the discrepancy invariant: it will
lead to a distortion (and possibly a dissection) of the various bins,
but this influences neither $w_n$ nor (by definition) $\si_n$.  Owing to the
finiteness of $M$, a finite point set can, in fact, have zero discrepancy in
this case, namely if every bin $\Aset_n$ contains precisely $w_nN$ points
(assuming this number to be integer for every $n$).

Because $\Cfam$ is $M$-dimensional, it is easiest to formulate everything in 
$\Real^M$. We define
\begin{equation}
   \phiv \;\df\; (\phi_1,\phi_2,\ldots,\phi_n) \quad,\quad\quad
   \phi_n = \frac{1}{w_n}\,\inp{\vt_n}{\phi} \;\;, 
\end{equation}
and divide $\Hilb$ into equivalence classes by the prescription that 
$\phi\sim\vhi$ if
$\phiv=\vhiv$. This space is $\Cfam$, and it is isomorphic to $\Real^M$
with inner product $\inp{\phiv}{\vhiv}\df\sum_{n=1}^Mw_n\phi_n\vhi_n$. 
The operator $\Prim$ restricted to $\Real^M$ is given by 
\begin{equation}
   (\Prim\phiv)_n = \frac{\si_n}{\sqrt{w_n}}\,\phi_n \;\;.
\end{equation}
Notice that $\Prim$ is self adjoint. The Gaussian measure $\mu$ can now be
defined rigorously in terms of a finite-dimensional path integral. If $F$ is 
a functional on $C$, then  
\begin{equation}
   \int_{\Cfam}F[\phi]\,d\mu[\phi]
   \;=\; \int[d\phiv]\,F\Big[\sum_{n=1}^M\phi_n\vt_n\Big]\,
         \exp(-S_0[\phiv])  \;\;,
\end{equation}
with
\begin{equation}
   [d\phiv] \;=\; \prod_{n=1}^M\frac{d\phi_n}{\sqrt{2\pi\si_n^2}}
   \quad\quad\textrm{and}\quad\quad
   S_0[\phiv]\;=\; \frac{1}{2}\sum_{n=1}^M\frac{1}{\si_n^2}\,\phi_n^2\;\;.
\end{equation}
The two-point function and the reduced two-point function can be written in terms
of matrices as
\begin{equation}
   \twoC(x_1,x_2) = \sum_{n,m=1}^M\twoC_{n,m}\vt_n(x_1)\vt_m(x_2) \;\;,\quad 
   \twoB(x_1,x_2) = \sum_{n,m=1}^M\twoB_{n,m}\vt_n(x_1)\vt_m(x_2) \;\;,
\end{equation}
with
\begin{equation}
   \twoC_{n,m} = \si_n^2\de_{n,m} \;\;,\quad
   \twoB_{n,m} = \sum_{k=1}^M(\si_n\de_{n,k}-\si_kw_k)(\si_m\de_{m,k}-\si_kw_k)
   \;\;.
\end{equation}
In the path integral formulation of the generating function,
$\leb{\,e^{g\hphi}\,}$ occurs, and the series expansion of $\exp$ and the
properties of the characteristic functions tell us that 
$\leb{e^{g\phi}}=\sum_{n=1}^Mw_ne^{g\phi_n}$, so that
the generating function is given by 
\begin{equation}
   G(z)
   \;=\; \int[d\phiv]\,
         \exp\left(- S_0[\phiv] - gN\sum_{n=1}^Mw_n\phi_n   
	           + N\log\Big(\sum_{n=1}^Mw_ne^{g\phi_n}\Big) \right)  \;\;.
\label{defEq009}	 
\end{equation}
The discrepancy itself can be written as
\begin{equation}
   D_N
   = \frac{1}{N}\sum_{k,l=1}^N\twoB(x_k,x_l)
   = \frac{1}{N}\sum_{n,m=1}^M\Smath_n\twoB_{n,m}\Smath_m \;\;,
   \quad\textrm{where}\quad
   \Smath_n \df \sum_{k=1}^N\vt_n(x_k)
\end{equation}
is the number of points in bin $\Aset_n$. 

\subsubsection{The $\chi^2$-statistic}
We did not yet specify the strengths $\si_n$, but we will in particular look at
the choice for which $\si_n^2w_n=1$ for all $n=1,2,\ldots,M$. In this case,
$\Cfam$ consists of functions in which the largest fluctuations appear over the
smallest intervals. Although not a priori attractive in many cases,
this choice is actually quite appropriate for, e.g. particle physics
where cross sections display precisely this kind of behavior. The reduced
two-point function attains the simple form $\twoB_{n,m} =
\frac{1}{w_n}\,\de_{n,m}-1$ and the discrepancy becomes
\begin{equation}
   D_N \;=\; \sum_{n=1}^M\frac{(\Smath_n-w_nN)^2}{w_nN} \;\;,
\end{equation}
which is nothing but the $\chi^2$-statistic for $N$ data points distributed
over $M$ bins with expected number of points $w_nN$ (cf.~\cite{Knuth}).

\section{Appendices}\vspace{-1.2\baselineskip}
\newcommand{\Hilt}{\tilde{\Hilb}}
\newcommand{\HP}{\Hilb^{}_{\Prim}}
\newcommand{\HPt}{\Hilb_{\Prit}}
\newcommand{\muP}{\mu}
\newcommand{\Noi}{\mathcal{N}}
\newcommand{\inmu}[2]{\langle #1,#2\rangle_{\muP}}
\Appendix{\label{App3A}}
\noindent
Let $\Hilb\df L_2(\Kube)$ be the Hilbert space of (equivalence classes of
almost everywhere equal) real quadratically integrable functions on $\Kube$,
with inner product and norm
\begin{equation}
   \langle f,g\rangle \;\df\; \int_\Kube f(x)g(x)\,dx \quad,\qquad
   \|f\|^{}_2 \;\df\; \sqrt{ \langle f,f\rangle}  \;\;.
\end{equation}
A Hilbert space is self-dual, i.e.~there is an isomorphism between $\Hilb$ and
its dual space $\Hilt$ of continuous linear functions $\Hilb\mapsto\Real$. It
induces an invertible mapping $\Hilt\ni\eta\mapsto f_\eta\in\Hilb$ such that
$\langle f_\eta,g\rangle=\eta[g]$ for all $g\in\Hilb$, and we write
$\|\eta\|^{}_2\df\|f_\eta\|^{}_2$. 

Let $\Prim$ be a Hilbert-Schmidt operator on $\Hilb$, and $\Prit$ its
transposed which acts on $\Hilt$ through the definition
$\Prit\Pspc\eta\df\eta\circ\Prim$. It is easy to see that
$\Prit$ is a Hilbert-Schmidt
operator on $\Hilt$ and that 
$\|\Prit\Pspc\eta\|^{}_2=\|\Prim^\dagger f_\eta\|_2$ 
exists for every
$\eta\in\Hilt$. Furthermore, it is well known (cf.~\cite{Choquet}) that there
exists a Gaussian measure $\muP$ on $\Hilb$ with Fourier transform
\begin{equation}
  \int_{\Hilb}\exp(i\eta[f])\,d\muP[f]
  \;=\; \exp(-\half\|\Prit\Pspc\eta\|^2_2)  \;\;.
\end{equation}
By inserting $\la\eta$ where $\la$ is a real variable, and differentiating the 
above equation twice with respect to $\la$ before putting it to zero, one 
obtains the relation
\begin{equation}
   \int_{\Hilb}\eta[f]^2\,d\muP[f] \;=\; \|\Prit\Pspc\eta\|^2_2 \;\;.
\end{equation}
With $\muP$, a Hilbert space $L_2(\Hilb,\muP)$ can be defined, where the norm
is given by 
\begin{equation}
   \|\eta\|_{\muP} 
   \;\df\; \left(\int_{\Hilb}\eta[f]^2\,d\muP[f]\right)^{1/2} \;\;.
\end{equation}
It is clear that a mapping $\Noi:\Hilt\mapsto L_2(\Hilb,\muP)$ can directly be 
defined on the whole of $\Hilt$, but we need more: we want to apply it to 
Dirac-measures, which $\Hilt$ does not contain. 
Consider therefore the Hilbert space $\HPt$, which is the completion of 
$\Hilt$ under the norm
\begin{equation}
   \|\eta\|_{\Prit} \;\df\; \|\Prit\Pspc\eta\|^{}_2 \;\;.
\end{equation} 
$\Prit$ can be interpreted as a continuous mapping from $\HPt$ to $\Hilt$, with
$\Prit\HPt=\Hilt$.
Furthermore, $\Noi$ can
be extended to the whole of $\HPt$ since it is an isometry: 
\begin{equation}
   \forall\,\eta\in\HPt:\;\;\;
   \|\Noi\eta\|^2_{\muP} 
   \;=\; \int_{\Hilb}(\Noi\eta)[f]^2\,d\muP[f]
   \;=\; \|\Prit\Pspc\eta\|^2_2 \;=\; \|\eta\|^2_{\Prit} \;\;.
\end{equation}
Now, suppose that $\Prim$ maps $\Hilb$ continuously onto a space $\Cfam$ of
continuous functions, such that 
\begin{equation}
   \|\Prim\Pspc f\|_\infty\df\sup_{x\in\Kube}|(\Prim\Pspc f)(x)|\leq p\|f\|^{}_2
   \quad \textrm{for some $p>0$.}
\label{ForApEq1}   
\end{equation}
The dual space $\tilde{\Cfam}$ of
$\Cfam$ consists of bounded measures on $\Kube$, and containes the
Dirac-measures.  Then we have for every $\eta\in\tilde{\Cfam}$ and every
$f\in\Hilb$:
\begin{equation}
   \big|(\Prit\Pspc\eta)[f]\big| \;=\;
   \big|\eta[\Prim\Pspc f]\big| 
   \;\leq\; \|\eta\|\!\cdot\!\|\Prim\Pspc f\|_\infty 
                           \;\leq\; p\|\eta\|\!\cdot\!\|f\|^{}_2 \;\;,
\end{equation}
so that $\Prit$ maps $\tilde{\Cfam}$ continuously onto $\Hilt$. Therefore,  
$\tilde{\Cfam}\subset\HPt$, and $\Noi$ can be applied to $\tilde\Cfam$.

\Appendix{\label{App3B}}
\noindent
For the proof that (\ref{ForApEq1}) holds in case of the Fourier diaphony, we 
use that there is obviously a number $p$ such that 
\begin{equation}
  \left|\frac{k_s(x\ominus y)}{\sqrt{(1+\pi^2/3)^s-1}}\right| \leq p 
  \quad\textrm{for all $x,y\in\Kube$,}
\end{equation}
so that 
\begin{equation}
   |(\Prim\Pspc f)(x)| \;\leq\; p\intk|f(y)|\,dy
   \quad\textrm{for all $x\in\Kube$.}
\end{equation}
Now, we can apply the Cauchy-Schwarz inequality, with the result that
for all $x\in\Kube$
\begin{equation}
   |(\Prim\Pspc f)(x)| 
   \;\leq\; p\left(\intk 1\,dy\right)^{1/2}\left(\intk|f(y)|^2\,dy\right)^{1/2}
		  \;=\; p\,\|f\|^{}_2  \;\;.
\end{equation}

\clearemptydoublepage

\chapter{Instantons for discrepancies\label{ChapInst}}

\thispagestyle{empty}

It is mentioned in \Sec{SecPath} that an expansion of the path
integral representation of the generating function of quadratic discrepancies
around the trivial solution of the field equation is only correct, if this
solution gives the minimum of the action. Furthermore, it is suggested that the
non-trivial solutions, called {\em instantons}, might spoil the perturbation
expansion if they exist for real values of the order parameter $z$ of the
generating function that are arbitrarily close to $0$. In this chapter, we
take a closer look at the issue for the Lego discrepancy and the
$L_2^*$-discrepancy in one dimension, and show that instantons exist but do 
not threaten the perturbative expansion.

For the the $L_2^*$-case, a method had to be developed to analyze the
singularity structures of the solutions of implicit function equations with 
numerical help of a computer which is presented in \Sec{CAARSS}.

\vspace{\baselineskip}

\minitoc

\section{An alternative derivation of the path integral formulation}
We start with an alternative derivation of the representation of the generating
function as a path integral. For the Lego discrepancy, this goes as follows.
We consider the case for which $\si_n^2w_n=1$ for all $n=1,\ldots,M$, so that
the discrepancy is just the $\chi^2$-statistic 
\begin{equation}
   D_N
   \;=\; \frac{1}{N}\sum_{n=1}^{M}\frac{\Smath_n^2}{w_n} - N \;\;,
   \quad\textrm{where}\quad \Smath_n\df\sum_{k=1}^{N}\vt_n(x_k)
\end{equation}
is the number of points $x_k$ in
bin $n$. If the points $x_k$ are truly randomly distributed, the variables
$\Smath_n$ are distributed according to a multinomial distribution, so that the
generating function is given by
\begin{equation}
   \Exp(\,e^{zD_N}\,)
   \;=\; \sum_{\{\Smath_n\}}\frac{N!}{\Smath_1!\cdots \Smath_M!}\,
         w_1^{\Smath_1}\cdots w_M^{\Smath_M}\,
	 \exp\left(\frac{z}{N}
	 \sum_{n=1}^{M}\frac{\Smath_n^2}{w_n} - zN\right) \;\;,
\label{FieEq033}	 
\end{equation}
where the summation is over all configurations $\{\Smath_n\}$ which satisfy 
$\sum_{n=1}^{M} \Smath_n=N$. Notice that 
$\Exp(\,e^{zD_N}\,)>w_n^N\exp(zN/w_n-zN)$ 
for every $n$, so that the generating function is not defined if $N\ra\infty$ 
for the values of $z$ with $\textrm{Re}~z>\frac{w_n}{w_n-1}\log w_n$. 
Using Gaussian integration rules and the generalized binomial 
theorem, it is easy to see that \eqn{FieEq033} can be written as 
\begin{equation}
   \Exp(\,e^{zD_N}\,)
   \;=\; e^{-zN}\left(\prod_{n=1}^{M}\frac{w_n}{2\pi}\right)^{\frac{1}{2}}
         \int_{\Real^M} \exp\left(-\frac{1}{2}\sum_{n=1}^{M} w_ny_n^2\right)
	      \left(\sum_{n=1}^{M} w_ne^{gy_n}\right)^N d^M\!y \;\;,        
\end{equation}
with $g=\sqrt{2z/N}$.  By writing the $N$-th power as a power of $e$ and
substituting $y_n=\phi_n+Ng$, the path integral of \eqn{defEq009} is obtained.

For the $L_2^*$-discrepancy in one dimension,
$\Prim^{-1}\phi=\phi'\df\frac{d\phi}{dx}$ with the boundary condition that
$\phi(1)=0$ (\Sec{SecDefL2}). The action is therefore given by 
\begin{equation}
   S[\phi] \;=\; \half\qnm{\phi'} + \half M\phi(1)^2 
                   - N\log\leb{\,e^{g\hphi}\,}\;\;,
\label{InstEq001}		   
\end{equation}
where $M\ra\infty$.
We show now that there is a na\"{\i}ve 
continuum limit with this result. We use the fact that the 
discrepancy can be defined as the na\"{\i}ve continuum limit of 
\begin{equation}
  D_N^{(M)} 
   \;=\; \frac{1}{N}\sum_{\rho=1}^M\si_\rho^2
         \left(\sum_{k=1}^{N}\sum_{n=1}^{M} 
	 K_n^\rho\left[\vt_n(x_k)-w_n\right]\right)^2 \;\;, 
	 \label{FieEq034}
\end{equation}
where
\begin{equation}
  \vt_n = \vt_{[\frac{n-1}{M},\frac{n}{M})} \quad,\quad
  w_n = \leb{\vt_n} \quad,\quad
  K_n^\rho = \theta(n-\rho) \quad,\quad\textrm{and}\quad
   \si^2_\rho = \frac{1}{M} \;\;.
\end{equation}
$D_N^{(M)}$ is the discretized version of the $L_2^*$-discrepancy, obtained
when in \eqn{defEq005} the average over a finite number of points $y_n$, 
$n=1,\ldots,M$ is taken, instead of the average over the whole of $\Kube$. 
Notice that a whole class of `discrete' discrepancies can be written as 
\eqn{FieEq034}, by choosing different expressions for the $K^\rho_n$ and the 
$\si^2_\rho$. Just like the Lego discrepancy, such a discrepancy can be written 
in terms of variables $\Smath_n$ that count the number of points $x_k$ in bin
$n$: 
\begin{equation}
  D_N^{(M)}
   \;=\; \frac{1}{N}\sum_{n,m=1}^{M} R_{nm}\Smath_n\Smath_m 
         - 2\sum_{n=1}^{M} T_n\Smath_n + NU \;\;,
\end{equation}
with
\begin{equation}
  R_{nm} = \sum_{\rho=1}^M \si^2_\rho K^\rho_n K^\rho_m \;\;,\quad
   T_n = \sum_{m=1}^M R_{nm}w_m \;\;,\quad\textrm{and}\quad
   U = \sum_{n,m=1}^{M} R_{nm}w_nw_m \;\;.
\end{equation}
In the case of the $L_2^*$-type discrepancy, the matrix $R$ is given by
$R_{nm}=\min(M-n,M-m)/M$. The generating function is again given as the
expectation value under the multinomial distribution. If we assume that the
matrix $R$ is invertible and positive definite, as it is for the $L_2^*$-type
discrepancy, use the Gaussian integration rules and the generalized binomial
theorem and do the appropriate coordinate transformations, we find 
\begin{equation}
  G(z) 
  \;=\; \sqrt{\frac{\det R^{-1}}{(2\pi)^M}}\int_{\Real^M}
        \exp\left(-S[\phi]\right)d^M\!\phi \;\;, 
\end{equation}
with 
\begin{equation}
  S[\phi] 
   \;=\;    \frac{1}{2}\sum_{n,m=1}^{M} R^{-1}_{nm}\phi_n\phi_m
          + Ng\sum_{n=1}^{M} w_n\phi_n
	  - N\log\Big(\sum_{n=1}^{M} w_ne^{g\phi_n}\Big) \;\;.
\label{FieEq035}	  
\end{equation}
For the $L_2^*$-type discrepancy the inverse $R^{-1}$ of the matrix $R$ is easy 
to find and we get 
\begin{equation}
   \sum_{n,m=1}^{M} R^{-1}_{nm}\phi_n\phi_m
   \;=\; M\phi_M^2 + M\sum_{n=1}^{M-1}(\phi_{n+1}-\phi_n)^2 \;\;,
\label{FieEq036}   
\end{equation}
so that a na\"{\i}ve continuum limit clearly produces \eqn{InstEq001}.

\section{Instantons for the Lego discrepancy\label{FieSec02}}
We start this section with a repetition of the statement that non-trivial
instanton solutions only exist if $z\in[0,\infty)$ (\Sec{SecPath}).  In
order to investigate the instantons, we analyze the action in terms of the
variables $y_n=g\phi_n+2z$, that is, we consider the integral
$\int_{\Real^M}\exp(-N\vSi[y])\,d^M\!y$\,, with 
\begin{equation}
   \vSi[y]
   \;=\; z + \frac{1}{4z}\sum w_ny_n^2 - \log\left(\sum w_ne^{y_n}\right) \;\;.
\end{equation}
The sum is over $n=1,\ldots,M$.
We are interested in the minima of $\vSi$. The `perturbative' minimum 
$\phi_n=0$, $n=1,\ldots,M$ corresponds to $y_n=2z$, $n=1,\ldots,M$, and general
extrema of $\vSi$ are situated at points 
$y$ which are solutions of the equations
\begin{equation}
   \frac{\partial\vSi}{\partial y_k}(y) = 0 
   \quad\Leftrightarrow\quad
   \frac{e^{y_k}}{y_k}
   \;=\; \frac{1}{2z}\sum w_ne^{y_n} \;\;,\quad k=1,\ldots,M \;\;.
\label{FieEq020}   
\end{equation}
If $z$ is positive, $e^{y_k}/y_k$, and therefore 
$y_k$, has to be positive for every $k$. 
The result is that the 
$y_k$ can take at most two values in one solution $y$ (\fig{FieFig01}). 
\begin{figure}[t]
\begin{center}
\begin{picture}(150,150)(0,0)
\LinAxis(0,0)(150,0)(1,1,3,0,1.5)
\LinAxis(0,0)(0,150)(1,1,-3,0,1.5)
\LinAxis(0,150)(150,150)(1,1,-3,0,1.5)
\LinAxis(150,0)(150,150)(1,1,3,0,1.5)
\Line(50,0)(50,150)
\Line(0,60)(150,60) 
\DashLine(50,90)(150,90){1.5}\Text(40,90)[]{$e^v$}
\DashLine(59,60)(59,90){1.5}\Text(60,52)[]{$y_-$}
\DashLine(103.8,60)(103.8,90){1.5}\Text(104.8,52)[]{$y_+$}
\Text(  0.0,-10)[]{$-2$}
\Text( 50.0,-10)[]{$0$}
\Text(150.0,-10)[]{$4$}
\Text(-12,  0.0)[]{$-8$}
\Text(-7, 60.0)[]{$0$}
\Text(-10,150)[]{$12$}
\Text(75,-20)[]{$t$}
\Text(-35,75)[]{$f(t)$}
\Text(80,25)[l]{$f(t)=\displaystyle\frac{e^t}{t}$}
\Curve{(  0.0, 59.5) (  1.5, 59.4) (  3.0, 59.4) (  4.5, 59.3) (  6.0, 59.3)
       (  7.5, 59.2) (  9.0, 59.1) ( 10.5, 59.0) ( 12.0, 58.9) ( 13.5, 58.8)
       ( 15.0, 58.7) ( 16.5, 58.5) ( 18.0, 58.4) ( 19.5, 58.2) ( 21.0, 58.0)
       ( 22.5, 57.7) ( 24.0, 57.5) ( 25.5, 57.1) ( 27.0, 56.8) ( 28.5, 56.3)
       ( 30.0, 55.8) ( 31.5, 55.2) ( 33.0, 54.4) ( 34.5, 53.5) ( 36.0, 52.3)
       ( 37.5, 50.9) ( 39.0, 49.0) ( 40.5, 46.5) ( 42.0, 43.0) ( 43.5, 37.8)
       ( 45.0, 29.3) ( 46.5, 13.4) (47,0)}
\Curve{( 52,150)
       ( 53.0,130.5) ( 54.5,109.9) ( 56.0, 99.7) ( 57.5, 93.7) ( 59.0, 89.9)
       ( 60.5, 87.2) ( 62.0, 85.3) ( 63.5, 83.8) ( 65.0, 82.8) ( 66.5, 82.0)
       ( 68.0, 81.4) ( 69.5, 81.0) ( 71.0, 80.7) ( 72.5, 80.5) ( 74.0, 80.4)
       ( 75.5, 80.4) ( 77.0, 80.4) ( 78.5, 80.6) ( 80.0, 80.8) ( 81.5, 81.0)
       ( 83.0, 81.3) ( 84.5, 81.6) ( 86.0, 82.0) ( 87.5, 82.4) ( 89.0, 82.9)
       ( 90.5, 83.4) ( 92.0, 84.0) ( 93.5, 84.6) ( 95.0, 85.2) ( 96.5, 85.9)
       ( 98.0, 86.6) ( 99.5, 87.4) (101.0, 88.3) (102.5, 89.2) (104.0, 90.1)
       (105.5, 91.1) (107.0, 92.2) (108.5, 93.3) (110.0, 94.4) (111.5, 95.7)
       (113.0, 97.0) (114.5, 98.4) (116.0, 99.8) (117.5,101.3) (119.0,102.9)
       (120.5,104.6) (122.0,106.4) (123.5,108.3) (125.0,110.2) (126.5,112.3)
       (128.0,114.4) (129.5,116.7) (131.0,119.1) (132.5,121.6) (134.0,124.3)
       (135.5,127.0) (137.0,130.0) (138.5,133.0) (140.0,136.2) (141.5,139.6)
       (143.0,143.2) (144.5,146.9) (145.6,150)}
\end{picture}    
\vspace{25pt}
\caption{$y_-$ and $y_+$.}
\label{FieFig01}
\end{center}
\end{figure}
If they all take the same value, this value is $2z$, and we get the
perturbative solution.  If they take two values, one of them, $y_+$, is larger
that $1$ and the other, $y_-$, is smaller than $1$.  With these results, and
the fact that \eqn{FieEq020} implies that 
\begin{equation}
   \sum w_ny_n 
   \;=\; 2z \;\;,
\label{FieEq021}   
\end{equation}
we see that there are no solutions but the perturbative one if 
$2z<w_{\rmin}$, where $w_{\rmin}=\min_nw_n$. 

In the next section, the other extremal points will be analyzed and it will
appear that minima occur with $\vSi[y]<0$. This means that, in the limit of
$N\ra\infty$, the integral of $\exp(-N\vSi)$ is not defined; there is a `wall'
in the complex $z$ plane along the positive real side of the imaginary axis, to
the right of which the generating function is not defined. That this is not an
artifact of our approach, can be seen in the expression of the generating
function given by \eqn{FieEq033}. It is shown there that the generating
function is not defined if $\textrm{Re}~z>\frac{w_n}{w_n-1}\log w_n$ for any
one of the $w_n$.

We know (\Sec{CorSec2}) that, at the perturbative level, the
generating function has a singularity at $z=\half$, but the instanton
contributions cannot correspond with it, because they will appear already for
$\textrm{Re}~z<\half$.  However, in order to calculate the probability density
$H$ with the Laplace transform using the perturbative expression of $G(z)$, we
can just calculate the contribution of the singularity at $z=\half$, for {\em
that is} the contribution to the perturbative expansion of $H(t)$.

\subsection{The wall}
To expose the nature of the extrema of $\vSi$, we have to investigate the 
eigenvalues $\la$ of the second derivative matrix $\vDe$ of $\vSi$ in the 
extremal points. This matrix is given by 
\begin{equation}
   \vDe_{k,l}(y)
   \;\df\; \frac{\partial^2\vSi}{\partial y_k\partial y_l}(y) 
   \;=\; a_k(y)\de_{k,l}+b_k(y)b_l(y)\;\;, 
\end{equation}
with
\begin{equation}
   a_k(y)=\frac{w_k}{2z}(1-y_k)\quad\textrm{and}\quad
   b_k(y)=\frac{w_ky_k}{2z}\;\;.
\end{equation}
To show that $\vSi$ becomes negative, we only use its minima, 
and these correspond with 
extremal points in which all eigenvalues of $\vDe$ are positive. According to 
\App{App4A}, 
we are therefore only interested in cases where the degeneracy of 
negative $a_k$ is one, for else $\la=a_k$ would be a solution. We further are 
only interested in cases where there is only one negative $a_k$, for if there 
where more, say $a_k$ and $a_{k+1}$ with $a_k<a_{k+1}$, then there would be 
a solution $a_k<\la<a_{k+1}<0$. So we see that the only extremal points we 
are interested in have all co-ordinates $y_k$ equal, or have one 
$y_k=y_+$ and the others equal to $y_-$. If they are all
equal, then they have to be equal to $2z$, and for the extremal point to be a 
minimum $2z$ has to be smaller than $1$. This is the perturbative minimum. 
Whether the other extremal points are minima depends on whether 
$\det\vDe$ is positive in these points. The determinant can be written as 
\begin{equation}
   \det\vDe(t)
   \;=\; \frac{\prod_n w_n}{(2z)^{M+1}}\,(1-y_-)^{M-1}(y_+-1)
         \left(\frac{w_+y_+}{y_+-1} + \frac{(1-w_+)y_-}{y_--1}\right) \;\;.
\end{equation}
Now we notice that all extremal points can be labeled with a parameter $v$ by 
defining
\begin{equation}
   \frac{e^{y_\pm(v)}}{y_\pm(v)}
   \;=\; e^v \quad\textrm{with}\quad v\in(1,\infty) \;\;.
\end{equation}
We see that $y_\pm$ is a continuous and differentiable function of $v$ and we 
have that $dy_\pm/dv=y_\pm/(y_\pm-1)$. This parameterization induces 
a parameterization of $2z$, and with the help of \eqn{FieEq021} we see that
\begin{equation}
   \frac{d(2z)}{dv}
   \;=\; \frac{w_+y_+}{y_+-1} + \frac{(1-w_+)y_-}{y_--1} \;\;. 
\end{equation}
So we see that the sign of 
$\det\vDe$ is the same as the sign of $d(2z)/dv$: if an extremal point is a 
minimum, then $d(2z)/dv>0$. 
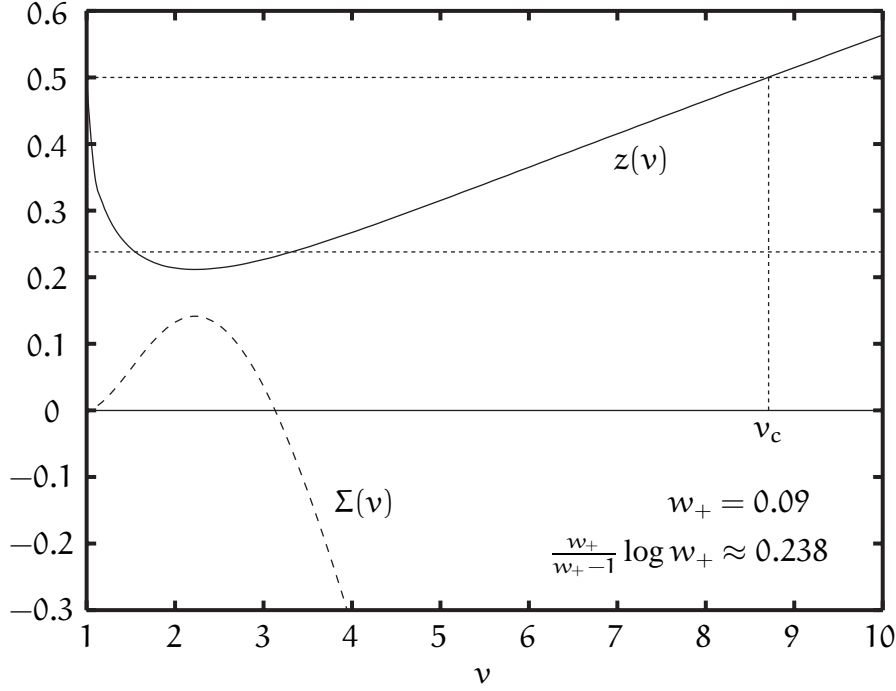
\begin{figure}[t]
\begin{center}
\begin{picture}(300,226)(0,0)
\LinAxis(0,0)(300,0)(9,1,3,0,1.5)
\LinAxis(0,0)(0,226)(9,1,-3,0,1.5)
\LinAxis(0,226)(300,226)(9,1,-3,0,1.5)
\LinAxis(300,0)(300,226)(9,1,3,0,1.5)
\Text(0,-10)[]{$1$}
\Text(33.3,-10)[]{$2$}
\Text(66.7,-10)[]{$3$}
\Text(100,-10)[]{$4$}
\Text(133.3,-10)[]{$5$}
\Text(166.7,-10)[]{$6$}
\Text(200,-10)[]{$7$}
\Text(233.3,-10)[]{$8$}
\Text(266.7,-10)[]{$9$}
\Text(300,-10)[]{$10$}
\Text(-17,0)[]{$-0.3$}
\Text(-17,25.1)[]{$-0.2$}
\Text(-17,50.2)[]{$-0.1$}
\Text(-13,75.3)[]{$0$}
\Text(-15,100.4)[]{$0.1$}
\Text(-15,125.6)[]{$0.2$}
\Text(-15,150.7)[]{$0.3$}
\Text(-15,175.8)[]{$0.4$}
\Text(-15,200.9)[]{$0.5$}
\Text(-15,226)[]{$0.6$}
\Text(210,170)[]{$z(v)$}
\Text(105,40)[]{$\vSi(v)$}
\Text(220,40)[l]{$w_+=0.09$}
\Text(175,20)[l]{$\frac{w_+}{w_+-1}\log w_+\approx 0.238$}
\Text(150,-25)[]{$v$}
\Line(0,75.3)(300,75.3)
\DashLine(0,135.1)(300,135.1){1.5}
\DashLine(0,200.9)(300,200.9){1.5}
\DashLine(257,75.3)(257,200.9){1.5}\Text(258,67.3)[]{$v_c$}
\Curve{(  0.0,200.9) (  3.0,164.5) (  6.0,153.5) (  9.0,146.6)
       ( 12.0,141.7) ( 15.0,138.1) ( 18.0,135.3) ( 21.0,133.3)
       ( 24.0,131.7) ( 27.0,130.5) ( 30.0,129.6) ( 33.0,129.1)
       ( 36.0,128.7) ( 39.0,128.5) ( 42.0,128.5) ( 45.0,128.6)
       ( 48.0,128.9) ( 51.0,129.2) ( 54.0,129.6) ( 57.0,130.1)
       ( 60.0,130.7) ( 63.0,131.4) ( 66.0,132.1) ( 69.0,132.8)
       ( 72.0,133.6) ( 75.0,134.5) ( 78.0,135.3) ( 81.0,136.2)
       ( 84.0,137.1) ( 87.0,138.1) ( 90.0,139.1) ( 93.0,140.1)
       ( 96.0,141.1) ( 99.0,142.1) (102.0,143.1) (105.0,144.2)
       (108.0,145.2) (111.0,146.3) (114.0,147.4) (117.0,148.5)
       (120.0,149.6) (123.0,150.7) (126.0,151.8) (129.0,152.9)
       (132.0,154.0) (135.0,155.1) (138.0,156.2) (141.0,157.4)
       (144.0,158.5) (147.0,159.6) (150.0,160.7) (153.0,161.9)
       (156.0,163.0) (159.0,164.1) (162.0,165.3) (165.0,166.4)
       (168.0,167.5) (171.0,168.7) (174.0,169.8) (177.0,171.0)
       (180.0,172.1) (183.0,173.2) (186.0,174.4) (189.0,175.5)
       (192.0,176.6) (195.0,177.8) (198.0,178.9) (201.0,180.0)
       (204.0,181.1) (207.0,182.3) (210.0,183.4) (213.0,184.5)
       (216.0,185.7) (219.0,186.8) (222.0,187.9) (225.0,189.0)
       (228.0,190.2) (231.0,191.3) (234.0,192.4) (237.0,193.5)
       (240.0,194.7) (243.0,195.8) (246.0,196.9) (249.0,198.0)
       (252.0,199.1) (255.0,200.2) (258.0,201.4) (261.0,202.5)
       (264.0,203.6) (267.0,204.7) (270.0,205.8) (273.0,206.9)
       (276.0,208.0) (279.0,209.2) (282.0,210.3) (285.0,211.4)
       (288.0,212.5) (291.0,213.6) (294.0,214.7) (297.0,215.8)
       (300.0,216.9) }
\DashCurve{(  0.0, 75.3) (  3.0, 76.4) (  6.0, 78.6) (  9.0, 81.6)
       ( 12.0, 85.0) ( 15.0, 88.8) ( 18.0, 92.7) ( 21.0, 96.6)
       ( 24.0,100.2) ( 27.0,103.4) ( 30.0,106.2) ( 33.0,108.4)
       ( 36.0,109.9) ( 39.0,110.7) ( 42.0,110.8) ( 45.0,110.1)
       ( 48.0,108.7) ( 51.0,106.6) ( 54.0,103.7) ( 57.0,100.2)
       ( 60.0, 95.9) ( 63.0, 91.1) ( 66.0, 85.7) ( 69.0, 79.7)
       ( 72.0, 73.2) ( 75.0, 66.2) ( 78.0, 58.8) ( 81.0, 51.0)
       ( 84.0, 42.9) ( 87.0, 34.4) ( 90.0, 25.6) ( 93.0, 16.5)
       ( 96.0,  7.1) ( 98.2, 0)}{3} 
\end{picture}    
\vspace{25pt}
\caption{$\vSi$ and $z$ for instanton solutions parameterized with $v$.}
\label{FieFig02}
\end{center}
\end{figure}
The minimal value that 
$v$ can take to represent a solution is $1$, which corresponds to 
$y_+=y_-=1$ and $2z=1$. 
It is easy to see that $d(2z)/dv\ra-\infty$ if $v\da 1$ and $w_+<\half$, where  
$w_+$ is the value of the weight belonging to the co-ordinate with the value 
$y_+$.
This means 
that if $v$ starts from $v=1$ and increases, then it will represent solutions 
with $d(2z)/dv<0$, which are local maxima. We know that, if $v\ra\infty$, then
$y_-\ra0$, $y_+\ra\infty$ and $2z=w_+y_++(1-w_+)y_-\ra\infty$, so that 
$d(2z)/dv$ has to become larger than $0$ at some point. The first point where 
$2z$ becomes equal to $1$ again we call $v_c$ (\fig{FieFig02}), so 
\begin{equation}
   z(v_c)=z(1)=\half \;\;.
\end{equation}
Also the function $\vSi$ itself can be written in terms of $z(v)$ in the 
extremal points. We use that
\begin{equation}
   \frac{d}{dv}[w_+y_+^2 + (1-w_+)y_-^2]
   \;=\; 4z + 4\frac{dz}{dv} \;\;
\end{equation}
and that $w_+y_+^2 + (1-w_+)y_-^2=1$ if $v=1$, so that 
\begin{equation}
   \vSi(v)
   \;=\; z(v) + \frac{1}{z(v)}\int_1^v z(x)\,dx + 1 - \frac{1}{4z(v)} - v
         - \log(\,2z(v)\,) \;\;.
\end{equation}
Now the problem arises. From the previous analysis of $z(v)$ we know that, 
if $1\leq v\leq v_c$, then $z(v)<\half$ so that 
\begin{equation}
   \vSi(v_c)
   \;=\; 1 - v_c + 2\int_1^{v_c}z(x)\,dx \;<\; 0 \;\;. 
\end{equation}
Furthermore, we find that 
\begin{equation}
  \frac{d\vSi}{dv} 
  = \left[1-\frac{1}{4z^2}\left(w_+y_+^2+(1-w_+)y_-^2\right)\right]\frac{dz}{dv}
  = -\frac{w_+(1-w_+)(y_+-y_-)^2}{4z^2}\,\frac{dz}{dv} \;\;,
\end{equation}
so that also $d\vSi/dv<0$ in $v_c$. So there clearly is a region in $[1,v_c]$ 
where $dz/dv>0$ and $\vSi(v)<0$. This means that in the region 
$\half w_{\rmin} <z<\half$ 
there are instanton solutions with negative action. The situation is shown in 
\fig{FieFig02} for $w_{\rmin}=0.09$. A region where $dz/dv>0$ and $S(v)<0$ is 
clearly visible in $[1,v_c]$.

\section{Instantons for the $L_2^*$-discrepancy\label{InstL2Sec}}
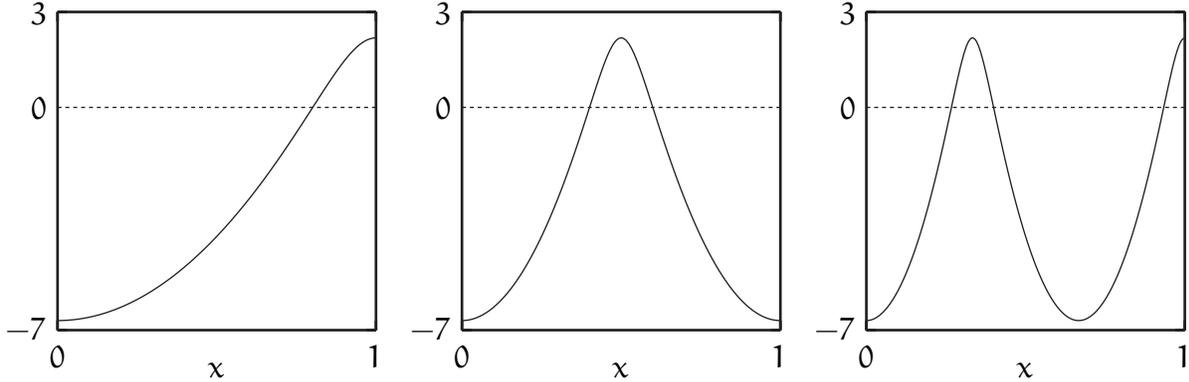
\begin{figure}
\begin{center}
\begin{picture}(120.0,120.0)(23,0)
\LinAxis(  0.0,  0.0)(120.0,  0.0)(1.0,1,3,0,1)
\LinAxis(  0.0,  0.0)(  0.0,120.0)(1,1,-3,0,1)
\LinAxis(  0.0,120.0)(120.0,120.0)(1.0,1,-3,0,1)
\LinAxis(120.0,  0.0)(120.0,120.0)(1,1,3,0,1)
\Text(  0.0,-10)[]{$  0$}
\Text(120.0,-10)[]{$  1$}
\Text(-12,  0.0)[]{$ -7$}
\Text(-7, 84.0)[]{$  0$}
\Text(-7,120.0)[]{$  3$}
\DashLine(0,84)(120,84){1.5}
\Text(60,-15)[]{$x$}
\Curve{(  0.00,  3.61) (  1.20,  3.63) (  2.40,  3.67) (  3.60,  3.73)
       (  4.80,  3.82) (  6.00,  3.93) (  7.20,  4.07) (  8.40,  4.24)
       (  9.60,  4.43) ( 10.80,  4.64) ( 12.00,  4.89) ( 13.20,  5.15)
       ( 14.40,  5.45) ( 15.60,  5.76) ( 16.80,  6.11) ( 18.00,  6.47)
       ( 19.20,  6.87) ( 20.40,  7.29) ( 21.60,  7.73) ( 22.80,  8.20)
       ( 24.00,  8.70) ( 25.20,  9.22) ( 26.40,  9.77) ( 27.60, 10.34)
       ( 28.80, 10.94) ( 30.00, 11.56) ( 31.20, 12.21) ( 32.40, 12.88)
       ( 33.60, 13.58) ( 34.80, 14.30) ( 36.00, 15.05) ( 37.20, 15.83)
       ( 38.40, 16.63) ( 39.60, 17.45) ( 40.80, 18.31) ( 42.00, 19.18)
       ( 43.20, 20.08) ( 44.40, 21.01) ( 45.60, 21.96) ( 46.80, 22.94)
       ( 48.00, 23.94) ( 49.20, 24.97) ( 50.40, 26.03) ( 51.60, 27.11)
       ( 52.80, 28.21) ( 54.00, 29.34) ( 55.20, 30.49) ( 56.40, 31.67)
       ( 57.60, 32.88) ( 58.80, 34.11) ( 60.00, 35.36) ( 61.20, 36.64)
       ( 62.40, 37.95) ( 63.60, 39.28) ( 64.80, 40.63) ( 66.00, 42.01)
       ( 67.20, 43.42) ( 68.40, 44.85) ( 69.60, 46.30) ( 70.80, 47.78)
       ( 72.00, 49.28) ( 73.20, 50.80) ( 74.40, 52.35) ( 75.60, 53.93)
       ( 76.80, 55.52) ( 78.00, 57.14) ( 79.20, 58.79) ( 80.40, 60.45)
       ( 81.60, 62.14) ( 82.80, 63.85) ( 84.00, 65.57) ( 85.20, 67.32)
       ( 86.40, 69.09) ( 87.60, 70.88) ( 88.80, 72.68) ( 90.00, 74.50)
       ( 91.20, 76.33) ( 92.40, 78.17) ( 93.60, 80.03) ( 94.80, 81.89)
       ( 96.00, 83.76) ( 97.20, 85.62) ( 98.40, 87.48) ( 99.60, 89.34)
       (100.80, 91.18) (102.00, 92.99) (103.20, 94.78) (104.40, 96.54)
       (105.60, 98.24) (106.80, 99.89) (108.00,101.47) (109.20,102.96)
       (110.40,104.35) (111.60,105.64) (112.80,106.79) (114.00,107.80)
       (115.20,108.65) (116.40,109.32) (117.60,109.81) (118.80,110.11)
       (120.00,110.21)}
\end{picture}
\begin{picture}(120.0,120.0)(-7,0)
\LinAxis(  0.0,  0.0)(120.0,  0.0)(1.0,1,3,0,1)
\LinAxis(  0.0,  0.0)(  0.0,120.0)(1,1,-3,0,1)
\LinAxis(  0.0,120.0)(120.0,120.0)(1.0,1,-3,0,1)
\LinAxis(120.0,  0.0)(120.0,120.0)(1,1,3,0,1)
\Text(  0.0,-10)[]{$  0$}
\Text(120.0,-10)[]{$  1$}
\Text(-12,  0.0)[]{$ -7$}
\Text(-7, 84.0)[]{$  0$}
\Text(-7,120.0)[]{$  3$}
\DashLine(0,84)(120,84){1.5}
\Text(60,-15)[]{$x$}
\Curve{(  0.00,  3.61) (  1.20,  3.67) (  2.40,  3.82) (  3.60,  4.07)
       (  4.80,  4.43) (  6.00,  4.89) (  7.20,  5.45) (  8.40,  6.11)
       (  9.60,  6.87) ( 10.80,  7.73) ( 12.00,  8.70) ( 13.20,  9.77)
       ( 14.40, 10.94) ( 15.60, 12.21) ( 16.80, 13.58) ( 18.00, 15.05)
       ( 19.20, 16.63) ( 20.40, 18.31) ( 21.60, 20.08) ( 22.80, 21.96)
       ( 24.00, 23.94) ( 25.20, 26.03) ( 26.40, 28.21) ( 27.60, 30.49)
       ( 28.80, 32.88) ( 30.00, 35.36) ( 31.20, 37.95) ( 32.40, 40.63)
       ( 33.60, 43.42) ( 34.80, 46.30) ( 36.00, 49.28) ( 37.20, 52.35)
       ( 38.40, 55.52) ( 39.60, 58.79) ( 40.80, 62.14) ( 42.00, 65.57)
       ( 43.20, 69.09) ( 44.40, 72.68) ( 45.60, 76.33) ( 46.80, 80.03)
       ( 48.00, 83.76) ( 49.20, 87.48) ( 50.40, 91.18) ( 51.60, 94.78)
       ( 52.80, 98.24) ( 54.00,101.47) ( 55.20,104.35) ( 56.40,106.79)
       ( 57.60,108.65) ( 58.80,109.81) ( 60.00,110.21) ( 61.20,109.81)
       ( 62.40,108.65) ( 63.60,106.79) ( 64.80,104.35) ( 66.00,101.47)
       ( 67.20, 98.24) ( 68.40, 94.78) ( 69.60, 91.18) ( 70.80, 87.48)
       ( 72.00, 83.76) ( 73.20, 80.03) ( 74.40, 76.33) ( 75.60, 72.68)
       ( 76.80, 69.09) ( 78.00, 65.57) ( 79.20, 62.14) ( 80.40, 58.79)
       ( 81.60, 55.52) ( 82.80, 52.35) ( 84.00, 49.28) ( 85.20, 46.30)
       ( 86.40, 43.42) ( 87.60, 40.63) ( 88.80, 37.95) ( 90.00, 35.36)
       ( 91.20, 32.88) ( 92.40, 30.49) ( 93.60, 28.21) ( 94.80, 26.03)
       ( 96.00, 23.94) ( 97.20, 21.96) ( 98.40, 20.08) ( 99.60, 18.31)
       (100.80, 16.63) (102.00, 15.05) (103.20, 13.58) (104.40, 12.21)
       (105.60, 10.94) (106.80,  9.77) (108.00,  8.70) (109.20,  7.73)
       (110.40,  6.87) (111.60,  6.11) (112.80,  5.45) (114.00,  4.89)
       (115.20,  4.43) (116.40,  4.07) (117.60,  3.82) (118.80,  3.67)
       (120.00,  3.61)}
\end{picture}
\begin{picture}(120.0,120.0)(-37,0)
\LinAxis(  0.0,  0.0)(120.0,  0.0)(1.0,1,3,0,1)
\LinAxis(  0.0,  0.0)(  0.0,120.0)(1,1,-3,0,1)
\LinAxis(  0.0,120.0)(120.0,120.0)(1.0,1,-3,0,1)
\LinAxis(120.0,  0.0)(120.0,120.0)(1,1,3,0,1)
\Text(  0.0,-10)[]{$  0$}
\Text(120.0,-10)[]{$  1$}
\Text(-12,  0.0)[]{$ -7$}
\Text(-7, 84.0)[]{$  0$}
\Text(-7,120.0)[]{$  3$}
\DashLine(0,84)(120,84){1.5}
\Text(60,-15)[]{$x$}
\Curve{(  0.00,  3.61) (  1.20,  3.73) (  2.40,  4.07) (  3.60,  4.64)
       (  4.80,  5.45) (  6.00,  6.47) (  7.20,  7.73) (  8.40,  9.22)
       (  9.60, 10.94) ( 10.80, 12.88) ( 12.00, 15.05) ( 13.20, 17.45)
       ( 14.40, 20.08) ( 15.60, 22.94) ( 16.80, 26.03) ( 18.00, 29.34)
       ( 19.20, 32.88) ( 20.40, 36.64) ( 21.60, 40.63) ( 22.80, 44.85)
       ( 24.00, 49.28) ( 25.20, 53.93) ( 26.40, 58.79) ( 27.60, 63.85)
       ( 28.80, 69.09) ( 30.00, 74.50) ( 31.20, 80.03) ( 32.40, 85.62)
       ( 33.60, 91.18) ( 34.80, 96.54) ( 36.00,101.47) ( 37.20,105.64)
       ( 38.40,108.65) ( 39.60,110.11) ( 41.20,109.32) ( 42.40,106.79)
       ( 43.60,102.96) ( 44.80, 98.24) ( 46.00, 92.99) ( 47.20, 87.48)
       ( 48.40, 81.89) ( 49.60, 76.33) ( 50.80, 70.88) ( 52.00, 65.57)
       ( 53.20, 60.45) ( 54.40, 55.52) ( 55.60, 50.80) ( 56.80, 46.30)
       ( 58.00, 42.01) ( 59.20, 37.95) ( 60.40, 34.11) ( 61.60, 30.49)
       ( 62.80, 27.11) ( 64.00, 23.94) ( 65.20, 21.01) ( 66.40, 18.31)
       ( 67.60, 15.83) ( 68.80, 13.58) ( 70.00, 11.56) ( 71.20,  9.77)
       ( 72.40,  8.20) ( 73.60,  6.87) ( 74.80,  5.76) ( 76.00,  4.89)
       ( 77.20,  4.24) ( 78.40,  3.82) ( 79.60,  3.63) ( 81.20,  3.73)
       ( 82.40,  4.07) ( 83.60,  4.64) ( 84.80,  5.45) ( 86.00,  6.47)
       ( 87.20,  7.73) ( 88.40,  9.22) ( 89.60, 10.94) ( 90.80, 12.88)
       ( 92.00, 15.05) ( 93.20, 17.45) ( 94.40, 20.08) ( 95.60, 22.94)
       ( 96.80, 26.03) ( 98.00, 29.34) ( 99.20, 32.88) (100.40, 36.64)
       (101.60, 40.63) (102.80, 44.85) (104.00, 49.28) (105.20, 53.93)
       (106.40, 58.79) (107.60, 63.85) (108.80, 69.09) (110.00, 74.50)
       (111.20, 80.03) (112.40, 85.62) (113.60, 91.18) (114.80, 96.54)
       (116.00,101.47) (117.20,105.64) (118.40,108.65) (119.60,110.11)}
\end{picture}
\vspace{20pt}
\caption{Instanton solutions $\phi_k(x)$ with $E=5.7$ and number of 
         bending points $k=1,2,3$.}
\label{FieFig03}	 
\end{center}
\end{figure}

In order to investigate the instantons for the $L_2^*$-discrepancy in 
one dimension, we analyze 
$\vSi[\phi]\df S[\phi/g]/N$, with $S$ as \eqn{InstEq001}, 
because this new action does not depend on $N$: 
\begin{equation}
   \vSi[\phi]
   \;=\; \frac{1}{4z}\qnm{\phi'} + \frac{1}{4z}M\phi(1)^2
                                 - \log\leb{\,e^{\hphi}\,} \;\;,
\end{equation}
where $2z=Ng^2$ and $M\ra\infty$.
Extremal points of this action are solutions of the field equation 
\begin{equation}
   -\frac{1}{2z}\,\phi''(x) + 1 
   - \frac{e^{\phi(x)}}{\leb{\,e^\phi\,}}
   \;=\; 0 \;\;
\end{equation}
that also satisfy the boundary conditions, which are
$\phi(0)=\phi'(1)=0$ at this point. We proceed however by applying the gauge
transformation $\Tran:\phi\mapsto\phi-\log\leb{\,e^{\phi}\,}$, so that
$\leb{\,e^{\Tran\phi}\,}=1$ and, in this gauge, the equation becomes
\begin{equation}
   -\frac{1}{2z}\,\phi''(x) + 1 - e^{\phi(x)} \;=\; 0 
   \;\;,\quad\textrm{with}\;\;
   \phi'(1)=0\;\;\textrm{and}\;\;\leb{\,e^{\phi}\,}=1\;\;.
\end{equation}
Integration over $\Kube$ of this equation leads to the identity $\phi'(0)=0$. 
The problem is now reduced to that of the motion of a classical 
particle with a mass $1/\sqrt{4z}$\, in 
a potential 
\begin{equation}
\raisebox{-55pt}{
\begin{picture}(125,115)(0,-13)
\LinAxis(  0.0,  0.0)(125.0,  0.0)(1.0,1,3,0,1.5)
\LinAxis(  0.0,  0.0)(  0.0,100.0)(1.0,1,-3,0,1.5)
\LinAxis(  0.0,100.0)(125.0,100.0)(1.0,1,-3,0,1.5)
\LinAxis(125.0,  0.0)(125.0,100.0)(1.0,1,3,0,1.5)
\Text(  0.0,-10)[]{ -1.5}
\Text(75,-10)[]{0}
\Text(125.0,-10)[]{  1.0}
\Text(-12,  2)[]{  0.0}
\Text(-12,100.0)[]{  0.7}
\Curve{(  4.2, 91.2) (  5.0, 89.4) (  5.8, 87.7) (  6.7, 86.0) (  7.5, 84.3)
       (  8.3, 82.5) (  9.2, 80.9) ( 10.0, 79.2) ( 10.8, 77.5) ( 11.7, 75.8)
       ( 12.5, 74.2) ( 13.3, 72.6) ( 14.2, 70.9) ( 15.0, 69.3) ( 15.8, 67.7)
       ( 16.7, 66.1) ( 17.5, 64.5) ( 18.3, 63.0) ( 19.2, 61.4) ( 20.0, 59.9)
       ( 20.8, 58.3) ( 21.7, 56.8) ( 22.5, 55.3) ( 23.3, 53.8) ( 24.2, 52.3)
       ( 25.0, 50.9) ( 25.8, 49.4) ( 26.7, 48.0) ( 27.5, 46.6) ( 28.3, 45.2)
       ( 29.2, 43.8) ( 30.0, 42.4) ( 30.8, 41.0) ( 31.7, 39.7) ( 32.5, 38.4)
       ( 33.3, 37.1) ( 34.2, 35.8) ( 35.0, 34.5) ( 35.8, 33.2) ( 36.7, 32.0)
       ( 37.5, 30.8) ( 38.3, 29.5) ( 39.2, 28.4) ( 40.0, 27.2) ( 40.8, 26.0)
       ( 41.7, 24.9) ( 42.5, 23.8) ( 43.3, 22.7) ( 44.2, 21.6) ( 45.0, 20.6)
       ( 45.8, 19.5) ( 46.7, 18.5) ( 47.5, 17.6) ( 48.3, 16.6) ( 49.2, 15.7)
       ( 50.0, 14.7) ( 50.8, 13.8) ( 51.7, 13.0) ( 52.5, 12.1) ( 53.3, 11.3)
       ( 54.2, 10.5) ( 55.0,  9.7) ( 55.8,  9.0) ( 56.7,  8.3) ( 57.5,  7.6)
       ( 58.3,  6.9) ( 59.2,  6.3) ( 60.0,  5.6) ( 60.8,  5.1) ( 61.7,  4.5)
       ( 62.5,  4.0) ( 63.3,  3.5) ( 64.2,  3.0) ( 65.0,  2.6) ( 65.8,  2.2)
       ( 66.7,  1.8) ( 67.5,  1.5) ( 68.3,  1.2) ( 69.2,  0.9) ( 70.0,  0.7)
       ( 70.8,  0.5) ( 71.7,  0.3) ( 72.5,  0.2) ( 73.3,  0.1) ( 74.2,  0.0)
       ( 75.0,  0.0) ( 75.8,  0.0) ( 76.7,  0.1) ( 77.5,  0.2) ( 78.3,  0.3)
       ( 79.2,  0.5) ( 80.0,  0.7) ( 80.8,  1.0) ( 81.7,  1.3) ( 82.5,  1.6)
       ( 83.3,  2.0) ( 84.2,  2.5) ( 85.0,  3.0) ( 85.8,  3.5) ( 86.7,  4.1)
       ( 87.5,  4.7) ( 88.3,  5.4) ( 89.2,  6.1) ( 90.0,  6.9) ( 90.8,  7.7)
       ( 91.7,  8.6) ( 92.5,  9.6) ( 93.3, 10.5) ( 94.2, 11.6) ( 95.0, 12.7)
       ( 95.8, 13.9) ( 96.7, 15.1) ( 97.5, 16.4) ( 98.3, 17.7) ( 99.2, 19.1)
       (100.0, 20.6) (100.8, 22.1) (101.7, 23.7) (102.5, 25.3) (103.3, 27.1)
       (104.2, 28.9) (105.0, 30.7) (105.8, 32.6) (106.7, 34.6) (107.5, 36.7)
       (108.3, 38.9) (109.2, 41.1) (110.0, 43.4) (110.8, 45.8) (111.7, 48.2)
       (112.5, 50.8) (113.3, 53.4) (114.2, 56.1) (115.0, 58.8) (115.8, 61.7)
       (116.7, 64.7) (117.5, 67.7) (118.3, 70.8) (119.2, 74.1) (120.0, 77.4)
       (120.8, 80.8) (121.7, 84.3) (122.5, 87.9) (123.3, 91.6)}
\Line(75,0)(75,100)
\DashLine(0,30)(125,30){3}
\Text(-10,30)[]{$E$}
\Text(32,77)[]{$U(\phi)$}
\end{picture}
}
\hspace{100pt}
   U(\phi) \;=\; e^\phi-\phi-1\;\;, 
\hspace{53pt}
\end{equation}
and the solution can be written implicitly as 
\begin{equation}
   \sqrt{4z}\,\frac{dx}{d\phi} \;=\; \frac{1}{\sqrt{E-U(\phi)}} \;\;,
\label{FieEq025}   
\end{equation}
where the integration constant $E$, the {\em energy}, has to be larger than 
zero for solutions to exist.
It is easy to see that the solutions are 
oscillatory and that, if $\phi(x)$ is a solution with one bending point, then 
also 
\begin{equation}
   \phi_k(x) \;=\; 
   \begin{cases}
      \phi(kx-p)   & \frac{p}{k}\leq x \leq\frac{p+1}{k}\;\;\textrm{$p$ even}\\
      \phi(1+p-kx) & \frac{p}{k}\leq x \leq\frac{p+1}{k}\;\;\textrm{$p$ odd}
   \end{cases}
   \;,\quad p = 0,1,\ldots,k-1\;\;,
\end{equation}
is a solution for $k=2,3,\ldots$.
These new solutions have the same energy, but a larger number of bending 
points, namely $k$, and the value of $z$ increases by a factor $k^2$. 
Hence, we can classify the solutions 
according to the energy and the number bending points.  
This classification in terms of the number of bending points is quite natural 
and this can best be understood by looking at the limit of $N\ra\infty$. 
Then, the equation becomes 
\begin{equation}
   -\phi''(x)-2z\hphi(x) \;=\; 0 \;\;,
\end{equation}
with $\phi(0)=\phi'(1)=0$, and the solutions are given by 
\begin{equation}
   \phi_k(x)=\textstyle{\sqrt{\frac{2}{3}}}\,[1-\cos(k\pi x)]\;\;,\;\;
   2z=k^2\pi^2\;\;,\;\;
   k=1,2,\ldots \;\;,
\end{equation}
so that the instantons are completely classified with the number of bending 
points $k$. 
If $N$ becomes finite, these solutions are deformed but keep the same value 
of $k$ (\fig{FieFig03}). For given $k$ there are infinitely many solutions 
classified by $E$. 

\subsection{Existence of instantons}
We now concentrate on the instantons with one bending point, because 
the numerical value of the action is independent of the number of bending 
points. 
Those instantons are completely characterized by their energy. The values of 
$z$ for which these instantons exist are defined as a function of $E$ by 
\eqn{FieEq025}, which states that 
\begin{equation}
   T(E)
   \;\df\; \sqrt{4z}
   \;=\; \int\limits_{\phi_-}^{\phi_+}\frac{d\phi}{\sqrt{E-U(\phi)}} \;\;,
\label{FieEq026}   
\end{equation}
where $\phi_-$ and $\phi_+$ are the classical turning points. 
They are solutions of $U(\phi_\pm)=E$ with $\phi_-<0<\phi_+$.
In classical mechanics, $T(E)$ is proportional to the period of 
a particle in the potential $U$ (cf.~\cite{LandL}).

\begin{figure}[t]
\begin{center}
\begin{picture}(300,226)(-15,0)
\LinAxis(0,0)(300,0)(6,1,3,0,1.5)
\LinAxis(0,0)(0,226)(5.5,1,-3,0,1.5)
\LinAxis(0,226)(300,226)(6,1,-3,0,1.5)
\LinAxis(300,0)(300,226)(5.5,1,3,0,1.5)
\Text(118,195)[r]{numerical integration}\Line(123,195)(148,195)
\Text(118,180)[r]{expansion}\DashLine(123,180)(148,180){3}
\Text(118,165)[r]{large $E$}\DashLine(123,165)(148,165){1.5}
\Text(0,-10)[]{$0$}\Text(50,-10)[]{$2$}\Text(100,-10)[]{$4$}
\Text(150,-10)[]{$6$}\Text(200,-10)[]{$8$}\Text(250,-10)[]{$10$}
\Text(300,-10)[]{$12$}\Text(-10,0)[]{$4$}\Text(-10,41.1)[]{$5$}
\Text(-10,82.2)[]{$6$}\Text(-10,123.3)[]{$7$}
\Text(-10,164.4)[]{$8$}\Text(-10,205.5)[]{$9$}
\Text(150,-25)[]{$E$}\Text(-35,123.3)[]{$T(E)$}
\DashLine(157.1,0)(157.1,226){6}
\Curve{(  0.2, 18.4) (  3.2, 20.2) (  6.2, 22.0) (  9.2, 23.9) ( 12.2, 25.7)
       ( 15.2, 27.6) ( 18.2, 29.4) ( 21.2, 31.3) ( 24.2, 33.2) ( 27.2, 35.0)
       ( 30.2, 36.9) ( 33.2, 38.8) ( 36.2, 40.7) ( 39.2, 42.5) ( 42.2, 44.4)
       ( 45.2, 46.3) ( 48.2, 48.1) ( 51.2, 50.0) ( 54.2, 51.9) ( 57.2, 53.7)
       ( 60.2, 55.6) ( 63.2, 57.4) ( 66.2, 59.2) ( 69.2, 61.1) ( 72.2, 62.9)
       ( 75.2, 64.7) ( 78.2, 66.5) ( 81.2, 68.3) ( 84.2, 70.1) ( 87.2, 71.9)
       ( 90.2, 73.7) ( 93.2, 75.5) ( 96.2, 77.2) ( 99.2, 79.0) (102.2, 80.7)
       (105.2, 82.5) (108.2, 84.2) (111.2, 85.9) (114.2, 87.6) (117.2, 89.3)
       (120.2, 91.0) (123.1, 92.7) (126.1, 94.3) (129.1, 96.0) (132.1, 97.7)
       (135.1, 99.3) (138.1,100.9) (141.1,102.6) (144.1,104.2) (147.1,105.8)
       (150.1,107.4) (153.1,109.0) (156.1,110.5) (159.1,112.1) (162.1,113.7)
       (165.1,115.2) (168.1,116.8) (171.1,118.3) (174.1,119.8) (177.1,121.3)
       (180.1,122.8) (183.1,124.3) (186.1,125.8) (189.1,127.3) (192.1,128.8)
       (195.1,130.3) (198.1,131.7) (201.1,133.2) (204.1,134.6) (207.1,136.1)
       (210.1,137.5) (213.1,138.9) (216.1,140.4) (219.1,141.8) (222.1,143.2)
       (225.1,144.6) (228.1,146.0) (231.1,147.3) (234.1,148.7) (237.1,150.1)
       (240.1,151.5) (243.0,152.8) (246.0,154.2) (249.0,155.5) (252.0,156.8)
       (255.0,158.2) (258.0,159.5) (261.0,160.8) (264.0,162.1) (267.0,163.5)
       (270.0,164.8) (273.0,166.1) (276.0,167.4) (279.0,168.6) (282.0,169.9)
       (285.0,171.2) (288.0,172.5) (291.0,173.7) (294.0,175.0) (297.0,176.3)
       (300.0,177.5) }
\DashCurve{(  0.2, 18.4) (  3.2, 20.2) (  6.2, 22.0) (  9.2, 23.9) ( 12.2, 25.7)
       ( 15.2, 27.6) ( 18.2, 29.4) ( 21.2, 31.3) ( 24.2, 33.2) ( 27.2, 35.0)
       ( 30.2, 36.9) ( 33.2, 38.8) ( 36.2, 40.7) ( 39.2, 42.5) ( 42.2, 44.4)
       ( 45.2, 46.3) ( 48.2, 48.1) ( 51.2, 50.0) ( 54.2, 51.9) ( 57.2, 53.7)
       ( 60.2, 55.6) ( 63.2, 57.4) ( 66.2, 59.2) ( 69.2, 61.1) ( 72.2, 62.9)
       ( 75.2, 64.7) ( 78.2, 66.5) ( 81.2, 68.3) ( 84.2, 70.1) ( 87.2, 71.9)
       ( 90.2, 73.7) ( 93.2, 75.5) ( 96.2, 77.2) ( 99.2, 79.0) (102.2, 80.7)
       (105.2, 82.5) (108.2, 84.2) (111.2, 85.9) (114.2, 87.6) (117.2, 89.3)
       (120.2, 91.0) (123.1, 92.7) (126.1, 94.3) (129.1, 96.0) (132.1, 97.7)
       (135.1, 99.3) (138.1,100.9) (141.1,102.6) (144.1,104.2) (147.1,105.8)
       (150.1,107.4) (153.1,109.1) (156.1,111.0) (159.1,113.2) (162.1,116.4)
       (165.1,122.2) (168.1,133.8) (171.1,159.6) (174.1,218.2) }{3}
\DashCurve{( 33.2,  0.0) ( 36.2,  7.1) ( 39.2, 13.0) ( 42.2, 18.2) ( 45.2, 22.8)
       ( 48.2, 26.9) ( 51.2, 30.7) ( 54.2, 34.2) ( 57.2, 37.6) ( 60.2, 40.7)
       ( 63.2, 43.7) ( 66.2, 46.6) ( 69.2, 49.3) ( 72.2, 52.0) ( 75.2, 54.5)
       ( 78.2, 57.0) ( 81.2, 59.5) ( 84.2, 61.8) ( 87.2, 64.1) ( 90.2, 66.4)
       ( 93.2, 68.6) ( 96.2, 70.8) ( 99.2, 72.9) (102.2, 75.0) (105.2, 77.1)
       (108.2, 79.1) (111.2, 81.1) (114.2, 83.1) (117.2, 85.0) (120.2, 86.9)
       (123.1, 88.8) (126.1, 90.7) (129.1, 92.6) (132.1, 94.4) (135.1, 96.2)
       (138.1, 98.0) (141.1, 99.8) (144.1,101.5) (147.1,103.2) (150.1,105.0)
       (153.1,106.7) (156.1,108.3) (159.1,110.0) (162.1,111.7) (165.1,113.3)
       (168.1,115.0) (171.1,116.6) (174.1,118.2) (177.1,119.8) (180.1,121.4)
       (183.1,122.9) (186.1,124.5) (189.1,126.0) (192.1,127.6) (195.1,129.1)
       (198.1,130.6) (201.1,132.1) (204.1,133.6) (207.1,135.1) (210.1,136.6)
       (213.1,138.0) (216.1,139.5) (219.1,141.0) (222.1,142.4) (225.1,143.8)
       (228.1,145.3) (231.1,146.7) (234.1,148.1) (237.1,149.5) (240.1,150.9)
       (243.0,152.3) (246.0,153.7) (249.0,155.0) (252.0,156.4) (255.0,157.7)
       (258.0,159.1) (261.0,160.4) (264.0,161.8) (267.0,163.1) (270.0,164.4)
       (273.0,165.8) (276.0,167.1) (279.0,168.4) (282.0,169.7) (285.0,171.0)
       (288.0,172.3) (291.0,173.5) (294.0,174.8) (297.0,176.1) (300.0,177.4) 
       }{1.5}
\end{picture} 
\vspace{25pt}
\caption{$T(E)$ computed by numerical integration, as an expansion around $E=0$
         and as an approximation for large $E$. The expansion is up to and 
	 including $\Ord(E^{49})$.}
\label{FieFig04}
\end{center}
\end{figure}
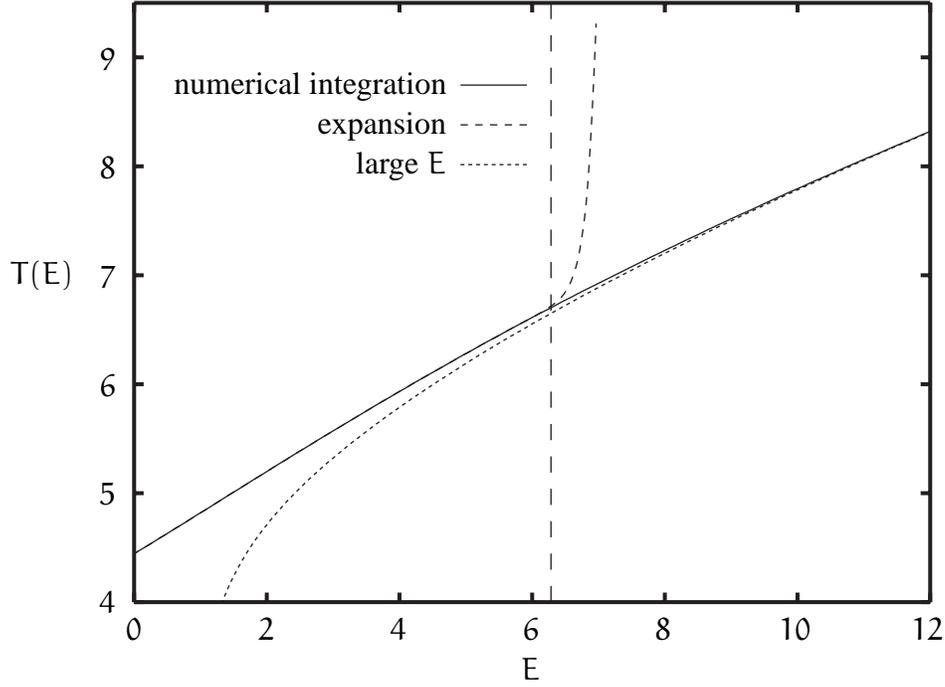

The function $T$ cannot be expressed in terms of elementary functions, but a 
number of its properties can be derived, as we shall now discuss.
For small $E$, a quadratic approximation of the potential can be made with 
$\phi_\pm=\pm\sqrt{2E}$ with the result that 
\begin{equation}
   \lim_{E\da0}T(E)
   \;=\; \pi\sqrt{2}  \quad\Longrightarrow\quad
   \lim_{E\da0}z(E)
   \;=\; \half\pi^2  \;\;.
\end{equation}
The question is now whether $z$ is 
increasing as a function of $E$. 
To calculate $T(E)$ for large $E$, $U(\phi)$ can be approximated by 
$-1-\phi$ for $\phi<0$ and by $e^\phi$ for $\phi>0$, so that 
\begin{equation}
   T(E\ra\infty)
   \;\approx\; 2\sqrt{E+1} 
         + \frac{2}{\sqrt{E}}\,\log\left(\sqrt{E}+\sqrt{E-1}\right) \;\;,
\label{FieEq027}	 
\end{equation}
so $T(E)$ is clearly increasing for large $E$.
To analyze $T(E)$ for small $E$, we make an 
expansion in powers of $E$. Therefore, we write 
\begin{equation}
   T(E)
   \;=\; \int\limits_0^{\sqrt{2E}}\left(E-\half v^2\right)^{-1/2}
         \frac{d}{dv}\left[f(v)-f(-v)\right]\,dv \;\;,
\label{FieEq028}	 
\end{equation}
where $f$ is a continuous solution of the implicit equation 
\begin{equation}
   e^{f(v)} - f(v) - 1 = \half v^2 \;\;,
\label{ImplFunc}   
\end{equation}
with $f(v)\sim v$ for small $v$. In \Sec{CAARSS} it is shown that it is given by 
the function values on the principal Riemann sheet of the general continuous 
solution and that is has an expansion $f(v)=\sum_{n=0}^\infty\al_nv^n$ 
with the coefficients $\al_n$ given by 
\begin{equation}
   \al_1=1\quad\textrm{and}\quad
   \al_n =-\frac{1}{n+1}\left[
          \half(n-1)\al_{n-1}+\sum_{k=2}^{n-1}k\al_k\al_{n+1-k}\right]\;\;
   \textrm{for}\;\; n>1 \;\;,
\end{equation}
and with the radius of convergence equal to $\sqrt{4\pi}$. If we substitute 
the power series into \eqn{FieEq028} and integrate term by term, we obtain 
the following power series for $|E|<2\pi$:
\begin{equation}
   T(E)
   \;=\; \sum_{n=1,~\textrm{$n$ odd}}^\infty
         \frac{\Gamma(\half)\Gamma(\frac{n}{2})}{\Gamma(\frac{n+1}{2})}\,
	 n\al_n\, 2^\frac{n}{2} E^\frac{n-1}{2} \;\;.
\end{equation}
The first few terms in this expansion are 
\begin{equation}
   T(E) 
   \;=\; \pi\sqrt{2}\left[1 + \frac{E}{12}
                            +\frac{1}{4}\left(\frac{E}{12}\right)^2
			    -\frac{139}{180}\left(\frac{E}{12}\right)^3
			    -\frac{571}{2880}\left(\frac{E}{12}\right)^4
			    +\Ord(E^5)\right] \;\;.
\end{equation}
The asymptotic behavior of the coefficients $\al_n$ will be determined in 
\Sec{CAARSS}, with the result that, for large and integer $k$,  
\begin{equation}
  \al_n \;\sim\; \frac{1}{(4\pi)^\frac{n}{2}\,n^\frac{3}{2}}\times
  \begin{cases}
      -2(-)^k         & \textrm{if $n=4k$}\\
      0               & \textrm{if $n=4k+1$} \\
      -2(-)^k         & \textrm{if $n=4k+2$}\\
      +2\sqrt{2}(-)^k & \textrm{if $n=4k+3$}
  \end{cases}    
\end{equation}
The results are summarized in \fig{FieFig04}. Depicted are the behavior for 
large $E$, the expansion for small $E$ and a numerical evaluation of the 
integral of \eqn{FieEq026}. Notice the strong deviation of the expansion from the 
other curves for $E>2\pi$, the radius of convergence. For this plot the first  
$50$ terms were used. It appears that $T$ is indeed an increasing function 
of $E$.

\subsection{The wall}
We now turn to the analysis of the value of the action for an instanton. 
In the foregoing, we have shown for which positive values of $z$ no instantons 
exist. Now we will show that the action indeed becomes negative for $z$ 
positive and large enough. For an instanton solution with one bending point,
the action is given by 
\begin{align}
  &S(E)
   \;=\; \frac{1}{4z(E)}\int_0^1\phi''(x)\,dx + \int_0^1\phi(x)\,dx
   \;=\; E + 2\,\frac{T_1(E)}{T(E)} \;\;,\label{FieEq029}\\
  &T_1(E)
  \;=\; \int\limits_{\phi_-}^{\phi_+}\frac{\phi\,d\phi}{\sqrt{E-U(\phi)}} 
        \label{FieEq030}\;\;.
\end{align}
With the use of the same approximations for $U(\phi)$ as in the derivation of 
\eqn{FieEq027}, it is easy to see that, for large $E$, $T_1(E)$ is bounded by  
\begin{equation}
   - \frac{4}{3}(E+1)^{3/2} 
         + \frac{2\log E}{\sqrt{E}}\,\log\left(\sqrt{E}+\sqrt{E-1}\right) \;\;,
	 \label{FieEq031}
\end{equation}
so that $S(E)$ clearly becomes negative for large $E$.

\begin{figure}[t]
\begin{center}
\begin{picture}(300.0,226.0)(-15,0)
\LinAxis(  0.0,  0.0)(300.0,  0.0)(6.0,1,3,0,1.5)
\LinAxis(  0.0,  0.0)(  0.0,226.0)(7.0,1,-3,0,1.5)
\LinAxis(  0.0,226.0)(300.0,226.0)(6.0,1,-3,0,1.5)
\LinAxis(300.0,  0.0)(300.0,226.0)(7.0,1,3,0,1.5)
\Text(  0.0,-10)[]{$  0$}
\Text( 50.0,-10)[]{$  2$}
\Text(100.0,-10)[]{$  4$}
\Text(150.0,-10)[]{$  6$}
\Text(200.0,-10)[]{$  8$}
\Text(250.0,-10)[]{$ 10$}
\Text(300.0,-10)[]{$ 12$}
\Text(-20,  0.0)[]{$ -3.0$}
\Text(-20, 32.3)[]{$ -2.5$}
\Text(-20, 64.6)[]{$ -2.0$}
\Text(-20, 96.9)[]{$ -1.5$}
\Text(-20,129.1)[]{$ -1.0$}
\Text(-20,161.4)[]{$ -0.5$}
\Text(-17,193.7)[]{$  0.0$}
\Text(-17,226.0)[]{$  0.5$}
\Text(150,-25)[]{$E$}
\Text(-50,113)[]{$S(E)$}
\Text(118,65)[r]{numerical integration}\Line(123,65)(148,65)
\Text(118,50)[r]{expansion}\DashLine(123,50)(148,50){3}
\Text(118,35)[r]{large $E$}\DashLine(123,35)(148,35){1.5}
\Line(0,193.7)(300,193.7)
\DashLine(157.1,0)(157.1,226){6.0}
\Curve{(  0.2,193.7) (  3.2,193.7) (  6.2,193.5) (  9.2,193.4) ( 12.2,193.1)
       ( 15.2,192.8) ( 18.2,192.3) ( 21.2,191.9) ( 24.2,191.3) ( 27.2,190.7)
       ( 30.2,190.1) ( 33.2,189.4) ( 36.2,188.6) ( 39.2,187.7) ( 42.2,186.8)
       ( 45.2,185.9) ( 48.2,184.9) ( 51.2,183.9) ( 54.2,182.8) ( 57.2,181.7)
       ( 60.2,180.5) ( 63.2,179.3) ( 66.2,178.0) ( 69.2,176.7) ( 72.2,175.4)
       ( 75.2,174.0) ( 78.2,172.6) ( 81.2,171.2) ( 84.2,169.7) ( 87.2,168.2)
       ( 90.2,166.7) ( 93.2,165.2) ( 96.2,163.6) ( 99.2,162.0) (102.2,160.4)
       (105.2,158.7) (108.2,157.1) (111.2,155.4) (114.2,153.7) (117.2,151.9)
       (120.2,150.2) (123.1,148.4) (126.1,146.6) (129.1,144.8) (132.1,143.0)
       (135.1,141.2) (138.1,139.3) (141.1,137.5) (144.1,135.6) (147.1,133.7)
       (150.1,131.8) (153.1,129.9) (156.1,128.0) (159.1,126.0) (162.1,124.1)
       (165.1,122.1) (168.1,120.2) (171.1,118.2) (174.1,116.2) (177.1,114.2)
       (180.1,112.2) (183.1,110.2) (186.1,108.1) (189.1,106.1) (192.1,104.1)
       (195.1,102.0) (198.1,100.0) (201.1, 97.9) (204.1, 95.8) (207.1, 93.8)
       (210.1, 91.7) (213.1, 89.6) (216.1, 87.5) (219.1, 85.4) (222.1, 83.3)
       (225.1, 81.2) (228.1, 79.1) (231.1, 76.9) (234.1, 74.8) (237.1, 72.7)
       (240.1, 70.5) (243.0, 68.4) (246.0, 66.3) (249.0, 64.1) (252.0, 61.9)
       (255.0, 59.8) (258.0, 57.6) (261.0, 55.4) (264.0, 53.3) (267.0, 51.1)
       (270.0, 48.9) (273.0, 46.7) (276.0, 44.5) (279.0, 42.3) (282.0, 40.1)
       (285.0, 37.9) (288.0, 35.7) (291.0, 33.5) (294.0, 31.3) (297.0, 29.1)
       (300.0, 26.9)}
\DashCurve{(  0.2,193.7) (  3.2,193.7) (  6.2,193.5) (  9.2,193.4) ( 12.2,193.1)
       ( 15.2,192.8) ( 18.2,192.3) ( 21.2,191.9) ( 24.2,191.3) ( 27.2,190.7)
       ( 30.2,190.1) ( 33.2,189.3) ( 36.2,188.6) ( 39.2,187.7) ( 42.2,186.8)
       ( 45.2,185.9) ( 48.2,184.9) ( 51.2,183.9) ( 54.2,182.8) ( 57.2,181.7)
       ( 60.2,180.5) ( 63.2,179.3) ( 66.2,178.0) ( 69.2,176.7) ( 72.2,175.4)
       ( 75.2,174.0) ( 78.2,172.6) ( 81.2,171.2) ( 84.2,169.7) ( 87.2,168.2)
       ( 90.2,166.7) ( 93.2,165.2) ( 96.2,163.6) ( 99.2,162.0) (102.2,160.4)
       (105.2,158.7) (108.2,157.1) (111.2,155.4) (114.2,153.6) (117.2,151.9)
       (120.2,150.2) (123.1,148.4) (126.1,146.6) (129.1,144.8) (132.1,143.0)
       (135.1,141.2) (138.1,139.3) (141.1,137.5) (144.1,135.6) (147.1,133.7)
       (150.1,131.9) (153.1,130.1) (156.1,128.6) (159.1,127.5) (162.1,127.8)
       (165.1,131.1) (168.1,141.5) (171.1,168.0) (173.9,226.0)}{3.0}
\DashCurve{( 27.2,115.3) ( 30.2,127.3) ( 33.2,135.2) ( 36.2,141.1) ( 39.2,145.8)
       ( 42.2,149.6) ( 45.2,152.7) ( 48.2,155.3) ( 51.2,157.4) ( 54.2,159.1)
       ( 57.2,160.5) ( 60.2,161.7) ( 63.2,162.5) ( 66.2,163.1) ( 69.2,163.6)
       ( 72.2,163.8) ( 75.2,163.9) ( 78.2,163.9) ( 81.2,163.7) ( 84.2,163.4)
       ( 87.2,163.0) ( 90.2,162.5) ( 93.2,161.9) ( 96.2,161.2) ( 99.2,160.4)
       (102.2,159.6) (105.2,158.6) (108.2,157.7) (111.2,156.6) (114.2,155.5)
       (117.2,154.4) (120.2,153.2) (123.1,151.9) (126.1,150.6) (129.1,149.3)
       (132.1,147.9) (135.1,146.5) (138.1,145.0) (141.1,143.6) (144.1,142.0)
       (147.1,140.5) (150.1,138.9) (153.1,137.3) (156.1,135.7) (159.1,134.1)
       (162.1,132.4) (165.1,130.7) (168.1,129.0) (171.1,127.2) (174.1,125.5)
       (177.1,123.7) (180.1,121.9) (183.1,120.1) (186.1,118.3) (189.1,116.4)
       (192.1,114.6) (195.1,112.7) (198.1,110.8) (201.1,108.9) (204.1,107.0)
       (207.1,105.1) (210.1,103.1) (213.1,101.2) (216.1, 99.2) (219.1, 97.2)
       (222.1, 95.2) (225.1, 93.2) (228.1, 91.2) (231.1, 89.2) (234.1, 87.2)
       (237.1, 85.1) (240.1, 83.1) (243.0, 81.1) (246.0, 79.0) (249.0, 76.9)
       (252.0, 74.8) (255.0, 72.8) (258.0, 70.7) (261.0, 68.6) (264.0, 66.5)
       (267.0, 64.3) (270.0, 62.2) (273.0, 60.1) (276.0, 58.0) (279.0, 55.8)
       (282.0, 53.7) (285.0, 51.5) (288.0, 49.4) (291.0, 47.2) (294.0, 45.1)
       (297.0, 42.9) (300.0, 40.7)}{1.5}
\end{picture}
\vspace{25pt}
\caption{$S(E)$ computed by numerical integration and as an expansion around 
         $E=0$. The expansion is up to and including $\Ord(E^{48})$ and its 
	 radius of convergence is $2\pi$.
	 The curve for large $E$ is the upper bound of \eqn{FieEq031}.} 
\label{FieFig05}
\end{center}
\end{figure}
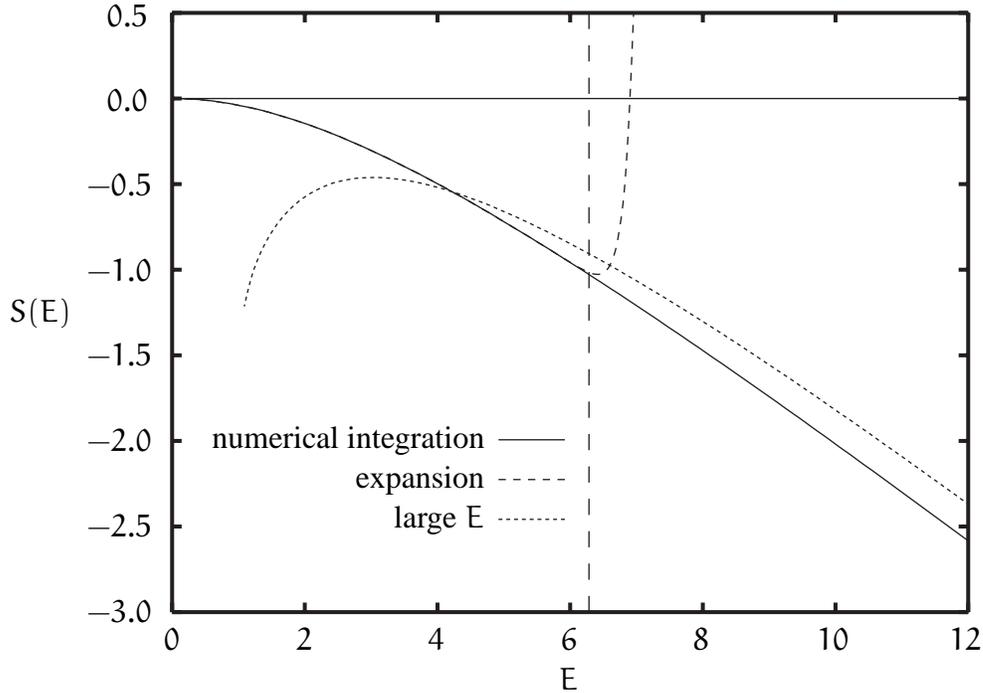
To investigate the behavior of $S(E)$ for small $E$, we use an expansion again.
It can be obtained using \eqn{FieEq029} and the relation 
\begin{equation}
   \frac{dT_1}{dE}(E)
   \;=\; -\frac{1}{2}\,T(E) - E\,\frac{dT}{dE}(E) \;\;.
\label{FieEq032}   
\end{equation}
A derivation of this relation is given in \App{App4B}. For $E\da0$ a 
quadratic approximation of the potential $U(\phi)$ can be used in \eqn{FieEq030} 
and we find that $T_1(0)=0$, so that the expansion of $T(E)$ can be substituted 
in \eqn{FieEq032} and an expansion of $T_1(E)$ can be obtained by integrating 
term by term. The expansions of $T(E)$ and $T_1(E)$ can then be used to find 
the expansion of $S(E)$ using \eqn{FieEq029}. The first few terms are 
\begin{align}
  &T_1(E)
  \;=\; \pi\sqrt{2}\left[-\frac{E}{2}-\frac{E^2}{16}-\frac{5\,E^3}{3456}
                         +\frac{973\,E^4}{2488320}+\Ord(E^5)\right]  \;\;,\\
  &S(E) 
  \;=\; -\frac{E^2}{24}+\frac{E^3}{432}+\frac{89\,E^4}{414720}+\Ord(E^5) \;\;.
\end{align}
In \fig{FieFig05}, we plot $S(E)$ as obtained from the series expansion, from 
the asymptotic behavior, and from numerical integration. The conclusion is 
that $S(E)$ is always negative.

\section{Computer-aided analysis of Riemann sheet structures\label{CAARSS}}
In the previous section, we encountered the problem of finding solutions to the
implicit function equation (\ref{ImplFunc}), or at least series expansions of
solutions. It can be classified as a particular case of slightly more general
problems one encounters in theoretical physics that are formulated as follows:
consider an entire function $F:\Comp\mapsto\Comp$ such that
\begin{equation}
F(y) \sim y^m \quad\textrm{as $y\to0$}\;\;,
\end{equation}
with nonnegative integer $m$ (in practice, we have met cases with
$m=1$ and $m=2$). The task at hand is then to find 
information about $y:\Comp\mapsto\Comp$ such that 
\begin{equation}
   F(\,y(x)\,) = x^m\;\;.
\label{NewEq001}
\end{equation}
In general, both the form of the series expansion of $y(x)$ around $x=0$ and
the nature of its singularities are of interest.  Apart from \Sec{InstL2Sec},
such questions arise, for instance, in the combinatorial problem of determining
the number of Feynman diagrams contributing to given scattering amplitudes in
various quantum field theories \cite{counting}, in the statistical bootstrap
model for hot hadronic matter (refs. in \cite{hagedorn}), and in
renormalization theory connected with the 't Hooft transformation \cite{khuri}.
An important and interesting example, studied in detail in \cite{hagedorn}, is
the so-called {\em bootstrap equation}:
\begin{equation}
   F_{\textrm{b}}(y) = 2y + 1 - e^y\;\;,
\label{NewEq002}
\end{equation}
which obviously has $m=1$.
We shall consider functions $F$ of the more general form
\begin{equation}
   F(y) = P(y) + Q(y)\,e^y \;\;,
\end{equation}
where $P$ and $Q$ are polynomials of finite degree $d_P >0$ 
and $d_Q \ge0$, respectively, with
real coefficients. As our working example, taken from \Sec{InstL2Sec},
we shall consider the function $F_{\textrm{w}}$ defined as
\begin{equation}
   F_{\textrm{w}}(y) = - 2 - 2y + 2e^y\;\;,
\label{NewEq003}
\end{equation}
for which $m=2$.
It is, in fact, closely related to the bootstrap equation (\ref{NewEq002}): by
substituting, in \eqn{NewEq002}, $y\to \log2+y$ and $x\to2\log2 - 1 -x^2$, we
obtain \eqn{NewEq003}. Its Riemann sheet structure, however, is quite
different, as we shall see.  We shall concentrate on the analysis of the
Riemann sheet structure of those solutions of these equations that have a
series expansion around $x=0$. To determine the asymptotic behavior of these
expansions, the nature of the singularities will be analyzed numerically.  The
results are justified by the fact that, in our calculations, only finite
computer accuracy is required, as we shall demonstrate.

\subsection{Identification of the Riemann sheets}
As a first step we identify the various Riemann sheets by their value of
$y(0)$: the sheet labeled $s$ will have $y(0)=Y_s$ for that sheet.
Obviously, $y(0)=0$ is a solution with multiplicity $m$. In general,
there will be $d_P$ solutions if $Q(y)=0$, and infinitely many if $Q$
is non-vanishing. It will be helpful if we can identify the Riemann sheet
on which pairs $(x,y(x))$ lie when $x$ is small but nonzero. 
This is indeed possible, and
we shall illustrate it using $F_{\textrm{w}}$. Let us write $y=\xi + i\eta$ with 
$\xi$ and $\eta$ real numbers. We are then looking for solutions of
$F_{\textrm{w}}(\xi+i\eta)=0$, or
\begin{align}
 \xi &= \log\left(\frac{\eta}{\sin\eta}\right) \;\;,\\
   0 &= 1 + \log\left(\frac{\eta}{\sin\eta}\right) - \frac{\eta}{\tan\eta} \;\;.
\end{align}
Inspecting the left-hand side of the last equation, we can immediately
see that its zeroes are quite nicely distributed. We can usefully
enumerate them as Im$(Y_s) = u_s$, where  the sheet number $s$
takes only the odd integer values $\pm1,\pm3,\pm5,\ldots$.
For positive $s$, the zero $u_s$ is certainly located in the interval
where $\sin u_s>0$, i.e.~$(s-1)\pi\le u_s<s\pi$,
and $u_{-s}=-u_s$. We have $u_1=u_{-1}=0$, and
for increasing $s$ the zero $u_s$ moves upwards in its interval, until
asymptotically we have 
$u_s \sim a_s - (\log a_s)/a_s$ with $a_s = (s-1/2)\pi$.
In \tab{NewTab01} we give the values of $Y_s$ for $F_{\textrm{w}}$, for the first
few values of $s$.
\begin{table}
\begin{center}
\begin{tabular}[b]{|l|c|}
\hline
$s$ & $Y_s/\pi$ \\ \hline
    1 &  (    0.0000,    0.0000 )\\
    3 &  (    0.6649,    2.3751 )\\
    5 &  (    0.8480,    4.4178 )\\
    7 &  (    0.9633,    6.4374 )\\
    9 &  (    1.0478,    8.4490 )\\
   11&   (    1.1145,   10.4567 )\\
\hline
\end{tabular}
\caption[.]{The first few Riemann sheet solutions for $F_{\textrm{w}}(Y_s)=0$.}
\label{NewTab01}
\end{center}
\vspace{-20pt}
\end{table}
Because the values $Y_s$ fall in disjoint intervals, 
for small $x$ we need to know $y(x)$ only to a limited
accuracy in order to be able to identify its Riemann sheet. The only
nontrivial case is that of sheets $-1$ and $1$, where it is sufficient to
consider the complex arguments:
for $\arg(x)-\arg(y) = 0$ we are on sheet $1$, 
for $|\arg(x)-\arg(y)|=\pi$ we are on sheet $-1$.
Again, limited computer accuracy is acceptable here, and for larger $m$
we simply have $m$ different values of the argument, distinguished in
an analogous manner. 
Note that of course the
labeling of the sheets is rather arbitrary: we have chosen the odd integers
in order to emphasize that both sheet $1$ and $-1$ can be considered the
principal Riemann sheet. For the bootstrap equation (\ref{NewEq002})
it is more natural to label the single principal Riemann sheet with
$y(0)=0$ as sheet number zero.

\subsection{Series expansion}
We want to compute $y(x)$ as a Taylor series around $x=0$:
\begin{equation}
y(x) = \sum_{n\ge0}\alpha_nx^n\;\;.
\label{NewEq004}
\end{equation}
Obviously, $\alpha_0$ can be chosen as one of the $u_s$ above. On principal
sheets, with $\alpha_0=0$, we also have immediately that $\alpha_1$ must be
chosen out of the $m$ possibilities with $\alpha_1^m=1$. The other
coefficients must then be computed (algebraically or numerically) by some
recursive method, which we shall now discuss.

It would be straightforward to plug the expansion (\ref{NewEq004}) into
\eqn{NewEq001} and equate the powers of $x$ on both sides, but notice that,
for $Q$ non-vanishing, the number of possible products of coefficients
grows very rapidly, so that the computer time
needed to find the first $N$ coefficients grows exponentially
with $N$. As already mentioned in \cite{hagedorn}, 
the better way is to differentiate
\eqn{NewEq001} with respect to $x$ so that we obtain the
nonlinear differential equation
\begin{equation}
y'(x)\left[P'(y)Q(y) + (Q(y)+Q'(y))(x^m-P(y))\right] = mx^{m-1}Q(y)\;\;.
\end{equation}
This equation yields a recursion relation involving 
products of at most $d_P+d_Q+1$ coefficients, 
so that a truncated power series can be computed in polynomial time.
As an example, for $F_{\textrm{w}}$ we find the following differential equation:
\begin{equation}
y'(x)(x^2+ 2y(x)) = 2x\;\;,
\end{equation}
and the following recursion relation:
\begin{align}
&\alpha_0\alpha_1 =  0\;\;\;,\;\;\;
2\alpha_0\alpha_2 + \alpha_1^2 - 1  =  0\;\;,\notag\\
&n\alpha_0\alpha_n + (n-2)\alpha_{n-2} + 
2\sum_{p=1}^{n-1}p\alpha_p\alpha_{n-p}  =  0\;\;,\;\;n\ge3\;\;.
\label{NewEq005}
\end{align}
We see immediately that $y(x)$ is necessarily even in $x$ if
$\alpha_0\ne0$, i.e.~on the non-principal Riemann sheets.
In that case, we also see that if $\al_n$, $n=0,2,\ldots$ is a solution, 
then also $\al^*_n$, $n=0,2,\ldots$ is a solution, where the asterix 
stands for complex conjugation.  This is a result of the fact that 
if $y(x)$ is a solution of \eqn{NewEq003}, then also $y^*(x^*)$ is a solution. 
In practice, these solutions 
give the function values on the different Riemann sheets of one solution. The 
analysis of the previous section proves that $y_s(0)=y_{-s}(0)^*$ so that 
the solutions satisfy $y_s^*(x)=y_{-s}(x^*)$ and the expansion coefficients 
satisfy
\begin{equation}
   \al^{(s)}_n=(\al^{(-s)}_n)^* \;\;.
\label{NewEq006}   
\end{equation}
On the principal Riemann sheets we have 
$\alpha_0=0$ and $\alpha_1^2=1$ as mentioned,
and the two solutions on sheet $1$ and sheet $-1$ are related by
$y_{-1}(x) = y_1(-x)$. 
For $y_1(x)$ we find, finally:
\begin{equation}
\alpha_n = -\frac{1}{2(n+1)}
\left[(n-1)\alpha_{n-1} +
 2\sum_{p=2}^{n-1}p\alpha_p\alpha_{n+1-p}\right]\;\;,
\end{equation}
for $n\ge2$. Using this relation we have been able to compute many thousands
of terms. The recursion appears
to be stable in the forward direction, but we
have not tried to prove this or examine the stability in the general case.

In series expansions it is of course always important to know the
convergence properties or, equivalently, the asymptotic behavior
of $\alpha_n$ as $n$ becomes very large. In the next section, we therefore
turn to the singularity structure of $y(x)$.

\subsection{Singularities and branches}
In order to find information about the singularity structure of $y(x)$, we
employ the techniques developed in \cite{counting}, which we recapitulate here.
Singularities are situated at those values $y_k$ of $y$ where
\begin{equation}
F'(y_k) = 0\;\;.
\end{equation}
Since $F$ is entire we also know that these singular points must form an
enumerable set, i.e.~we can find, and label, them as distinct points.
We shall assume that these singularities are square-root branch points,
for which it is necessary that
\begin{equation}
F''(y_k) \ne 0\;\;,
\end{equation}
If $F''$ vanishes at $y_k$ but $F'''$ does not, 
we have a cube-root branch point,
and so on. If, for non-vanishing $Q$, 
all derivatives vanish (as for instance when $F(y)=e^y$) we
have, of course, a logarithmic branch point. We know that $y=-\infty$ 
corresponds to a logarithmic branch point, and it is 
to remove this to infinity
in the $x$ plane that we have required $d_P>0$. In our
examples all the singularities at finite $x$ will be square-root
branch points. 
The position of the singularity in the $x$ plane, $x_k$, is of course
given by
\begin{equation}
F(y_k) = x_k^m\;\;,
\end{equation}
so that there are $m$ different possible positions, lying
equally spaced on a circle around the origin. 
We shall denote them by $x_{k,p}$ with $p=1,2,\ldots,m$.
Note that, in first instance, it is not clear at all whether $x_{k,p}$ for 
certain $k$ and $p$ is indeed a singular point on a specific Riemann sheet. 
Later on, we shall describe how to determine this numerically.
For values of $x$ close to an observed singular point
$x_{k,p}$ we may expand the left-hand and 
right-hand side of \eqn{NewEq001} to obtain
\begin{equation}
\frac{1}{2}(y-y_k)^2F''(y_k) \sim mF(y_k)\left(\frac{x}{ x_{k,p}}-1\right)\;\;,
\label{NewEq007}
\end{equation}
where we have dropped the higher derivative terms. Very close to the
branch point we may therefore approximate $y(x)$ by
\begin{equation}
y(x)  \sim  y_k + \beta_{k,p}\;\left(1-\frac{x}{ x_{k,p}}\right)^{1/2}
\;\;,\quad
\beta_{k,p}^2  \df  -\frac{2mF(y_k)}{ F''(y_k)}\;\;.
\end{equation}
Note that there are only two possible values for $\beta_{k,p}$,
and each singular point $x_{k,p}$ goes with one or the other of these.
Again numerical methods will help in determining which one of the two is 
the correct choice.

We are now in a position to compute the asymptotic behavior of the
coefficients $\alpha_n$. To find it, we first determine, {\em for a
given Riemann sheet}, which are the $x_{k,p}$ that lie closest to the
origin: this gives us the radius of convergence of the expansion
of $y(x)$ in that Riemann sheet. We then have to determine those $p$
for which $x_{k,p}$ is actually a singular point. We shall do this
numerically, in the way described in the following section. Let us denote
the set of values of $p$ for which this is the case by ${\bf P}$.
Now, we may use the fact that
\begin{align}
   \sqrt{1-x}  =  1 - \sum_{n\ge1}\gamma_n x^n\;\;,\quad
   \gamma_n = \frac{(2n-2)!}{2^{2n-1}(n-1)!n!}
   \;\overset{n\to\infty}{\sim}\;
   \frac{1}{\sqrt{4\pi}}\,n^{-3/2} + {\cal O}(n^{-5/2})\;,
\end{align}
where we have chosen that square root that is real and 
positive for $1-x$ real and 
positive. The asymptotic behavior of $\alpha_n$ as $n\to\infty$ 
must therefore be given by
\begin{equation}
\alpha_n \sim \frac{-1}{ n^{3/2}\sqrt{4\pi}}
\sum_{p\in{\bf P}} \frac{\beta_{k,p}}{ x_{k,p}^n}\;\;.
\label{NewEq008}
\end{equation}
Amongst other things, this provides a powerful numerical check on the
accuracy of the $\alpha_n$ as computed by the recursive technique.
We shall now discuss how the singularity structure of our problem can be
investigated numerically.

\subsection{Computer searches for sheet structures}
The main tool we use for our computer studies is a method for
taking small steps over a Riemann sheet, that is, given the fact that
for some value $x_1$ the point
$y_1=y(x_1)$ is determined to belong to a certain Riemann sheet, we perform
a small step $\Delta x$ to a point $x_2$
and find the point $y_2 = y(x_2)$
on the same Riemann sheet. Our method to do this is nothing but
Newton-Raphson iteration: we simply iterate the mapping
\begin{equation}
y \leftarrow y - \frac{F(y) - x_2^m}{ F'(y)}\;\;,
\end{equation}
until satisfactory convergence is obtained. The starting value
for this iteration is just the point $y_1$. A few remarks are in
order here. In the first place, it must be noted that for this method to
work, $y_1$ must be in the basin of attraction of $y_2$. Since, except
at the branch points, which we shall expressly avoid, $y(x)$ is
a continuous and differentiable function of $x$, this can always be
arranged by taking $\Delta x$ small enough. In the second place, the
accuracy with which $y_1$ is actually a solution of \eqn{NewEq001} is
not important as long as it is in the basin of attraction of $y_2$: 
therefore, there is no buildup of numerical errors in this method
if we restrict ourselves to just keeping track of which Riemann sheet
we are on. Finally, problems could arise if two Riemann sheet values
of $y$ for the same $x$ are very close. But, since $F$ is
an entire function, we know that the solutions of \eqn{NewEq001}
must either completely coincide or be separated by a finite
distance, any inadvertent jump from one sheet to another can be detected
and cured by, again, taking a small enough $\Delta x$.

We have applied the following method for detecting and characterizing the
various singular points. We start on a Riemann sheet $s_1$ at a value
$x$ close to zero, and determine $y(x)$ on that Riemann sheet. We then let
the parameter $x$ follow a prescribed contour that circles a selected
would-be singularity $x_{k,p}$ once (and no other singularities), and then
returns to the starting point close to the origin. We then determine
to which Riemann sheet the resulting $y$ belongs. In this way we can find 
whether $x_{k,p}$ is, in fact, a singular point for the starting sheet,
and, if so, which two sheets are connected there. It is also possible, of
course, to certify the square-root branch point nature of a singular point
by circling twice around it, and checking that one returns to the
original Riemann sheet.

One important remark is in order here. In our tracking over the Riemann
sheet, it is necessary that we do not cross branch cuts (except of course
the one connected to the putative singularity). Since these branch cuts
can be moved around in the complex $x$ plane, {\em the contour
chosen defines the (relative) position of the branch cuts}. 
The sheets that are said to be connected at a particular branch cut
are therefore also determined by the choice of contour. 
Of course, choosing a different contour will change the
whole system of interconnected sheets in a consistent
manner, so that in fact, given one choice of contour and its system
of sheets, we can work out what system of sheets will
correspond to another choice of contour.
We shall illustrate this in the following.

Suppose, now, that $x_{k,p}$ is one of the singular points on a certain
sheet that is closest to the origin. We can then follow, on that sheet,
a straight line running from $x_1$ close to the origin to a point 
$x_2$ for which $x_2/x_{k,p}$ is real and just a bit smaller than one.
Since $x_{k,p}$ is by assumption closest to the origin, there is then
no ambiguity involved in determining which one of the two possible
complex arguments of $\beta_{k,p}$ we have to take. Thus, we can find
all the information needed to compute the asymptotic behavior of
$\alpha_n$ on that sheet.

\subsection{An example}
Having established the necessary machinery, we shall now discuss a concrete
example of our method. For this, we have taken the function $F_{\textrm{w}}$ of
\eqn{NewEq003}, which is closely related to the very well-understood bootstrap
equation (\ref{NewEq002}), as we have shown. Note that the origin $x=0$, $y=0$
for $F_{\textrm{w}}$ corresponds to the first singularity in $F_{\textrm{b}}$.

\subsubsection{The singularities}
\begin{figure}
\begin{center}
\begin{picture}(120.0,120.0)(0,0)
\LinAxis(  0.0,  0.0)(120.0,  0.0)(1.0,1,3,0,1)
\LinAxis(  0.0,  0.0)(  0.0,120.0)(1.0,1,-3,0,1)
\LinAxis(  0.0,120.0)(120.0,120.0)(1.0,1,-3,0,1)
\LinAxis(120.0,  0.0)(120.0,120.0)(1.0,1,3,0,1)
\Text(  0.0,-10)[]{$-5$} \Text( 60.0,-10)[]{$0$} \Text(120.0,-10)[]{$5$}
\Text(-10,  0.0)[]{$-5$} \Text(-10, 60.0)[]{$0$} \Text(-10,120.0)[]{$5$}
\Vertex( 90.08, 90.08){1} \Vertex(102.54,102.54){1} \Vertex(112.10,112.10){1}
\Vertex( 29.92, 90.08){1} \Vertex( 17.46,102.54){1} \Vertex(  7.90,112.10){1}
\Vertex( 29.92, 29.92){1} \Vertex( 17.46, 17.46){1} \Vertex(  7.90,  7.90){1}
\Vertex( 90.08, 29.92){1} \Vertex(102.54, 17.46){1} \Vertex(112.10,  7.90){1}
\Text( 86.08, 89.08)[r]{$\sss{(-1,1)}$} \Text( 98.54,101.54)[r]{$\sss{(-2,1)}$}
\Text(108.10,111.10)[r]{$\sss{(-3,1)}$} \Text( 33.92, 89.08)[l]{$\sss{(1,1)}$}
\Text( 21.46,101.54)[l]{$\sss{(2,1)}$} \Text( 11.90,111.10)[l]{$\sss{(3,1)}$}
\Text( 33.92, 30.92)[l]{$\sss{(-1,2)}$} \Text( 21.46, 18.46)[l]{$\sss{(-2,2)}$}
\Text( 11.90,  8.90)[l]{$\sss{(-3,2)}$} \Text( 86.08, 30.92)[r]{$\sss{(1,2)}$}
\Text( 98.54, 18.46)[r]{$\sss{(2,2)}$} \Text(108.10,  8.90)[r]{$\sss{(3,2)}$}
\DashLine(60,0)(60,120){1.5}
\DashLine(0,60)(120,60){1.5}
\end{picture}
\vspace{10pt}
\caption{The numbering $(k,p)$ of the singularities.}
\label{NewFig04}
\end{center}
\vspace{-20pt}
\end{figure}
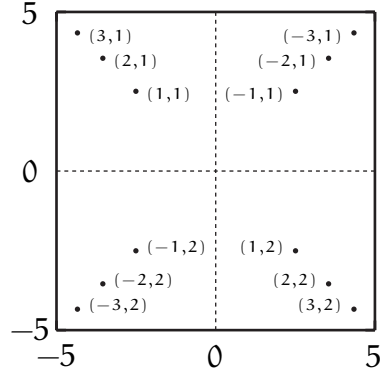
The values of $y(0)$ on the different Riemann sheets for $F_{\textrm{w}}$,
namely $Y_s$ for $s=\pm1,\pm3,\ldots$ have already been discussed above. 
The singular values $y_k$ are simply given
by
\begin{equation}
F_{\textrm{w}}'(y_k) = 2e^{y_k} - 2 =0\;\;\;\Rightarrow\;\;\; y_k = 2i\pi k\;\;,
\end{equation}
so that the possible singular points $x_{k,p}$ satisfy
\begin{equation}
x_{k,p}^2 = -4i\pi k\;\;.
\end{equation}
Note that $k=0$ does not correspond to a singular point. 
The positions of the possible singularities in the complex $x$ plane
are therefore as follows. For positive integer $k$:
\begin{align}
   x_{k,1}  &=  iz_k\;\;\;,\;\;\;x_{k,2} = -iz_k\;\;\;,\notag\\
   x_{-k,1}  &=  z_k\;\;\;,\;\;\;x_{-k,2} = -z_k\;\;\;,\qquad
   z_k  =  (1+i)\sqrt{2\pi k}\;\;.
\end{align}
At all these various possible singularities, we have
\begin{equation}
\beta_{k,p}^2 = 8i\pi k\;\;,
\end{equation}
and therefore we may write
\begin{align}
&\textrm{for $k>0$}:\quad \beta_{k,p} = \ep_{k,p}(1+i)\sqrt{4\pi|k|}\;\;,\notag\\
&\textrm{for $k<0$}:\quad \beta_{k,p} = \ep_{k,p}(1-i)\sqrt{4\pi|k|}\;\;,
\end{align}
where the only number to be determined is $\ep_{k,p}\in\{-1,1\}$. It must be 
kept in mind
that the value of $\ep$ depends of course on the sheet: we take the
convention that we work on the sheet with the lowest number
(in absolute value). When viewed from the other sheet, the value of
$\ep$ is simply opposite.

\subsubsection{The Riemann sheet structure}
We now have to discuss how the branch cuts should run in the complex $x$
plane. There are two simple options (and an infinity of more  complicated
ones): in the first option (I), we choose to let the branch cuts
extend away from the origin parallel to the real axis. This corresponds
to tracking a contour that, say, first moves in the imaginary direction,
and then in the real direction, to arrive close to the chosen singularity.
The other option (II) is to take the cuts parallel to the imaginary
axis, so that a contour that does not cross branch cuts {\it en route\/}
first goes in the real direction, and then in the imaginary direction.
Note that these two alternatives do, indeed, correspond to different
implied relative positionings of the branch cuts.

In \fig{NewFig01}.a we show the contour used in examining singularity $x_{2,1}$
under option I. 
\begin{figure}
\begin{center} 
\begin{picture}(120.0,120.0)(20,0)
\LinAxis(  0.0,  0.0)(120.0,  0.0)(1.0,1,3,0,1)
\LinAxis(  0.0,  0.0)(  0.0,120.0)(1.0,1,-3,0,1)
\LinAxis(  0.0,120.0)(120.0,120.0)(1.0,1,-3,0,1)
\LinAxis(120.0,  0.0)(120.0,120.0)(1.0,1,3,0,1)
\Text(  0.0,-10)[]{$-5$} \Text(100.0,-10)[]{$0$} \Text(120.0,-10)[]{$1$}
\Text(-10,  0.0)[]{$-1$} \Text(-10, 20.0)[]{$0$} \Text(-10,120.0)[]{$5$}
\Text(60,-30)[]{a: the $x$-plane}
\DashLine(100,0)(100,120){1.5} \DashLine(0,20)(120,20){1.5}
\Line( 99.00, 21.00)( 99.00, 90.90) \Line( 99.00, 90.90)( 39.10, 90.90)
\CArc( 29.10, 90.90)(10,0,360)
\Vertex( 49.87, 70.13){1}\DashLine(0, 70.13)( 49.87, 70.13){6}
\Vertex( 29.10, 90.90){1}\DashLine(0, 90.90)( 29.10, 90.90){6}
\Vertex( 13.17,106.83){1}\DashLine(0,106.83)( 13.17,106.83){6}
\end{picture}   
\begin{picture}(120.0,120.0)(-35,0)  
\LinAxis(  0.0,  0.0)(120.0,  0.0)(1.0,1,3,0,1)
\LinAxis(  0.0,  0.0)(  0.0,120.0)(1.0,1,-3,0,1)
\LinAxis(  0.0,120.0)(120.0,120.0)(1.0,1,-3,0,1)
\LinAxis(120.0,  0.0)(120.0,120.0)(1.0,1,3,0,1)
\Text(  0.0,-10)[]{$-2$} \Text( 34.3,-10)[]{$0$} \Text(120.0,-10)[]{$5$}
\Text(-10,  0.0)[]{$-1$} \Text(-10, 17.1)[]{$0$} \Text(-10,120.0)[]{$6$}
\Text(60,-30)[]{b: sheet 1, single loop}
\Curve{( 33.43, 18.01) ( 33.45, 18.99) ( 33.49, 19.90) ( 33.55, 20.81)
       ( 33.62, 21.72) ( 33.82, 23.53) ( 34.07, 25.33) ( 34.39, 27.12) 
       ( 34.77, 28.89) ( 35.21, 30.63) ( 35.71, 32.35) ( 36.27, 34.04) 
       ( 36.88, 35.70) ( 37.55, 37.33) ( 38.27, 38.92) ( 39.03, 40.47) 
       ( 39.87, 42.03) ( 40.79, 43.59) ( 41.75, 45.09) ( 42.75, 46.54) 
       ( 43.79, 47.93) ( 44.90, 49.30) ( 46.08, 50.65) ( 47.28, 51.93) 
       ( 48.53, 53.18) ( 49.84, 54.38) ( 51.17, 55.51) ( 52.57, 56.62) 
       ( 53.98, 57.65) ( 55.45, 58.65) ( 56.91, 59.58) ( 57.67, 60.03) 
       ( 58.43, 60.46) ( 59.21, 60.90) ( 59.99, 61.31)}
\Curve{( 55.15, 79.81) ( 55.17, 78.93) ( 55.20, 78.07) ( 55.26, 77.17)
       ( 55.34, 76.29) ( 55.56, 74.56) ( 55.87, 72.80) ( 56.25, 71.11) 
       ( 56.71, 69.40) ( 57.24, 67.72) ( 57.84, 66.08) ( 58.50, 64.45) 
       ( 58.85, 63.66) ( 59.22, 62.86) ( 59.60, 62.07) ( 59.99, 61.31)}
\Curve{( 55.15, 79.81) ( 55.16, 80.69) ( 55.20, 81.58) ( 55.26, 82.48)
       ( 55.34, 83.33) ( 55.45, 84.19) ( 55.58, 85.06) ( 55.74, 85.92)
       ( 55.93, 86.78) ( 56.15, 87.64) ( 56.39, 88.50) ( 56.67, 89.35)
       ( 56.97, 90.19) ( 57.30, 91.02) ( 57.66, 91.84) ( 58.04, 92.64)
       ( 58.45, 93.43) ( 58.89, 94.19) ( 59.35, 94.94) ( 59.83, 95.66)
       ( 60.36, 96.39) ( 60.90, 97.10) ( 61.46, 97.77) ( 62.03, 98.42)
       ( 62.65, 99.07) ( 63.27, 99.69) ( 63.93,100.30) ( 64.59,100.88)
       ( 65.03,101.18) ( 65.64,100.55) ( 66.36,100.01) ( 67.12, 99.60)
       ( 67.94, 99.31) ( 68.80, 99.15) ( 69.68, 99.12) ( 70.56, 99.22)
       ( 71.42, 99.46) ( 72.20, 99.82) ( 72.93,100.30) ( 73.58,100.89)
       ( 74.11,101.55) ( 74.55,102.31) ( 74.85,103.11) ( 75.03,103.97)
       ( 75.07,104.85)}
\Curve{( 63.73,104.64) ( 63.79,105.52) ( 63.98,106.38) ( 64.31,107.19)
       ( 64.75,107.93) ( 65.32,108.61) ( 65.97,109.17) ( 66.72,109.63)
       ( 67.53,109.98) ( 68.39,110.20) ( 69.28,110.28) ( 70.16,110.23)
       ( 71.03,110.04) ( 71.86,109.72) ( 72.63,109.27) ( 73.30,108.73)
       ( 73.89,108.09) ( 74.38,107.36) ( 74.74,106.57) ( 74.98,105.72)
       ( 75.07,104.85)}
\Curve{( 63.73,104.64) ( 63.82,103.77) ( 64.04,102.93) ( 64.40,102.12)
       ( 64.87,101.38)}
\DashLine(34.3,0)(34.3,120){1.5} \DashLine(0, 17.1)(120, 17.1){1.5}       
\end{picture}  \\ 
\begin{picture}(120.0,120.0)(20,60)
\LinAxis(  0.0,  0.0)(120.0,  0.0)(1.0,1,3,0,1)
\LinAxis(  0.0,  0.0)(  0.0,120.0)(1.0,1,-3,0,1)
\LinAxis(  0.0,120.0)(120.0,120.0)(1.0,1,-3,0,1)
\LinAxis(120.0,  0.0)(120.0,120.0)(1.0,1,3,0,1)
\Text(  0.0,-10)[]{$ -4$} \Text( 53.3,-10)[]{$  0$} \Text(120.0,-10)[]{$  5$}
\Text(-10,  0.0)[]{$  7$} \Text(-10,120.0)[]{$ 16$}
\Text(60,-30)[]{c: sheet 3, single loop}
\DashLine( 53.3,0)( 53.3,120){1.5}
\Curve{( 80.53, 10.09) ( 80.58,  8.75) ( 80.79,  7.43) ( 81.18,  6.15)}
\Curve{( 80.53, 10.09) ( 80.61, 11.43) ( 80.82, 12.76) ( 81.16, 14.05)
       ( 81.62, 15.32)(82.43,16.2)}
\Curve{( 52.70, 42.16) ( 52.74, 41.74) ( 53.26, 40.45) ( 53.84, 39.14)
       ( 54.43, 37.91) ( 55.09, 36.67) ( 55.76, 35.50) ( 56.49, 34.33)
       ( 57.27, 33.17) ( 58.07, 32.09) ( 58.91, 31.03) ( 59.79, 30.00)
       ( 60.72, 29.00) ( 61.68, 28.04) ( 62.67, 27.12) ( 63.69, 26.25)
       ( 64.76, 25.41) ( 65.87, 24.59) ( 66.98, 23.83) ( 68.11, 23.11)
       ( 69.29, 22.42) ( 70.47, 21.76) ( 71.66, 21.14) ( 72.86, 20.55)
       ( 74.07, 19.98) ( 75.29, 19.42) ( 76.52, 18.87) ( 77.75, 18.33)
       ( 78.98, 17.78) ( 80.21, 17.22) ( 81.43, 16.64) (82.43,16.2)}
\Curve{( 52.70, 42.16) ( 53.98, 42.67) ( 55.29, 43.24) ( 56.53, 43.85)
       ( 57.71, 44.47) ( 58.91, 45.17) ( 60.04, 45.89) ( 61.18, 46.67)
       ( 62.26, 47.47) ( 63.34, 48.34) ( 64.36, 49.21) ( 65.37, 50.15)
       ( 66.31, 51.11) ( 67.24, 52.12) ( 68.11, 53.14) ( 68.96, 54.23)
       ( 69.74, 55.32) ( 70.50, 56.47) ( 71.19, 57.62) ( 71.85, 58.83)
       ( 72.44, 60.04) ( 72.99, 61.30) ( 73.48, 62.57) ( 73.91, 63.88)
       ( 74.29, 65.18) ( 74.59, 66.48) ( 74.84, 67.83) ( 75.03, 69.16)
       ( 75.16, 70.54) ( 75.23, 71.90) ( 75.24, 73.24)}
\Curve{( 68.99, 93.02) ( 69.26, 92.48) ( 69.94, 91.31) ( 70.60, 90.11)
       ( 71.22, 88.88) ( 71.79, 87.66) ( 72.33, 86.41) ( 72.83, 85.16)
       ( 73.29, 83.87) ( 73.71, 82.58) ( 74.08, 81.28) ( 74.40, 79.98)
       ( 74.67, 78.66) ( 74.90, 77.33) ( 75.07, 75.98) ( 75.18, 74.62)
       ( 75.24, 73.24)}
\Curve{( 68.99, 93.02) ( 70.16, 93.72) ( 71.36, 94.39) ( 72.57, 95.00)
       ( 73.79, 95.55) ( 75.03, 96.06) ( 76.30, 96.52) ( 77.58, 96.93)
       ( 78.88, 97.28) ( 80.20, 97.58) ( 81.52, 97.81) ( 82.85, 97.97)
       ( 84.18, 98.05) ( 85.52, 98.05) ( 86.86, 97.95) ( 88.18, 97.75)
       ( 89.2,97.6)}
\Line( 88.52, 93.84)( 88.50, 95.17) \Line( 88.50, 95.17)( 88.86, 96.50)
\Line( 88.86, 96.50)( 89.2,97.6) \Line( 88.52, 93.84)( 88.66, 92.51)
\end{picture}  
\begin{picture}(120.0,120.0)(-35,60)
\LinAxis(  0.0,  0.0)(120.0,  0.0)(1.0,1,3,0,1)
\LinAxis(  0.0,  0.0)(  0.0,120.0)(1.0,1,-3,0,1)
\LinAxis(  0.0,120.0)(120.0,120.0)(1.0,1,-3,0,1)
\LinAxis(120.0,  0.0)(120.0,120.0)(1.0,1,3,0,1)
\Text(  0.0,-10)[]{$ -4$} \Text( 53.3,-10)[]{$  0$} \Text(120.0,-10)[]{$  5$}
\Text(-10,  0.0)[]{$  7$} \Text(-10,120.0)[]{$ 16$}
\Text(60,-30)[]{d: sheet 3, double loop}
\DashLine( 53.3,0)( 53.3,120){1.5}
\Curve{( 80.53, 10.09) ( 80.58,  8.75) ( 80.79,  7.43) ( 81.18,  6.15)}
\Curve{( 80.53, 10.09) ( 80.61, 11.43) ( 80.82, 12.76) ( 81.16, 14.05)
       ( 81.62, 15.32) (82.43,16.2)}
\Curve{( 52.70, 42.16) ( 52.74, 41.74) ( 53.26, 40.45) ( 53.84, 39.14)
       ( 54.43, 37.91) ( 55.09, 36.67) ( 55.76, 35.50) ( 56.49, 34.33)
       ( 57.27, 33.17) ( 58.07, 32.09) ( 58.91, 31.03) ( 59.79, 30.00)
       ( 60.72, 29.00) ( 61.68, 28.04) ( 62.67, 27.12) ( 63.69, 26.25)
       ( 64.76, 25.41) ( 65.87, 24.59) ( 66.98, 23.83) ( 68.11, 23.11)
       ( 69.29, 22.42) ( 70.47, 21.76) ( 71.66, 21.14) ( 72.86, 20.55)
       ( 74.07, 19.98) ( 75.29, 19.42) ( 76.52, 18.87) ( 77.75, 18.33)
       ( 78.98, 17.78) ( 80.21, 17.22) ( 81.43, 16.64) (82.43,16.2)}
\Curve{( 52.70, 42.16) ( 53.98, 42.67) ( 55.29, 43.24) ( 56.53, 43.85)
       ( 57.71, 44.47) ( 58.91, 45.17) ( 60.04, 45.89) ( 61.18, 46.67)
       ( 62.26, 47.47) ( 63.34, 48.34) ( 64.36, 49.21) ( 65.37, 50.15)
       ( 66.31, 51.11) ( 67.24, 52.12) ( 68.11, 53.14) ( 68.96, 54.23)
       ( 69.74, 55.32) ( 70.50, 56.47) ( 71.19, 57.62) ( 71.85, 58.83)
       ( 72.44, 60.04) ( 72.99, 61.30) ( 73.48, 62.57) ( 73.91, 63.88)
       ( 74.29, 65.18) ( 74.59, 66.48) ( 74.84, 67.83) ( 75.03, 69.16)
       ( 75.16, 70.54) ( 75.23, 71.90) ( 75.24, 73.24)}
\Curve{(  6.82, 76.65) (  6.88, 78.20) (  7.00, 79.74) (  7.19, 81.28)
       (  7.40, 82.60) (  7.96, 85.21) (  8.72, 87.79) (  9.67, 90.30) 
       ( 10.80, 92.74) ( 12.11, 95.08) ( 13.58, 97.31) ( 15.51, 99.75) 
       ( 17.62,101.99) ( 19.91,104.00) ( 22.35,105.78) ( 24.89,107.29) 
       ( 27.51,108.54) ( 30.16,109.51) ( 32.81,110.21) ( 35.61,110.68) 
       ( 38.49,110.88) ( 41.23,110.81) ( 44.12,110.46) ( 46.93,109.86) 
       ( 49.60,109.02) ( 52.23,107.94) ( 54.66,106.71) ( 56.99,105.30) 
       ( 59.27,103.68) ( 61.38,101.94) ( 63.37,100.04) ( 65.24, 98.02) 
       ( 66.95, 95.89) ( 68.54, 93.61) ( 69.94, 91.31) ( 71.22, 88.88) 
       ( 72.33, 86.41) ( 73.29, 83.87) ( 74.08, 81.28) ( 74.40, 79.98) 
       ( 74.67, 78.66) ( 74.90, 77.33) ( 75.07, 75.98) ( 75.18, 74.62) 
       ( 75.24, 73.24)}
\Curve{(  6.82, 76.65) (  6.83, 75.12) (  6.90, 73.59) (  7.03, 72.07)
       (  7.22, 70.56) (  7.79, 67.60) (  8.58, 64.73) (  9.58, 61.96) 
       ( 10.78, 59.31) ( 12.17, 56.80) ( 13.72, 54.45) ( 15.42, 52.26) 
       ( 17.24, 50.24) ( 19.33, 48.27) ( 21.66, 46.41) ( 24.08, 44.80) 
       ( 26.56, 43.43) ( 29.07, 42.31) ( 31.74, 41.38) ( 34.53, 40.68) 
       ( 37.26, 40.24) ( 40.03, 40.03) ( 42.81, 40.07) ( 45.56, 40.34) 
       ( 48.24, 40.83) ( 49.62, 41.17) ( 50.93, 41.56) ( 52.30, 42.01)
       ( 52.70, 42.16)}
\end{picture}
\vspace{90pt}
\caption{Loops around $x_{2,1}$ under option I.}
\label{NewFig01}
\end{center}
\end{figure}

\vspace*{\fill}

\begin{figure}  
\begin{center}  
\begin{picture}(120.0,120.0)(25,0)
\LinAxis(  0.0,  0.0)(120.0,  0.0)(1.0,1,3,0,1)
\LinAxis(  0.0,  0.0)(  0.0,120.0)(1.0,1,-3,0,1)
\LinAxis(  0.0,120.0)(120.0,120.0)(1.0,1,-3,0,1)
\LinAxis(120.0,  0.0)(120.0,120.0)(1.0,1,3,0,1)
\Text(  0.0,-10)[]{$-5$} \Text(100.0,-10)[]{$0$} \Text(120.0,-10)[]{$1$}
\Text(-10,  0.0)[]{$-1$} \Text(-10, 20.0)[]{$0$} \Text(-10,120.0)[]{$5$}
\Text(60,-30)[]{a: the $x$-plane}
\DashLine(100,0)(100,120){1.5} \DashLine(0,20)(120,20){1.5}
\Line( 99.00, 21.00)( 29.10, 21.00) \Line( 29.10, 21.00)( 29.10, 80.90)
\CArc( 29.10, 90.90)(10,0,360)
\Vertex( 49.87, 70.13){1}\DashLine( 49.87, 70.13)( 49.87,120){6}
\Vertex( 29.10, 90.90){1}\DashLine( 29.10, 90.90)( 29.10,120){6}
\Vertex( 13.17,106.83){1}\DashLine( 13.17,106.83)( 13.17,120){6}
\end{picture} 
\begin{picture}(120.0,120.0)(-5,0)
\LinAxis(  0.0,  0.0)(120.0,  0.0)(1,1,3,0,1)
\LinAxis(  0.0,  0.0)(  0.0,120.0)(1,1,-3,0,1)
\LinAxis(  0.0,120.0)(120.0,120.0)(1,1,-3,0,1)
\LinAxis(120.0,  0.0)(120.0,120.0)(1,1,3,0,1)
\Text(  0.0,-10)[]{$-10$} \Text( 75.0,-10)[]{$  0$} \Text(120.0,-10)[]{$  6$}
\Text(-10,  7.5)[]{$  0$} \Text(-10,120.0)[]{$ 15$}
\Text(60,-30)[]{b: sheet $1$, single loop}       
\DashLine(75.0,0)(75.0,120){1.5} \DashLine(0,7.5)(120,7.5){1.5}
\Line(20.39,  8.99)(74.63,  7.88)       
\Curve{( 20.39,  8.99) ( 20.43, 10.58) ( 20.50, 12.18) ( 20.59, 13.77) 
       ( 20.71, 15.36) ( 20.85, 16.96) ( 21.02, 18.55) ( 21.22, 20.14) 
       ( 21.45, 21.73) ( 21.70, 23.33) ( 21.98, 24.92) ( 22.29, 26.51) 
       ( 22.62, 28.10) ( 22.98, 29.69) ( 23.37, 31.28) ( 23.78, 32.87) 
       ( 24.23, 34.46) ( 24.70, 36.06) ( 25.17, 37.57) ( 25.64, 39.00) 
       ( 26.13, 40.43) ( 26.64, 41.86) ( 27.18, 43.29) ( 27.74, 44.73) 
       ( 28.32, 46.16) ( 28.92, 47.59) ( 29.54, 49.03) ( 30.19, 50.46) 
       ( 30.86, 51.90) ( 31.55, 53.34) ( 32.26, 54.77) ( 32.99, 56.21) 
       ( 33.74, 57.65) ( 34.51, 59.09) ( 35.31, 60.53) ( 36.12, 61.97) 
       ( 36.96, 63.42) ( 37.81, 64.86) ( 38.68, 66.30) ( 39.47, 67.58) 
       ( 40.28, 68.86) ( 41.10, 70.13) ( 41.94, 71.41) ( 42.79, 72.68) 
       ( 43.66, 73.94) ( 44.54, 75.21) ( 45.44, 76.46) ( 46.36, 77.72) 
       ( 47.30, 78.96) ( 48.25, 80.20) ( 49.23, 81.43) ( 50.23, 82.65) 
       ( 51.25, 83.86) ( 52.31, 85.06) ( 53.39, 86.25) ( 54.51, 87.44) 
       ( 55.08, 87.88) ( 55.63, 87.36) ( 56.27, 86.81) ( 56.93, 86.29) 
       ( 57.60, 85.81) ( 58.29, 85.36) ( 59.68, 84.56) ( 61.09, 83.91) 
       ( 62.59, 83.36) ( 64.16, 82.93) ( 65.70, 82.66) ( 67.28, 82.53) 
       ( 68.85, 82.53) ( 70.41, 82.66) ( 71.94, 82.93) ( 73.47, 83.32)
       ( 74.93, 83.82) ( 76.35, 84.44) ( 77.74, 85.17) ( 79.06, 86.00)
       ( 80.31, 86.93) ( 81.49, 87.95) ( 82.58, 89.05) ( 83.58, 90.23)
       ( 84.48, 91.49) ( 85.27, 92.81) ( 85.94, 94.20) ( 86.48, 95.62)
       ( 86.89, 97.09) ( 87.17, 98.60) ( 87.25, 99.35) ( 87.30,100.12) 
       ( 87.32,100.88)}
\Curve{( 86.75,105.39) ( 86.92,104.66) ( 87.06,103.91) ( 87.18,103.16)
       ( 87.26,102.40) ( 87.31,101.63) ( 87.32,100.88)}
\Curve{( 86.75,105.39) 
       ( 87.51,105.64) ( 88.25,105.83)
       ( 88.98,106.02) ( 89.71,106.20) ( 90.45,106.40) ( 91.17,106.60)
       ( 91.89,106.82) ( 92.62,107.05) ( 93.33,107.30) ( 94.03,107.57)
       ( 94.72,107.88) ( 95.39,108.22) ( 96.04,108.61)} 
\Curve{( 94.99,111.42) ( 95.16,110.69) ( 95.41,109.98) ( 95.74,109.30) 
       ( 96.04,108.61)}
\end{picture} 
\begin{picture}(120.0,120.0)(-35,0)
\LinAxis(  0.0,  0.0)(120.0,  0.0)(1.0,1,3,0,1)
\LinAxis(  0.0,  0.0)(  0.0,120.0)(1.0,1,-3,0,1)
\LinAxis(  0.0,120.0)(120.0,120.0)(1.0,1,-3,0,1)
\LinAxis(120.0,  0.0)(120.0,120.0)(1.0,1,3,0,1)
\Text(  0.0,-10)[]{$  0.5$} \Text(120.0,-10)[]{$  3.5$}
\Text(-12,  2.0)[]{$  4.5$} \Text(-12,120.0)[]{$  7.5$}
\Text(60,-30)[]{c: sheet $3$, single loop}
\Curve{( 63.54,118.45) ( 64.53,114.57) ( 66.07,110.87) ( 68.01,107.34)
       ( 70.30,104.02) ( 72.89,100.93) ( 75.73, 98.09) ( 78.78, 95.48)}
\Curve{( 56.52, 49.31) ( 56.69, 53.36) ( 57.23, 57.39) ( 58.11, 61.31)
       ( 59.31, 65.14) ( 60.81, 68.87) ( 62.57, 72.51) ( 64.55, 76.01)
       ( 66.71, 79.41) ( 69.01, 82.72) ( 71.39, 85.94) ( 73.84, 89.10)
       ( 76.34, 92.26) ( 78.78, 95.48)}
\Curve{( 56.52, 49.31) ( 56.71, 45.31) ( 57.26, 41.27) ( 58.13, 37.34)
       ( 58.18, 36.81)}
\Curve{( 48.71, 24.70) ( 48.92, 26.75) ( 49.46, 28.73) ( 50.30, 30.58)
       ( 51.46, 32.33) ( 52.85, 33.84) ( 54.50, 35.12) ( 56.27, 36.10)
       ( 58.18, 36.81)}
\Curve{( 48.71, 24.70) ( 48.83, 22.66) ( 49.26, 20.68) ( 50.03, 18.73)
       ( 51.06, 16.97) ( 52.39, 15.36) ( 53.90, 14.01) ( 55.63, 12.89)
       ( 57.52, 12.04) ( 59.45, 11.51) ( 61.45, 11.27) ( 63.48, 11.33)
       ( 65.48, 11.71) ( 67.41, 12.39) ( 69.23, 13.36) ( 70.82, 14.57)
       ( 72.23, 16.03) ( 73.41, 17.71) ( 74.29, 19.52) ( 74.89, 21.50)
       ( 75.16, 23.53)}
\Curve{( 58.18, 36.81) ( 58.97, 37.01) ( 60.96, 37.29) ( 62.98, 37.28)
       ( 64.98, 36.96) ( 66.91, 36.34) ( 68.73, 35.45) ( 70.39, 34.28)
       ( 71.87, 32.86) ( 73.09, 31.27) ( 74.06, 29.48) ( 74.73, 27.59)
       ( 75.16, 23.53)}
\end{picture}
\vspace{30pt}
\caption{Loops around $x_{2,1}$ under option II.}
\vspace{-30pt}
\label{NewFig02}
\end{center}
\end{figure}
The contour starts on sheet number 1 close to the origin (so that $y$ is close
to $Y_1$), moves upwards and then to the left, circles the singularity once
anti-clockwise, and returns to its starting point by the same route in order to
enable us to determine the resulting Riemann sheet number.
\fig{NewFig01}.b shows the corresponding path in the $y$ plane. 
It ends again close to $Y_1$ so that, {\em for this choice of contour and its
induced branch structure\/} (indicated in the figure), sheet 1 does not have a
branch point at $x_{2,1}$.
\fig{NewFig01}.c shows what happens if, instead of sheet number 1, we start at
sheet number 3: the $y$ track starts then at close to $Y_3$, but ends up close
to $Y_5$, so that we conclude that sheets 3 and 5 are connected at $x_{2,1}$.
If we run through the whole contour twice, we get the $y$ track presented in
\fig{NewFig01}.d, where the $y$ track ends up again at $Y_3$ as expected for a
square root branch cut.

Under option II, we rather use the contour indicated in \fig{NewFig02}.a, which
first moves to the left and then upwards. 
\fig{NewFig02}.b shows the resulting
$y$ path, which does not return to $Y_1$ but rather to $Y_5$, indicating
that under this choice of contour the sheets labeled 1 and 5 are connected
at $x_{2,1}$. \fig{NewFig02}.c shows that, now, sheet 3 is insensitive to
this singularity.

In this way we have mapped the various singularities around the origin.
\begin{table}
\begin{center}
\begin{tabular}{|l|c|c|c|c|c|c|c|c|}
\hline 
$k$   & $x_{k,1}$ & $\ep_{k,1}$ & $x_{k,2}$ & $\ep_{k,2}$ &
$x_{-k,1}$ & $\ep_{-k,1}$ & $x_{-k,2}$ & $\ep_{-k,2}$ \\
\hline
1     & (1,3)  & -1  & (-1,3)  & -1  & (-1,-3) & -1  & (1,-3)  & -1  \\
2     & (3,5)  & -1  & (3,5)   & -1  & (-3,-5) & -1  & (-3,-5) & -1  \\
3     & (5,7)  & -1  & (5,7)   & -1  & (-5,-7) & -1  & (-5,-7) & -1  \\
4     & (7,9)  & -1  & (7,9)   & -1  & (-7,-9) & -1  & (-7,-9) & -1  \\
5     & (9,11) & -1  & (9,11)  & -1  & (-9,-11)& -1  & (-9,-11)& -1  \\
\hline 
\end{tabular}
\caption[.]{The first few sheets and singularities (option I), and
the corresponding value for $\epsilon$.}
\label{NewTab02}
\end{center}
\end{table}
In \tab{NewTab02} we present the pairs of sheets that are pairwise connected
at the first few singularities, under option I, and the observed value for
$\ep$, which turns out to be $-1$ in all cases.
We point out that at each singularity
only two sheets out of all infinitely many are connected. Note the
somewhat atypical situation at the lowest-lying singularities $x_{1,\pm1}$
and $x_{-1,\pm1}$.
\begin{table}
\begin{center}
\begin{tabular}{|l|c|c|c|c|c|c|c|c|}
\hline 
$k$   & $x_{k,1}$ & $\ep_{k,1}$ & $x_{k,2}$ & $\ep_{k,2}$ &
$x_{-k,1}$ & $\ep_{-k,1}$ & $x_{-k,2}$ & $\ep_{-k,2}$ \\
\hline
1     & (1,3) & -1   & (-1,3) & -1  & (-1,-3) & -1  & (1,-3) & -1   \\
2     & (1,5) & -1   & (-1,5) & -1  & (-1,-5) & -1  & (1,-5) & -1   \\
3     & (1,7) & -1   & (-1,7) & -1  & (-1,-7) & -1  & (1,-7) & -1   \\
4     & (1,9) & -1   & (-1,9) & -1  & (-1,-9) & -1  & (1,-9) & -1   \\
5     & (1,11)& -1   & (-1,11)& -1  & (-1,-11)& -1  & (1,-11)& -1   \\
\hline
\end{tabular}
\caption[.]{The first few sheets and singularities (option II), and
the corresponding value for $\epsilon$.}
\label{NewTab03}
\end{center}
\end{table}
The alternative option II results in \tab{NewTab03}. Note that the higher-lying
singularities now show a sheet structure similar to the lowest ones.  In fact,
this is the choice that corresponds most directly to the analysis of the sheet
structure of the bootstrap equation in \cite{hagedorn}, with of course the
extra complication in the fact that the bootstrap equation (\ref{NewEq002}) has
$m=1$ while for $F_{\textrm{w}}$, $m=2$. Note that, once again, $\ep=-1$ in all
cases.

\subsubsection{Asymptotic behavior of the series expansion coefficients}
We shall now illustrate how the information on the $x_{k,p}$ and $\beta_{k,p}$
allows us to compute the asymptotic behavior of the series expansion
coefficients $\alpha_n$.

\paragraph{First Riemann sheet.} In this sheet, the singularities closest
to the origin, and their corresponding $\beta$'s are
\begin{align}
&x_{1,1}  =  \sqrt{4\pi}\exp(3i\pi/4)\;\;\;,\;\;\;
\beta_{1,1} = \sqrt{8\pi}\exp(-3i\pi/4)\;\;,\notag\\
&x_{-1,2}  =  \sqrt{4\pi}\exp(-3i\pi/4)\;\;\;,\;\;\;
\beta_{-1,2} = \sqrt{8\pi}\exp(3i\pi/4)\;\;.
\end{align}
Using \eqn{NewEq008}, we see that the asymptotic form of the
coefficients on sheet 1 is given by
\begin{align}
&\alpha_n^{(1)}  \sim  \alasym_n\;\;,\notag\\
&\alasym_n  =  \frac{2}{ n^{3/2}(4\pi)^{n/2}}\,c_n\;\;,\notag\\
&c_n  =  -\sqrt{2}\cos\left(\frac{3n\pi}{4}+\frac{3\pi}{4}\right)
     =  \left\{\mbox{\begin{tabular}{ll}
 $(-)^{p}$           & $n=4p$ \\
 $0$                 & $n=4p+1$ \\
 $(-)^{p+1}$         & $n=4p+2$ \\
 $(-)^{p}\sqrt{2}$   & $n=4p+3$ \end{tabular}}\right.\;\;,
\end{align}
with integer $p$.
\begin{figure}
\begin{center}
\begin{picture}(200.0,151.0)(0,0)
\LinAxis(  0.0,  0.0)(200.0,  0.0)(6.0,1,3,0,1.5)
\LinAxis(  0.0,  0.0)(  0.0,151.0)(9.0,1,-3,0,1.5)
\LinAxis(  0.0,151.0)(200.0,151.0)(6.0,1,-3,0,1.5)
\LinAxis(200.0,  0.0)(200.0,151.0)(9.0,1,3,0,1.5)
\Text(  0.0,-10)[]{$0$}
\Text( 33.3,-10)[]{$1$}
\Text( 66.7,-10)[]{$2$}
\Text(100.0,-10)[]{$3$}
\Text(133.3,-10)[]{$4$}
\Text(166.7,-10)[]{$5$}
\Text(200.0,-10)[]{$6$}
\Text(-10,  0.0)[]{$-7$}
\Text(-10, 16.8)[]{$-6$}
\Text(-10, 33.6)[]{$-5$}
\Text(-10, 50.3)[]{$-4$}
\Text(-10, 67.1)[]{$-3$}
\Text(-10, 83.9)[]{$-2$}
\Text(-10,100.7)[]{$-1$}
\Text(-5,117.4)[]{$0$}
\Text(-5,134.2)[]{$1$}
\Text(-5,151.0)[]{$2$}
\Text(100,-30)[]{$\log n$}
\Text(-30,75.5)[]{$r_n$}
\Line(0,117)(200,117)
\Vertex(0.00,138.68){0.5} \Vertex(23.10,140.38){0.5}
\Vertex(36.62,138.94){0.5} \Vertex(46.21,133.98){0.5}
\Vertex(53.65,123.66){0.5} \Vertex(59.73, 96.41){0.5}
\Vertex(64.86,105.76){0.5} \Vertex(69.31,112.94){0.5}
\Vertex(73.24,108.29){0.5} \Vertex(76.75, 86.98){0.5}
\Vertex(79.93, 92.66){0.5} \Vertex(82.83,102.48){0.5}
\Vertex(85.50, 99.23){0.5} \Vertex(87.97, 78.81){0.5}
\Vertex(90.27, 85.47){0.5} \Vertex(92.42, 95.81){0.5}
\Vertex(94.44, 92.92){0.5} \Vertex(96.35, 72.07){0.5}
\Vertex(98.15, 80.89){0.5} \Vertex(99.86, 91.09){0.5}
\Vertex(101.48, 88.25){0.5} \Vertex(103.03, 66.66){0.5}
\Vertex(104.52, 77.53){0.5} \Vertex(105.93, 87.49){0.5}
\Vertex(107.30, 84.61){0.5} \Vertex(108.60, 62.31){0.5}
\Vertex(109.86, 74.81){0.5} \Vertex(111.07, 84.56){0.5}
\Vertex(112.24, 81.63){0.5} \Vertex(113.37, 58.76){0.5}
\Vertex(114.47, 72.49){0.5} \Vertex(115.52, 82.09){0.5}
\Vertex(116.55, 79.12){0.5} \Vertex(117.55, 55.78){0.5}
\Vertex(118.51, 70.46){0.5} \Vertex(119.45, 79.95){0.5}
\Vertex(120.36, 76.94){0.5} \Vertex(121.25, 53.23){0.5}
\Vertex(122.12, 68.65){0.5} \Vertex(122.96, 78.06){0.5}
\Vertex(123.79, 75.02){0.5} \Vertex(124.59, 51.01){0.5}
\Vertex(125.37, 67.02){0.5} \Vertex(126.14, 76.36){0.5}
\Vertex(126.89, 73.29){0.5} \Vertex(127.62, 49.04){0.5}
\Vertex(128.34, 65.53){0.5} \Vertex(129.04, 74.81){0.5}
\Vertex(129.73, 71.73){0.5} \Vertex(130.40, 47.27){0.5}
\Vertex(131.06, 64.16){0.5} \Vertex(131.71, 73.40){0.5}
\Vertex(132.34, 70.31){0.5} \Vertex(132.97, 45.66){0.5}
\Vertex(133.58, 62.89){0.5} \Vertex(134.18, 72.10){0.5}
\Vertex(134.77, 68.99){0.5} \Vertex(135.35, 44.19){0.5}
\Vertex(135.92, 61.71){0.5} \Vertex(136.48, 70.89){0.5}
\Vertex(137.03, 67.77){0.5} \Vertex(137.57, 42.84){0.5}
\Vertex(138.10, 60.61){0.5} \Vertex(138.63, 69.76){0.5}
\Vertex(139.15, 66.64){0.5} \Vertex(139.65, 41.59){0.5}
\Vertex(140.16, 59.57){0.5} \Vertex(140.65, 68.71){0.5}
\Vertex(141.14, 65.58){0.5} \Vertex(141.62, 40.42){0.5}
\Vertex(142.09, 58.60){0.5} \Vertex(142.56, 67.71){0.5}
\Vertex(143.02, 64.58){0.5} \Vertex(143.47, 39.32){0.5}
\Vertex(143.92, 57.68){0.5} \Vertex(144.36, 66.78){0.5}
\Vertex(144.79, 63.63){0.5} \Vertex(145.22, 38.29){0.5}
\Vertex(145.65, 56.80){0.5} \Vertex(146.07, 65.89){0.5}
\Vertex(146.48, 62.74){0.5} \Vertex(146.89, 37.32){0.5}
\Vertex(147.29, 55.97){0.5} \Vertex(147.69, 65.05){0.5}
\Vertex(148.09, 61.89){0.5} \Vertex(148.48, 36.41){0.5}
\Vertex(148.86, 55.18){0.5} \Vertex(149.24, 64.24){0.5}
\Vertex(149.62, 61.08){0.5} \Vertex(149.99, 35.54){0.5}
\Vertex(150.36, 54.43){0.5} \Vertex(150.73, 63.48){0.5}
\Vertex(151.09, 60.31){0.5} \Vertex(151.44, 34.71){0.5}
\Vertex(151.80, 53.70){0.5} \Vertex(152.15, 62.74){0.5}
\Vertex(152.49, 59.58){0.5} \Vertex(152.83, 33.92){0.5}
\Vertex(153.17, 53.01){0.5} \Vertex(153.51, 62.04){0.5}
\Vertex(153.84, 58.87){0.5} \Vertex(154.17, 33.16){0.5}
\Vertex(154.49, 52.34){0.5} \Vertex(154.81, 61.37){0.5}
\Vertex(155.13, 58.20){0.5} \Vertex(155.45, 32.44){0.5}
\Vertex(155.76, 51.70){0.5} \Vertex(156.07, 60.72){0.5}
\Vertex(156.38, 57.54){0.5} \Vertex(156.68, 31.75){0.5}
\Vertex(156.98, 51.09){0.5} \Vertex(157.28, 60.10){0.5}
\Vertex(157.58, 56.92){0.5} \Vertex(157.87, 31.08){0.5}
\Vertex(158.16, 50.49){0.5} \Vertex(158.45, 59.50){0.5}
\Vertex(158.74, 56.31){0.5} \Vertex(159.02, 30.44){0.5}
\Vertex(159.30, 49.92){0.5} \Vertex(159.58, 58.92){0.5}
\Vertex(159.86, 55.73){0.5} \Vertex(160.13, 29.82){0.5}
\Vertex(160.41, 49.36){0.5} \Vertex(160.68, 58.35){0.5}
\Vertex(160.94, 55.17){0.5} \Vertex(161.21, 29.23){0.5}
\Vertex(161.47, 48.82){0.5} \Vertex(161.73, 57.81){0.5}
\Vertex(161.99, 54.62){0.5} \Vertex(162.25, 28.65){0.5}
\Vertex(162.51, 48.30){0.5} \Vertex(162.76, 57.29){0.5}
\Vertex(163.01, 54.10){0.5} \Vertex(163.26, 28.10){0.5}
\Vertex(163.51, 47.80){0.5} \Vertex(163.76, 56.78){0.5}
\Vertex(164.00, 53.58){0.5} \Vertex(164.24, 27.56){0.5}
\Vertex(164.48, 47.30){0.5} \Vertex(164.72, 56.28){0.5}
\Vertex(164.96, 53.09){0.5} \Vertex(165.19, 27.04){0.5}
\Vertex(165.43, 46.83){0.5} \Vertex(165.66, 55.80){0.5}
\Vertex(165.89, 52.61){0.5} \Vertex(166.12, 26.53){0.5}
\Vertex(166.35, 46.36){0.5} \Vertex(166.57, 55.33){0.5}
\Vertex(166.80, 52.14){0.5} \Vertex(167.02, 26.04){0.5}
\Vertex(167.24, 45.91){0.5} \Vertex(167.46, 54.88){0.5}
\Vertex(167.68, 51.68){0.5} \Vertex(167.90, 25.56){0.5}
\Vertex(168.11, 45.47){0.5} \Vertex(168.33, 54.44){0.5}
\Vertex(168.54, 51.24){0.5} \Vertex(168.75, 25.10){0.5}
\Vertex(168.96, 45.05){0.5} \Vertex(169.17, 54.01){0.5}
\Vertex(169.38, 50.81){0.5} \Vertex(169.59, 24.64){0.5}
\Vertex(169.79, 44.63){0.5} \Vertex(170.00, 53.59){0.5}
\Vertex(170.20, 50.38){0.5} \Vertex(170.40, 24.20){0.5}
\Vertex(170.60, 44.22){0.5} \Vertex(170.80, 53.18){0.5}
\Vertex(171.00, 49.97){0.5} \Vertex(171.19, 23.77){0.5}
\Vertex(171.39, 43.82){0.5} \Vertex(171.58, 52.77){0.5}
\Vertex(171.78, 49.57){0.5} \Vertex(171.97, 23.36){0.5}
\Vertex(172.16, 43.43){0.5} \Vertex(172.35, 52.38){0.5}
\Vertex(172.54, 49.18){0.5} \Vertex(172.73, 22.95){0.5}
\Vertex(172.91, 43.06){0.5} \Vertex(173.10, 52.00){0.5}
\Vertex(173.28, 48.80){0.5} \Vertex(173.47, 22.55){0.5}
\Vertex(173.65, 42.68){0.5} \Vertex(173.83, 51.63){0.5}
\Vertex(174.01, 48.42){0.5} \Vertex(174.19, 22.16){0.5}
\Vertex(174.37, 42.32){0.5} \Vertex(174.55, 51.26){0.5}
\Vertex(174.72, 48.06){0.5} \Vertex(174.90, 21.78){0.5}
\Vertex(175.08, 41.97){0.5} \Vertex(175.25, 50.91){0.5}
\Vertex(175.42, 47.70){0.5} \Vertex(175.60, 21.41){0.5}
\Vertex(175.77, 41.62){0.5} \Vertex(175.94, 50.55){0.5}
\Vertex(176.11, 47.35){0.5} \Vertex(176.28, 21.04){0.5}
\Vertex(176.44, 41.28){0.5} \Vertex(176.61, 50.21){0.5}
\Vertex(176.78, 47.00){0.5} \Vertex(176.94, 20.69){0.5}
\Vertex(177.11, 40.94){0.5} \Vertex(177.27, 49.88){0.5}
\Vertex(177.43, 46.67){0.5} \Vertex(177.60, 20.34){0.5}
\Vertex(177.76, 40.61){0.5} \Vertex(177.92, 49.55){0.5}
\Vertex(178.08, 46.34){0.5} \Vertex(178.24, 20.00){0.5}
\Vertex(178.40, 40.29){0.5} \Vertex(178.55, 49.22){0.5}
\Vertex(178.71, 46.01){0.5} \Vertex(178.87, 19.66){0.5}
\Vertex(179.02, 39.98){0.5} \Vertex(179.18, 48.91){0.5}
\Vertex(179.33, 45.69){0.5} \Vertex(179.48, 19.33){0.5}
\Vertex(179.64, 39.67){0.5} \Vertex(179.79, 48.59){0.5}
\Vertex(179.94, 45.38){0.5} \Vertex(180.09, 19.01){0.5}
\Vertex(180.24, 39.36){0.5} \Vertex(180.39, 48.29){0.5}
\Vertex(180.54, 45.08){0.5} \Vertex(180.68, 18.69){0.5}
\Vertex(180.83, 39.06){0.5} \Vertex(180.98, 47.99){0.5}
\Vertex(181.12, 44.77){0.5} \Vertex(181.27, 18.38){0.5}
\Vertex(181.41, 38.77){0.5} \Vertex(181.56, 47.69){0.5}
\Vertex(181.70, 44.48){0.5} \Vertex(181.84, 18.08){0.5}
\Vertex(181.99, 38.48){0.5} \Vertex(182.13, 47.40){0.5}
\Vertex(182.27, 44.19){0.5} \Vertex(182.41, 17.78){0.5}
\Vertex(182.55, 38.20){0.5} \Vertex(182.69, 47.12){0.5}
\Vertex(182.83, 43.90){0.5} \Vertex(182.96, 17.49){0.5}
\Vertex(183.10, 37.92){0.5} \Vertex(183.24, 46.84){0.5}
\Vertex(183.38, 43.62){0.5} \Vertex(183.51, 17.20){0.5}
\Vertex(183.65, 37.65){0.5} \Vertex(183.78, 46.56){0.5}
\Vertex(183.91, 43.35){0.5} \Vertex(184.05, 16.91){0.5}
\Vertex(184.18, 37.38){0.5} \Vertex(184.31, 46.29){0.5}
\Vertex(184.45, 43.08){0.5} \Vertex(184.58, 16.63){0.5}
\Vertex(184.71, 37.11){0.5} \Vertex(184.84, 46.03){0.5}
\Vertex(184.97, 42.81){0.5} \Vertex(185.10, 16.36){0.5}
\Vertex(185.23, 36.85){0.5} \Vertex(185.36, 45.76){0.5}
\Vertex(185.48, 42.55){0.5} \Vertex(185.61, 16.09){0.5}
\Vertex(185.74, 36.59){0.5} \Vertex(185.87, 45.51){0.5}
\Vertex(185.99, 42.29){0.5} \Vertex(186.12, 15.82){0.5}
\Vertex(186.24, 36.34){0.5} \Vertex(186.37, 45.25){0.5}
\Vertex(186.49, 42.03){0.5} \Vertex(186.61, 15.56){0.5}
\Vertex(186.74, 36.09){0.5} \Vertex(186.86, 45.00){0.5}
\Vertex(186.98, 41.78){0.5} \Vertex(187.10, 15.30){0.5}
\Vertex(187.23, 35.84){0.5} \Vertex(187.35, 44.75){0.5}
\Vertex(187.47, 41.53){0.5} \Vertex(187.59, 15.05){0.5}
\Vertex(187.71, 35.60){0.5} \Vertex(187.83, 44.51){0.5}
\Vertex(187.95, 41.29){0.5} \Vertex(188.06, 14.80){0.5}
\Vertex(188.18, 35.36){0.5} \Vertex(188.30, 44.27){0.5}
\Vertex(188.42, 41.05){0.5} \Vertex(188.53, 14.55){0.5}
\Vertex(188.65, 35.13){0.5} \Vertex(188.77, 44.03){0.5}
\Vertex(188.88, 40.81){0.5} \Vertex(189.00, 14.31){0.5}
\Vertex(189.11, 34.89){0.5} \Vertex(189.23, 43.80){0.5}
\Vertex(189.34, 40.58){0.5} \Vertex(189.45, 14.07){0.5}
\Vertex(189.57, 34.66){0.5} \Vertex(189.68, 43.57){0.5}
\Vertex(189.79, 40.35){0.5} \Vertex(189.90, 13.83){0.5}
\Vertex(190.01, 34.44){0.5} \Vertex(190.13, 43.34){0.5}
\end{picture}
\vspace{30pt}
\caption{$r_n$, defined in \eqn{NewEq009}, as function of $\log n$.}
\label{NewFig03}
\end{center}
\end{figure}
In \fig{NewFig03} we have plotted the observed behavior of 
\begin{equation}
r_n = \log\left(\frac{(4\pi)^{n/2}n^{3/2}}{2}\,
\big|\alpha_n-\alasym_n\big|\right)
\label{NewEq009}
\end{equation}
on the first Riemann sheet, against $\log n$. The coefficients clearly
converge to the computed behavior, and we can even distinguish that the
leading corrections go as $n^{-5/2}$; the four separate lines that emerge
are just the four different forms of $c_n$. The series expansion for Riemann
sheet $-1$ are simply obtained from
\begin{equation}
\alpha_n^{(-1)} = (-)^n\alpha_n^{(1)}\;\;.
\end{equation}
\paragraph{Higher Riemann sheets.} We first consider positive
sheet label $s=3,5,7,\ldots$ and put $k=(s-1)/2$.
We then have
\begin{equation}
x_{k,1} = -x_{k,2} = \sqrt{4\pi k}\exp(i\pi/4)\;\;\;,\;\;\;
\beta_{k,1} = \beta_{k,2} = (1+i)\sqrt{4\pi k}\;\;.
\end{equation}
As we have already seen $\alpha_n$ vanishes for odd $n$, and for
even $n$ we have the following asymptotic form:
\begin{equation}
   \alpha_{4p}^{(s)}  \sim  {2(1+i)\sqrt{k}\over(4\pi k)^{2p}}
   (-)^{p+1}\;\;,\quad
   \alpha_{4p+2}^{(s)}  \sim  {2(1-i)\sqrt{k}\over(4\pi k)^{2p+1}}
   (-)^{p}\;\;,
\end{equation}
for integer $p$. For negative $s$, we use \eqn{NewEq006}, which also holds 
asymptotically.

\newpage
\section{Conclusions}
For the $L_2^*$-discrepancy and the Lego discrepancy, we have addressed the
problem that non-trivial extremal points of the the action in the path integral
representation of the generating function of quadratic discrepancies, called
instantons, might spoil the saddle point approximation around the trivial
extremal point. We have shown that instantons appear in both cases, but only if
the order parameter $z$ of the generating function $G$ is larger then a certain
positive value. In the Lego-case this value is half of the size of the smallest
bin, and in the $L_2^*$-case it is $\half\pi^2$, the smallest positive value of
$z$ at which $G(z)$ in the limit of an infinite number of random points $N$ has
a singularity. Although the instantons do not threaten the perturbation
expansion, they cause $G(z)$ to be undefined for asymptotically large $N$ when
the real part of $z$ is larger then the mentioned values.

For the analyses in the $L_2^*$-case, a numerical method to investigate the
Riemann sheet structure of the solution of certain algebraic complex equations
is used, which is treated in \Sec{CAARSS}.  The method is in particular
suitable for the determination of  the series expansions around the origin on
the different sheets and the asymptotic behavior of their coefficients. The
results of the numerical analyses have been justified by the fact that only
finite computer accuracy was required in the specific calculations.

\section{Appendices}
\vspace{-1.2\baselineskip}
\Appendix{:\; Matrices of the form $A_{n,m}=a_n\de_{n,m}+\epsilon b_nb_m$
          \label{App4A}}
The eigenvalues $\la$ of a real-valued matrix $A$ are given by the zeros of the 
characteristic polynomial $P_A$. If $A$ is an $M\times M$ matrix with
matrix elements
\begin{equation}
   A_{n,m} \;=\; a_n\de_{nm} +\epsilon\, b_nb_m \;\;,\quad 
   a_n,b_n\in\Real\;,\;\;n=1,\ldots,M\;\;,\quad\epsilon=\pm 1\;\;,
\label{FieEq037}   
\end{equation}
then the characteristic polynomial $P_A$ is given by 
\begin{equation}
   P_A(x) \;=\; Q_A(x)\prod_{n=1}^M(a_n-x) \;\;,\quad
   Q_A(x) \;=\; 1+\epsilon\sum_{m=1}^M\frac{b_m^2}{a_m-x} \;\;,
\end{equation}
which is easily derived using $P_A(x)=\sum_{\pi\in
S_M}\prod_{n=1}^M[A_{n,\pi(n)}-x\de_{n,\pi(n)}]$, where the sum is over all
permutations of $(1,\ldots,M)$.  Without loss of generality, we assume that the
coefficients $a_n$ are ordered such that $a_1\leq a_2\leq\cdots\leq a_M$.
If a number of $d_n$ coefficients $a_n$ take the same value, that is, if 
$a_n$ is $d_n$-fold degenerate, then a $(d_n-1)$-fold degenerate 
eigenvalue of $A$ is given by $\la=a_n$. The remaining eigenvalues are given by the zeros of the 
function $Q_A$.
Except of the poles at $x=a_n$, $n=1,\ldots,M$, this function is continuous 
and differentiable on the whole of $\Real$. Furthermore, the sign of the 
derivative is equal to $\epsilon$. This
means that for each zero $\la$ of $Q_A$ except one, there is an $n$, such 
that $a_n<\la<a_{n+m}$ for the nearest and non-equal neighbor $a_{n+m}$ of 
$a_n$. The one other zero is smaller than $a_1$ if $\epsilon=-1$, 
and larger than $a_M$ if $\epsilon=1$. This 
is easy to see because $\lim_{x\ra\infty}Q_A(x)=\lim_{x\ra-\infty}Q_A(x)=1$.

\Appendix{:\; Derivation of \eqn{FieEq032}\label{App4B}}
We use the definitions of $T(E)$ as the r.h.s.~of \eqn{FieEq026} and $T_1(E)$ 
as given in \eqn{FieEq030}:
\begin{equation}
   T(E) 
   \,=\, \int\limits_{\phi_-}^{\phi_+}\frac{d\phi}{\sqrt{E-U(\phi)}} \;,\quad
   T_1(E) 
   \,=\, \int\limits_{\phi_-}^{\phi_+}\frac{\phi\,d\phi}{\sqrt{E-U(\phi)}} 
   \;,\quad
   U(\phi) 
   = e^\phi - \phi - 1 \;.
\end{equation}
Because the end points $\phi_+$ and $\phi_-$ depend on $E$ such that 
$E-U(\phi_\pm)=0$, we can use Leibnitz's rule for differentiation under the 
integral sign to write
\begin{equation}
   T(E) \;=\; 
   2\,\frac{d}{dE}\int\limits_{\phi_-}^{\phi_+}\sqrt{E-U(\phi)}\,d\phi \;\;.
\end{equation}
Now we write $\sqrt{E-U(\phi)}=(E-e^\phi+\phi+1)(E-U(\phi))^{-1/2}$ and use 
that $1-e^\phi=dU/d\phi$, so that 
\begin{equation}
   T(E) \;=\; 
   2\,\frac{d}{dE}\left(
                   ET(E) + T_1(E) - \int\limits_{\phi_-}^{\phi_+}
                   \frac{1}{\sqrt{E-U(\phi)}}\frac{dU}{d\phi}\,d\phi\right)\;\;.
\end{equation}
But the last integral is equal to zero, and as a result, we obtain 
\eqn{FieEq032}.

\clearemptydoublepage

\chapter{Gaussian limits for discrepancies\label{GausChap}}

This chapter deals with the calculation of the generating function of the
probability densities of quadratic discrepancies in the limit of a large number
of truly random points. These densities depend on the dimension
$s$ of the integration region, or, in the case of the Lego discrepancy,
on the number of bins $M$ the integration region is dissected in.  We will
derive a `Law of Large Number of Modes', which describes the conditions under
which these densities approaches a normal density if $s$ or $M$ become large.
Throughout this discussion, we shall only consider the asymptotic limit of a
very large number of random points. This implies that, in this chapter, we
cannot make any statements on how the number of points has to approach infinity
with respect to $s$ or $M$, as was for instance done in \cite{Leeb}. 

\vspace{\baselineskip}

\minitoc

\section{The generating function}
We want to calculate the the probability density $H$ of quadratic discrepancies
in the limit of $N\ra\infty$, where $N$ is the number of points in the point
set. We will use the generating function in this limit, that we denote by 
\begin{equation}
   G_0(z) \;\df\; \lim_{N\ra\infty}\Exp(\,e^{zD_N}\,) \;\;, 
\end{equation}
so that the probability density is given by the inverse Laplace transform of
$G_0$ (\Sec{ProSecGen}). It results in the weak limit of $H$. 
Starting from \eqn{FieEq005}, it is easy to see that
$G_0$ is given by  
\begin{equation}
   G_0(z) \;=\; \int\exp(\,z\leb{\phi^2}+z\leb{\phi}^2\,)\,d\mu[\phi] \;\;.
\end{equation}
In the Landau gauge, the boundary conditions on the functions $\phi$ that give
a contribution are such that $\leb{\phi}=0$ (notice that $G_0$ contains the
same gauge freedom as $G$).  We can apply the formalism of \Sec{GmFdSec3}, 
and conclude that $\log G_0(z)$ is equal to the sum of the
contributions of all possible connected diagrams consisting only of vertices
with two legs. Consequently, they are of the form 
\begin{equation}
   \diagram{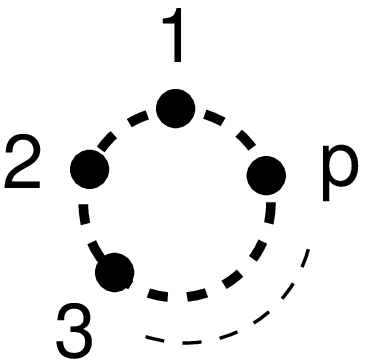}{50}{16} \quad\quad,
\end{equation}
and carry a symmetry factor equal to $1/2p$, where $p$ is the number of
vertices. Every vertex contributes with a factor $2z$, and represents a 
convolution of reduced two-point functions $\twoB$, so that
\begin{equation}
   \log G_0(z) 
   \;=\; \sum_{p=1}^\infty\frac{(2z)^p}{2p}\,R_p \;\;,
\label{GauEq010}			 
\end{equation}
with
\begin{equation}
   R_p 
   \;\df\; \int_{\Kube^p}\twoB(x_1,x_2)\twoB(x_2,x_3)\cdots
                         \twoB(x_{p-1},x_p)\twoB(x_p,x_1)\,
                         dx_1dx_2\cdots dx_{p-1}dx_p\;\;.
\label{GauEq009}			 
\end{equation}
The coefficients $R_p$ can be written in terms of the eigenfunctions $u_n$ and
the eigenvalues $\si^2_n$ of the two-point function $\twoC$ interpreted as an
integration kernel (\Sec{GMOACB}). We can use the expression of
\eqn{defEq008} for $\twoB$, which tells us that we have to repeatedly calculate
the integral
\begin{equation}
   \vGa_{n,m}
   \;\df\; \intk\si_n(u_n(x)-\leb{u_n})\si_m(u_m(x)-\leb{u_m})\,dx
   \;=\;   \si_n^2\de_{n,m} - \si_n\leb{u_n}\si_m\leb{u_m}  \;\;. 
\end{equation}
$\vGa$ is an infinite dimensional matrix, and the coefficients $R_p$ can be 
written in terms of $\vGa$ through
\begin{equation}
   R_p = \Tr(\vGa^p) \;\;,
\end{equation}
where $\vGa^p$ denotes the $p$-fold matrix product, and $\Tr$ the
trace, i.e., the sum over the diagonal elements.  The generating function
itself can also be expressed directly in terms of $\vGa$, since
\begin{equation}
   \log G_0(z)
   \;=\; \sum_{p=1}^\infty\frac{(2z)^p}{2p}\,\Tr(\vGa^p)
   \;=\; \Tr\left(\sum_{p=1}^\infty\frac{(2z\vGa)^p}{2p}\right)
   \;=\; -\Tr(\,\half\log(1-2z\vGa)\,) \;\;,
\end{equation}
so that 
\begin{equation}
   G_0(z) = (\,\det(1-2z\vGa)\,)^{-1/2} \;\;,
\label{GauEq007}   
\end{equation}
where we used the well known rule that, for a general matrix $A$, 
$\det(e^A)=e^{\Tr(A)}$. In \App{App5A}, it is shown how $G_0$ can be written 
in terms of the strengths $\si_n$ and the weights $\leb{u_n}$, with the result 
that 
\begin{equation}
   G_0(z) \;=\; \frac{e^{\psi(z)}}{\sqrt{\chi(z)}} \;\;,
\label{GauEq003}   
\end{equation}
where
\begin{equation}
   \psi(z) \;=\; -\sum_{n}\half\log(1-2z\si_n^2)\;\;,\qquad\textrm{and}\qquad
   \chi(z) \;=\; 1 + \sum_{n}\frac{\leb{u_n}^2}{1-2z\si_n^2} \;\;.
\label{GauEq004}   
\end{equation}
Notice that if the basis is in the Landau gauge, then $\leb{u_n}=0$ for all
functions and $\chi(z)=1$, so that the generating function is just given by
$\prod_n(1-2z\si_n^2)^{-1/2}$. In the Landau gauge, the matrix $\vGa$ is
diagonal, so that this result follows directly from \eqn{GauEq007}. If the
eigenvalues of $\vGa$, in a general gauge, are denoted $\la_n$, then the
generating function is given by $\prod_n(1-2z\la_n)^{-1/2}$.

\subsection{Standardized variables and the Gaussian limit}
We have now derived the expression for $G_0(z)$ in the large-$N$ limit. Given
the form of $\vGa$, we can now compute $H(t)$ for given discrepancy $t$, if
only numerically; in fact this was done for the $L_2^*$-discrepancy in
\cite{jhk} for several dimensionalities. In some special cases, $H(t)$ can
even be given as an analytic expression \cite{hk1,hk2}. Here, however, we are
interested in possible Gaussian limits, and therefore it is useful use the
standardized variable instead of the discrepancy itself 
(\Sec{SecProbRV}). 
Notice that the expectation value and the variance of the
discrepancy are just given by 
\begin{equation}
   \Exp(D_N) = R_1 \qquad\textrm{and}\qquad 
   \Var(D_N) \overset{N\ra\infty}{\longrightarrow} 2R_2 \;\;.
\end{equation}
The generating function $\Hat{G}$ of the standardized variable is given by 
\begin{equation}
   \Hat{G}_0(z)
   \;=\; \exp\left(\half z^2 
                   + \sum_{p=3}^\infty\frac{(\sqrt{2}\,z)^p}{2p}\,\sqrt{\ga_p}
                                      \right)\;\;,\qquad\textrm{with}\qquad
   \ga_p \df \frac{R_p^2}{R_2^p}\;\;.				      
\end{equation}
All information on the particulars of the discrepancy are now
contained in the constants $\ga_p$, and we have that {\it the standardized 
  probability density approaches the normal density whenever
  $\ga_p\to 0$ for all $p\ge3$.\/} It remains to examine under what
circumstances this can happen.

\subsection{A Law of Large Number of Modes}
Since we know that $G_0(z)$ has no
singularities for negative values of $\textrm{Re}\,z$, the eigenvalues of
$\vGa$ are also nonnegative, and we may write
\begin{equation}
  \Tr(\vGa^k) 
  \;=\; 
  \sum_n\la_n^k
  \quad,\quad
  \ga_k 
  \;=\; 
  \left(
    \sum_n\la_n^k\right)^2\left(\sum_n\la_n^2
  \right)^{-k}
  \quad,\quad
  \la_n\ge0
  \;\;,
\label{GauEq011}
\end{equation}
where the various eigenvalues have been denoted by $\la_n$. Note that
the sum may run over a finite or an infinite number of eigenvalues,
but all these sums must converge since $\Exp(D_N)$ is finite. Note,
moreover, that $\ga_k$ is homogeneous of degree zero in the $\la_n$:
therefore, any scaling of the eigenvalues by a constant does not
influence the possible Gaussian limit (although it will, of course,
affect the mean and variance of $D_N$).

We now proceed by noting that $\ga_{k+1}\le\ga_k$, because
\begin{equation}
  \left( \sum_n\la_n^{k+1} \right)^2 
  \;\le\;
  \left( \sum_n\la_n^{2k} \right)
  \left( \sum_n\la_n^2 \right) 
  \;\le\;
  \left( \sum_n\la_n^k \right)^2
  \left( \sum_n\la_n^2 \right)
  \;\;,
\label{GauEq012}
\end{equation}
where the first inequality is simply the Schwarz inequality, and the
second one holds because the $\la_n$ are nonnegative. This means that
$\ga_k$ will approach zero for $k>3$, whenever $\ga_3$ approaches
zero. To see when this happens we define
\newcommand{\rhosup}{\rho_{\textrm{s}}}
\begin{equation}
  \rho_n 
  \;\df\; 
  \frac{\la_n}{\sqrt{\sum_m\la_m^2}}
  \qquad\textrm{and}\qquad
  \rhosup \df \sup_{n} \rho_n
  \;\;,
\end{equation}
so that $\sum_n\rho_n^2=1$. It is then trivial to see that
\begin{equation}
  \rhosup^3 \;\le\; \ga_3^{1/2} \;\le\; \rhosup
  \;\;,
\end{equation}
from which we derive that 
{\sl the necessary and sufficient condition for the discrepancy distribution, 
in the limit of an infinite number of points in the point set, to
approach a Gaussian is that}
\newcommand{\lasup}{\lambda_{\textrm{s}}}
\begin{equation}
  \frac{\lasup^2}{\sum_n\la_n^2} 
  \;\to\; 0
  \quad,\quad
  \lasup \df \sup_n \la_n
  \;\;.
  \label{GauEq005}
\end{equation} 
The Gaussian limit is thus seen to be equivalent to the statement that
even the largest eigenvalue becomes unimportant. 

Clearly, a necessary
condition for this is that the total number of non-vanishing
eigenvalues (number of {\it modes\/}) approaches infinity.
Incidentally, the condition (\ref{GauEq005}) also implies that
\begin{equation}
  \lasup \;\to\; 0
  \;\;,\qquad
  \sum_n\la_n^2 \;\to\; 0  
  \;\;,
\end{equation}
for all those discrepancies that have $\Exp(D_N)=\sum_n\la_n=1$.  This
is eminently reasonable, since a distribution centered around 1 and
(by construction) vanishing for negative argument can only approach a
normal distribution if its variance approaches zero.  On the other
hand, the condition $\lasup\to0$ is by itself {\it not\/} sufficient, as
proven by a counterexample given in \App{App5B}.

Another piece of insight can be obtained if we allow the eigenvalues
to take on random values. We may introduce the rather dizzying concept
of {\it an ensemble of different definitions of discrepancy,\/} each
characterized by its set of eigenvalues (all nonnegative)
$\vec{\la}=\{\la_1,\la_2,\ldots,\la_M\}$, with the usual constraint
that they add up to 1; we keep $M$ finite for simplicity.  A natural
probability measure on this ensemble is given by the probability
density $P_{\la}(\vec{\la})$ of the random vector $\la$:
\begin{equation}
  P_{\la}(\vec{\la}) 
  \;\df\; 
  \Ga(M)\,
  \delta\left(\sum_{n=1}^M\la_n - 1\right)
  \;\;.
\end{equation}
Here $\Ga$ denotes Euler's gamma-function and $\de$ stands for the Dirac
delta-distribution.  It is easily computed that the expectation and variance of
$R_k=\sum_n\la_n^k$ are given, for large $M$, by
\begin{equation}
  \Exp(R_k) \sim \frac{k!}{M^{k-1}}
  \quad,\quad
  \Var(R_k) \sim \frac{(2k)!-(1+k^2)(k!)^2}{M^{2k-1}}
  \;\;,
\end{equation}
so that the $R_k$ become sharply peaked around their expectation for
large $M$. In that case, we have
\begin{equation}
  \ga_3 \;\sim\; \frac{9}{2M}
  \;\;,
\end{equation}
and we see that, in the above sense, almost all discrepancies have a
Gaussian distribution in the limit where $M$, the number of modes,
approaches infinity.

\section{Applications to different examples}
\subsection{Fastest approach to a Gaussian limit}
We now examine the various definitions of discrepancies, and assert their
approach to a Gaussian limit. Usually this is envisaged, for instance in
\cite{Leeb}, as the limit where the dimensionality of the integration
region becomes very large. But, as we have shown, this is only a special case
of the more general situation where the number of relevant modes becomes very
large: another possible case is that where, in one dimension, the number of
modes with essentially equal strength $\si_n$ becomes very large. As an
illustration, consider the case where the basis functions with the Gaussian
measure are orthonormal and $M$ of the nontrivial modes have equal strength
$\si_n^2=1/M$, and the rest have strength zero. The moment-generating function
then takes on a particularly simple form, and so does the discrepancy
distribution \cite{hk2}:
\begin{equation}
  \log G_0(z) 
  \;=\; 
  -\frac{M}{2}\log\left(1-\frac{2z}{M}\right)
  \quad,\quad
  H(t)
  \;=\; 
  \frac{(M/2)^{M/2}}{\Ga(M/2)}\,t^{M/2-1}e^{-tM/2}
  \;\;.
  \label{GauEq006}
\end{equation}
It is easily seen that the gamma-distribution $H(t)$ approaches a normal one
when $M$ becomes very large. At the same time, we see the `physical' reason
behind this: it is the fact that the singularity of $G_0(z)$ in the complex
plane (in the more general case, the singularity nearest to $z=0$) moves away
to infinity.  One observation is relevant here: in the inverse Laplace
transform, to go from $G_0$ to $H$, we have kept the integration along the
imaginary axis $\textrm{Re}\,z=0$.  We might consider performing a saddle-point
integration, with a non-vanishing value of $\textrm{Re}\,z$. That may give us,
for a finite number of modes, a good approximation to the actual form of
$H(t)$.  It is quite possible, and, indeed, it happens in the above
equal-strength model, that this approximation is already quite similar to a
Gaussian.  In the equal-strength model, a saddle-point approximation for $H(t)$
gives precisely the form of \eqn{GauEq006}, the only difference being that
$\Ga(M/2)$ is replaced by its Stirling approximation.  On the other hand, for
not-so-large $M$, this form is not too well approximated by a Gaussian centered
around $t=1$, since the true maximum resides at $t=1-2/M$. Nevertheless, in
this chapter we are only interested in the limiting behavior of $H(t)$, and we
shall stick to the use of condition (\ref{GauEq005}) as an indicator of the
Gaussian limit.

One interesting remaining observation is the following.  For any
finite number $M$ of eigenvalues $\la_n$ $(n=1,2,\ldots,M)$, the {\em
  smallest\/} value of the indicator $\lasup^2/\sum_n\la_n^2$ is obtained
when $\la_n=1/M$ for all $n$. In this sense, the equal-strengths model
gives, for finite $M$, that discrepancy distribution that is closest
to a Gaussian.

\subsection[The $L_2^*$-discrepancy]
           {The $L_2^*$-discrepancy\label{GauSec01}}
Here we shall discuss the standard $L_2^*$-discrepancy (\Sec{SecDefL2}).
The eigenfunctions $u_{\vn}$ are equal to
$2^{s/2}\prod_{\nu=1}^s\cos\left((n_\nu+\half)\pi x^\nu\right)$ so that
$\leb{u_{\vn}}=2^{s/2}\si_{\vn}$ where the strengths, and the matrix $\vGa$,
are given by
\begin{equation}
   \vGa_{\vm,\vn}
   \;=\; \si_{\vm}^{2}\delta_{\vm,\vn}-2^s\si_{\vm}^{2}\si_{\vn}^{2} 
   \;\;,\qquad
   \si_{\vn}^{2} 
   \;=\; \left(\frac{4}{\pi^2}\right)^s
         \prod_{\nu=1}^s\frac{1}{(2n_\nu+1)^2} \;\;. 
\end{equation}
The components $n_\nu$ of the integer vector $\vn$ can take all
non-negative integer values, including zero. The eigenvalue equation for the
eigenvalues $\la$ of $\vGa$ can be written down easily:
\begin{equation}
  \prod_{\vn}(\si_{\vn}^{2}-\la)
  \left[
    1-2^s\sum_{\vm}\frac{\si_{\vm}^{4}}{\si_{\vm}^{2}-\la}
  \right] 
  \;=\; 0
  \;\;.
\end{equation}
The strengths $\si_{\vn}$ are degenerate in the values they take. Labeling the
strengths with different values by $\si_{p}$ with
$p=\prod_{\nu=1}^s(2n_\nu+1)$, the degeneracy is given by
\begin{equation}
  \Qw(p) 
  \;\df\; 
  \sum_{\vn\geq0} 
  \theta\left(p=\prod_{\nu=1}^s (2n_\nu+1)\right)\;\;.
\end{equation}
We introduced the logical step function here, which is simply defined by 
\begin{equation}
   \theta(\Pi) \;\df\; \begin{cases}
                        1 &\textrm{if $\Pi$ is true} \;,\\
			0 &\textrm{if $\Pi$ is false}\;.
		     \end{cases}	
\end{equation}
So $\la=\si_{p}^{2}$ is a solution to the eigenvalue equation with
a $(\Qw(p)-1)$-fold degeneracy. If we factorize these solutions we
obtain the following equation for the remaining eigenvalues: 
\begin{equation}
  1-2^s\sum_{p}\Qw(p)\,\frac{\si_{p}^{4}}{\si_{p}^{2}-\la}
  \;=\;
  0
  \;\;.
\end{equation}
Some assertions concerning the remaining eigenvalues can be made using
this equation. On inspection, it can be seen that there are no
negative solutions, nor solutions larger than $\si_{1}^{2}$, so that
$\si_1^2$ can be used as an upper bound of the eigenvalues of $\vGa$.
If we order the $\la$ such that $\la_1\geq\la_3\geq\ldots$, then
$\si^2_1\geq\la_1\geq\si^2_3\geq\la_3\geq\ldots$. This implies that
$\Tr(\vGa^k)=\sum_{p}\Qw(p)\,\si_{p}^{2k}-\epsilon$ where
$0\leq\epsilon\leq\si_{1}^{2k}$. Now we have
\begin{equation}
   \sum_{p}\Qw(p)\,\si_{p}^{2k}
  \;=\; 
  \left(\frac{4}{\pi^2}\right)^{ks}\xi(2k)^s
  \quad,\quad
  \xi(p) 
  \;=\; 
  \sum_{n\ge0}{1\over(2n+1)^p}
  \;\;,  
\end{equation}
and therefore, for $k\ge3$, that 
\begin{equation}
  \ga_k 
  \;\le\; 
  \left(\frac{\xi(2k)^2}{\xi(4)^k}\right)^s
  \left(1-2\brfr{4}{5}^s+\brfr{2}{3}^s\right)^{-k}
  \;\;.
\end{equation}
The second factor decreases monotonically from $(15)^k$ for $s=1$
to one as $s\to\infty$; for the first factor, we note that
$1<\xi(2k)<\xi(4)$ for all $k>2$. Therefore $\ga_k$ can be made
arbitrarily small by choosing $s$ large enough, and the Gaussian limit
of high dimensionality is proven. Note, however, that the approach is
not particularly fast: for large $s$, we have
$\ga_3\sim(24/25)^s\sim\exp(-s/25)$, so that $s$ has to become of the
order of one hundred or so to make the Gaussian behavior manifest. In
fact, this was already noted by explicit numerical computation in \cite{jhk}.

\subsection{The Fourier diaphony\label{GauSecFou}}
In the case of the Fourier diaphony (\Sec{SecDefDia}), the eigenfunctions are in the Landau 
gauge by definition, so that the matrix $\vGa$ is just given by 
\begin{equation}
  \vGa_{\vm,\vn} 
  \;=\; 
  \si_{\vn}^{2}\delta_{\vm,\vn} 
  \;\;,
\end{equation}
with the strengths $\si_{\vn}$ as in \eqn{defEq010}.  The normalization of the
strengths ensures that $\Exp(D_N)=1$, independent of $s$. In this case, keeping
in mind that sines and cosines occur in the eigenfunctions with equal strength,
we have to consider the multiplicity function
\begin{equation}
  \Qfp(p) 
  \;\df\; 
  \sum_{\vn\ge0}\theta\left(p=\prod_{\nu}r(k_{\nu}(\vn))\right)
  \;\;,
\end{equation}
Actually, before assigning a strength $\si_{\vn}$, or rather $\si^2_p$,
we have to know the behavior of $\Qfp(p)$ in order to ensure
convergence of $\Exp(D_N)$. In order to do so, we introduce the
Dirichlet generating function for $\Qfp(p)$:
\begin{equation}
  F^{[1]}_s(x) 
  \;\df\; 
  \sum_{p>0}\frac{\Qfp(p)}{p^x} \;=\; \left(1+2\zeta(x)\right)^s
  \;\;,
\end{equation}
where we use the Riemann $\zeta$ function. Since this function (and,
therefore, $F^{[1]}_s(x)$ as well), converges for all $x>1$, we are
ensured that $\Qfp(p)$ exceeds the value $cp^{1+\epsilon}$ at
most for a finite number of values of $p$, for all positive $c$ and
$\epsilon$. This is proven in \App{App5C}. It is therefore sufficient
that $\si^2_p$ decreases as a power (larger than 1) of $p$. In fact,
taking 
\begin{equation}
  \si^2_p \;=\; c p^{-\be}
  \quad,\quad
  \be\;>\; 1
  \;\;,
\end{equation}
we immediately have that
\begin{equation}
  R_k 
  \;=\; 
  \sum_{\vn>0}\si_{\vn}^{2k} 
  \;=\; 
  \sum_{p>0}\Qfp(p)\si^{2k}_p - \si^{2k}_1
  \;=\; 
  c^k\left[\left(1+2\zeta(k\be)\right)^s - 1\right]
  \;\;,
\end{equation}
which, for given $\be$, fixes $c$ such that $R_1=\Exp(D_N)= 1$, and,
moreover, gives 
\begin{equation}
  \ga_3 \sim a(\be)^s\quad\textrm{as}\quad s\to\infty
  \quad,\quad
  a(\be) 
  \;=\; 
  {\left(1+2\zeta(3\be)\right)^2\over \left(1+2\zeta(2\be)\right)^{-3}}
  \;\;.
\end{equation}
In \Sec{SecDefDia}, the value $\be=2$ is used,
with $a(2)\sim 0.291$.  The supremum of $a(\be)$ equals $1/3$, as
$\be\to\infty$, and the (more interesting) infimum is $a(1)$, about
$0.147$.  We conclude that, for all diaphonies of the above type, the
Gaussian limit appears for high dimensionality. For large $\be$, where
the higher modes are greatly suppressed, the convergence is slowest,
in accordance with the observation that the `equal-strength' model
gives the fastest convergence; however, the convergence is still much
faster than for the $L_2^*$-discrepancy, and the Gaussian
approximation is already quite good for $s\sim4$. The {\it fastest\/}
approach to the Gaussian limit occurs when we force all modes to have
as equal a strength as is possible within the constraints on the
$\be$. The difference between the supremum and infimum of $a(\be)$ is,
however, not much more than a factor of $2$.

Another possibility would be to let $\si^2_p$ depend exponentially on
$p$. In that way one can ensure convergence of the $R_k$ while at the
same time enhancing as many low-frequency modes as possible.  It is
proven in \App{App5C} that the function
\begin{equation}
  F^{[2]}_s(x) 
  \;\df\; 
  \sum_{p>0}\Qfp(p)\,x^p
\label{GauEq008}  
\end{equation}
has radius of convergence equal to one, and therefore we may take
$\si^2_p = (\be')^p$ with $\be'$ between zero and one. If we choose
$\be'$ to be very small, we essentially keep only the modes with
$p=1$, and therefore in that case we have $\ga_3\sim1/(3^s-1)$.  This
is of course in reality the same type of discrepancy as the above one,
with $\be\to\infty$. On the other hand, taking $\be'\to1$ we arrive at
$\ga_3\to0$ (see, again, \App{App5C}).  The difference with the first
model is, then, that we can approach the Gaussian limit arbitrarily
fast, at the price, of course, of having a function $\twoB(x_k,x_l)$
that is indistinguishable from a Dirac $\delta$-distribution in
$x_k-x_l$, and hence meaningless for practical purposes.

\subsubsection{Fourier diaphony with sum clustering}
In the above, we have let the strength $\si_{\vn}$ depend on the {\it
  product\/} of the various $r(n_{\nu})$. This can be seen as mainly a
matter of expediency, since the generalization to $s>1$ is quite
simple in that case. From a more `physical' point of view, however,
this grouping of the $\si$ is not so attractive, if we keep in mind
that each $\vn$ corresponds to a mode with wave vector $\vk(\vn)$.
Under the product rule, wave vectors differing only in their direction
but with equal length may acquire vastly different weights: for
instance, $\vk = (m\sqrt{s},0,0,\ldots)$ and $\vk=(m,m,m,\ldots)$ have
equal Euclidean length, $m\sqrt{s}$, but their strengths under the
product rule are $1/(sm^2)$ and $1/(m^{2s})$, respectively. This lack
of `rotational' symmetry could be viewed as a drawback in a
discrepancy distinguished by its nice `translational' symmetry. One
may attempt to soften this problem by grouping the strengths
$\si_{\vn}$ in another way, for instance by taking
\begin{equation}
  \si_{\vn} 
  \;=\; 
  \si\left(\sum_{\nu}k(n_{\nu})\right)
  \;\;,
\end{equation}
so that $\si$ depends on the sum of the components rather than on
their product. The multiplicity of a given strength now becomes, in
fact, somewhat simpler:
\begin{equation}
  \Qfs(p) 
  \;\df\; 
  \sum_{\vn>0}\theta\left(p=\sum_{\nu=1}^sk(n_{\nu})\right)
  \;=\; 
  \sum_{m\ge0}\binom{s}{m}\binom{s-1+p-m}{p-m}
  \;\;,
\end{equation}
where the last identity follows from the generating function
\begin{equation}
  F^{[3]}_s(x) 
  \;\df\; 
  \sum_{p\ge0}\Qfs(p)\,x^p 
  \;=\; 
  \left(\frac{1+x}{1-x}\right)^s
  \;\;.
\end{equation}
This also immediately suggests the most natural form for the strength:
$\si^2_{\vn} = \be^p$, where $p$ is $\sum_{\nu}k(n_{\nu})$ as above.
We see that $R_1$ converges as long as $\be<1$, and moreover,
\begin{equation}
  \ga_3 
  \;=\; 
  \frac{\left[\left(\frac{1+\be^3}{1-\be^3}\right)^s-1\right]^2}
  {\left[\left(\frac{1+\be^2}{1-\be^2}\right)^s-1\right]^3}
  \;\sim\; 
  a(\be)^s
  \;\;,
\end{equation}
where $a(\be)$ has supremum $a(0)=1$, and decreases monotonically with
increasing $\be$. For $\be$ close to one, we have
$a(\be)\sim4(1-\be)/9$, so that the Gaussian limit can be reached as
quickly as desired (again with the reservations mentioned above). At
the other extreme, note that for very small $\be$ we shall have
\begin{equation}
  \ga_3 
  \;\sim\; 
  \frac{1}{2s}\quad\textrm{if}\quad s\be^2 \ll 1
  \;\;.
\end{equation}
This just reflects the fact that, for extremely small $\be$, only the
$2s$ lowest nontrivial modes contribute to the discrepancy; and even
in that case the Gaussian limit is attained, although much more
slowly.  The criterion that determines whether the behavior of
$\ga_3$ with $s$ and $\be$ is exponential or of type $1/(2s)$ is seen
to be whether $s\be^2$ is considered to be large or small, respectively.

Another alternative might be a power-law-like behavior of the
strengths, such as $\si^2_p = 1/p^\al$. Also in this case we may
compute the $R_k$, as follows:
\begin{equation}
  R_k 
  \;=\; 
  \sum_{p>0}\Qfs(p)\,\frac{1}{p^{k\al}} 
  \;=\;
  \frac{1}{\Ga(k\al)}\int\limits_0^{\infty}\;
  z^{k\al-1}\left(F^{[3]}_s(e^{-z})-1\right)\,dz
  \;\;,
\end{equation}
from which it follows that $\al>s$ to ensure convergence of
$\Exp(D_N)$. In the large-$s$ limit, we therefore find that, also in
this case, $\ga_3\to1/(2s)$.

\subsubsection{Fourier diaphony with spherical clustering}
A clustering choice which is, at least in principle, even more
attractive from the symmetry point of view than sum clustering, is to
let $\si_{\vn}$ depend on $|\vk(\vn)|^2$, hence assuring the maximum
possible amount of rotational invariance under the constraint of
translational invariance.  We therefore consider the choice
\begin{equation}
  \si_{\vn}^2 
  \;=\; 
  \exp\left(-\al\sum_{\nu}k(n_{\nu})^2\right)
  \;\;.
\end{equation}
For the function $\twoB(x_1,x_2) = \twoB(x_1-x_2)$ we now have the
following two alternative forms, related by Poisson summation:
\begin{align}
  \twoB(x) 
  & \;=\; 
  -1 + \prod_{\nu=1}^s\left(
    \sum_{k=-\infty}^{+\infty}e^{-\al k^2}\cos(2\pi
    kx^{\nu})\right)
  \notag\\
  & \;=\; 
  -1 + \left(\frac{\pi}{\al}\right)^{s/2}
  \sum_{\vec{m}}\exp\left(-\frac{\pi^2(\vec{x}+\vec{m})^2}{\al}\right)
  \;\;,
\end{align}
of which the first converges well for large, and the second for small,
values of $\al$; the sum over $\vec{m}$ extends over the whole integer
lattice. The $R_k$ are, similarly, given by
\begin{align}
  R_k  
  & \;=\; 
  \left(
    \sum_{q=-\infty}^{+\infty}e^{-k\al q^2}
  \right)^s-1
  \notag\\
  & \;=\; 
  \left(\frac{\pi}{k\al}\right)^{s/2}
  \left(
    \sum_{m=-\infty}^{+\infty}e^{-\pi^2m^2/k\al}
  \right)^s-1
  \;\;.
\end{align}
For large $\al$ (where, again, only the first few modes really
contribute) we recover, again, the limit $\ga_3\to1/(2s)$ as
$s\to\infty$: for small $\al$ we have, again, an exponential approach
to the Gaussian limit:
\begin{equation}
  \ga_3 \sim \left(\frac{8\al}{9\pi}\right)^{s/2}\quad
  \textrm{as}\quad s\to\infty
  \;\;.
\end{equation}
The distinction between the two limiting behaviors is now the
magnitude of the quantity $se^{-2\al}$, which now takes over the
r\^{o}le of the $s\be^2$ of the previous paragraph.

\subsection{The Walsh diaphony}
Another type of diaphony is based on Walsh functions, which are
defined as follows. Let, in one dimension, the real number $x$ be
given by the decomposition
\begin{equation}
  x 
  \;=\; 
  2^{-1}x_1 + 2^{-2}x_2 + 2^{-3}x_3 + \cdots
  \quad,\quad
  x_i\in\{0,1\}
  \;\;,
\end{equation}
and let the nonnegative integer $n$ be given by the decomposition
\begin{equation}
  n 
  \;=\; 
  n_1 + 2n_2 + 2^2n_3 + 2^3n_4 + \cdots
  \quad,\quad
  n_i\in\{0,1\}
  \;\;.
\end{equation}
Then, the $n^{\textrm{\scriptsize th}}$ Walsh function $W_n(x)$ is defined as
\begin{equation}
  W_n(x) 
  \;\df\; 
  (-1)^{(n_1x_1 + n_2x_2 + n_3x_3 + \cdots)}
  \;\;.
\end{equation}
The extension to the multidimensional case is of course
straightforward, and it is easily seen that the Walsh functions form
an orthonormal set. The Walsh diaphony is then given by
\begin{equation}
  D_N 
  \;=\; 
  \frac{1}{N}\sum_{\vn>0}\si_{\vn}^2
  \left|\sum_{k=1}^N W_{\vn}(x_k)\right|^2
  \;\;.
\end{equation}
In \cite{Dyadic}, the following choice is made:
\begin{align}
  &\si_{\vn}^2 
  \;\df\; 
  \frac{1}{3^s-1}\prod_{\nu=1}^s {1\over r(n_\nu)^2}
  \;\;,\notag\\
  &r(n) 
  \;\df\; 
  \theta(n=0) +
  \theta(n>0)
  \sum_{p\ge0} 2^{p}\,\theta\left(2^p\le n < 2^{p+1}\right)
  \;\;.
\end{align}
Note that, in contrast to the Fourier case where each mode of
frequency $n$ contains two basis functions (one sine and one cosine),
the natural requirement of `translational invariance' in this case
requires that the Walsh functions from $2^p$ up to $2^{p+1}$ get equal
strength.  The clusterings are therefore quite different from the
Fourier case.  We slightly generalize the notions of \cite{Leeb},
and write
\begin{align}
  &\si_{\vn}^2 
  \;=\; 
  \prod_{\nu=1}^s {1\over r(n_{\nu})^2}
  \;\;,\notag\\
  &r(n) 
  \;=\; 
  \theta(n=0) + 
  \theta(n>0)\sum_{p\ge0}
  \left(\al\be^p\right)^{-1/2}\theta\left(2^p\le n<2^{p+1}\right)
  \;\;.
\end{align}
Here, we have disregarded the overall normalization of the $\si$'s
since it does not influence the Gaussian limit.  It is an easy matter
to compute the $R_k$; we find
\begin{equation}
  R_k 
  \;=\; 
  \sum_{\vn>0}\si_{\vn}^{2k} 
  \;=\;
  \left(1+\frac{\al^k}{1-2\be^k}\right)^s - 1
  \;\;,
\end{equation}
so that the requirement $\Exp(D_N) = R_1 <\infty$ implies that we must
have $\be<1/2$. Therefore, for not too small values of $\al$, we have
\begin{equation}
  \ga_3 
  \;\sim\; 
  a(\al,\be)^s 
  \quad,\quad
  a(\al,\be) 
  \;=\; 
  \frac{(1+\al^3/(1-2\be^3))^2}
  {(1+\al^2/(1-2\be^2))^3}
  \;\;.
\end{equation}
The choice made in \cite{Dyadic} corresponds to $\al=1$ and
$\be=1/4$, for which we find $a(1,1/4)\sim 0.4197$. The Gaussian limit
should, therefore, be a good approximation for $s$ larger than 6 or
so.  An interesting observation is that for fixed $\be$, $a(\al,\be)$
attains a minimum at $\al =(1-2\be^3)/(1-2\be^2)$, so that the choice
$\be=1/4$ could in principle lead to $a(31/28,1/4)=0.4165$ with a
marginally faster approach to the Gaussian. The overall infimum is
seen to be $a(3/2,1/2) = 2/11 \sim 0.182$. As in the Fourier case with
product clustering and a power-law strength, there is a limit on the
speed with which the Gaussian is approached: in both cases this is
directly related to the type of clustering.

At the other extreme, for very small $\al$ we find the limiting
behavior
\begin{equation}
  \ga_3 
  \;\sim\; 
  \frac{(1-2\be^2)^3}{(1-2\be^3)^2}\,\frac{1}{s}\quad
  \textrm{if}\quad s\al^2 \ll 1\;\;.
\end{equation}
Again in this case, the slowest possible approach to the Gaussian
limit is like $1/s$, directly related to the symmetry of the
discrepancy definition with respect to the various coordinate axes.

\subsection{The Lego discrepancy}
In the case of the Lego discrepancy (\Sec{SecDefLego}), the matrix
$\vGa_{m,n}$ has indices that label the bins $\Aset_n$ $(n=1,2,\ldots M)$ the
hypercube is dissected into, where $M$ is the total number of bins. Because the
characteristic functions of the bins are not normalized, the matrix looks a bit
different:
\begin{equation}
  \vGa_{m,n} 
  \;=\; 
  \si_m\si_n\left(w_m\delta_{m,n} - w_mw_n\right)
  \;\;,
\end{equation}
where $w_n\df\leb{\vt_n}$ is the volume of bin $\Aset_n$. This matrix satisfies
$\Tr(\vGa^p)=R_p$ for all $p>0$.
We shall now examine under what circumstances the criterion
(\ref{GauEq005}) for the appearance of the Gaussian limit is fulfilled.
The eigenvalues $\la_i$ of the matrix $\vGa_{m,n}$ are
given as the roots of the eigenvalue equation
\begin{equation}
  \left(
    \prod_{m=1}^M(\la_i-\si_m^2w_m)
  \right)
  \left(
    \sum_{n=1}^M\frac{w_n\la_i}{\la_i-\si_n^2w_n}
  \right) \;=\; 0
  \;\;.
\end{equation}
It is seen that there is always one zero eigenvalue (the corresponding
eigenvector has $1/\si_m$ for its $m^{\textrm{\scriptsize th}}$
component).  Furthermore the eigenvalues are bounded by
$\max_m(\si_m^2w_m)$, and this bound is an eigenvalue if there is more
than one $m$ for which the maximum is attained. At any rate, we have
for our criterion, that
\begin{equation}
  \rhosup 
  \;\df\; 
  \frac{\lasup^2}{\sum_i\la_i^2} 
  \;\le\;
  \frac{\max_m(\si_m^2w_m)^2}{\Tr(\vGa^2)} 
  \;=\; 
  \frac{\max_m(\si_m^2w_m)^2}{\sum_{m}\si_m^4w_m^2(1-2w_m) + 
    (\sum_{m}\si_m^2w_m)^2}
  \;\;.
\end{equation}
Since the generality of the Lego discrepancy allows us to choose from
a multitude of possibilities for the $\si$'s and $w$'s, we now
concentrate on a few special cases.
\begin{enumerate}
\item {\sl All $w_m$ equal.} This models integrands whose local
      details are not resolved within areas smaller than $1/M$, but whose
      magnitude may fluctuate. In that case, we have
      \begin{equation}
        \rhosup \;<\; \frac{1}{1-2/M}\frac{(\max_m\si_m)^4}{\sum_n\si_n^4}
        \;\;,
      \end{equation}
      and a sufficient condition for the Gaussian limit is for this bound to
      approach zero. Note that here, as in the general case, only bins
      $m$ with $\si_m\ne0$ contribute to the discrepancy as well as to the
      criterion $\rhosup$, so that one has to be careful with models in which
      the integrand is fixed at zero in a large part of the integration
      region: this type of model was, for instance, examined in \cite{Schlier}.
\item {\sl All $\si_m$ equal.} In this case, the underlying integrands
      have more or less bounded magnitude, but show finer detail in some
      places (with small $w$) than in other places (with larger $w$). 
      Now, it is simple to prove that 
      \begin{equation}
        \rhosup \;\le\; \frac{M\bar{w}^2}{1-2\bar{w}+1/M}
        \quad,\quad
        \bar{w} \df \max_mw_m
        \;\;,
      \end{equation}
      so that a sufficient condition is that $M\bar{w}^2$ should approach zero.
\item {\sl All $\si_m^2w_m$ equal.} In this case, the discrepancy is the
      $\chi^2$-statistic for the data points distributed over the bins with 
      expected fraction of points $w_n$. We simply have
      \begin{equation}
        \rhosup \;=\; \frac{1}{(M+2)(M-1)}
        \;\;,
      \end{equation}
      and the Gaussian limit follows whenever $M\to\infty$.
\end{enumerate}

\section{Conclusions}
We derived the probability
distribution, in the limit of a large number of points, over the
ensemble of truly random point-sets of quadratic discrepancies.  We have shown under what
conditions this distribution tends to a Gaussian.  In particular, the
question of the limiting behavior of a given distribution can be
reduced to solving an eigenvalue problem.  Using the knowledge
of the eigenvalues for a given function class it is possible to
determine under which conditions and how fast the Gaussian limit is
approached.  Finally, we have investigated the limiting behavior of
the probability distribution for the discrepancy of several function
classes explicitly.

The discrepancy that fastest approaches the Gaussian limit is obtained
for the model in which the number of modes with non-zero equal strength goes
to infinity, while the sum of the strengths is fixed. In fact, we
give an argument why we cannot improve much on this limit. However, a
drawback of this model is that the discrepancy itself becomes a sum of
Dirac $\delta$-distributions in this limit: it only measures whether points
coalesce, and is therefore not very useful in practice.

Secondly we looked at the $L_2^*$-discrepancy. Here a Gaussian
distribution appears in the limit of a large number of dimensions. It
is however a very slow limit: only when the number of dimensions
becomes of the order ${\mathcal O}\left(10^2\right)$ does the Gaussian
behavior become manifest.

For the different diaphonies the choice of the mode-strengths is more
arbitrary. The strengths we discuss are chosen on the basis of some
preferred global properties of the diaphony, such as translation-
and/or rotation-invariance. Again for large dimensions the Gaussian
limit is attained, either as a power-law or inverse of the number of
dimension. It is possible to choose the strengths in such a way that
the Gaussian limit is approached arbitrarily fast. But the diaphony
corresponding to that case again consists of a sum of Dirac
$\delta$-distributions.

Finally, for the Lego-discrepancy, we can assign strengths to the
different modes in several ways. One example is to keep the product of
the squared strength and volume of the modes fixed, then the Gaussian
limit is reached for a large number of modes.

All these results have been derived in the limit of large number of
points. It remains to be seen however whether this is reasonable in
practice. To determine when the asymptotic regime sets in, i.e.~for
which value of $N$, it is necessary to take into account the
next-to-leading contributions, which will be calculated in the following 
chapter.


\section{Appendices}
\vspace{-1.2\baselineskip}
\Appendix{:\; The form of $G_0(z)$\label{App5A}}
In this Appendix, we derive the result (\ref{GauEq003}) for the form of
$G_0(z)$. We introduce the notation $[BA^kB]\df\sum_{m,n}B_m(A^k)_{m,n}B_n$ for
matrices $A$ and vectors $B$, and the general form of the matrix $\vGa$:
\begin{equation}
  \vGa_{m,n} \;=\; A_{m,n} - B_mB_n
  \;\;.
\end{equation}
The $k^{\textrm{\scriptsize th}}$ power of this matrix has the general
form
\begin{equation}
  (\vGa^k)_{m,n} 
  \;=\; 
  (A^k)_{m,n} -
  \sum_{p,q,\nu_{0,1,2,\ldots}\ge0}
  \frac{(\,\sum_{r\ge0}\nu_r\,)!}{\nu_0!\nu_1!\nu_2!\cdots}
  (A^pB)_m(BA^q)_n\prod_{r\ge0}(-[BA^rB])^{\nu_r}
  \;\;,
\end{equation}
with the constraint $k-1=p+q+\nu_0+2\nu_1+3\nu_2+\cdots$. The
combinatorial factor follows directly from the possible positionings
of the dyadic factors $-B_mB_n$. Multiplying by $(2t)^{k-1}$ and
summing over the $k$ then gives us immediately
\begin{equation}
  \Tr\left(\frac{\vGa}{1-2t\vGa}\right) 
  \;=\;
  \sum_{k\ge1}(2t)^{k-1}\Tr(A^k) -
  \frac{\sum_{r\ge0}(r+1)(2t)^r[BA^rB]}{1+\sum_{n\ge1}(2t)^n[BA^{n-1}B]}
  \;\;,
\end{equation}
where the factor with $r+1$ comes from the double sum over $p$
and $q$ with $p+q=r$. Upon integration of this result over $t$ from 0
to $z$ we find
\begin{align}
  \log G_0(z) 
  \;&=\; 
  \sum_{n>0}\frac{(2z)^n}{2n}\Tr(\vGa^n)\notag\\
  \;&=\; 
  \sum_{n>0}\frac{(2z)^n}{2n}\Tr(A^n) -
  \frac{1}{2}\log\left(1+\sum_{n>0}(2z)^n[BA^{n-1}B]\right)
  \;\;.
\end{align}
If we now take $A_{m,n}=\si_n^2\de_{m,n}$ and $B_n=\si_n\leb{u_n}$, we obtain
(\ref{GauEq003}) with (\ref{GauEq004}).  This result has, in fact, already been
obtained for the case of the $L_2^*$-dis\-crep\-an\-cy in \cite{jhk}, but here
we demonstrate its general validity for more general discrepancy measures. In
those cases where $B_m=0$, the second term vanishes of course.

\Appendix{:\; A counterexample\label{App5B}}
In this Appendix we prove that the condition (\ref{GauEq005}) for the
occurrence of a Gaussian limit is, in a sense, the best possible.  Namely,
consider a set of eigenvalues $\la_n$, again adding up to unity as usual,
defined as follows: let $\la$ be a positive number, and take 
\begin{align}
  \la_1 = \la \;\;,\quad
  \la_n \;=\; \begin{cases}
                 (1-\la)/(M-1) &\textrm{for $n=2,3,\ldots,M$} \;,\\
		 0             &\textrm{for $n>M$} \;.
              \end{cases}
\end{align}
Clearly, $\la$ will indeed be the maximal eigenvalue as long as
$M>1/\la$.  Now,
\begin{equation}
  \frac{\la^2}{\sum_n\la_n^2} 
  \;=\; 
  \frac{\la^2}{\la^2 + (1-\la)^2/(M-1)}
  \;\;,
\end{equation}
and this ratio can be driven as close to unity as desired by choosing
$M$ sufficiently large. This shows that the simple condition $\la\to0$
is not always enough to ensure the Gaussian limit.

\Appendix{:\; The magnitude of $\Qfp(p)$\label{App5C}}
Here we present the proofs of our various
statements about the multiplicity function $\Qfp(p)$ of \eqn{GauEq008}.
In the first place, we know that its Dirichlet
generating function, $F^{[1]}(x)$, converges for all $x>1$. Now
suppose that $\Qfp(p)$ exceeded $cp^{\al}$ an infinite number of
times, with $c>0$ and $\al>1$. The Dirichlet generating function would
then contain an infinite number of terms all larger than $c$, for
$1<x<\al$, and therefore would diverge, in contradiction with its
convergence for all $x>1$.

In the second place, consider the `standard' generating function,
$F^{[2]}_s(x)$. By inspecting how many of the vector components
$n_{\nu}$ of $\vn$ are zero, we see that we may write, for $p>1$,
\begin{equation}
  \Qfp(p) 
  \;=\; 
  \sum_{t=1}^s\binom{s}{t}2^td_t(p)
  \quad,\quad
  d_t(p) 
  \;\df\; 
  \sum_{\vn\ge0}\theta\left(p=\prod_{\nu=1}^tn_{\nu}\right)
  \;\;,
\end{equation}
so that $d_t(p)$ counts in how many ways the integer $p$ can be
written as a product of $t$ factors, including ones; this function is
discussed, for instance, in \cite{Hardy}. Now, for $p$ prime, we
have $d_t(p)=t$, and therefore
\begin{equation}
  \Qfp(p) 
  \;\ge\; 
  2s(3^{s-1})
  \quad,\quad\textrm{equality for $p$ prime}.
\end{equation}
The radius of convergence of $F^{[2]}_s(x)$ is therefore {\it at
  most\/} equal to unity.  On the other hand, we can obtain a very
crude, but sufficient, upper bound on $\Qfp(p)$ as follows.
Since $d_t(p)$ is a nondecreasing function of $t$, we may bound
$\Qfp(p)$ by $(3^s-1)d_s(p)$. Now let $k_p$ be the number of
prime factors in $p$; then $k_p$ cannot exceed $\log(p)/\log(2)$, and
only is equal to this when $p$ is a pure power of 2. Also, the number
of ways to distribute $k$ object in $s$ groups (which may be empty) is
at most $s^k$, and is smaller if some of the objects are equal.
Therefore, $d_s(p)$ is at most $s^{k_p}$, and we see that
\begin{equation}
  \Qfp(p) 
  \;<\; 
  (3^s-1)p^{\log(s)/\log(2)}
  \;\;,
\end{equation}
or, in short, is bounded\footnote{Note that equality cannot occur in
  this case since the two requirements are mutually exclusive.}  by a
polynomial in $p$.  Therefore, the radius of convergence of
$F^{[2]}_s(x)$ is also {\it at least\/} unity, and we have proven the
assertion in \Sec{GauSecFou}.

Finally, we consider the limit
\begin{equation}
  \lim_{\beta'\to1}\ga_3 
  \;=\; 
  \lim_{x\to1}
  \frac{\left(F^{[2]}_s(x^3)\right)^2}{\left(F^{[2]}_s(x^2)\right)^3}
  \;\;.
\end{equation}
The same reasoning that led us to the radius of convergence shows
that, for $x$ approaching 1 from below, the function $F^{[2]}_s(x)$
behaves as $(1-x)^{-c}$, with $c\ge1$. Therefore, $\ga_3$ will behave
as $(8(1-x)/9)^c$, and approach zero as $x\to1$.  Note that the upper
bound on $\Qfp(p)$ is extremely loose: but it is enough.

\clearemptydoublepage

\chapter{Finite-sample corrections to discrepancy distributions\label{ChapCorr}}

\newcommand{\Lf}{L_{\textrm{f}}}
\newcommand{\Lb}{L_{\textrm{b}}}
\newcommand{\Lm}{L_{\textrm{m}}}
\newcommand{\If}{I_{\textrm{f}}}
\newcommand{\Ib}{I_{\textrm{b}}}

This chapter deals with the calculation of the $1/N$-corrections to the
asymptotic probability distributions of quadratic discrepancies in the limit of
an infinite number of random points $N$. In \Sec{CorSec1}, the explicit
diagrammatic expansion of the logarithm of the generating function up to
and including $\Ord(1/N^2)$ will be given. For the Lego discrepancy, the
$L_2^*$-discrepancy in one dimension and the Fourier diaphony in one dimension,
the explicit $1/N$-correction is calculated.

In \Chp{GausChap}, criteria were given for the asymptotic probability
distribution of several quad\-ratic discrepancies to become Gaussian when a
certain free parameter becomes infinitely large. This parameter often is the
dimension $s$ of the integration region. In the case of the Lego discrepancy,
it is the number of bins $M$. In \cite{Leeb}, it is shown that for the {\em
Fourier diaphony} a Gaussian limit is obtained when both $N$ and $s$ go to
infinity such that $c^s/N\ra0$, where $c$ is some constant larger than $1$.
This theorem clearly gives more information about the behavior of the
probability distribution, for it relates $s$ and $N$, whereas in
Chapter~\ref{GausChap} the limit of $N\ra\infty$ is assumed before considering
the behavior with respect to $s$ or $M$.  However, the techniques of this
chapter to calculate $1/N$-corrections to the asymptotic distributions {\em
give} the opportunity to relate $s$ or $M$ with $N$. In \Sec{CorSec2}, this
leads to limits for the Lego discrepancy, which is equivalent with a
$\chi^2$-statistic for $N$ data points distributed over $M$ bins, if $M$ as
well as $N$ become infinite. In \Sec{CorSec3}, a Gaussian limit is derived for
the Fourier diaphony, which is stronger than the one in \cite{Leeb} in the
sense that it provides convergence of the moments of the distribution, whereas
the limit in \cite{Leeb} is weak.

\vspace{\baselineskip}

\minitoc

\section{The first few orders\label{CorSec1}}
The Feynman diagrams that contribute to the first few orders in the
$1/N$-expansion of the generating function $G(z)$ of the probability
distribution of quadratic discrepancies are determined,  and are used in a few 
examples. 

\subsection{The diagrammatic expansion}
To calculate a term in the $1/N$-expansion of $G$, the contribution of all
diagrams that can be drawn using the Feynman rules, as given in \Sec{FormSec1}, and carry the right power of $1/N$ has to be included.  We
want to stress again that we only need to calculate the connected diagrams
without external lines. The sum of the contributions of all these diagrams
gives 
\begin{equation}
   W(z) 
   \;\df\; \log G(z) 
   \;=\; W_0(z) + \frac{1}{N}W_1(z) + \frac{1}{N^2}W_2(z) + \cdots \;\;.
\end{equation}
Usually, a Feynman diagram is a mnemonic representing a certain contribution to
a term in a series expansion, i.e.~a label. We will use the same drawing
for the contribution itself, apart of the symmetry factor of the diagram.
For example, the contribution of the diagram
$\diagram{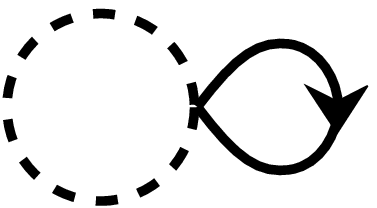}{30}{5.0}$ is equal to $\half Ng^2\intk\twoB(x,x)\,dx$ and its 
symmetry factor is equal to $\half$, so that we write 
\begin{equation}
   \frac{1}{2}\,\diagram{dcZ1}{30}{6} 
   \;=\; \frac{Ng^2}{2}\intk\twoB(x,x)\,dx \;\;.
\end{equation}

\subsubsection{The zeroth order}
The contribution to the zeroth order in $1/N$ can only come from diagrams in
which the power of $1/N$ coming from the vertices cancels the power of $N$,
coming from the fermion loops. This only happens in diagrams with vertices with
two bosonic legs only, and in which the fermion lines begin and end on the same
vertex. To write down their contribution, we introduce the two-point functions
$\twoB_p$, $p=1,2,\ldots$, defined by 
\begin{equation}
   \twoB_1(x_1,x_2) \df \twoB(x_1,x_2) \;\;,\quad
   \twoB_{p+1}(x_1,x_2) \df \intk\twoB_p(x_1,y,)\twoB(y,x_2)\,dy \;\;.
\end{equation}
The zeroth order term $W_0(z)$ is given by 
\begin{align}
   \frac{1}{2}\,\diagram{dcZ1}{30}{6} \;+\; \frac{1}{4}\,\diagram{dcZ2}{45}{6} 
                       \;+\; \frac{1}{6}\,\diagram{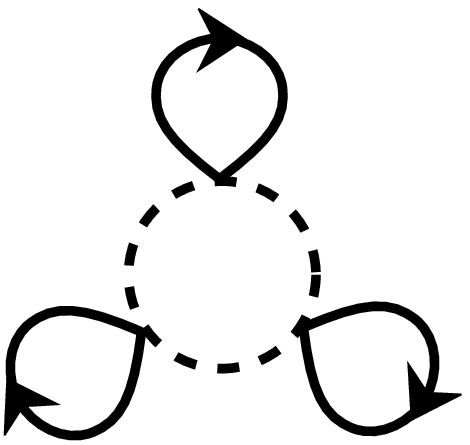}{38}{15} \;+\; \cdots 
   \;=\; \sum_{p=1}^{\infty}\frac{(Ng^2)^p}{2p}\intk\twoB_p(x,x)\,dx \;\;.
\label{CorEq025}   
\end{align}
The factor $1/2p$ is the symmetry factor of this type of diagram with $p$
fermion ``leaves''. If we substitute $g=\sqrt{2z/N}$ in this expression, we
find exactly the result of \eqn{GauEq010}.

\subsubsection{The first order\label{FiOrSec}}
As we have seen before, bosonic two-point vertices with a closed single fermion
line contribute with a factor $2z$, and without any dependence on $N$.
Therefore, it is useful to introduce the following effective vertex
\begin{equation}
   \diagram{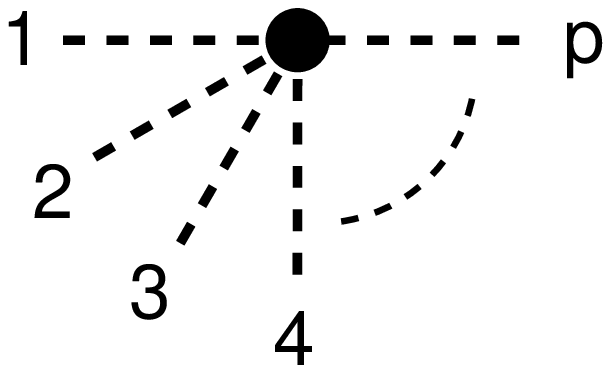}{50}{23} \;\df\; \diagram{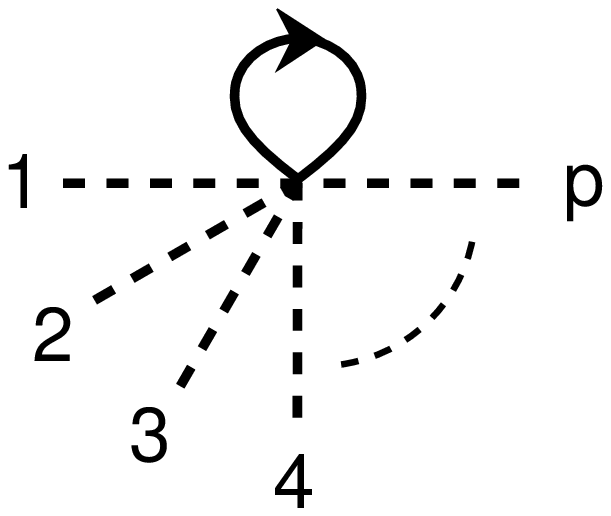}{50}{23}
                        \;=\; Ng^p\times\textrm{convolution}\;\;,
\label{CorEq026}			
\end{equation}
and the following {\em dressed} boson propagator
\begin{align}
   x\,\diagram{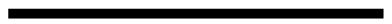}{30}{-1}\,y 
   \;&\df\;     x\,\diagram{dcP2}{30}{-1}\,y \;+\; x\,\diagram{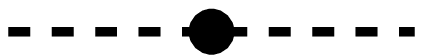}{30}{0}\,y 
          \;+\; x\,\diagram{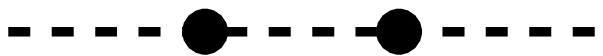}{39}{0}\,y 
	  \;+\; x\,\diagram{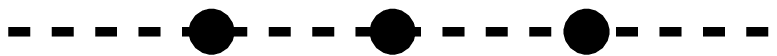}{60}{0}\,y \;+\; \cdots \;\;,\\
   \prop_z(x,y)	  
   \;&\df\; \sum_{p=1}^\infty(2z)^{p-1}\twoB_p(x,y) \;\;.
\end{align}
In terms of the basis in the Landau gauge, it is given by 
\begin{equation}
   \prop_z(x,y) 
   \;=\; \sum_n\frac{\si_{\textrm{L},n}^2}{1-2z\si_{\textrm{L},n}^2}\,
               u^{(\textrm{L})}_n(x)u^{(\textrm{L})}_n(y)\;\;,
\end{equation}
which is, apart of a factor $2z$, the same expression as in Eq.~(67) in
\cite{hk1}. Notice that $\prop_z$ and $\twoB$ satisfy the relation 
\begin{equation}
   \lim_{z\ra0}\prop_z(x,y)
   \;=\; \prop_{z=0}(x,y) 
   \;=\; \twoB(x,y) \;\quad \forall\,x,y,\in\Kube  \;\;.
\end{equation}
Furthermore, notice that $\prop_z$ and $W_0$ satisfy 
\begin{equation}
   \frac{\partial}{\partial z}\,W_0(z) 
   \;=\; \intk\prop_z(x,x)\,dx \;\;,
\label{CorEq008}   
\end{equation}
and that this relation determines $W_0$ uniquely, because we know that 
$W_0(0)$ has to be equal to $0$ in order for the asymptotic probability 
distribution to be normalized to $1$.

The first order term in the expansion of $W(z)$ is 
\begin{equation}
   \frac{1}{N}W_1(z) 
   \;=\; \frac{1}{8}\,\diagram{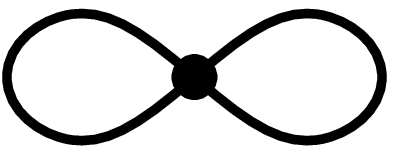}{31}{3.5}
   + \frac{1}{8}\,\diagram{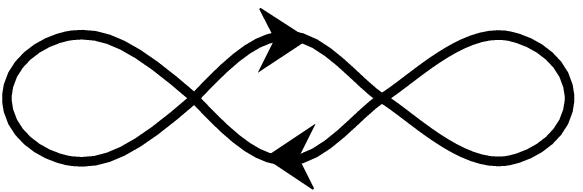}{45}{4.5}
   + \frac{1}{4}\,\diagram{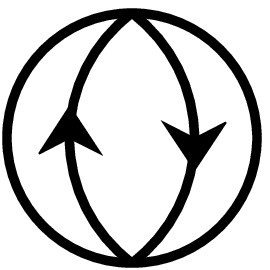}{22}{9}  
   + \frac{1}{8}\,\diagram{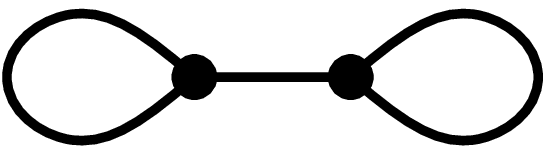}{43}{3.5}
   + \frac{1}{12}\,\diagram{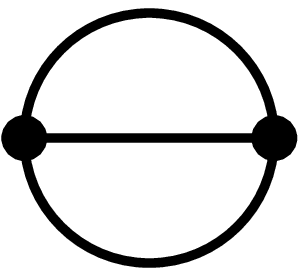}{24}{8.5} \;\;,
\label{CorEq009}
\end{equation}
or, more explicitly, 
\begin{align}
   W_1(z) 
   &\;=\;
   \frac{z^2}{2}\intk\prop_z(x,x)^2\,dx
             \;-\; \frac{z^2}{2}\left(\intk\prop_z(x,x)\,dx\right)^2
	     \;-\; z^2\intkk\prop_z(x,y)^2\,dxdy \notag\\ 
 &\hspace{22pt}+ z^3\intkk\prop_z(x,x)\prop_z(x,y)\prop_z(y,y)\,dxdy
             \;+\; \frac{2z^3}{3}\intkk\prop_z(x,y)^3\,dxdy  \;\;.
\end{align}

\subsubsection{The second order}
The second order term in the expansion of $W(z)$ is denoted 
$\sfrac{1}{N^2}W_2(z)$ and is given by
\begin{align}
&\hspace{20pt}\frac{1}{48}\,\diagram{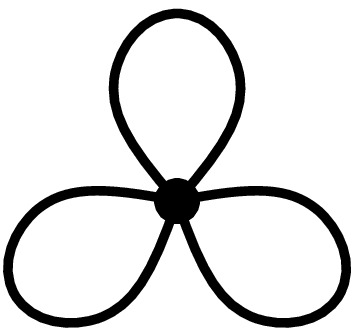}{28}{9}
\;+\;\frac{1}{48}\,\diagram{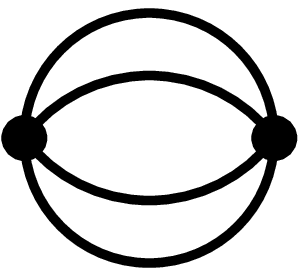}{24}{9}
\;+\;\frac{1}{16}\,\diagram{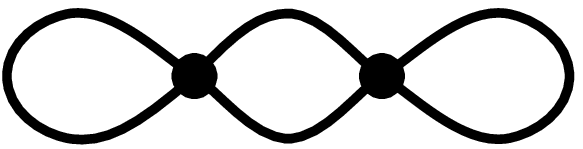}{46}{4}
\;+\;\frac{1}{12}\,\diagram{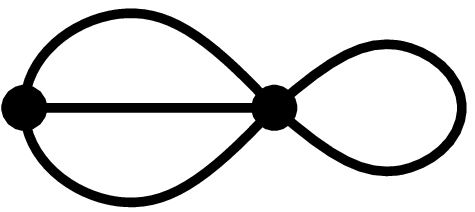}{38}{7}
\;+\;\frac{1}{24}\,\diagram{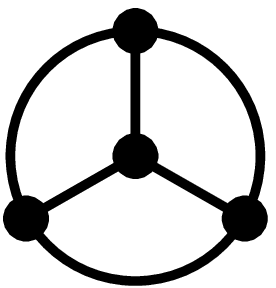}{23}{9.5}
\;+\;\frac{1}{16}\,\diagram{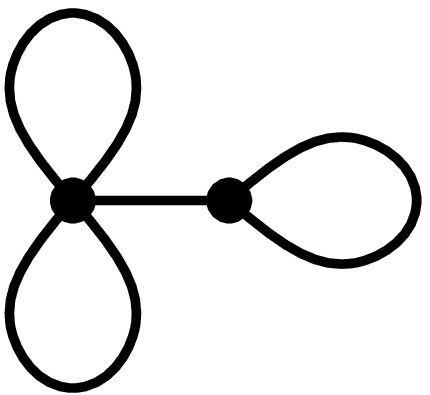}{34}{13} \notag \\
&\;+\;\frac{1}{8}\,\diagram{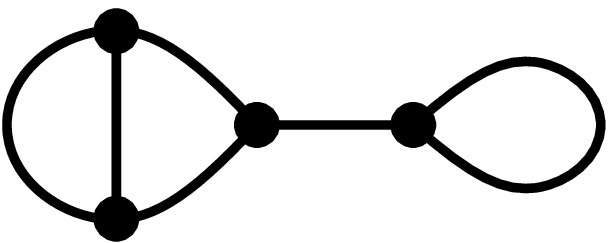}{49}{7}
\;+\;\frac{1}{8}\,\diagram{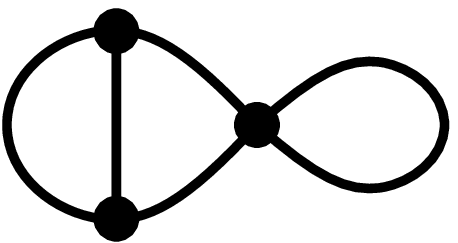}{37}{8}
\;+\;\frac{1}{8}\,\diagram{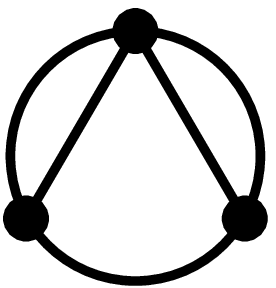}{24}{11}
\;+\;\frac{1}{16}\,\diagram{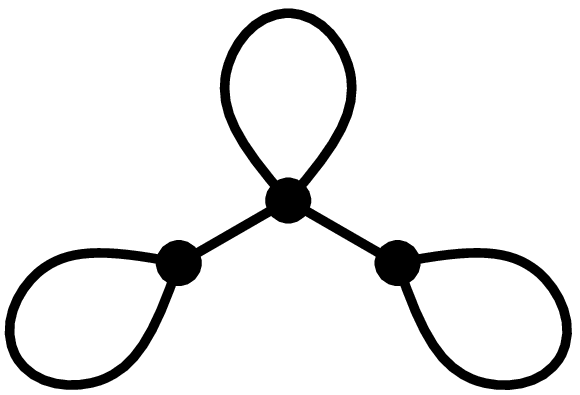}{45}{12}
\;+\;\frac{1}{16}\,\diagram{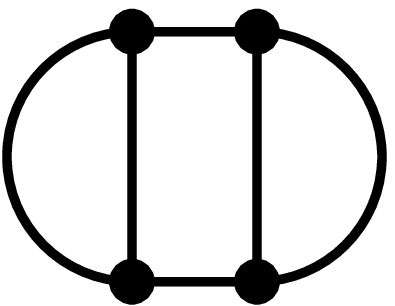}{32}{10}  
\;+\;\frac{1}{48}\,\diagram{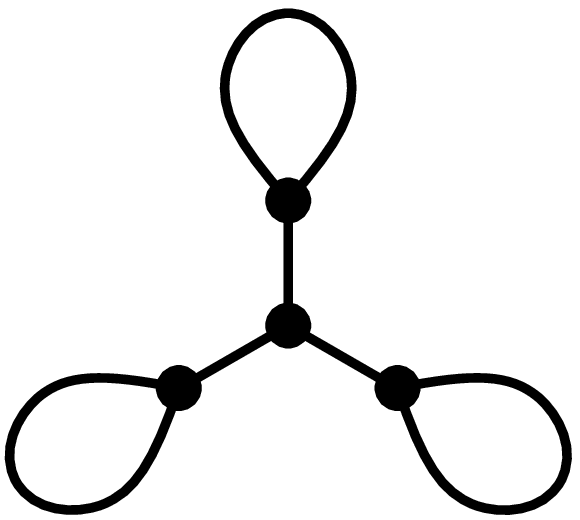}{45}{17} \notag\\
&\;+\;\frac{1}{8}\,\diagram{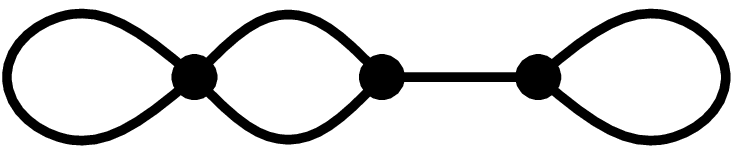}{58}{4}
\;+\;\frac{1}{16}\,\diagram{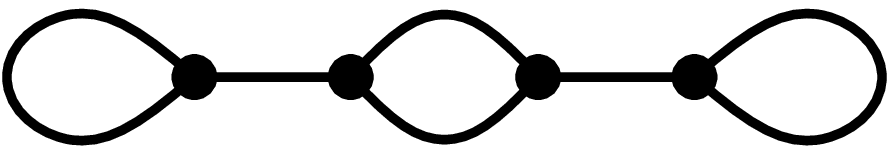}{72}{4}
\;+\;\frac{1}{12}\,\diagram{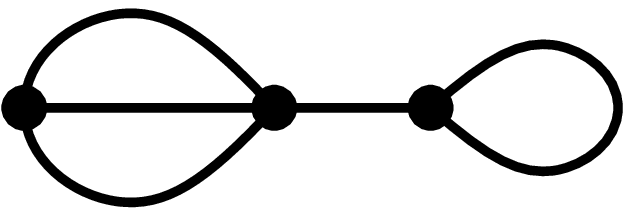}{49}{6}
\;+\;\frac{1}{4}\,\diagram{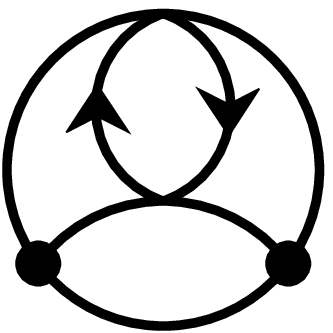}{23}{9}
\;+\;\frac{1}{8}\,\diagram{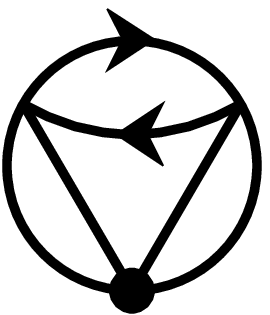}{23}{12}
\;+\;\frac{1}{4}\,\diagram{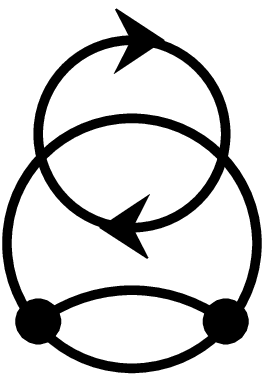}{23}{12} \notag\\  
&\;+\;\frac{1}{4}\,\diagram{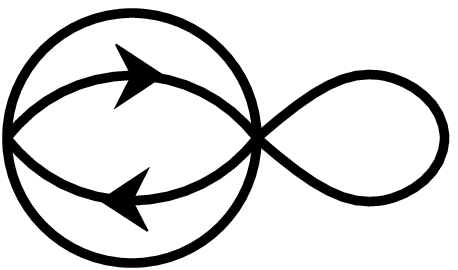}{38}{9}
\;+\;\frac{1}{4}\,\diagram{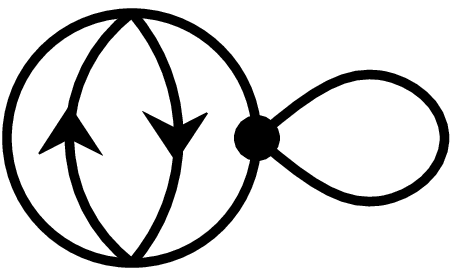}{38}{9}
\;+\;\frac{1}{2}\,\diagram{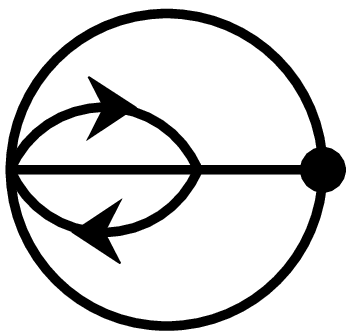}{25}{10}
\;+\;\frac{1}{16}\,\diagram{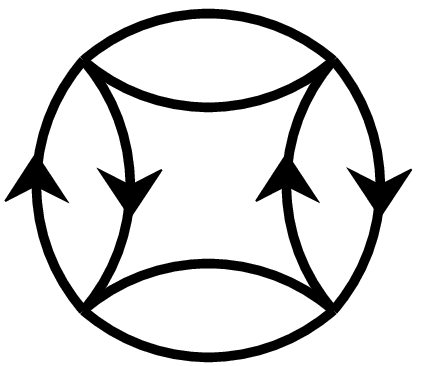}{32}{12}
\;+\;\frac{1}{8}\,\diagram{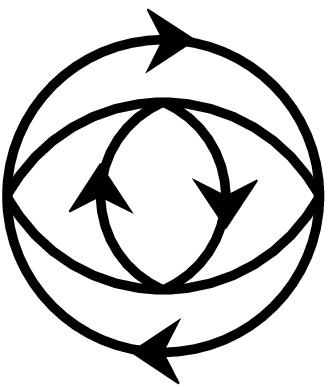}{26}{14}
\;+\;\frac{1}{4}\,\diagram{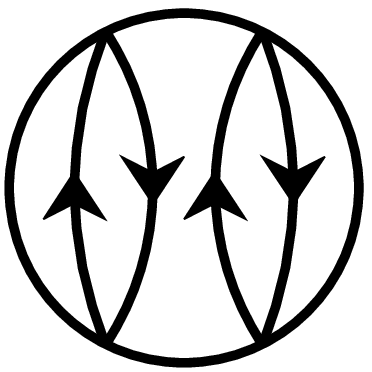}{28}{12}
\;+\;\frac{1}{12}\,\diagram{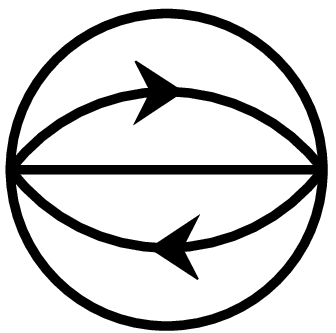}{23}{9} \notag\\
&\;+\;\frac{1}{8}\,\diagram{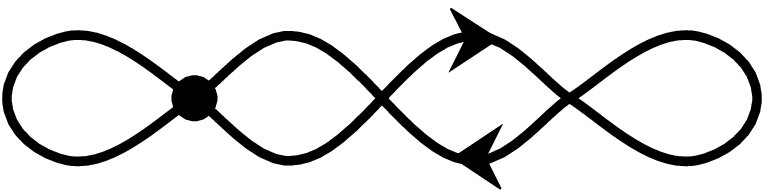}{60}{6}
\;+\;\frac{1}{16}\,\diagram{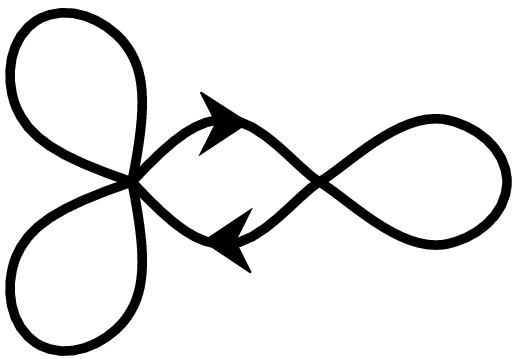}{41}{13}
\;+\;\frac{1}{8}\,\diagram{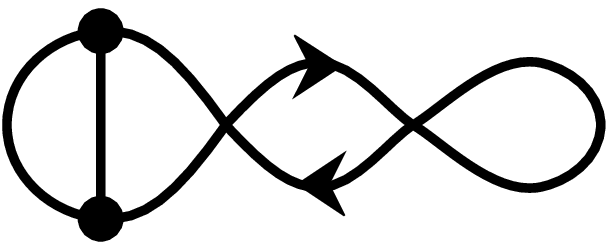}{49}{8}
\;+\;\frac{1}{12}\,\diagram{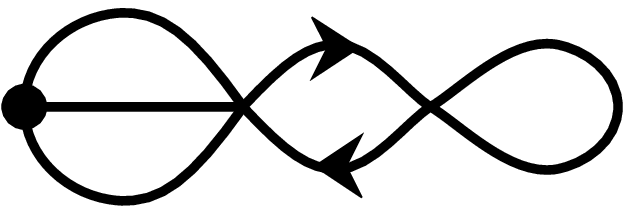}{53}{8}
\;+\;\frac{1}{8}\,\diagram{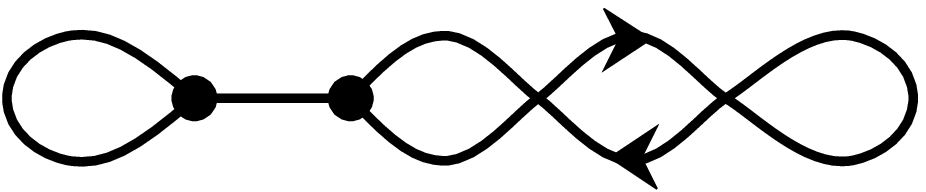}{71}{6} \notag\\
&\;+\;\frac{1}{16}\,\diagram{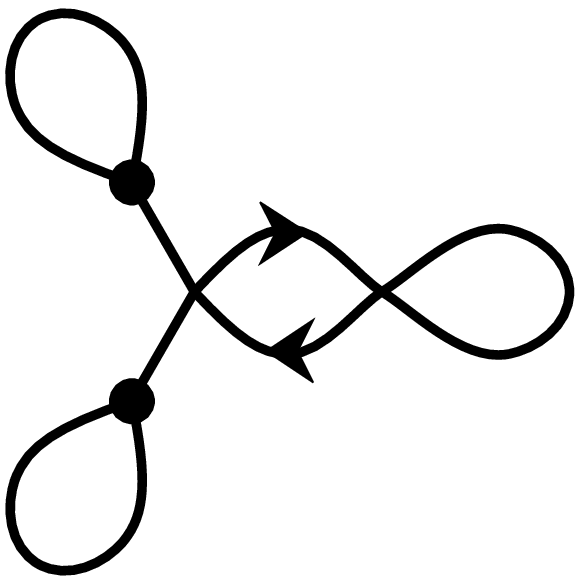}{45}{21}
\;+\;\frac{1}{8}\,\diagram{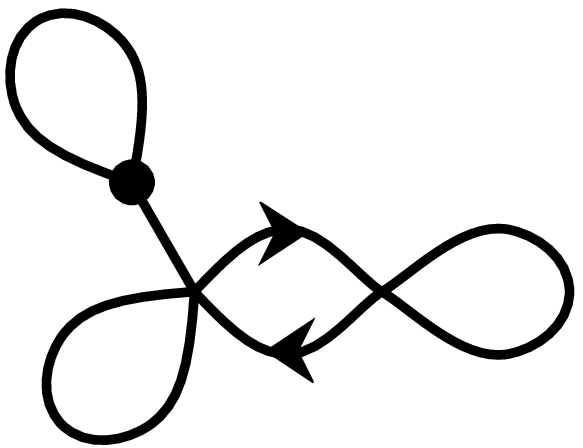}{45}{11}
\;+\;\frac{1}{4}\,\diagram{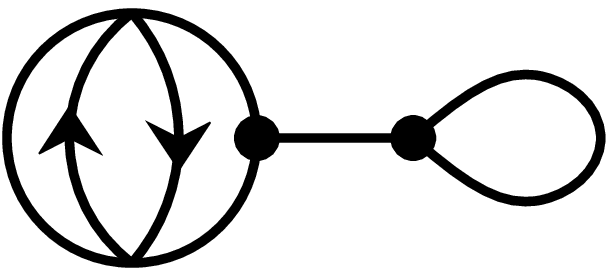}{49}{8}
\;+\;\frac{1}{4}\,\diagram{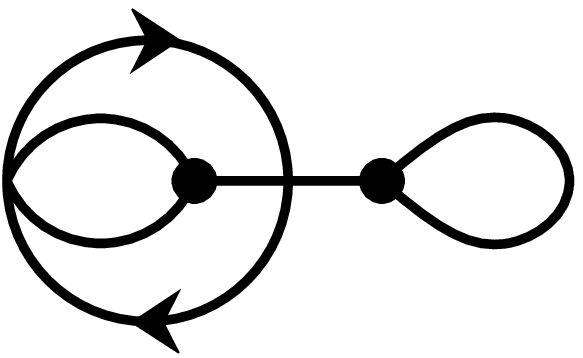}{48}{13} 
\;+\;\frac{1}{8}\,\diagram{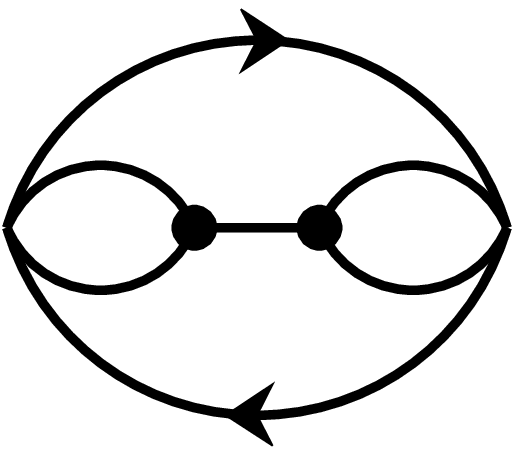}{39}{15}
\;+\;\frac{1}{3}\,\diagram{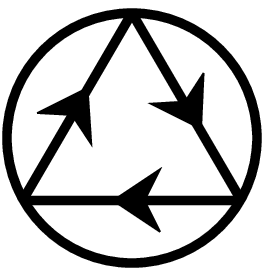}{22}{9}\notag\\
&\;+\;\frac{1}{4}\,\diagram{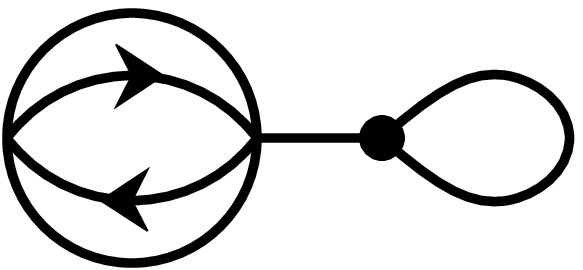}{48}{9}
\;+\;\frac{1}{4}\,\diagram{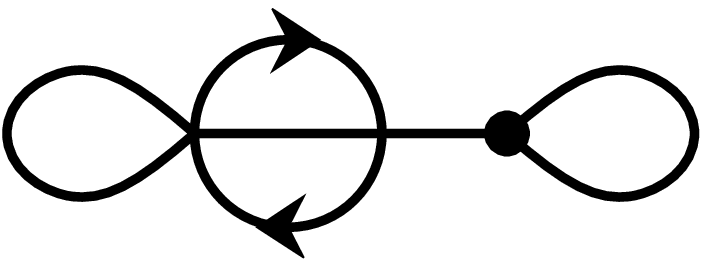}{56}{9} 
\;+\;\frac{1}{8}\,\diagram{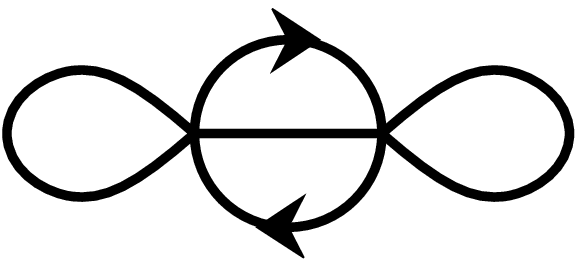}{45}{9}
\;+\;\frac{1}{4}\,\diagram{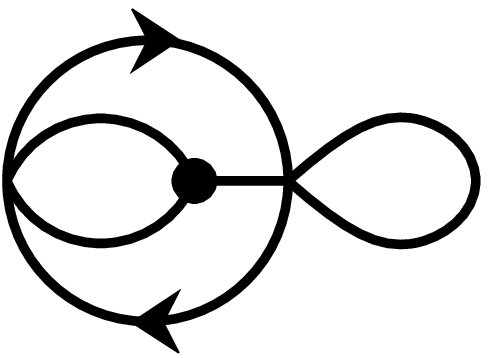}{41}{14}
\;+\;\frac{1}{8}\,\diagram{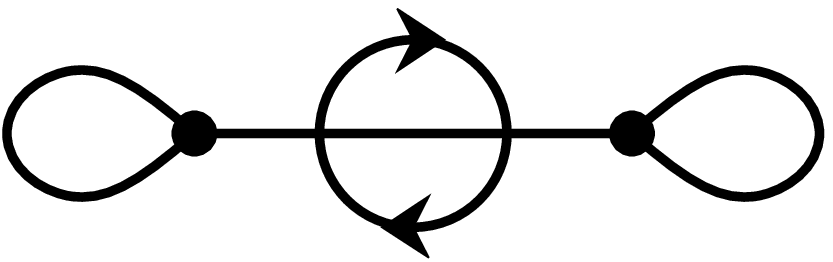}{64}{8}\notag \\
&\;+\;\frac{1}{24}\,\diagram{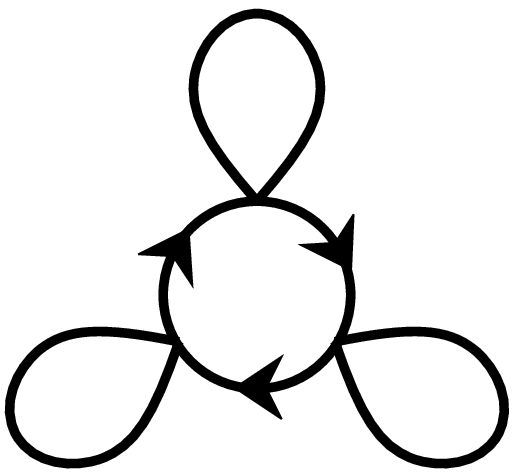}{41}{14}
\;+\;\frac{1}{4}\,\diagram{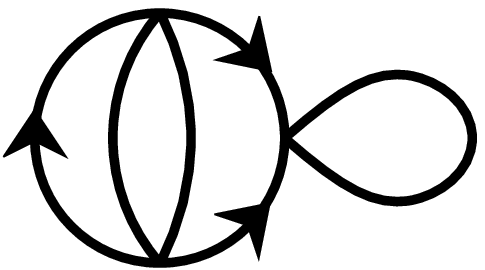}{38}{9}
\;+\;\frac{1}{16}\,\diagram{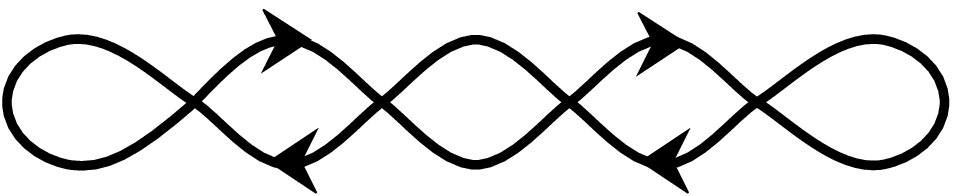}{74}{5}
\;+\;\frac{1}{4}\,\diagram{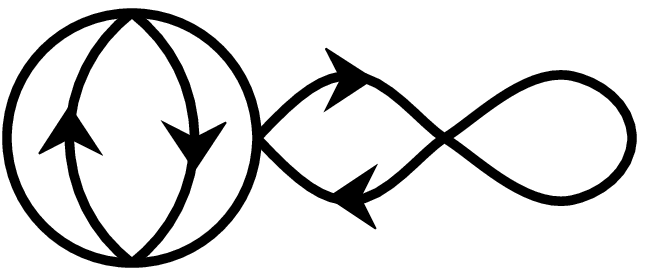}{54}{9}
\;\;.
\end{align}

\subsubsection{One-vertex decomposability\label{OVRSec}}
For some discrepancies, the contribution of a bosonic part of a diagram that
consists of two pieces connected by {\em only one} vertex, is equal to
the product of the contribution of those pieces. Such diagrams we call {\em
one-vertex reducible}, and discrepancies with this property we call {\em
one-vertex decomposable}. Examples of such discrepancies are those for which
$\twoB$ is translation invariant, i.e., $\twoB(x,y)=\twoB(x+a,y+a)$
$\forall\,x,y,a\in\Kube$, such as the Fourier diaphony. Also the Lego
discrepancy with equal bins is one-vertex decomposable. In contrast, the 
$L_2^*$-discrepancy is not one-vertex decomposable.

As a result of the one-vertex decomposability, many diagrams cancel or give 
zero. For example, the first and the second diagram in (\ref{CorEq009}) cancel, and the fourth gives zero, so that 
\begin{equation}
   \frac{1}{N}W_1(z) 
   \;=\;       \frac{1}{4}\,\diagram{dcA21}{22}{9}  
                \;+\; \frac{1}{12}\,\diagram{dcA12}{24}{8.5} \;\;.
\label{CorEq010}		
\end{equation}
To second order, only the following remains:
\begin{align}
\frac{1}{N^2}W_2&(z) \;=\;
     \frac{1}{48}\,\diagram{dcB12}{24}{9}
\;+\;\frac{1}{24}\,\diagram{dcB111}{23}{9.5}
\;+\;\frac{1}{8}\,\diagram{dcB18}{24}{11}
\;+\;\frac{1}{2}\,\diagram{dcB210}{25}{10}
\;+\;\frac{1}{8}\,\diagram{dcB24}{23}{12}  
\;+\;\frac{1}{4}\,\diagram{dcB28}{23}{12}  
\;+\;\frac{1}{4}\,\diagram{dcB29}{23}{9} \notag\\
&\;+\;\frac{1}{16}\,\diagram{dcB115}{32}{10}
\;+\;\frac{1}{16}\,\diagram{dcB36}{32}{12}
\;+\;\frac{1}{8}\,\diagram{dcB37}{26}{14}
\;+\;\frac{1}{4}\,\diagram{dcB38}{28}{12}
\;+\;\frac{1}{12}\,\diagram{dcB211}{23}{9} 
\;+\;\frac{1}{3}\,\diagram{dcB31}{22}{9} \;\;.
\label{CorEq015}
\end{align}
We now derive a general rule of diagram cancellation. First, we extend
the notion of one-vertex reducibility to complete diagrams, including the
fermionic part, with the rule that the two pieces both must contain a 
bosonic part. Consider the following diagram
\begin{equation}
   \diagram{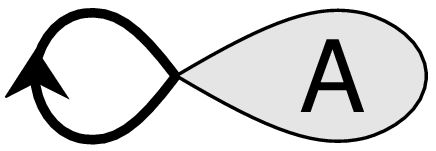}{49}{0}  \;\;.
\label{CorEq011}   
\end{equation}
The only restriction we put one the ``leave'' $A$ is that it must be one-vertex 
irreducible with respect to the vertex that connects it to the fermion loop.
For the rest, it may be anything. We define the contribution of the leave by
the contribution of the whole diagram divided by $-N$, and denote it with
$C(A)$. This contribution includes internal symmetry factors. Now consider a
diagram consisting of a fermion loop as in diagram (\ref{CorEq011}) with
attached to the one vertex $n_1$ leaves of type $A_1$, $n_2$ leaves of type
$A_2$, and so on, up to $n_p$ leaves of type $A_p$. The extra symmetry factor
of such a diagram is $(n_1!n_2!\cdots n_p!)^{-1}$, and, for one-vertex
decomposable discrepancies, the contribution is equal to the product of the
contributions of the leaves, so that the total contribution is given by
\begin{equation}
   -N\prod_{q=1}^p\frac{C(A_q)^{n_q}}{n_q!} \;\;.
\end{equation}
Now we sum the contribution of all possible diagrams of this kind that can
made with the $p$ leaves, and denote the result by
\begin{align}
   \diagram{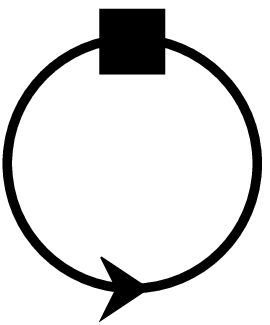}{22}{12} 
   \;=\; -N\sum_{n_1,n_2,\ldots\geq1}\prod_{q=1}^p
           \frac{C(A_q)^{n_q}}{n_q!} 
   \;=\; -N\Big(\exp\Big(\sum_{q=1}^pC(A_q)\Big) - 1\Big) \;\;.
\label{CorEq012}   
\end{align}
Because the little square in l.h.s.~of
\eqn{CorEq012} represents all possible ways to put the leaves together
onto one vertex, the sum of all possible ways to put the leaves onto one
fermion loop is given by
\begin{equation}
   \diagram{lcFl1}{22}{12} \;+\;
   \diagram{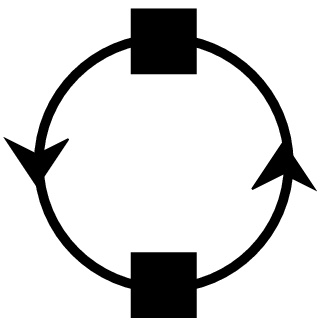}{26}{12} \;+\;
   \diagram{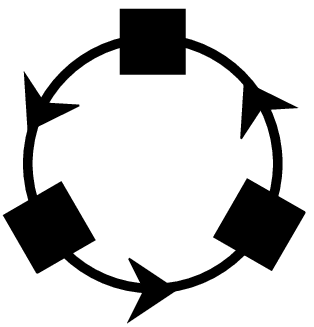}{25}{12} \;+\; \cdots 
   \;=\; -N\sum_{n=1}^{\infty}\frac{(-1)^{n-1}}{n}
         \Big(\exp\Big(\sum_{q=1}^pC(A_q)\Big) - 1\Big)^n    \;\;.
\label{CorEq013}	 
\end{equation}
The $(-1)^{n-1}$ in the sum comes from the vertices and $1/n$ is the extra
symmetry factor of such diagram with $n$ squares. The sum can be evaluated 
further and is equal to
\begin{equation}
    -N\log\Big(\exp\Big(\sum_{q=1}^pC(A_q)\Big)\Big) 
    \;=\; -N\sum_{q=1}^pC(A_q) \;\;,
\end{equation}
i.e., the sum of all possible ways to put $p$ different leaves onto one
fermion loop is equal to the sum of all leaves, each of them put onto its own 
fermion loop. This means that diagrams, consisting of two or more leaves put 
onto one fermion loop, cancel.

Now consider the following equation, which holds for 
every one-vertex decomposable discrepancy: 
\begin{equation}
   \diagram{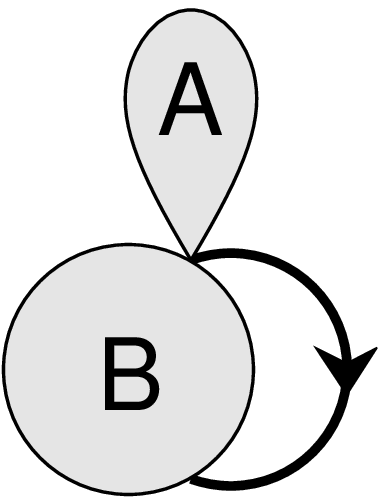}{38}{11} \;=\; -\; \diagram{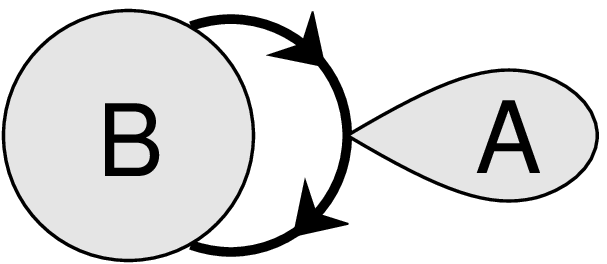}{60}{11} \;\;,
\label{CorEq014}   
\end{equation}
where we only assume that $B$ is not of the type on the l.h.s.~of
\eqn{CorEq013}. The minus sign comes from the fact that the first diagram has
one vertex less. Because the number of fermion lines a fermion loop consists
of is equal to the number of vertices it contains, we can always pair the
diagrams into one diagram of the l.h.s.~type and one of the r.h.s.~type so that
they cancel. We can summarize the result with the rule that 
\begin{equation}
\begin{tabular}{c}
   \textsl{for one-vertex decomposable discrepancies,}\\
   \textsl{only the one-vertex irreducible diagrams contribute.}
\end{tabular}   
\end{equation}

\subsection{Applications}
We apply the general formulae given above to the Lego discrepancy, the 
$L_2^*$-discrepancy in one dimension and the Fourier diaphony in one dimension.

\subsubsection{The Lego discrepancy}
We take the strengths $\si_n$ equal to $1/\sqrt{w_n}$\,, so that the
discrepancy is just the $\chi^2$-statistic that determines how well the points
are distributed over the bins (\Sec{SecDefLego}). The propagator is given by
\begin{equation}
   \twoB(x,y) \;=\; \sum_{n=1}^M\frac{\vt_n(x)\vt_n(y)}{w_n}-1 \;\;, 
\end{equation}
and it is easy to see that $\twoB_p(x,y)=\twoB(x,y)$, $p=2,3,\ldots$, so that
the dressed propagator is given by 
\begin{equation}
   \prop_z(x,y)  \;=\; \frac{1}{1-2z}\,\twoB(x,y) \;\;.
\end{equation}
The zeroth order term can be found with the relation of \eqn{CorEq008}, which
results in the following expression
\begin{equation}
   W_0(z) \;=\; -\frac{1}{2}\log(1-2z)\intk\twoB(x,x)\,dx
               \;=\; -\frac{M-1}{2}\log(1-2z) \;\;,
\label{CorEq017}	       
\end{equation}
which is exactly the logarithm of the generating function of the
$\chi^2$-distribution (notice that this is by definition the distribution of
the $\chi^2$-statistic in the limit of an infinite number of random data). To
write down the first order term, we introduce 
\begin{equation}
   M_2 = \sum_{n=1}^M\frac{1}{w_n} \;\;,
   \quad\textrm{and}\quad \eta(z) = \frac{2z}{1-2z}  \;\;,
\end{equation}
so that 
\begin{equation}
   W_1(z)
   \;=\; \frac{1}{8}\left(M_2-M^2-2M+2\right)\eta(z)^2
         +\frac{1}{24}\left(5M_2-3M^2-6M+4\right)\eta(z)^3  \;\;.
\label{CorEq018}	 
\end{equation}
If the bins are equal, so that $w_n=1/M$ $n=1,2,\ldots,M$, then only the 
contribution of the diagrams of \eqn{CorEq011} remains, and the result is 
\begin{equation}
   W_1(z)
   \;=\; -\frac{1}{4}\,E\eta(z)^2 + \frac{1}{12}(E^2-E)\eta(z)^3 \;\;,
\end{equation}
where we denote
\begin{equation}
   E = M-1 \;\;.
\end{equation}
To second order in $1/N$, the contribution comes from the diagrams in \eqn{CorEq015}, and is given by 
\begin{align}
   W_2(z)
   \;=\;       &(5E^3-12E^2+7E)\frac{\eta(z)^6}{48} 
         \;+\;  (E^3-6E^2+5E)\frac{\eta(z)^5}{8}    \notag\\
	 \;&+\; (E^3-28E^2+43)\frac{\eta(z)^4}{48}
	 \;+\;  (-E^2-5E)\frac{\eta(z)^3}{12}       \;\;.
\label{CorEq020}   
\end{align}
In \App{App6A}, we present the expansion of $G(z)$ in the case of equal bins,
up to and including the $1/N^4$ term. It is calculated using the path integral
expression (\ref{defEq009}) of $G(z)$ and computer algebra. The reader may
check that this expression for $G(z)$ and the above terms of $W(z)$ satisfy
$G(z)=e^{W(z)}$ up to the order of $1/N^2$.

\subsubsection[The $L_2^*$-discrepancy]
           {The $L_2^*$-discrepancy}
In one dimension, the basis in the Landau gauge is given by the set of
functions $\{\sqrt{2}\cos(n\pi x), n=1,2,\ldots\}$ (\Sec{SecDefL2}), so that
the propagator is given by 
\begin{align}
   \twoB(x,y) 
   \;=\; \sum_{n=1}^\infty\frac{2\cos(n\pi x)\cos(n\pi y)}{n^2\pi^2} 
   \;=\; \min(1-x,1-y) + \half x^2 + \half y^2 - \sfrac{2}{3}\;\;.
\end{align}
The dressed propagator is given by 
\begin{align}
   \prop_z(x,y) 
   \;&=\; \sum_{n=1}^\infty\frac{2\cos(n\pi x)\cos(n\pi y)}{n^2\pi^2-2z} \\
   \;&=\; \frac{1}{u^2} - \frac{1}{2u\sin u}
                          \{\cos[u(1-|x+y|)] \;+\; \cos[u(1-|x-y|)]\} \;\;,
\end{align}
with 
\begin{equation}
   u=\sqrt{2z}  \;\;.
\end{equation}
The zeroth order term can be obtained using \eqn{CorEq008}: 
\begin{equation}
   W_0(z) \;=\; -\frac{1}{2}\log\left(\frac{\sin u}{u}\right) \;\;,
\label{CorEq021}   
\end{equation}
which is the well-known result. After some algebra, also the first order term
follows:
\begin{equation}
   W_1(z) 
   \;=\; \frac{1}{288}\left(24-8\frac{u}{\sin u}-7\frac{u^2}{\sin^2u}
                          -7\frac{u}{\tan u}-2\frac{u^2}{\tan^2u}\right) \;\;.
\label{CorEq022}   
\end{equation}

\subsubsection{The Fourier diaphony}
We consider the one-dimensional case, with a slightly different definition than 
the one given in \Sec{SecDefDia}: we multiply the discrepancy with a 
factor $\pi^2/3$, so that the propagator is given by 
\begin{align}
   \twoB(x,y) 
   \;=\; \sum_{n=1}^\infty\frac{2\cos(2n\pi\{x-y\})}{n^2} 
   \;=\; \frac{\pi^2}{3}\left[1-6\{x-y\}(1-\{x-y\})\right]  \;\;,
\end{align} 
where we use the notation $\{x\}=x\mod1$. The dressed propagator is given by 
\begin{equation}
   \prop_z(x,y)
   \;=\; \sum_{n=1}^\infty\frac{2\cos(2n\pi\{x-y\})}{n^2-2z}
   \;=\; \frac{\pi^2}{v^2}\left(1-\frac{v\cos[v(2\{x-y\}-1)]}{2\sin v}\right)
   \;\;,
\end{equation}
where
\begin{equation}
   v=\sqrt{2\pi^2 z} \;\;.
\end{equation}
This two-point function is, apart of a factor $\pi^2/v^2$, the same as the one 
in Eq.~(26) in \cite{hk2}.
The zeroth order term can easily be obtained from the dressed propagator and 
is given by
\begin{equation}
   W_0(z) \;=\; -\log\left(\frac{\sin v}{v}\right) \;\;,
\label{CorEq023}   
\end{equation}
which is in correspondence with Eq.~(21) in \cite{hk2}.
Because the propagator is translation invariant, i.e.,
$\twoB(x+a,y+a)=\twoB(x,y)$ $\forall\,x,y,a\in\Kube$, the contributions
of the first two diagrams in \eqn{CorEq009} cancel, and the contribution of the
fourth diagram is zero. The contribution of the remaining diagrams gives
\begin{equation}
   W_1(z) 
   \;=\; \frac{1}{36}\left(3 + v^2 - 3\frac{v^2}{\sin^2v}\right) \;\;.
\label{CorEq024}   
\end{equation}

\section{Scaling limits for the Lego discrepancy\label{CorSec2}}
In this section, we take a closer look at the Lego discrepancy in the case that
it is equivalent with a $\chi^2$-statistic for $N$ data points distributed over
$M$ bins (\Sec{SecDefLego}).  First, we will show that the natural expansion
parameter in the calculation of the moment generating function is $M/N$, and
calculate a few terms. We will see, however, that a strict limit of
$M\ra\infty$ does not exist, and, in fact, this is well known because the
$\chi^2$-distribution, which gives the lowest order term in this expansion,
does not exist if the number of degrees of freedom becomes infinite. We
overcome this problem by going over to the standardized variable, which is
obtained from the discrepancy by shifting and rescaling it such that it has
zero expectation and unit variance. In fact, it is this variable for which the
results in \cite{Leeb} and Chapter~\ref{GausChap} were obtained. In this
section, we derive similar results for the Lego discrepancy, depending on the
behavior of the sizes of the bins if $M$ goes to infinity. We will see that
various asymptotic probability distributions occur if $M,N\ra\infty$ such that
$M^\al/N\ra\textsl{constant}$ with $\al\geq0$. If, for example, the bins become
asymptotically equal and $\al>\half$, then the probability distribution becomes
Gaussian. Notice that this includes limits with $\al<1$, which is in stark
contrast with the rule of thumb that, in order to trust the
$\chi^2$-distribution, each bin has to contain at least a few, say five (see
e.g. \cite{Knuth}), data points. Our result states that, for large $M$ and $N$,
the majority of bins is allowed to remain empty!

\subsection{Sequences and notation}
In the following, we will investigate limits in which the number of bins $M$
goes to infinity. Note that for each value of $M$, we have to decide on the
values of the volumes $w_n$ of the bins. They clearly have to scale with $M$,
because their sum has to be equal to one. There are, of course, many possible
ways for the measures to scale, i.e., many double-sequences $\{w_n^{[M]},1\leq
n\leq M,M>0\}$ of positive numbers with
\begin{equation}
   \sum_{n=1}^Mw_n^{[M]} \;=\; 1  \quad\forall\,M>0 
   \quad\quad\textrm{and}\quad\quad
   \lim_{M\ra\infty}\sum_{n=1}^Mw_n^{[M]} \;=\; 1  \;\;.
\end{equation}
We, however, want to restrict ourselves to discrepancies in which the relative 
sizes of the bins stay of the same order, i.e., sequences for which
\begin{equation}
     \inf_{n,M}Mw_n^{[M]}>0  \qquad\textrm{and}\qquad
     \sup_{n,M}Mw_n^{[M]}<\infty  \;\;.
\label{LegEq001}   
\end{equation}
It will appear to be appropriate to specify the sequences under consideration 
by another criterion, which is for example satisfied by the sequences mentioned 
above. It can be formulated in terms of the objects
\begin{equation}
   M_p \;\df\; \sum_{n=1}^{M}\left(w_n^{[M]}\right)^{1-p} 
   \;\;,\quad p\geq1  \;\;,
\label{LegEq022}   
\end{equation}
and is given by the demand that 
\begin{equation}
    h_p\in[1,\infty) \quad\forall\,p\geq1 \;\;,\quad\textrm{where}\quad
    h_p\df\lim_{M\ra\infty}\frac{M_p}{M^p} \;\;.
\label{LegEq018}   
\end{equation}
Within the set of sequences we consider, there are those with for which the bins become asymptotically equal, i.e., sequences with 
\begin{equation}
   w_n^{[M]} = \frac{1 + \vp_n^{[M]}}{M}          \qquad\textrm{with}\qquad
   \lim_{M\ra\infty}\max_{1\leq n\leq M}\left|\vp_n^{[M]}\right| = 0  \;\;,
\end{equation}
and $\vp_n^{[M]}>-1$, $1\leq n\leq M$ of course.
They belong to the set of sequences with $h_p=1$ $\forall\,p\geq1$, which will
allow for special asymptotic probability distributions.

In the following analysis, we will consider functions of $M$ and their
behavior if $M\ra\infty$. To specify relative behaviors, we will use the
symbols ``$\sim$'', ``$\asymp$'' and ``$\prec$''. The first one is used as
follows:
\begin{equation}
   f_1(M) \sim f_2(M) \quad\Longleftrightarrow\quad 
   \lim_{M\ra\infty} \frac{f_1(M)}{f_2(M)} \;=\; 1 \;\;.
\end{equation}
If a limit as 
above is not necessarily equal to one and not equal to zero, then we use the second symbol:
\begin{equation}
   f_1(M) \asymp f_2(M) \quad\Longleftrightarrow\quad 
   f_1(M) \sim cf_2(M) \;\;,\quad c\in(0,\infty) \;\;.
\end{equation}
We only use this symbol for those cases in which $c\neq0$. For the cases in
which $c=0$ we use the third symbol:
\begin{equation}
   f_1(M) \prec f_2(M) \quad\Longleftrightarrow\quad 
   \lim_{M\ra\infty} \frac{f_1(M)}{f_2(M)} \;=\; 0 \;\;.
\end{equation}
We will also use the $\Ord$-symbol, and do this in the usual sense.
We can immediately use the symbols to specify the behavior of $M_p$ with $M$, 
for the criterion of \eqn{LegEq018} tells us that
\begin{equation}
   M_p \asymp M^p  \;\;,
\label{LegEq019}   
\end{equation}
and that
\begin{equation}
   M_p \sim M^p  \quad\textrm{if}\quad h_p=1 \;\;.
\end{equation}

In our formulation, also the number of data points $N$ runs with $M$.
We will, however, never denote the dependence of $N$ on $M$ explicitly and
assume that it is clear from now on. Also the upper index at the measures $w_n$
we will omit from now on.

\subsection{Feynman rules}
The Feynman rules to calculate the generating function $G(z)$ in a diagrammatic
expansion are given in \Sec{FormSec1}. The boson propagator is a matrix in this
case, i.e., 
\begin{equation}
   \textrm{boson propagator:}\quad
    n\,\diagram{dcP2}{30}{-1}\,m \;=\; \twoB_{n,m}=\frac{\de_{n,m}}{w_n}-1
    \;\;, 
\end{equation}
and boson propagators are convoluted as
$\sum_{m=1}^Mw_m\twoB_{m,n_1}\twoB_{m,n_2}\cdots\twoB_{m,n_p}$ in the vertices.
Only {\em connected} diagrams have to be calculated, since 
\begin{equation}
   \log G(z) \;=\; \textsl{the sum of the connected vacuum diagrams.}
   \tag*{rule 1}
\end{equation}
Furthermore, the bosonic part of each diagram decouples completely from the
fermionic part, and the contribution of the fermionic part can easily be
determined, for 
\begin{equation}
   \textsl{every fermion loop only gives a factor $-N$.}\tag*{rule 2}
\end{equation}
Because of the rather simple expression for the bosonic
propagator, we are able to deduce from the basic Feynman rules some effective
rules for the bosonic parts of the Feynman diagrams. Remember that the bosonic
parts decouples completely from the fermionic parts. The following rules apply
after having counted the number of fermion loops and the powers of $g$ coming
from the vertices, and after having calculated the symmetry factor of the
original diagram. When we mention the {\em contribution} of a diagram in this
section, we refer to the contribution apart from the powers of $g$ and symmetry
factors. This contribution will be represented by the same drawing as the
diagram itself.

The first rule is a consequence of the fact that 
\begin{equation}
   \sum_{n=1}^Mw_m\twoB_{n_1,m}\twoB_{m,n_2}
   \;=\; \twoB_{n_1,n_2}
\end{equation}
and states that 
\begin{equation}
  \textsl{all vertices with only two legs that do not form a single loop
          can be removed.}
\tag*{rule 3}	  
\end{equation}
The second rule is a consequence of the fact that for any $M\times M$-matrix
$f$ 
\begin{equation}
   \sum_{n,m=1}^Mw_nw_mf_{n,m}\twoB_{n,m}
   \;=\; \sum_{n=1}^Mw_nf_{n,n} 
         - \sum_{n,m=1}^Mw_nw_mf_{n,m}   \;\;,
\end{equation}
and states that the contribution of a diagram is the same as that of
the diagram in which a boson line is contracted and the two vertices,
connected to that line, are fused together to form one vertex, minus the
contribution of the diagram in which the line is simply removed and the
vertices replaced by vertices with one boson leg less. This rule
can be depicted as follows
\begin{equation}
   \diagram{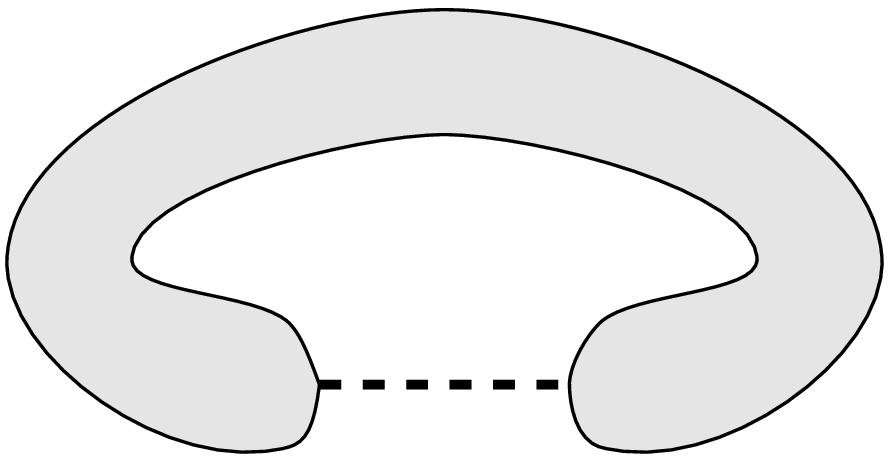}{56}{13} 
   \;=\; \diagram{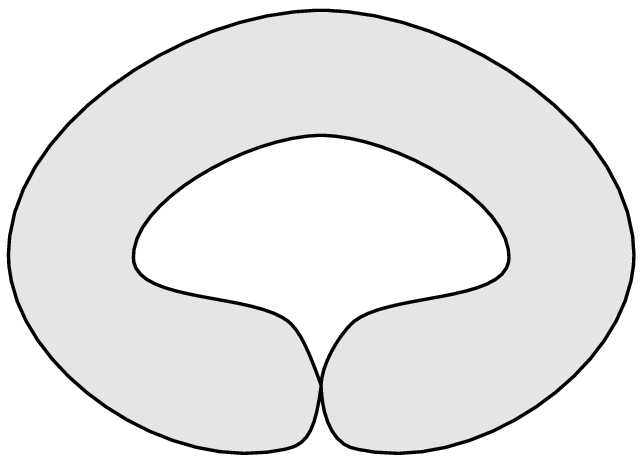}{39}{13} - \diagram{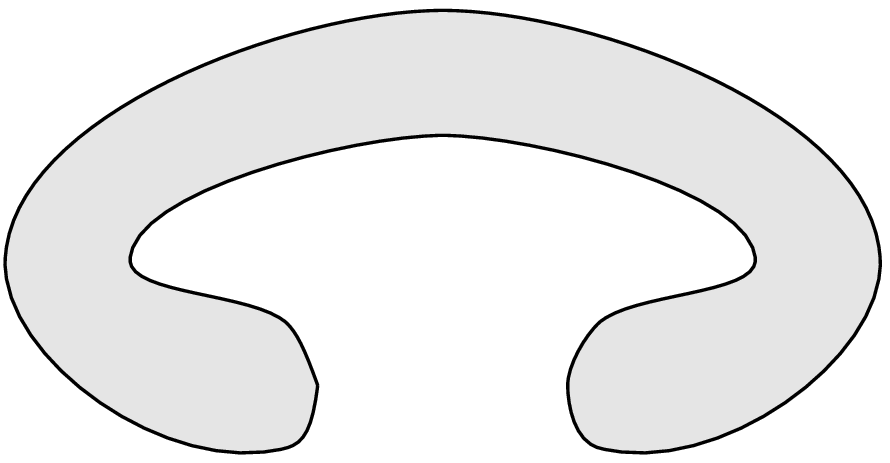}{56}{13} \;\;.\tag*{rule 4}
\end{equation}
By repeated
application of these rules, we see that the contribution of a connected
bosonic diagram is equal to the contribution of a sum of products
of so called {\em daisy} diagrams~\footnote{For example 
$\diagram{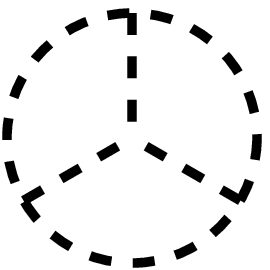}{20}{8}=\diagram{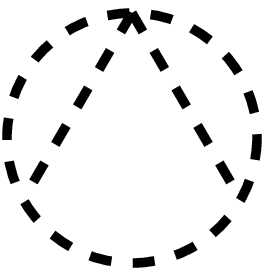}{20}{8}-\diagram{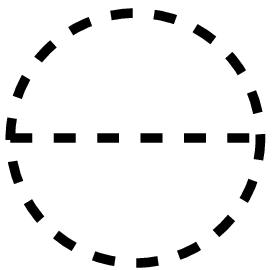}{20}{8}
                   =\diagram{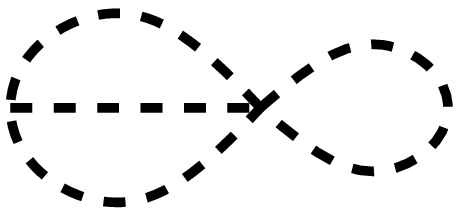}{34}{6}-2\,\diagram{lcBr3}{20}{8}
		   =\diagram{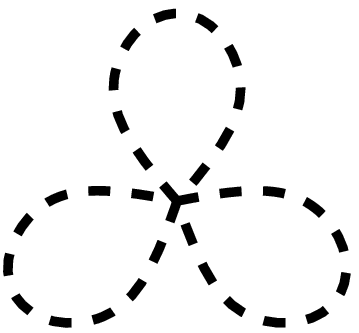}{26}{10}-\diagram{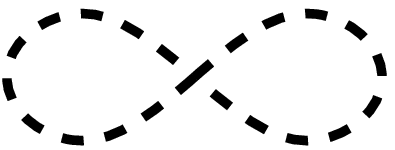}{30}{4}
	       - 2\left(\diagram{lcBr6}{30}{4}-\diagram{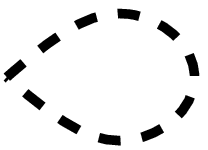}{15}{4}\right)$},
which are of the type 
\begin{equation}
   \diagram{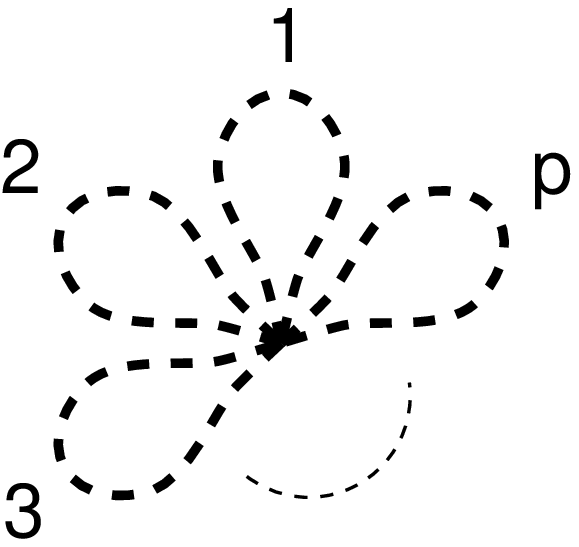}{50}{18} \;\;.
\label{LegEq003}   
\end{equation} 
They are characterized by the fact that all lines begin and end on the same
vertex and form single loops. The contribution of such a diagram is given by 
\begin{equation}
   d_p(M) 
   \;=\; \sum_{n=1}^Mw_n\twoB_{n,n}^p
   \;=\; \sum_{q=0}^p\binom{p}{q}(-1)^{p-q}M_q  
   \;=\; M_p[1+\Ord(M^{-1})]\;\;,
\label{LegEq021}   
\end{equation}
where the last equation follows from \eqn{LegEq019}.
The maximal number of leaves in a product in the sum of
daisy diagrams is equal to the number of loops $\Lb$ in the original
diagram, so that 
\begin{equation}
   \textsl{the contribution of a diagram with $\Lb$ boson loops 
           is $M_{\Lb}[1+\Ord(M^{-1})]$.}  \tag*{rule 5}
\end{equation}
The leading order contribution of a diagram with $\Lb$ boson loops is thus 
of the order of $M^{\Lb}$.

\subsubsection{Extra rule if ${h_p=1}$}
If $h_p=1$ $\forall\,p\geq1$, then all kind of cancellations between diagrams 
occur, because in those cases $M_p\sim M^p$ $\forall\,p\geq1$.
As a result of this, the contribution of a daisy diagram is $d_p(M)\sim M^p$,
and we can deduce the following rule: the contribution of a diagram that falls
apart in disjunct pieces if a vertex is cut, is equal to the product of the
contributions of those disjunct pieces times one plus vanishing corrections.
Diagrammatically, the rule looks like
\begin{equation}
   \diagram{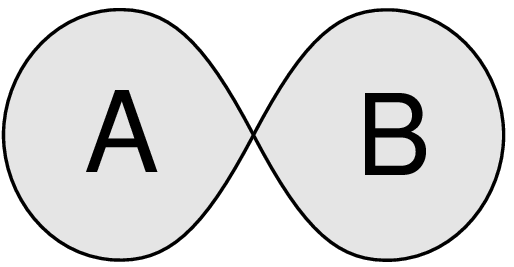}{45}{10} 
   \;\sim\; \diagram{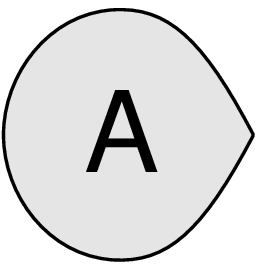}{23}{10}\times\diagram{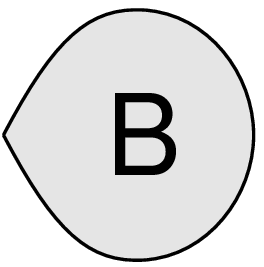}{23}{10} \;\;,
\label{LegEq016}
\end{equation}
In \Sec{OVRSec} we called discrepancies for which \eqn{LegEq016} is exact 
one-vertex decomposable, and have shown that for those discrepancies only the 
{\em one-vertex irreducible} diagrams contribute, i.e., diagrams that do not 
fall apart in pieces containing bosonic parts if a vertex is cut. The previous 
rule tells us that, if $h_p=1$ $\forall\,p\geq1$, then 
\begin{equation}
   \log G(z) 
   \;\sim\; \textsl{sum of all connected one-vertex irreducible
                      diagrams}. \tag*{rule 6}
\end{equation}
The connected one-vertex irreducible diagrams we call {\em relevant} and the 
others {\em irrelevant}.

\subsection{Loop analysis}
We want to determine the contribution of the diagrams in this section, and in 
order to do that, we need to introduce some notation:
\begin{align}
   \Lb  \;&\df\; \textrm{the number of boson loops} \;\;;\\
   \Lf  \;&\df\; \textrm{the number of fermion loops} \;\;;\\
   L    \;&\df\; \textrm{the total number of loops}  \;\;;\\
   \Ib  \;&\df\; \textrm{the number of bosonic lines} \;\;;\\
   \If  \;&\df\; \textrm{the number of fermionic lines} \;\;;\\
   v    \;&\df\; \textrm{the number of vertices} \;\;;\\
   \Lm  \;&\df\; L-\Lb-\Lf \;=\; \textrm{number of mixed loops}\;\;.
\end{align}
These quantities are in principle functions of the diagrams, but we will never 
denote this dependence explicitly, for it will always be clear which diagram 
we are referring to when we use the quantities. 

With the foregoing, we deduce that the contribution $C_\Delta$ of 
a connected diagram $\Delta$ with no external legs satisfies
\begin{equation}
    C_\Delta \;\asymp\; M^{\Lb}N^{\Lf}g^{2\Ib}  \;\;.
\end{equation}
The Feynman rules and basic graph theory tell us that, for connected diagrams 
with no external legs, $v=\If$ and $L=\Ib+\If-v+1$, so that 
\begin{equation}
   \Ib \;=\; L-1 \;=\; \Lb+\Lf+\Lm-1 \;\;.
\end{equation}
If we furthermore use that $g=\sqrt{2z/N}\,$, we find that the contribution is 
given by 
\begin{equation}
   C_\Delta \;\asymp\; \frac{M^{\Lb}}{N^{\Lm}N^{\Lb-1}}\,(2z)^{\Ib} \;\;.
\label{LegEq005}   
\end{equation}
Notice that this expression does not depend on $\Lf$. Furthermore, it is clear
that, for large $M$ and $N$, the largest contribution comes from diagrams with
$\Lm=0$. Moreover, we see that we must have $N=\Ord(M)$, for else the
contribution of higher-order diagrams will grow with the number of boson loops,
and the perturbation series becomes completely senseless. If, however, $N\asymp
M$, then the contribution of each diagram with $\Lm=0$ is more important than
the contribution of each of the diagrams with $\Lm>0$. Finally, we also see
that the contribution of the $\Ord(M^{-1})$-corrections of a diagram
(\eqn{LegEq021}) is always negligible compared to the leading contribution of
each diagram with $\Lm=0$.  These observations lead to the conclusion that, if
$N$ and $M$ become large with $N\asymp M$, then the leading contribution to
$\log G(z)$ comes from the diagrams with $\Lm=0$, and that there are no
corrections to these contributions. If we assume that $M/N$ is small, then the
importance of these diagrams decreases with the number of boson loops $\Lb$ as
$(M/N)^{\Lb}$.

\subsubsection{The loop expansion of ${\log G(z)}$}
Now we calculate the first few terms in the loop expansion of $\log G(z)$. We
start with the diagrams with one loop (remember that it is an expansion in
boson loops and that we only have to calculate {\em connected} diagrams for
$\log G(z)$). The sum of all $1$-loop diagrams with $\Lm=0$ is given by the
l.h.s.~of \eqn{CorEq025}, resulting in the r.h.s.~of \eqn{CorEq017}.  To
calculate the higher loop diagrams, we introduce the effective vertex of
\eqn{CorEq026} and the partly re-summed propagator
\begin{align}   
   n\,\diagram{dcP6}{30}{-1}\,m 
   \;&\df\;     n\,\diagram{dcP2}{30}{-1}\,m \;+\; n\,\diagram{dcP3}{30}{0}\,m 
          \;+\; n\,\diagram{dcP4}{39}{0}\,m 
	  \;+\; n\,\diagram{dcP5}{60}{0}\,m \;+\; \cdots \notag\\
   \;&=\; \sum_{p=0}^{\infty}(Ng^2)^p\times n\,\diagram{dcP2}{30}{-1}\,m 
   \;=\; \frac{1}{1-2z}\times n\,\diagram{dcP2}{30}{-1}\,m \;\;.
\end{align}
The contribution of the $2$-loop diagrams with $\Lm=0$ is given by 
\begin{align}
   & \frac{1}{8}\,\diagram{dcA11}{31}{3.5}
   \;+\; \frac{1}{8}\,\diagram{dcA22}{45}{4.5}
   \;+\; \frac{1}{8}\,\diagram{dcA13}{43}{3.5}
   \;+\; \frac{1}{12}\,\diagram{dcA12}{24}{8.5} 
    \notag\\
   \;&=\; \left[
          - \frac{1}{8}\,\frac{Ng^2M^2}{(1-2z)^2}
	  + \frac{1}{8}\,\frac{Ng^2M_2}{(1-2z)^2}
          + \frac{1}{8}\,\frac{(Ng^3)^2(M_2-M^2)}{(1-2z)^3}
          + \frac{1}{12}\,\frac{(Ng^3)^2M_2}{(1-2z)^3}\right]
	  [1+\Ord(M^{-1})] \notag\\
   \;&=\; \frac{1}{N}
         \left[\frac{1}{8}(M_2-M^2)\eta(z)^2 
	  + \left(\frac{5M_2}{24}-\frac{M^2}{8}\right)\eta(z)^3\right]
	  [1+\Ord(M^{-1})] \;\;,
\end{align}
where we define
\begin{equation}
   \eta(z) \;=\; \frac{2z}{1-2z} \;\;.
\end{equation}
Notice that the first three diagrams vanish if $h_p=1$
$\forall\,p\geq1$. The contribution of the $3$-loop diagrams with $\Lm=0$ is 
given by
\begin{align}
&    \frac{1}{48}\,\diagram{dcB12}{24}{9}
\;+\;\frac{1}{24}\,\diagram{dcB111}{23}{9.5}
\;+\;\frac{1}{8}\,\diagram{dcB18}{24}{11}
\;+\;\frac{1}{16}\,\diagram{dcB115}{32}{10} 
\;+\;\frac{1}{48}\,\diagram{dcB11}{28}{9}
\;+\;\frac{1}{12}\,\diagram{dcB14}{38}{7}  
\;+\;\frac{1}{8}\,\diagram{dcB16}{58}{4}\notag\\
&\;+\;\frac{1}{16}\,\diagram{dcB13}{46}{4}
\;+\;\frac{1}{8}\,\diagram{dcB17}{37}{8}
\;+\;\frac{1}{16}\,\diagram{dcB19}{45}{12}
\;+\;\frac{1}{8}\,\diagram{dcB212}{71}{6}
\;+\;\frac{1}{16}\,\diagram{dcB114}{72}{4}\notag\\
&\;+\;\frac{1}{8}\,\diagram{dcB113}{49}{7}
\;+\;\frac{1}{12}\,\diagram{dcB110}{49}{6}  
\;+\;\frac{1}{8}\,\diagram{dcB21}{60}{6}
\;+\;\frac{1}{16}\,\diagram{dcB22}{41}{13}
\;+\;\frac{1}{8}\,\diagram{dcB26}{49}{8}\notag\\
&\;+\;\frac{1}{24}\,\diagram{dcB32}{41}{14} 
\;+\;\frac{1}{48}\,\diagram{dcB112}{45}{17}
\;+\;\frac{1}{16}\,\diagram{dcB213}{45}{21}
\;+\;\frac{1}{8}\,\diagram{dcB214}{45}{11}
\;+\;\frac{1}{16}\,\diagram{dcB15}{34}{13}\notag\\
&\;+\;\frac{1}{12}\,\diagram{dcB27}{53}{8}
\;+\;\frac{1}{16}\,\diagram{dcB34}{74}{5} \;\;.
\end{align}
If $h_p=1$ $\forall\,p\geq1$, then only the first four diagrams are relevant, 
and their contribution $C$ satisfies
\begin{equation}
   C \;\sim\; \frac{M^3}{N^2}\left[      \frac{1}{48}\eta(z)^4
                        \;+\; \frac{1}{8}\eta(z)^5
			\;+\; \frac{5}{48}\eta(z)^6\right] \;\;.
\end{equation}

\subsection{Various limits}
In the previous calculations, $M/N$ was the expansion parameter and the
expansion of the generating function only makes sense if it is considered to be
small. Furthermore, a limit in which $M\ra\infty$ does not exist, because
the zeroth order term is proportional to $M$. In order to analyze limits in 
which $M$ as well as $N$ go to infinity, we can go over to the standardized 
variable $(D_N-E)/\sqrt{V}\,$ of the discrepancy (\Sec{SecProbRV}), where
\begin{align}
   E   \;&\df\; \Exp(D_N) \;=\; M-1 \\   
   V \;&\df\; \Var(D_N) \;=\; 2(M-1) + \frac{M_2-M^2-2(M-1)}{N}  \;\;.
\end{align}
The generating function of the probability distribution of the standardized
variable is given by 
\begin{equation}
   \Hat{G}(\xi) 
   \;=\; \Exp\left(\,\exp\left(\xi\,\frac{D_N-E}{\sqrt{V}}\right)\,\right) 
   \;=\; \exp\left(-\frac{E\xi}{\sqrt{V}}\right)\,
         G\left(\frac{\xi}{\sqrt{V}}\right)  \;\;. 
\label{LegEq008}
\end{equation}
Instead of the parameter $z$, the parameter
$\xi=z\sqrt{V}$ is considered to be of 
$\Ord(1)$ in this perspective, in the sense that it are these values of $\xi$ 
that give the important contribution to the inverse Laplace transform to get 
from the generating function to the probability density, and the 
contribution of a diagram changes from (\ref{LegEq005}) to 
\begin{equation}
   C_\Delta \;\asymp\; 
   \frac{M^{\Lb}}{N^{\Lm}N^{\Lb-1}V^{\shalf(\Lb+\Lf+\Lm-1)}}\,
   (2\xi)^{\Ib} \;\;.
\end{equation}
In the following we will investigate limits of $M\ra\infty$ with, at first
instance, the criterion of \eqn{LegEq018} as only restriction. The fact that
the variance $V$ shows up explicitly in the contribution of the diagrams,
forces us to specify the behavior of $M_2$ more precisely. We will take
\begin{equation}
   M_2-M^2 \asymp M^\ga \;\;,\quad 0\leq\ga\leq2 \;\;.
\label{LegEq023}   
\end{equation}
Notice that $h_2=1$ if $\ga<2$ and that $h_2$ does not exist if $\ga>2$. 
Furthermore, we cannot read off the natural expansion parameter from the 
contribution of the diagrams anymore, and have to specify the behavior of $N$. 
We will only consider limits in which 
\begin{equation}
   N \asymp M^\al \;\;,\quad \al>0  \;\;.
\label{LegEq024}
\end{equation}
Although they are a small subset of possible limits, those that can be
specified by a pair $(\al,\ga)$ show an interesting picture. We will derive the
results in the next section, but present them now in the following phase
diagram:
\begin{center}
\begin{picture}(160,160)(0,0)
\LongArrow(30,10)(160,10)
\LongArrow(30,140)(160,140)
\Line(30,10)(30,139.5)
\Line(30,105)(33,105)
\Line(61.5,105)(64.5,105)
\DashLine(65,10)(65,105){3}
\DashLine(65,105)(30,140){3}
\Text(30,0)[]{$0$}
\Text(65.5,0)[]{$\frac{1}{2}$}
\Text(20,10)[]{$0$}
\Text(20,106)[]{$\frac{3}{2}$}
\Text(20,140)[]{$2$}
\LongArrow(10,75)(10,85)
\Text(2,80)[]{$\gamma$}
\Text(150,0)[]{$\alpha$}
\Text(48,80)[]{$\bs{T}$}
\Text(110,100)[]{$\bs{U}$}
\LongArrow(105,40)(70,50)
\Text(110,40)[]{$\bs{\ell}$}
\end{picture}    
\end{center}
\vspace{20pt}
It shows the region
$\bs{S}=\{(\al,\ga)\in\Real^2\,|\,\al\in[0,\infty),\,\ga\in[0,2]\}$ of the
real $(\al,\ga)$-plane. In this region, there is a {\em critical line}
$\bs{\ell}$, given by 
\begin{equation}
   \bs{\ell}\df\{(f_{\bs{\ell}}(t),t)\in\bs{S} \mid t\in[0,2]\}  
   \quad\textrm{with}\quad
   f_{\bs{\ell}}(t) \df \begin{cases}
             \half &\textrm{if $0\leq t\leq\frac{3}{2}$}\;\;,\\
	     2-t   &\textrm{if $\frac{3}{2}\leq t\leq 2$}\;\;.
          \end{cases}
\end{equation}
It separates $\bs{S}$ into two regions $\bs{T}$ and $\bs{U}$, neither of which
contains $\bs{\ell}$.
Our results are the following. Firstly, 
\begin{equation}
   \textsl{in the region $\bs{T}$, the limit of $M\ra\infty$ is not defined.}
\label{LegRes01}   
\end{equation}
In this region, the standardized variable is not appropriate, and we see that 
there are too many diagrams that grow indefinitely with $M$. Secondly, 
\begin{equation}
   \textsl{in the region $\bs{U}$, the limit of $M\ra\infty$ gives a Gaussian 
           distribution.}
\label{LegRes02}   
\end{equation}
Because we used the standardized variable, this distribution has necessarily 
zero expectation and unit variance. Finally,  
\begin{equation}
   \textsl{on the line $\bs{\ell}$, various limits exist, depending on the 
           behaviour of $M_p$, $p>2$.}
\label{LegRes03}   
\end{equation}
One of these limits we were able to calculate explicitly. It appears if
$M_p-M^p\prec M^{p-\frac{1}{2}}$ $\forall p\geq1$, which is, for example,
satisfied in the case of equal binning. In this limit, the generating function
is given by 
\begin{equation}
   \log\Hat{G}(\xi) \;=\; 
   \frac{1}{\la^2}\left(e^{\la\xi}-1-\la\xi\right) \;\;,\quad 
   \la\df\lim_{M\ra\infty}\frac{\sqrt{2M}}{N} \;\;.
   \label{LegEq014}
\end{equation}
In \App{App6B}, we show that the probability distribution $\Hat{H}$ belonging
to this generating function, which is the inverse Laplace transform, is given
by 
\begin{equation}
   \Hat{H}(\tau) \;=\; 
   \sum_{n\in\Natu}
   \de\left(\tau-\left[n\la-\frac{1}{\la}\right]\right)
   \frac{1}{n!}\left(\frac{1}{\la^2}\right)^n
   \exp\left(-\frac{1}{\la^2}\right) \;\;.
\label{LegEq020}   
\end{equation}
It consists of an infinite number of Dirac delta-distributions, weighed with a
Poisson distribution. 
The delta-distributions reveal the fact that, for finite
$N$ and $M$, the Lego discrepancy, and also the $\chi^2$-statistic, can only
take a finite number of values, so that the probability density {\em should}
consist of a sum of delta-distributions. In the usual limit of $N\ra\infty$, the
discrete nature of the random variable disappears, and the
$\chi^2$-distribution is obtained. In our limit, however, the discrete nature
does not yet disappear. A continuous distribution is obtained if $\la\ra0$,
which corresponds with going over from $\al=\half$ to $\al>\half$. Then 
$\Hat{G}(\xi)\ra\exp(\half\xi^2)$.

\subsection{Derivation of the various limits}
We will deal with the cases $\ga=2$, $\ga-\al\leq1$  and $\ga-\al>1$
separately.

\subsubsection{${\ga=2}$}
We distinguish the three cases $0<\al<1$, $\al=1$ and $\al>1$. 

If $\al>1$, then $V\asymp M$, and the contribution $C_\Delta$ of a diagram $\Delta$
satisfies $C_\Delta\asymp M^\beta$, with 
\begin{equation}
   \beta \;=\; (\half -\al)\Lb - \half \Lf + (\al + \half)(1-\Lm) \;\;.
\end{equation}
A short analysis shows that only diagrams with $(\Lb,\Lf,\Lm)=(1,1,0)$ or 
$(\Lb,\Lf,\Lm)=(1,2,0)$ give a non-vanishing contribution, and those diagrams 
are
\begin{align}
   \frac{1}{2}\,\diagram{dcZ1}{30}{6} 
   \;&=\; \frac{1}{2}\,\frac{N(M-1)2\xi}{N\sqrt{V}} 
                        \;=\;  \frac{E\xi}{\sqrt{V}}\label{LegEq009}\\
   \frac{1}{4}\,\diagram{dcZ2}{45}{6} 
   \;&=\; \frac{1}{4}\frac{N^2(M-1)4\xi^2}{N^2V}
                          \;=\; \frac{\xi^2}{2} + \Ord(M^{-1}) 
			  \label{LegEq010}  \;\;.
\end{align}
The first diagram gives a contribution that is linear in $\xi$ and cancels 
with the exponent in \eqn{LegEq008}. This has to happen for every value of 
$\al$, and as we will see, this diagram will occur always. Notice that the 
diagrams above are the first two diagrams in the series on the l.h.s~of 
\eqn{CorEq025}. The logarithm of the generating function becomes 
quadratic, so that the probability distribution becomes Gaussian.

If $\al=1$, then again $V\asymp M$, so that
$\beta=-\half(\Lb+\Lf)+\sfrac{3}{2}(1-\Lm)$, and we have to add the diagrams
with $(\Lb,\Lf,\Lm)=(2,1,0)$:
\begin{equation}
   \frac{1}{8}\,\diagram{lcF1B2b}{30}{11} \;+\; 
   \frac{1}{8}\,\diagram{lcF1B2}{45}{8} 
   \;=\; \frac{1}{8}\,\frac{N(M_2-M^2)4\xi^2}{N^2V} 
   \;=\;  \frac{(M_2-M^2)\xi^2}{2NV}  \;\;.
\label{LegEq011}   
\end{equation}

If $0<\al<1$, then $V\asymp M^{2-\al}$ and 
$\beta=-\sfrac{\al}{2}\Lb-(1-\sfrac{\al}{2})\Lf-(\sfrac{\al}{2}+1)\Lm+\sfrac{\al}{2}+1$, so that, besides the diagram of \eqn{LegEq009}, only the diagrams of 
\eqn{LegEq011} give a non-vanishing contribution, and this contribution is 
equal to $\xi^2/2$.

\subsubsection{${\ga-\al\leq1}$}
In this case, $V\asymp M$, and the contribution $C_\Delta$ of a diagram $\Delta$
satisfies $C_\Delta\asymp M^\beta$ with 
\begin{equation}
   \beta \;=\; (\half -\al)\Lb - \half \Lf + (\al + \half)(1-\Lm) \;\;.
\end{equation}

If $\al<\half$, then $\beta$ increases with the number of boson loops $\Lb$,
and we are not able to calculate the limit of $M\ra\infty$. 

If $\al>\half$,
then the only diagrams that have a non-vanishing contribution are those with
$(\Lb,\Lf,\Lm)=(1,1,0)$, $(1,2,0)$ or $(2,1,0)$. These are exactly the
diagrams of \eqn{LegEq009}, \eqn{LegEq010} and \eqn{LegEq011}. Notice, however,
that the diagrams of \eqn{LegEq011} cancel if $\ga-\al<0$: then they are {\em
irrelevant}. The resulting asymptotic distribution is Gaussian again. 

If $\al=\half$, then $\Lb$ disappears from the equation for $\beta$, and we
obtain a non-Gaussian asymptotic distribution. The diagrams that contribute are
those with $(\Lf,\Lm)=(1,0)$ or $(2,0)$. There is, however, only one {\em
relevant} diagram with $(\Lf,\Lm)=(1,0)$, namely the diagram of \eqn{LegEq009}
that gives the linear term. We have to be careful here, because the other
diagrams with $(\Lf,\Lm)=(1,0)$ still might be non-vanishing. 
A short analysis shows that they are given by the sum of all ways to put 
daisy diagrams to one fermion loop, and that their contribution is given by 
\begin{equation}
   C_1(M) \;=\; 
   N\log\left(1+\sum_{p=1}^\infty\frac{(\half g^2)^pd_p(M)}{p!}\right) \;\;.
\end{equation}
We know that, if $h_p=1$, then  $d_p(M)=M^p[1+\vp_p(M)]$ with
$\lim_{M\ra\infty}\vp_p(M)=0$, so that 
\begin{equation}
   C_1(M) 
   \;=\; \half NMg^2 
         + N\log\left(1 + e^{-\shalf Mg^2}
	   \sum_{p=1}^\infty\frac{(\half Mg^2)^p\vp_p(M)}{p!}\right)  \;\;.
\label{LegEq004}	   
\end{equation}
The first term gives the leading contribution; the contribution of the relevant
diagram, which consists of a boson loop and a fermion loop attached to one
vertex. The second term is irrelevant with respect to the first, but can still
be non-vanishing, depending on the behavior of $\vp_p(M)$. Remember that
$\al=\half$ and $V\asymp M$, so that $Mg^2=2\xi
M/(N\sqrt{V}\,)\ra\textsl{constant}$, and we can see that the contribution is
only vanishing if 
\begin{equation}
   \lim_{M\ra\infty}N\vp_p(M)
   \;=\;0 \quad\forall\,p\geq1 \quad\Longleftrightarrow\quad
   M_p-M^p\prec M^{p-\frac{1}{2}} \quad\forall\,p\geq1 \;\;.
\end{equation}
For $p=1$ this relation is satisfied because $\vp_1(M)=0$. For $p=2$ this
relation is also satisfied if $\ga<\frac{3}{2}$. 

If the relation is also satisfied for the other values of $p$,
then the only diagrams that contribute to the generating function are the
relevant diagrams with $(\Lf,\Lm)=(2,0)$:
\begin{equation}
   \diagram{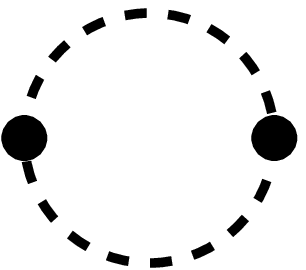}{30}{11} \;+\;
   \diagram{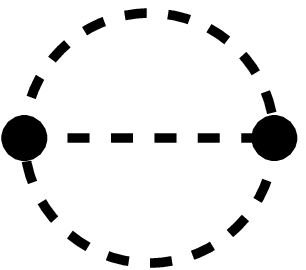}{30}{11} \;+\;
   \diagram{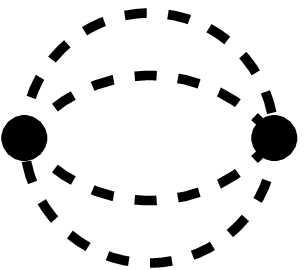}{30}{11} \;+\; \cdots  \;\;, 
\label{LegEq012}
\end{equation}
where we used the effective vertex (\ref{CorEq026}) again.
The contribution of a diagram of this type with $p$ boson lines is given by 
\begin{equation}
   \frac{1}{2\,p!}\left(\frac{2\xi}{N\sqrt{V}}\right)^p
         N^2M_p[1+\Ord(M^{-1})] 
   \;\sim\;\frac{N^2}{2M}\,\frac{1}{p!}
              \left(\frac{2M\xi}{N\sqrt{V}}\right)^p.
\label{LegEq013}	 
\end{equation}
The factor $1/2\,p!$ is the symmetry factor of this type of diagram. If we sum 
the contribution of these diagrams and use that $V\sim2M$, we obtain
\begin{equation}
   \log\Hat{G}(\xi) \;\sim\; 
   \frac{1}{\la^2}\left(e^{\la\xi}-1-\la\xi\right) \;\;,\quad 
   \la\df\lim_{M\ra\infty}\frac{\sqrt{2M}}{N} \;\;.
\end{equation}

\subsubsection{${\ga-\al>1}$}
In this case, $V\asymp M^{\ga-\al}$ and the contribution $C_\Delta$ of a diagram $\Delta$
satisfies $C_\Delta\asymp M^\beta$ with
\begin{equation}
   \beta \;=\; (1-\sfrac{\ga+\al}{2})\Lb - \sfrac{\ga-\al}{2}\,\Lf 
               + \sfrac{\ga+\al}{2}(1-\Lm) \;\;.
\end{equation}

If $\ga+\al<2$, then $\beta$ increases with the number of boson loops $\Lb$,
and we are not able to calculate the limit of $M\ra\infty$. 

If $\ga+\al>2$,
then the only diagrams that have a non-vanishing contribution are those with
$(\Lb,\Lf,\Lm)=(1,1,0)$, $(1,2,0)$ or $(2,1,0)$. These are exactly the
diagrams of \eqn{LegEq009}, \eqn{LegEq010} and \eqn{LegEq011}. Notice, however,
that the diagrams of \eqn{LegEq011} cancel if $\ga-\al<0$: then they are {\em
irrelevant}. The resulting asymptotic distribution is Gaussian. 

If $\ga+\al=2$, then $\beta=(\al-1)\Lf+1-\Lm$. Because $\ga-\al>1$, we have 
$\al<\half$, and non-vanishing diagrams have $(\Lf,\Lm)=(1,0)$. 
Their contribution is given by the r.h.s.~of \eqn{LegEq004}, the first term 
of which gives the term linear in $\xi$. The second term is non-vanishing, 
because $Mg^2\asymp M^{1-(\ga+\al)/2}\ra\textsl{constant}$ and
$N\vp_2(M)\asymp M^{\al+\ga-2}\ra\textsl{constant}$.

\section{Stronger-than-weak limits for diaphony\label{CorSec3}}
In \cite{Leeb}, it is proven that the standardized variable of the Fourier
diaphony (\Sec{SecDefDia}) converges in distribution to a Gaussian variable if
the number $N$ of points in the point set, and with it the number $s_N$ of
dimensions of the integration region, goes to infinity such that
$c_{\textrm{w}}^{s_N}/N$ goes to zero, where  
\begin{equation}
   c_{\textrm{w}} \;=\; 
   \frac{\sqrt{1+\frac{2\pi^4}{15}+\frac{8\pi^6}{945}+\frac{\pi^8}{945}}}
        {1+\frac{\pi^4}{45}} \;=\; 1.79218\cdots \;\;.
\end{equation}
To be more precise, if $s_1,s_2,\ldots$ is a nondecreasing sequence such that 
\begin{equation}
   \lim\sup_N\frac{c_{\textrm{w}}^{s_N}}{N}=0 \qquad\textrm{then}\qquad
   \frac{D_N-\Exp(D_N)}{\sqrt{\Var(D_N)}} \;\dcon\; \textrm{Normal} \;\;,
\label{weaklim}   
\end{equation}
where we include in the notation ``$D_N$'' the dependence on $N$ through $s_N$.
The proof makes use of the Central Limit Theorem as given in \Sec{CeLiTh}, and 
is, roughly speaking, based on the fact that the conditions in the theorem are
satisfied if $\Exp(D_N^4)/\Var(D_N)^2\ra0$. 

\subsection{The observation}
The Central Limit Theorem provides a weak limit for the distribution, which
becomes dramatically clear in a short calculation. 
The expectation value and the variance of the diaphony are given by  
$\Exp(D_N) = \intk\twoB(x,x)\,dx$ and 
$\Var(D_N)=2\frac{N-1}{N}\intk\twoB(x,y)^2\,dxdy$, leading to 
\begin{equation}
  E \df \Exp(D_N) = 1 \quad,\qquad
  V_N \df \Var(D_N) 
       = 2\,\frac{N-1}{N}\frac{\left(1+\frac{\pi^4}{45}\right)^{s_N}-1}
 	      {\left[\left(1+\frac{\pi^2}{3}\right)^{s_N}-1\right]^2} \;\;.
\end{equation}
The moments of the diaphony contain contributions of all kind of
convolutions of the reduced two-point function $\twoB$. One such convolution 
for the fifth moment $\Exp(D_N^5)$ is given by 
\begin{align}
  \intk\twoB(x,y)^5\,dxdy \;=\; 
   \Big[&  \left(1+\sfrac{2\pi^{4}}{9}+\sfrac{4\pi^{6}}{189} 
           +\sfrac{\pi^{8}}{189}+\sfrac{4\pi^{10}}{18711}\right)^{s_N}
        - 5\left(1+\sfrac{2\pi^{4}}{15}+\sfrac{8\pi^{6}}{945}
	          +\sfrac{\pi^{8}}{945}\right)^{s_N} \notag\\
        &+ 10\left(1 + \sfrac{\pi^{4}}{15}+\sfrac{2\pi^{6}}{945}\right)^{s_N} 
         - 10\left(1+\sfrac{\pi^{4}}{45}\right)^{s_N} + 4\Big]
	\left[\left(1+\sfrac{\pi^{2}}{3}\right)^{s_N} - 1\right]^{-5} \;.
\notag	
\end{align}
If $s_N$ becomes large, the leading contribution in the expression above is
given by the first term. The convolution contributes to $\Exp(D_N^5)$ with a
third power of $1/N$, and this means that the fifth moment of the standardized
variable behaves at least as
\begin{equation}
  \frac{\left(1+\sfrac{2\pi^{4}}{9}+\sfrac{4\pi^{6}}{189} 
          +\sfrac{\pi^{8}}{189}+\sfrac{4\pi^{10}}{18711}\right)^{s_N}}
 {\left[\left(1+\sfrac{\pi^{2}}{3}\right)^{s_N}-1\right]^{5}N_{}^{3}V_N^{5/2}}
 \;>\; \left(\frac{N}{2(N-1)}\right)^{5/2}\left(\frac{(1.85)^{s_N}}{N}\right)^3
\notag
\end{equation}
if $N$ becomes large, and calculation of the other contributions to
$\Exp(D_N^5)$ shows that this behavior is not canceled. So clearly, the
fifth moment of the standardized variable may explode in the weak limit of
(\ref{weaklim}). 

\subsection{The statement}
In this section, we derive a limit in which all moments of the standardized
variable converge to the moments of a normal variable, which is therefore
`stronger' than the weak limit, in the sense that if $s_N$ grows with $N$ such
that the `strong' limit appears, it grows such that the criterion of
(\ref{weaklim}) is certainly satisfied (\Sec{CPDSec}). Actually, we will see
that the `strong' limit appears under the same type of condition, but of course
with a constant $c_{\textrm{s}}>c_{\textrm{w}}$. Our exact statement shall be
that {\sl if $s_N\ra\infty$ as $N\ra\infty$ such that}
\begin{equation}
   \lim\sup_N\frac{c_{\textrm{s}}^{s_N}}{N}=0
\label{StrEq009}			   
\end{equation}
{\sl then all moments of the standardized variable converge to the moments of a
normal variable}.
We shall show that it works for $c_{\textrm{s}}\geq\alpha^{4/3}$ and probably
even for $\alpha\leq c_{\textrm{s}}<\alpha^{4/3}$, where
\begin{equation}
   \alpha = \left(1+\sfrac{\pi^2}{3}\right)
                           \left(1+\sfrac{\pi^4}{45}\right)^{-1/2}\;\;.
\end{equation}
Note that $\alpha=2.41146\cdots$ and $\alpha^{4/3}=3.23376\cdots$.

\subsection{The scenario}
The logarithm $\Hat{W}\df\log\Hat{G}$ of the generating function of the
standardized variable is given by 
\begin{equation}
   \Hat{W}(\xi)
   \;=\; \log\Exp\left(\exp\left(\xi\,\frac{D_N-E}{\sqrt{V_N}}\right)\right)
   \;=\; -\frac{E\xi}{\sqrt{V_N}} + W\left(\frac{\xi}{\sqrt{V_N}}\right) \;\;,
\label{StrEq001}	 
\end{equation}
where $W\df\log G$ is the logarithm of the generating function of the
probability distribution of $D_N$.  So $\Hat{W}$ can be calculated using the
Feynman rules of \Sec{FormSec1} if $g\df\sqrt{2z/N}$ is replaced by
$g\df\sqrt{2\xi/(N\sqrt{V_N}\,)}$. This is equivalent with using
$g\df\sqrt{2\xi/N}$ and replacing the propagator $\twoB$ by
$\Hat{\twoB}\df\twoB/\sqrt{V_N}\,$, and it is this what we shall do. Only
connected diagrams contribute to $\Hat{W}(\xi)$, and there is only one diagram
that gives a contribution linear in $\xi$, namely 
\begin{equation}
   \frac{1}{2}\,\diagram{dcZ1}{30}{6} 
   \;=\; \frac{Ng^2}{2}\intk\Hat{\twoB}(x,x)\,dx
   \;=\; \frac{E\xi}{\sqrt{V_N}} \;\;,
\label{StrEq002}   
\end{equation}
which cancels against the term
$-E\xi/\sqrt{V_N}$ in \eqn{StrEq001}. The second-order in $\xi$ is given
by
\begin{align}
\frac{1}{4}\,\raisebox{-20pt}{\epsfig{figure=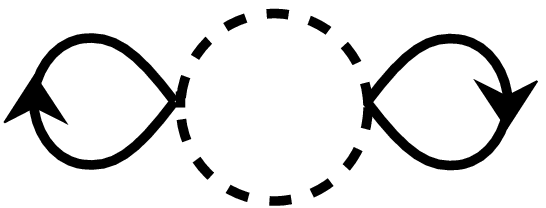,width=45pt,angle=90}} \;+\;
\frac{1}{4}\,\raisebox{-15pt}{\epsfig{figure=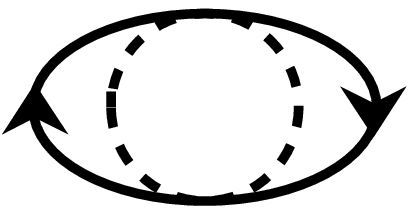,width=35pt,angle=90}} \;+\;
\frac{1}{8}\,\raisebox{-13pt}{\epsfig{figure=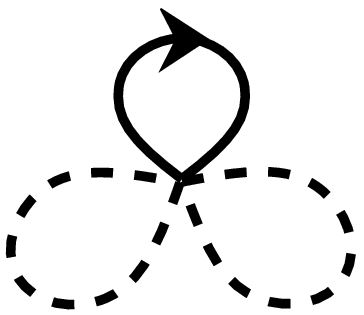,width=30pt,angle=90}} \;+\;
\frac{1}{8}\,\raisebox{-20pt}{\epsfig{figure=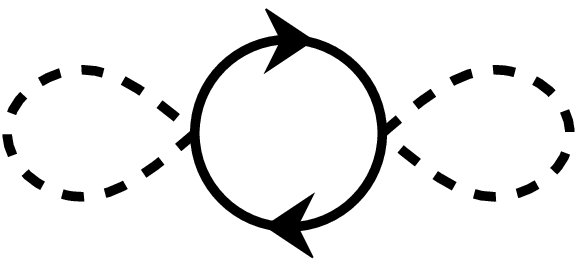,width=45pt,angle=90}}
   \;=\; \frac{(N^2-N)g^4}{4}\intk\Hat{\twoB}(x,y)^2\,dxdy 
   \;=\; \frac{\xi^2}{2} \;.
\label{StrEq003}   
\end{align}
Note that the third and the fourth diagram cancel each other. These results 
for the first two orders in $\xi$ are in correspondence with the fact that we
use the standardized variable. 
If we find a criterion dictating how $s_N\ra\infty$ as $N\ra\infty$ such that
the contribution of all other diagrams vanishes, regardless of the value of 
$\xi$, then this vanishing happens order by order in $\xi$ because each order
consists of a sum of diagrams. This 
then implies that all moments of the standardized variable converge to the 
moments of a normal variable.

\subsection{The calculation}
In order to calculate the diagrams, we expand $\Hat{\twoB}$ in terms of the
complex Fourier basis
\begin{equation}
   \Hat{\twoB}(x,y) 
   \;=\; \sum_{\vn}\Hat{\si}_{\vn}^2\,e^{2\pi i\vn\cdot(x-y)} \;\;,
\end{equation}
where the sum is over all $\vn\in\Zatu^s$ except the constant mode 
$\vn=(0,0,\ldots,0)$, and we denote $\vn\cdot x\df \sum_{\nu=1}^sn_\nu x^\nu$. 
The strengths $\Hat{\si}_{\vn}$ are given by
\begin{equation}
   \Hat{\si}_{\vn}^2 
   \;=\; \frac{1}{\sqrt{\tau_N}} 
	 \prod_{\nu=1}^s\frac{1}{r(n_\nu)^2}
   \quad,\qquad	  
   r(n_\nu)=\begin{cases}
           n_\nu &\textrm{if $n_\nu\neq0$ ,}\\
           1 &\textrm{if $n_\nu=0$ ,}
        \end{cases}
\end{equation}
where
\begin{equation}
   \tau_N \;\df\; \rho_N\left[\left(1+\sfrac{\pi^4}{45}\right)^{s}-1\right] 
   \quad,\qquad
   \rho_N \df 2\left(1-\frac{1}{N}\right)\;\;.
\end{equation}
Notice that, because we absorbed the factor $\sqrt{V_N}$ in $\Hat{\twoB}$, the
strengths depend on $N$.  
Because we expanded $\Hat{\twoB}$ in terms of the complex exponentials, 
convolutions of this two-point function can be calculated as sums over 
products of the strengths $\Hat{\si}_{\vn}^2$. As a result of this, we can go 
over to another boson propagator
\begin{equation}
   \vn\,\diagram{dcP2}{30}{-1}\,\vm 
   \;=\; \Hat{\si}_{\vn}^2\,\de_{\vn,-\vm} \;\;,
\end{equation}
and the rule that in a vertex with $k$ boson legs, boson propagators have to be convoluted as
\begin{equation}
   \sum_{\vn_1,\vn_2,\ldots,\vn_k}
   \Hat{\si}_{\vn_1}^2\,\de_{\vn_1,-\vm_1}
   \Hat{\si}_{\vn_2}^2\,\de_{\vn_2,-\vm_2}\cdots
   \Hat{\si}_{\vn_k}^2\,\de_{\vn_k,-\vm_k}\,
   \theta(\vn_1+\vn_2+\cdots+\vn_k=0) \;\;,
\end{equation}
where the logical $\theta$-function expresses that the sum of the labels has 
to be zero. 
These Feynman rules give the same result as the original rules. 

Because the Fourier diaphony is translation invariant, it is {\it one-vertex
decomposable}, so that we only have to consider {\it one-vertex irreducible}
(1VI) diagrams (\Sec{OVRSec}). The other diagrams cancel exactly. For each  1VI
diagram that has vertices that are not of the type of \eqn{CorEq026}, there
exists a diagram that has exactly the same bosonic part, but only effective
vertices of the type of \eqn{CorEq026}, and therefore carries a smaller power
of $1/N$. Combining this with the fact that we only have to consider connected
diagrams to calculate $\Hat{W}(\xi)$, we see that 
\begin{equation}
\begin{array}{c}
   \textsl{for the limit of large $N$,
           we only have to consider connected 1VI diagrams}  \\
   \textsl{with all vertices of the type of \eqn{CorEq026}, 
           which we call relevant}.
\end{array}	   
\label{StrEq004}	   
\end{equation}
The power of $1/N$ that is carried by a relevant diagram is given by
$1/N^{p/2-v}$, where $p$ is the sum of all bosonic legs of all vertices, and
$v$ the number of vertices.  Basic graph theory tells us that the number of
bosonic lines $I$ is equal to $p/2$, and the number of bosonic loops $L$ is
equal to $I-v+1$, so that 
\begin{equation}
\begin{array}{c}
   \textsl{the power of $1/N$ carried by a relevant diagram is $1/N^{L-1}$,}\\
   \textsl{where $L$ is the number of bosonic loops.}
\end{array}	   
\label{StrEq005}	   
\end{equation}
So the natural way to order the diagrams is by number of bosonic loops. 
From now on, the drawing of a diagram only represents the contribution coming 
from the bosonic part of the diagram, stripped from its factors of 
$g=\sqrt{2\xi/N}$, its factors of $N$ coming from the fermionic piece, and 
the symmetry factors, which we call the {\em bare} contribution.

\subsubsection{One loop}
Diagrams with no loops do not exist (because of the Landau gauge), and the 
relevant diagrams with only one loop contribute with 
\begin{equation}
    \diagram{dcZp}{40}{16}
    \;=\; \sum_{\vn}\Hat{\si}_{\vn}^{2p}
    \;=\; \tau_N^{-p/2}
          \Big[\Big(1+\sum_{n=1}^\infty\frac{2}{n^{2p}}\Big)^s-1\Big]
    \;\overset{s,N\ra\infty}{\longrightarrow}\;\gamma_p^{s}
\end{equation}
where
\begin{equation}
   \gamma_p \;=\; \Big(1+\sum_{n=1}^\infty\frac{2}{n^{2p}}\Big)
                 \Big(1+\sum_{n=1}^\infty\frac{2}{n^{4}}\Big)^{-p/2} \;\;.
\end{equation}
Using the line of argument with \eqn{GauEq011} and \eqn{GauEq012}, we derive
that $\gamma_{p+1}<\gamma_p$. Explicit calculation shows that
$\gamma_3=(\frac{2\pi^6}{945}+1)(\frac{\pi^4}{45}+1)^{-3/2}<1$, so that
$\gamma_{p}<1$ for all $p>2$, and 
\begin{equation}
\begin{array}{c}
   \textsl{the contribution of all one-loop diagrams with more than two 
           vertices}\\
   \textsl{vanishes if $N\ra\infty$ and $s\ra\infty$}.	   
\end{array}
\end{equation}
The diagrams with one or two vertices contribute 
to the first two powers in $\xi$ (\eqn{StrEq002} and \eqn{StrEq003}). 
If $N\ra\infty$ and $s$ stays finite, only one-loop diagrams do not vanish,  
and the analysis above was just a repetition of what was done in 
\Chp{GausChap}.

\subsubsection{More than one loop}
In order to estimate the contribution of the higher-loop diagrams, we observe
that, because $0<\Hat{\si}_{\vn}^2<1$ for all 
$\vn$ and all values of $N$, 
\begin{align}
   \sum_{\vm_1,\vm_2}\theta(\vn_1+\vm_1=0)\,
                     \Hat{\si}_{\vm_1}^2\,\de_{\vm_1,-\vm_2}\,
                     \theta(\vn_2+\vm_2=0)
   \;&<\; \sum_{\vm}\theta(\vn_1+\vm=0)\,\theta(\vn_2-\vm=0) \notag\\
   \;&=\; \theta(\vn_1+\vn_2=0) \;\;.
\label{StrEq007}   
\end{align}
As a result of this, the bare contribution of a relevant diagram can be 
estimated by repeated application of the operation 
\begin{equation}
   \diagram{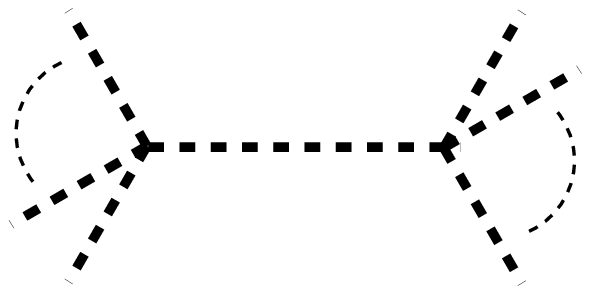}{64}{13} \;\;\ra\;\; \diagram{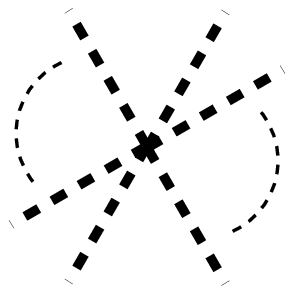}{31}{13} \;\;,
\end{equation}
until only one vertex remains. This operation leaves the number of loops $L$
invariant, so that the bare contribution of a relevant diagram is smaller than 
\begin{equation}
   \diagram{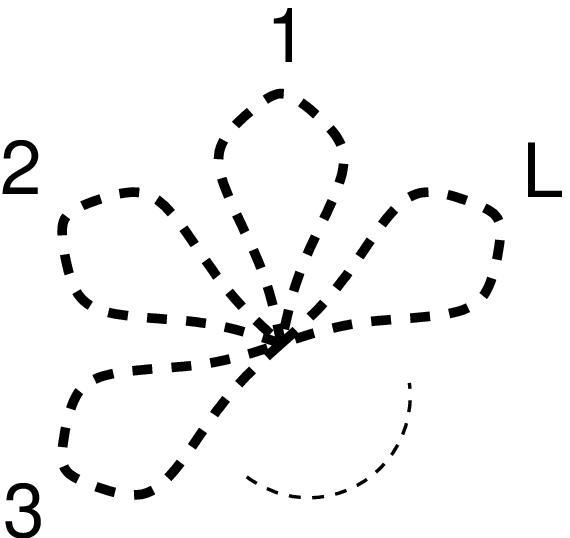}{56}{20}
   \;=\; \Big(\sum_{\vn}\Hat{\si}_{\vn}^2\Big)^L 
   \;=\; \left(\rho_N^{-1/2}\,d(N)\right)^L \;,
\end{equation}
where 
\begin{equation}
   d(N)
   \df \left(\frac{\rho_N}{\tau_N}\right)^{\!1/2}
           \Big[\Big(1+\sum_{n=1}^\infty\frac{2}{n^{2}}\Big)^{s}-1\Big]
   = \left[\left(1+\sfrac{\pi^2}{3}\right)^{s}-1\right]
         \left[\left(1+\sfrac{\pi^4}{45}\right)^{s}-1\right]^{-1/2} .
\end{equation}
Using this result, (\ref{StrEq004}) and (\ref{StrEq005}), we conclude that if
$s,N\ra\infty$, then the behavior of the contribution of any 1VI diagram with
$L>1$ loops is dominated by
\begin{equation}
   \frac{(\alpha^{s})^L}{N^{L-1}}
   \qquad\textrm{where}\qquad
   \alpha \df \left(1+\sfrac{\pi^2}{3}\right)
              \left(1+\sfrac{\pi^4}{45}\right)^{-1/2}\;\;,
   \end{equation}
\label{StrEq008}  
so that all diagrams with more than one loop vanish if $s=s_N\ra\infty$ 
as $N\ra\infty$, such that
\begin{equation}
   \lim\sup_N\frac{c_{\textrm{s}}^{s_N}}{N} = 0 
   \qquad\textrm{where}\qquad  
   c_{\textrm{s}}
   = \al^2\;\;.
\label{StrEq006}  
\end{equation}
We have already established that, in the expansion of $\Hat{W}(\xi)$ in $\xi$,
there is no linear term, and the quadratic term is given by $\xi^2/2$, as
demanded by the fact that we are dealing with the standardized variable.  The
one-loop diagrams contributing to the higher powers vanish if $s,N\ra\infty$,
and all other diagrams vanish if (\ref{StrEq006}) holds.

\subsubsection{Leading contributions}
In the previous section we have put a bound on the contribution of each
diagram, which resulted in (\ref{StrEq006}). This result comes from the bound
on the two-loop diagrams. For the lower-loop diagrams, however, the
determination of the {\em actual} leading behavior is attainable. There is, 
for example, only one relevant two-loop diagram, which has the following 
bosonic structure:
\begin{equation}
    \diagram{lcBr3}{25}{10}\;\;.
\end{equation}
Its bare contribution is
\begin{equation}
   \intk\Hat{\twoB}(x,y)^3\,dxdy
   \;=\; \frac{\left(1 + \sfrac{\pi^{4}}{15}+\sfrac{2\pi^{6}}{945}\right)^{s_N} 
               - 3\left(1+\sfrac{\pi^{4}}{45}\right)^{s_N} + 2}
	      {\rho_N^{3/2}\left[\left(1+\frac{\pi^4}{45}\right)^{s_N}
	                       -1\right]^{3/2}}
   \;\ra\; \frac{b_3^{s_N}}{2^{3/2}} \;\;,
\end{equation}
where
\begin{equation}
   b_3 
   \;\df\; \left(1+\sfrac{\pi^{4}}{15}+\sfrac{2\pi^{6}}{945}\right)
	   \left(1+\sfrac{\pi^4}{45}\right)^{-3/2}
   \;<\;   \alpha^{3/2} \;\;,		
\end{equation}
so that it suffices to take $c_{\textrm{s}}=\alpha^{3/2}$ in (\ref{StrEq006}).
The relevant three-loop diagrams have bosonic parts
\begin{equation}
   \diagram{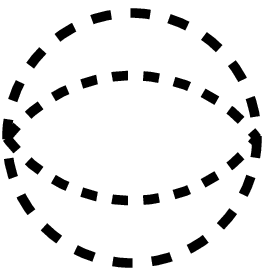}{25}{10}\;\;,\;\;\;
   \diagram{lcBr1}{25}{10}\;\;,\;\;\;
   \diagram{lcBr2}{25}{10}\;\;,\;\;\;
   \diagram{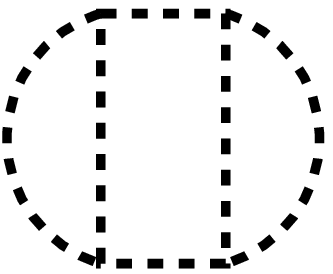}{31}{10}\;\;,
\end{equation}
and using (\ref{StrEq007}), we immediately see that the last two diagrams are
bounded by the first, which has a bare contribution 
\begin{align}
   \intk\Hat{\twoB}(x,y)^4\,dxdy
   \;&=\;\frac{  \left(1+\sfrac{2\pi^4}{15}+\sfrac{8\pi^6}{945}
                                          +\sfrac{\pi^8}{945}\right)^{s_N} 
              - 4\left(1+\sfrac{\pi^4}{15}+\sfrac{2\pi^6}{945}\right)^{s_N} 
	      + 6\left(1+\sfrac{\pi^4}{45}\right)^{s_N} - 3}
             {\rho_N^2\left[\left(1+\frac{\pi^4}{45}\right)^{s_N}-1\right]^2}
	     \notag\\
   \;&\ra\; \sfrac{1}{4}
           \left[\left(1+\sfrac{2\pi^4}{15}+\sfrac{8\pi^6}{945}
                                          +\sfrac{\pi^8}{945}\right)
		 \left(1+\sfrac{\pi^4}{45}\right)^{-2}\right]^{s_N}	
   \;\df\; \sfrac{1}{4}\,b_4^{s_N} \;\;.		 
\end{align}
Application of (\ref{StrEq007}) shows also that the bare contribution of the
second diagram is bounded by 
\begin{equation}
   \intk\Hat{\twoB}(x,y)^3\Hat{\twoB}(z,z)\,dxdydz
   \;\ra\; \sfrac{1}{4}
           \left[\left(1+\sfrac{\pi^{4}}{15}+\sfrac{2\pi^{6}}{945}\right)
	         \left(1+\sfrac{\pi^2}{3}\right)
		 \left(1+\sfrac{\pi^4}{45}\right)^{-2}\right]^{s_N}
   \;\df\; \sfrac{1}{4}m^{s_N} \;,
\end{equation}
and we have $b_4^{1/2}<\alpha^{4/3}$  and $m^{1/2}<\alpha^{4/3}$, so that
$c_{\textrm{s}}=\alpha^{4/3}$ also suffices in (\ref{StrEq006}). We suspect
that this works to all loop-orders, and that we can actually take
$c_{\textrm{s}}=\alpha$.

\section{Conclusions}
We have presented finite-sample corrections to the probability distributions of
quadratic discrepancies under sets of $N$ random points. The corrections are
terms in an $1/N$ expansion of the generating function of the probability
distribution, consisting of the contribution of a finite number of Feynman
diagrams. We presented the diagrams up to and including the order of $1/N^2$
for the general case, and derived a rule of diagram cancellation in the case of
special discrepancies, which we call one-vertex decomposable. 
We have applied the formalism to the Lego discrepancy, the $L_2^*$-discrepancy
in one dimension and the Fourier diaphony in one dimension, and calculated the
first two terms in the expansion. For the Lego discrepancy,
this resulted in \eqn{CorEq017} and \eqn{CorEq018}, for the $L_2^*$-discrepancy
in \eqn{CorEq021} and \eqn{CorEq022}, and for the Fourier diaphony in
\eqn{CorEq023} and \eqn{CorEq024}. The Fourier diaphony and the Lego
discrepancy with equal binning are one-vertex decomposable. For the latter, we
also calculated the $1/N^2$-term, which is in correspondence with the result
of an alternative calculation up to the order of $1/N^4$, given in \App{App6A}.

In the second part of the chapter, we focused on the variant of the Lego
discrepancy that is equivalent with a $\chi^2$-statistic of $N$ data points
distributed over $M$ bins. We have presented a procedure to calculate the
generating function perturbatively if $M$ and $N$ become large. The natural
expansion parameter we have identified to be $M/N$, and we have calculated the
first few terms in the series explicitly. 

In order to calculate limits for the Lego discrepancy in which $N,M\ra\infty$,
we have introduced the objects of \eqn{LegEq022} and restricted the behavior of
the size of the bins such that they satisfy \eqn{LegEq018}. Furthermore, we
have gone over to the standardized variable of the discrepancy. For this
variable, we have derived a phase diagram, representing the limits specified by
\eqn{LegEq023} and \eqn{LegEq024}. We have formulated the results in
(\ref{LegRes01}), (\ref{LegRes02}) and (\ref{LegRes03}).
On of these results is that there are non-trivial limits if $N,M\ra\infty$ such
that $M^\al/N\ra\textsl{constant}$ with $\al<1$. This result is in stark
contrast with the rule of thumb that, in order to trust the
$\chi^2$-distribution, each bin has to contain at least a few data points. 

Finally, we have derived a limit in which all the moments of the standardized 
variable of the Fourier diaphony converge to the moments of a normal variable,
which is given in (\ref{StrEq009}).

\section{Appendices}
\vspace{-1.2\baselineskip}
\Appendix{\label{App6A}}
If we define, for the Lego discrepancy with equal bins, $E=M-1$,
$\eta(z)=2z/(1-2z)$ and 
\begin{equation}
   (1-2z)^{E/2}G(z) 
   \;=\; \sum_{n,p\geq0}\frac{\eta(z)^p}{N^n}\,C^{(p)}_n(E) \;\;,
\end{equation}
then the only non-zero $C^{(p)}_n(E)$ up to $n=4$ are given by 
\begin{align}
   C_1^{(2)}(E) &= -\frac{1}{4}E\notag\\
   C_1^{(3)}(E) &= E\bigg(\frac{1}{12}E-\frac{1}{12}\bigg)\notag\\
   C_2^{(3)}(E) &= E\bigg(-\frac{1}{12}E+\frac{5}{12}\bigg)\notag\\
   C_2^{(4)}(E) &= E\bigg(\frac{1}{48}E^2-\frac{53}{96}E
                   +\frac{43}{48}\bigg)\notag\\
   C_2^{(5)}(E) &= E\bigg(\frac{5}{48}E^2-\frac{35}{48}E
                   +\frac{5}{8}\bigg)\notag\\
   C_2^{(6)}(E) &= E\bigg(\frac{1}{288}E^3+\frac{7}{72}E^2
                   -\frac{71}{288}E+\frac{7}{48}\bigg)\notag\\
   C_3^{(4)}(E) &= E\bigg(-\frac{1}{48}E^2+\frac{7}{12}E
                   -\frac{61}{48}\bigg)\notag\\
   C_3^{(5)}(E) &= E\bigg(\frac{1}{240}E^3-\frac{17}{30}E^2
                   +\frac{583}{120}E-\frac{1451}{240}\bigg)\notag\\
   C_3^{(6)}(E) &= E\bigg(\frac{53}{576}E^3-\frac{1153}{384}E^2
                   +\frac{7423}{576}E-\frac{527}{48}\bigg)\notag\\
   C_3^{(7)}(E) &= E\bigg(\frac{1}{576}E^4+\frac{461}{1152}E^3
                   -\frac{6581}{1152}E^2+\frac{8663}{576}E
                   -\frac{467}{48}\bigg)\notag\\
   C_3^{(8)}(E) &= E\bigg(\frac{11}{1152}E^4+\frac{85}{144}E^3
                   -\frac{5125}{1152}E^2+\frac{1555}{192}E
                   -\frac{17}{4}\bigg)\notag\\
   C_3^{(9)}(E) &= E\bigg(\frac{1}{10368}E^5+\frac{29}{3456}E^4
                   +\frac{955}{3456}E^3-\frac{12475}{10368}E^2
                   +\frac{953}{576}E-\frac{53}{72}\bigg)\notag\\
   C_4^{(5)}(E) &= E\bigg(-\frac{1}{240}E^3+\frac{37}{80}E^2
                   -\frac{337}{80}E+\frac{1397}{240}\bigg)\notag\\
   C_4^{(6)}(E) &= E\bigg(\frac{1}{1440}E^4-\frac{349}{960}E^3
                   +\frac{7193}{720}E^2-\frac{15283}{320}E
                   +\frac{67021}{1440}\bigg)\notag\\
   C_4^{(7)}(E) &= E\bigg(\frac{49}{960}E^4-\frac{29069}{5760}E^3
                   +\frac{372169}{5760}E^2-\frac{571727}{2880}E
                   +\frac{21503}{144}\bigg)\notag
\end{align}
\begin{align}
   C_4^{(8)}(E) &= E\bigg(\frac{13}{23040}E^5+\frac{13979}{23040}E^4
                   -\frac{2290601}{92160}E^3+\frac{1446743}{7680}E^2\notag\\
                &\phantom{= E\bigg(\frac{13}{23040}E^5}
		   -\frac{9583187}{23040}E+\frac{294773}{1152}\bigg)\notag\\
   C_4^{(9)}(E) &= E\bigg(\frac{73}{6912}E^5+\frac{35077}{13824}E^4
                   -\frac{781079}{13824}E^3+\frac{993515}{3456}E^2
                   -\frac{564301}{1152}E+\frac{24607}{96}\bigg)\notag\\
   C_4^{(10)}(E) &= E\bigg(\frac{1}{13824}E^6+\frac{139}{3072}E^5
                    +\frac{162721}{34560}E^4-\frac{596467}{9216}E^3\notag\\
		 &\phantom{= E\bigg(\frac{1}{13824}E^6}   
                    +\frac{1653251}{6912}E^2-\frac{253799}{768}E
                    +\frac{145199}{960}\bigg)\notag\\
   C_4^{(11)}(E) &= E\bigg(\frac{17}{41472}E^6+\frac{895}{13824}E^5
                    +\frac{55025}{13824}E^4-\frac{1505645}{41472}E^3\notag\\
		 &\phantom{= E\bigg(\frac{17}{41472}E^6}   
                    +\frac{19783}{192}E^2-\frac{137875}{1152}E
                    +\frac{1565}{32}\bigg)\notag\\
   C_4^{(12)}(E) &= E\bigg(\frac{1}{497664}E^7+\frac{11}{31104}E^6
                    +\frac{2431}{82944}E^5+\frac{155735}{124416}E^4\notag\\
		 &\phantom{= E\bigg(\frac{1}{497664}E^7}   
                    -\frac{3942431}{497664}E^3+\frac{249239}{13824}E^2
                    -\frac{250141}{13824}E+\frac{2575}{384}\bigg)\notag
\end{align}

\Appendix{\label{App6B}}
We want to calculate the integral 
\begin{equation}
   \Hat{H}(\tau) 
   \;=\; \frac{1}{2\pi i}\int\limits_{-i\infty}^{i\infty}e^{f_{\tau}(z)}\,dz 
   \quad,\quad 
   f_{\tau}(z)
   =\frac{1}{\lambda^2}\left(e^{\lambda z}-1-\lambda z\right) - z\tau \;\;.
\end{equation}
We will make use of the fact that 
\begin{equation}
   f_{\tau}\left(z+\frac{2\pi in}{\lambda}\right)
   \;=\; f_{\tau}(z) - 2\pi in\frac{1+\lambda\tau}{\lambda^2}  
\end{equation}
for all $n\in\Zatu$, so that
\begin{equation}
   \Hat{H}(\tau)
   \;=\; \frac{1}{2\pi i}\sum_{n\in\Zatu}\,
         \int\limits^{\frac{(2n+1)\pi i}{\lambda}}
	            _{\frac{(2n-1)\pi i}{\lambda}} e^{f_{\tau}(z)}\,dz
   \;=\; \frac{1}{2\pi i}\sum_{n\in\Zatu}
         e^{-2\pi in\frac{1+\lambda\tau}{\lambda^2}}
         \int\limits^{\frac{\pi i}{\lambda}}_{\frac{-\pi i}{\lambda}}
	 e^{f_{\tau}(z)}\,dz  \;\;.  
\label{CorEq019}  	 
\end{equation}
Notice that the integral is independent of $n$, so that the sum can be 
interpreted as a sum of Dirac delta-distributions:
\begin{equation}
   \sum_{n\in\Zatu}e^{-2\pi in\frac{1+\lambda\tau}{\lambda^2}}
   \;=\; \sum_{n\in\Zatu}
         \de\left(\frac{1+\lambda\tau}{\lambda^2} - n\right)   
   \;=\; \sum_{n\in\Zatu}
         \lambda\de\left(\tau-\left[n\lambda
	                              -\frac{1}{\lambda}\right]\right) \;\;.
\end{equation}
These delta-distributions restrict the values that $\tau$ can take.
If we use these restrictions and do the appropriate 
variable substitutions, the remaining integral in (\ref{CorEq019}) can be 
reduced to
\begin{equation}
   \int\limits^{\frac{\pi i}{\lambda}}_{\frac{-\pi i}{\lambda}}
   e^{f_{\tau}(z)}\,dz \;=\;
   \frac{e^{-\frac{1}{\lambda^2}}}{\lambda}
   \int\limits^{\pi i}_{-\pi i}\exp\left(\frac{e^\varphi}{\lambda^2}
                                         -n\varphi\right)\,d\varphi \;=\;
   \frac{e^{-\frac{1}{\lambda^2}}}{\lambda}
   \oint \frac{e^{\frac{1}{\lambda^2}w}}{w^{n+1}}\,dw \;\;,
\end{equation}
where $n\in\Zatu$ and the contour is closed around $w=0$. According to Cauchy's
theorem, the final integral is only non-zero if $n\in\Natu$, and in that case
its value is $2\pi i\frac{1}{n!}(\frac{1}{\lambda^2})^n$. The combination of
these results gives \eqn{LegEq020}.

\clearemptydoublepage

\chapter{Phase space integration\label{ChapPhSp}}

In particle physics, there is the need to integrate transition
probabilities of particle processes over phase space, the space of all possible
configurations of the final-state momenta (\Sec{IntroPSI}).
This is usually done with the Monte Carlo method, and the first sections of 
this chapter deal with its basics and some useful techniques. The formalism 
converges towards the application for phase space in \Sec{SecRambo}. 

\vspace{\baselineskip}

\minitoc

\section{Monte Carlo integration\label{SecPSMC}}
For the Monte Carlo (MC) method of numerical integration, the integral of a 
function $F$ over an integration region $\Mset$ has to be reduced to the 
integral of a function $f$ over an $s$-dimensional hypercube 
$\Kube\df[0,1]^s$. In order to do so, a 
suitable mapping $\vhi:\Kube\mapsto\Mset$ and the normalization function 
$g_\vhi$ have to be determined (\Sec{IntroIS}).
A conceptual help in the search for such mappings is 
considering them algorithms to generate random variables with a certain 
probability distribution. This probability 
distribution enters the integration problem as follows. Given a 
probability density $G$ on $\Mset$, the integral 
of $F$ over $\Mset$ can be written as
\begin{equation}
   \lebM{F}{\Mset}{} \;\df\; 
   \int_{\Mset}F(y)\,dy \;=\; \int_{\Mset}\frac{F(y)}{G(y)}\,G(y)dy  \;\;,
\end{equation}
so that the integral can be interpreted as the expectation value of $F/G$ under
the probability density $G$ on $\Mset$. The only restriction on $G$ is that
its support should contain the support of $F$. The Monte Carlo method can then
be directly applied to $\Mset$.  
Let us denote the average of a function $w$ over the first $N$ points of a
sequence $y_1,y_2,\ldots$ in $\Mset$ distributed following $G$ by
\begin{equation}
   G_N[w] \;\df\; \frac{1}{N}\sum_{k=1}^Nw(y_k) 
   \;\;,\quad\textrm{$y_1,y_2,\ldots$ distributed with density $G(y)$.}
\end{equation}
If $w$ is taken equal to $F/G$, then the expectation value
$\Exp(G_N[w])=\lebM{F}{\Mset}{}$ and the variance
$\Var(G_N[w])=\VAR_{\!G}[F]/N$, where
\begin{equation}
   \VAR_{\!G}[F] \;\df\;
  \lebM{F^2/G}{\Mset}{} - \lebM{F}{\Mset}{2}\;\;.
\end{equation}
If $\lebM{F^2/G}{\Mset}{}$ exists, so that $\VAR_{\!G}[F]$ is a finite
number, 
then $G_N[w]$ converges in probability to $\lebM{F}{\Mset}{}$, and
$\VAR_{\!G}[F]/N$ can be interpreted as the expected squared error
(\Sec{ProSecAMC}). This number is positive by definition, and extremalization
with respect to $G$ leads to $G=|F/\lebM{F}{\Mset}{}|$, which minimizes
$\VAR_{\!G}[F]$ to zero if $F$ is positive. Importance sampling can be
interpreted as the effort to make $G$ look like $|F|$ as much as possible. The
squared error can be estimated with 
\begin{equation}
   G_N^{(2)}[w] \;\df\; \frac{G_N[w^2] - G_N[w]^2}{N-1} \;\;,
\end{equation}
which satisfies $\Exp(G^{(2)}_N[w])=\Var(G_N[w])$. The integration is done
most efficiently if the numbers $w(y_i)$ fluctuate as little as possible, so
that $G^{(2)}_N[w]$ is as small as possible. That is what importance sampling
should take care of. 
The expected squared error on the error can be estimated
with the help of 
\begin{equation}
   G^{(4)}_N[w] \;\df\;
     \frac{G_N[w^4]-4G_N[w^3]G_N[w]+3G_N[w^2]^2}{N(N-2)(N-3)}
   - \frac{(4N-6)G_N^{(2)}[w]^2}{(N-2)(N-3)} \;\;,
\end{equation}
which satisfies $\Exp(G^{(4)}_N[w])=\Var(G^{(2)}_N[w])$. If
$\lebM{F^4/G^3}{\Mset}{}$ exists, then $G^{(4)}_N[F/G]$ is a good estimator for
the the expected squared error on the estimate of the expected squared
integration error.

The integration problem is now for large part reduced to that of generating the
sequence of points in $\Mset$ with the density $G$. In practice, the mapping
$\vhi$ still has to be made explicit, because algorithms usually start with the
generation of numbers between $0$ and $1$, but the analysis of the whole
algorithm can be done on a `higher level' by considering the piece that
generates the $y$-variables a given `black box'. The connection with
\Sec{IntroIS} can be established by taking $g_\vhi\df G\circ\vhi$.

\section{The unitary algorithm formalism}
In general, it is hard to find an efficient algorithm to generate sequences
with a given distribution. It is often even hard to determine the density under
which a sequence, generated with a given algorithm, is distributed. For a
certain class of algorithms the latter can be done analytically with the
unitary algorithm formalism (UAF). This is the class of {\em unitary}
algorithms, that is, algorithms which produce an output with probability one.
This may sound a bit mysterious, and in order to explain what we mean, we
introduce the class of {\em stepwise unitary} (SU) algorithms, of which all
{\em steps} produce an output with probability one. We illustrate this with an
example of a non-SU algorithm to generate a fair dice with five sides:
\begin{enumerate}
\item throw a fair dice, output $\leftarrow$ number of points;
\item if the number of points is less than six: 
      output $\leftarrow$ number of points, else throw again.
\end{enumerate}
The first step produces an output with probability one: if you throw a dice, it
generates a number. The second step, however, produces an output with
probability $\frac{5}{6}$, so that the algorithm is not SU.  The whole
algorithm, {\em is} unitary: the probability to produce no output is
$\lim_{n\ra\infty}\left(\frac{1}{6}\right)^n=0$. Consequently, stepwise
unitary is a relative concept. The steps may be ``black boxes'' that can be
trusted to produce an output with probability one. Consider the following
algorithm to generate numbers between $1$ and $11$:
\begin{enumerate}
\item throw a fair dice with five sides, output $\leftarrow$ number of points;
\item throw the dice again, output $\leftarrow$ number of points plus previous 
      number of points.
\end{enumerate}
This is a SU algorithm as long as we do not ask the question how the fair
dice with five sides is generated.

\subsection{Notation}
We shall introduce the UAF with the help of two examples, but first we
have to introduce some notation. We will frequently use the logical step
function $\theta$, which returns $1$ if a statement $\Pi$ is true, and $0$
otherwise:
\begin{equation}
   \theta(\Pi) \df \begin{cases}
                     1 &\textrm{if $\Pi$ is true}\;,\\
		     0 &\textrm{if $\Pi$ is false}\;.
		   \end{cases}
\end{equation}
A relation that is satisfied by this function and that we will often use is 
\begin{equation}
   \theta(\Pi) = 1-\theta(\textrm{not}\,\Pi) \;\;.
\label{PHEq002}   
\end{equation}
Also the Dirac delta-distribution will often appear, and we recall its most
important features (cf.~\cite{Choquet}): if $F$ is a sufficiently regular
function on $\Mset$, then
\begin{equation}
   \int_{\Mset}F(y)\de(y-y')\,dy \;=\; F(y') \;\;,
\end{equation}
and if $\vhi:x\mapsto y$ is an invertible and differentiable mapping, then
\begin{equation}
   \de(\vhi^{-1}(y)-x) \;=\; |J_\vhi(x)|\de(\vhi(x)-y) \;\;,
\end{equation}
where $J_\vhi$ is the determinant of the Jacobian matrix of $\vhi$. An integral
over many variables will from now on start with a single $\int$-symbol, and for
every variable $z$ a $dz$ means `integrate $z$ over the appropriate integration
region'. The order in which the variables appear is irrelevant. If it is not
evident what the `appropriate integration region' is, we shall make it explicit
with the help of $\theta$-functions.

\subsection{The UAF for SU algorithms}
The following is an example of the use of the UAF for a SU algorithm. It is an
algorithm to generate $n$ numbers $y_i$, $i=1,\ldots,n$ uniformly distributed in
$[0,1]$ such that their sum is equal to $1$, and we are going to prove that it
actually does. The algorithm goes as follows:
\begin{enumerate}
\item generate $n$ numbers $z_i$, $i=1,\ldots,n$ in $[0,\infty)$, distributed 
      independently\\ and with density $e^{-z_i}$;
\item put $L\lar\sum_{i=1}^nz_i$;
\item put $y_i\lar z_i/L$ for $i=1,\ldots,n$.      
\end{enumerate}
The algorithm clearly produces numbers the sum of which is equal to $1$.  The
question is whether they are distributed uniformly, i.e., whether the density
is, up to a normalization constant, equal to
$\de\left(1-\sum_{i=1}^ny_i\right)$. The UAF can answer the question as
follows. Write every generation of a variable in the algorithm as the integral
over the density with which it is generated, and interpret every assignment as
a generation with a density that is given by a Dirac delta-distribution. Only
the assignment of the final output should not be written as an integral, but
only with the delta-distributions. The integral obtained gives the generated
density $P$.  So in this case we have 
\begin{equation}
   P(y) \;=\; \int\Big(\prod_{i=1}^ndz_i\,e^{-z_i}\Big)\, 
            dL\,\de\Big(L-\sum_{i=1}^nz_i\Big)
            \Big(\prod_{i=1}^n\de(y_i-z_i/L)\Big) \;\;. 
\end{equation}
The unitarity of the algorithm is represented by the fact that integration 
over the $y$-variables of this equation gives the identity $1=1$.
To find the density, we have to eliminate the $z_i$-and 
$L$-integrals. Application of the rules for the delta-distributions gives
\begin{align}
   P(y) \;=\; \int\de\Big(1-\sum_{i=1}^ny_i\Big)\,
           dL\,L^{n-1}e^{-L}\,\theta(0\leq L\leq\infty)
     \;=\; \Gamma(n)\,\de\Big(1-\sum_{i=1}^ny_i\Big) \;,
\end{align}
and we see that the the $y_i$-variables are generated with the correct density. 
We even calculated the normalization factor, which is $\Gamma(n)=(n-1)!$. 
The step `generate $z$ with density $e^{-z}$' is a black box in this example, 
but can be made explicit. Such variable can be obtained by generating $x$ 
uniformly in $[0,1]$ and putting $z\lar-\log(x)$, since
\begin{equation}
   \int dx\,\de(z+\log(x))\,\theta(0\leq x\leq1) \;=\; e^{-z} \;\;.
\end{equation}

\subsection{The UAF for non-SU algorithms}
As an application of the UAF to a non-SU algorithm, we show the correctness 
of the ratio-of-uniforms method for the generation of a random variable with 
a given density $g$. The algorithm goes as follows \cite{Devroye}. 
Let $b\geq\sup_y\sqrt{g(y)}$, $a_-\leq\inf_yy\sqrt{g(y)}$ and 
$a_+\geq\sup_yy\sqrt{g(y)}$.  Then one has to
\begin{enumerate}
\item generate $x_1$ uniformly in $[0,b]$;
\item generate $x_2$ uniformly in $[a_-,a_+]$;
\item if $x_2^2\leq g(x_1/x_2)$ then put $y\lar x_1/x_2$, 
      else reject $x_1,x_2$ and start anew.
\end{enumerate}
Just as our algorithm for the fair dice with five sides, it uses the rejection 
method, and the third step is not unitary. For this algorithm, we can write 
down a recursive equation for the probability density $P$ that is generated. 
If we denote the volume of the space in which $x_1$ and $x_2$ are 
generated $V$, then this equation is then given by 
\begin{align}
   P(y) = \int dx_1dx_2\,\frac{1}{V}\,
              \Big[\theta\left(x_2^2\leq g(x_1/x_2)\right)
	                   \de(y-x_1/x_2) 
	          +\theta\left(x_2^2>g(x_1/x_2)\right)P(y) \Big] \;.
\end{align}  
Integration of the equation over $y$ gives the identity $1=1$ again, expressing
the unitarity of the algorithm. If we now use \eqn{PHEq002} and replace the
variables $x_1,x_2$ by $t\df x_1/x_2$ and $z\df x_2^2$, we get the equation 
\begin{equation}
   P(y) \;=\; 
   \int dtdz\,\frac{1}{V}\,
   \left\{\theta(z\leq g(t))\left[\de(y-t)-P(y)\right] + P(y)\right\} \;\;.
\end{equation}
The region in which $x_1$ and $x_2$ are generated are such that $\inf_y
g(y)\leq z\leq\sup_y g(y)$. Furthermore is $\int dtdz=\int dx_1dx_2=V$, so that
the equation becomes
\begin{equation}
   0 \;=\; \int dt\,\frac{1}{V}\,g(t)\left[\de(y-t)-P(y)\right]
   \quad\Longrightarrow\quad 
   P(y) = \frac{g(y)}{\int dt\,g(t)}\;\;, 
\end{equation}
and we see that the algorithm is correct. We even see that the function $g$, 
used in the algorithm, does not have to be normalized: the algorithm itself 
is unitary.

\section{Some useful techniques}
\subsection{Inversion\label{Inversion}}
The most straightforward way of generating random variables $y\in\Mset$ 
following a certain probability distribution $G$ is with an invertible mapping 
$\vhi:\Kube\mapsto\Mset$, by generating $x\in\Kube$ and putting $y\lar\vhi(x)$.
The generated density $P$ is given by
\begin{equation}
   P(y)
   \;=\; \intk dx\,\de(y-\vhi(x)) \;=\; |J_{\vhi^{-1}}(y)| \;\;,
\end{equation}
so that $|J_{\vhi^{-1}}(y)|$ should be equal to $G(y)$. The search for $\vhi$
is an integration and inversion problem, and is usually very hard to
solve in practice, even for one-dimensional variables.  

\subsection{Crude MC\label{CrudeMC}}
\newcommand{\oMset}{\overline{\Mset}}
Sometimes, part of the difficulty of the integration problem lies in the shape
of the integration region $\Mset$, which might be complicated.  Usually,
however, it can be seen as a subspace of a simpler manifold $\oMset$ with the
same dimension. One can look for a probability density $G$ on $\oMset$ then,
and integrate the function $F\vt_{\Mset}$, where $\vt_{\Mset}$ is the
characteristic function of $\Mset$. This just means that a density 
\begin{equation}
    \frac{G\vt_{\Mset}}{\lebM{G\vt_{\Mset}}{\oMset}{}}
\label{PHEq001}    
\end{equation}
is used on $\oMset$. The algorithm to generate a sequence of variables $y$
following this density is very simple:
\begin{enumerate}
\item generate $x$ in $\oMset$ following $G$;
\item if $x\in\Mset$ then put $y\lar x$, else reject $x$ and start anew. 
\end{enumerate}
This is called {\em crude} or {\em hit-and-miss} MC.
The proof of the correctness is also simple. In the UAF, the generated density 
$P$ satisfies
\begin{equation}
   P(y) \;=\; 
   \int dx\,G(x)[\theta(x\in\Mset)\de(y-x)+\theta(x\not\in\Mset)P(y)] \;\;.
\end{equation}
If we use \eqn{PHEq002} and evaluate the
integrals, (\ref{PHEq001}) is found as the solution to the equation. 

In principle, this method always works, but can be inefficient if the volume
of $\oMset$ is much larger than the volume of $\Mset$. If the integrand is as
simple as possible, i.e., $F(y)=1$ so that the original problem was that of
determining the volume $\Vol(\,\Mset\,)$ of $\Mset$, and if one would take for
$G$ the uniform distribution on $\oMset$, then
\begin{equation}
   \VAR_{\!G}[F] \;=\;  \Vol(\,\Mset\,)[\Vol(\,\oMset\,)-\Vol(\,\Mset\,)] \;\;.
\end{equation}
So if the difference between 
the volumes is large, one better chooses a density $G$ that is substantially 
larger on $\Mset$ than on $\oMset-\Mset$.

\subsection{Rejection\label{Rejection}}
Crude MC is a special case of the rejection method to generate variables $y$
following a density $F$ on $\Mset$. One needs an algorithm to generate
variables on $\Mset$ following some density $G$ and a number $c$ such that
$cG(y)>F(y)$ for all $y\in\Mset$. To obtain the density $F$, one should
\begin{enumerate}
\item generate $x$ following $G$, and $\rho$ uniformly in $[0,1]$;
\item if $\rho cG(x)\leq F(x)$ then put $y\lar x$, 
      else reject $x,\rho$ and start anew.
\end{enumerate}
The generated density $P$ satisfies
\begin{equation}
   P(y) 
   \;=\; \int dxG(x)\,d\rho\,[\theta(\,\rho cG(x)\leq F(x)\,)\,\de(y-x) + 
                            \theta(\,\rho cG(x)>F(x)\,)P(y)] \;,
\end{equation}
and if we use \eqn{PHEq002} again and the fact that
$\int_0^1d\rho\,\theta(a\rho\leq b)=b/a$, the solution
$P(y)=F(y)/\lebM{F}{\Mset}{}$ is found. In principle, this method works for any
bounded $F$, since there are always an easy to generate density $G$ and a $c$
that will do the job: $G=1/\Vol(\,\Mset\,)$ and
$c\geq\Vol(\,\Mset\,)\sup_{y\in\Mset}F(y)$. However, the algorithm can become
very inefficient. The efficiency can be expressed by
$\lebM{F}{\Mset}{}/(c\lebM{G}{\Mset}{})$, and if this number is small, the
variable $x$ will often be rejected in step 2.

\subsection{Sum of densities}
As an example in which integer random variables have to be generated, we
present a method to generate a density that is the normalized sum of a number
of positive functions $g_i$ with $i=1,\ldots,n$. To generate the density 
\begin{equation}
   G(y) \;=\;\sum_{j=1}^n\bar{g}_i(y)\quad,\qquad
   \bar{g}_i(y) \df \frac{g_i(y)}{\sum_{j=1}^n\lebM{g_j}{\Mset}{}} \;\;,
\end{equation}
one has to 
\begin{enumerate}
\item generate an integer $i$ with probability 
      $\lebM{\bar{g}_i}{\Mset}{}$
      and put $k\lar i$;
\item generate $x$ with density $g_k$ and put $y\lar x$.      
\end{enumerate}
To cast it into the UAF, a summation over the integer random variable has to 
be included into the equation for the density generated:
\begin{equation}
   P(y) \;=\;
   \int\sum_{i=1}^n\lebM{\bar{g}_i}{\Mset}{}\,
   \delta_{i,k}\, dx\, g_k(x)\delta(y-x) \;\;.
\end{equation}
The assignment `$k\lar i$' is represented by the Kronecker delta-symbol
$\de_{i,k}$. Evaluation of the integrals leads trivially to the correct
density.  To generate $i$, the unit interval $[0,1]$ can be dissected into $n$
bins of size $\lebM{\bar{g}_i}{\Mset}{}$, and $i$ becomes the number of the
bin a random number $z$, distributed uniformly in $[0,1]$, falls in.

\subsection{Adaptive MC\label{SecAMC}}
If one considers the actual calculation of a MC-integral a real random process,
it is a small step to the question whether the density $G$ may change during
the process. The answer is yes (\Sec{ProSecAMC}). If $y_1$ is generated
following a density $G_1$, and then $y_2$ following $G_2$ which depends on
$y_1$, and then $y_3$ following $G_3$ which depends on
$\{y\}_2\df\{y_1,y_2\}$ and so on, then $N^{-1}\sum_{k=1}^NF(y_k)/G_k(\{y\}_k)$
converges in probability to $\lebM{F}{\Mset}{}$, with an estimated squared
error given by $\VAR_{\!N}[F]/N$, where 
\begin{equation}
   \VAR_{\!N}[F]
   \;\df\;  \frac{1}{N}\sum_{k=1}^N\lebM{\bar{G}^{-1}_kF^2}{\Mset}{} 
                          \;-\; \lebM{F}{\Mset}{2}\;\;,
\label{PHEq003}
\end{equation}
with
\begin{equation}			  
   \bar{G}^{-1}_k(x_k) \;\df\;
   \int_{\Mset^{k-1}}\frac{G_1(y_1)\cdots G_{k-1}(\{y\}_{k-1})}
                          {G_k(\{y\}_{k})}\,dy_1\cdots dy_{k-1}  \;\;.
\end{equation}
The explicit dependence on $N$ of this expression shows that, by adapting the
density for each integration point in the right way, the error may be reduced
and the integration process optimized.

An example of a method to adapt the variance is weight optimization in
multichannel-MC \cite{Pittau1}, in which a density
$G_\al(y)\df\sum_{i=1}^n\al_ig_i(y)$ is generated, where each function $g_i$ is
a probability density itself, and the parameters $\al_i$ are positive and
satisfy $\sum_{i=1}^n\al_i=1$. Let us define
\begin{equation}
   W_i(\al) \;\df\; 
   \int_{\Mset}\frac{F(y)^2}{G_\al(y)^2}\,g_i(y)dy \;\;, 
\end{equation}
so that $\VAR_{\!G_\al}[F]=\sum_{i=1}^n\al_iW_i(\al) - \lebM{F}{\Mset}{2}$.  It
is not difficult to see that the variance, as function of the parameters
$\al_i$, has a (local) minimum if the values of these parameters are such that
$W_i(\al)$ has the same value for all $i=1,\ldots,n$. Of course, the problem of
finding these values is possibly even more difficult than the original
integration problem, but with adaptive MC the values might be found
approximately using an iterative procedure. The variance will then improve with
each step. 

In \cite{Pittau1}, the following is suggested: one starts with some
(sensible) values for the parameters $\al_i$ and, after generating a number of
$N$ points $y_k$ following the density $G_\al$, one estimates $W_i(\al)$ with
\begin{equation}
   E_i
   \;\df\; \frac{1}{N}\sum_{k=1}^N\frac{g_i(y_k)F(y_k)^2}{G_\al(y_k)^3}
\end{equation}
for all
$i=1,\ldots,n$. These numbers are then used to improve the values of the
parameters, for example through the prescription 
\begin{equation}
   \al_i\lar c\al_i\sqrt{E_i}
\end{equation}
where $c$ is some constant. The plausibility of this prescription is supported
by the example of stratified sampling, in which the functions $g_i$ are
normalized characteristic functions $\vt_i/\lebM{\vt_i}{\Mset}{}$ of
non-overlapping subspaces of the integration region. In that case,
$W_i(\al)=\al_i^{-2}\lebM{\vt_i}{\Mset}{}\lebM{\vt_iF^2}{\Mset}{}$, and we see
that putting $\al_i\lar c\al_i\sqrt{W_i(\al)}$ will give the local minimum
immediately, starting from any configuration for the parameters $\al_i$.

\section{Random momenta beautifully organized\label{SecRambo}}
As mentioned in \Sec{IntroPSI}, particle physicists often need to integrate
differential cross sections over phase space (PS), which is the space of all
physically possible final-state momentum configurations. Usually, it depends on
the transition amplitude which configurations are allowed, and here we mean by 
PS the space of all final-state momentum configurations for which the
separate momenta sum up to a given momentum, and for which the particles have
given masses. Because these restrictions reduce the dimension of the
integration region, it has measure zero in the space of all momentum
configurations so that the crude MC method is no option. 

One way to generate PS is by sequential two-body decays, i.e., by the recursive
splitting of each momentum generated so far into two momenta 
(cf.~\cite{Byckling}). The drawback of this method is that the efficiency is
poor if the number of momenta and the total energy become large 
(cf.~\cite{SKE}). The high-energy limit is equivalent with the limit in which
the masses of the particles become negligible, and for this situation, \rambo\
\cite{SKE} can be used. It generates any number of massless momenta with a
given total energy distributed uniformly in PS. We will not deal with the
algorithm adapted for the generation of massive momenta \cite{mambo}. Another
approach to PS generation, which we will also not address, uses the help of the
metropolis algorithm \cite{Kharraziha}.

\subsection{Notation}
The relativistic momentum of an elementary particle is a vector in $\Real^4$.
Its first component, also called the $0$-component, gives the energy of the
particle \footnote{We use units with which the speed of light is equal to
$1$.}, and the other three components give the real momentum in
three-dimensional space:
\begin{equation}
   p = (p^0,p^1,p^2,p^3) \df (p^0,\vec{p})  \;\;.
\end{equation}
The momentum with the opposite $3$-momentum is denoted by
\begin{equation}
   \tilde{p}\df(p^0,-\vec{p})  \;\;.
\end{equation}
The interpretation of $\Real^4$ as a real vector space can be carried forward 
in the sense that a system of non-interacting particles has 
a momentum that is equal to the sum of the momenta of the separate particles.
We shall need the first and the fourth canonical basis vectors, which we denote 
\begin{equation}
   \enul\df(1,0,0,0) \qquad\textrm{and}\qquad \ethr\df(0,0,0,1)  \;\;.
\end{equation}
$\Real^4$ becomes Minkowski space if it is endowed with the Lorentz invariant 
quadratic form 
\begin{equation}
   \invs{p}{} \df (p^0)^2 - |\vec{p}|^2  \quad,\qquad
   |\vec{p}|\df[(p^1)^2+(p^2)^2+(p^3)^2]^{1/2} \;\;.
\end{equation}
The same notation for the quadratic form and the $2$-component will not lead 
to confusion, because the $2$-component will not appear explicitly anymore
after this section.
The combination $\invs{(p+q)}{}-\invs{p}{}-\invs{q}{}$ defines a bi-linear 
product of two momenta, which is two times the scalar product
\begin{equation}
   \ipb{p}{q} \df
   \ip{p}{q} \df p^0q^0 - \ip{\vec{p}}{\vec{q}} \quad,\qquad
   \ip{\vec{p}}{\vec{q}} \df p^1q^1 + p^2q^2 + p^3q^3 \;.
\end{equation}
The notation with the parentheses shall be used in the next chapter. 
For physical particles, $\invs{p}{}$ has to be positive, and in that case, the
square root gives the invariant mass of the particle:
\begin{equation}
   \invm{p} \df \sqrt{\invs{p}{}} \quad\textrm{if}\;\; \invs{p}{}\geq0 \;\;.
\end{equation}
The group of linear transformations on $\Real^4$ that leave the quadratic form 
invariant, and the members of which have determinant $1$ and leave the sign of
the $0$-component of a momentum invariant, is called the Lorentz group. It is
generated by boosts, which are represented by symmetric matrices, and
rotations, which are represented by orthogonal matrices. A boost that
transforms a momentum $p$, with $\invs{p}{}>0$, to $\invm{p}\enul$ is
denoted $\Bo_p$, so 
\begin{equation}
   \Bo_pp = \invm{p}\enul \qquad\textrm{and}\qquad 
   \invm{p}\Bo_p\enul = \tilde{p}  \;\;.
\end{equation}
More explicitly, such a boost is given by
\begin{equation}
   \Bo_pq
   = (a,\vec{q}-b\vec{p})
   \qquad\textrm{where}\qquad
   a = \frac{\ip{p}{q}}{\invm{p}} \;\;,\quad 
   b = \frac{q^0+a}{p^0+\invm{p}}\;\;.
\end{equation}
A rotation that transforms $p$ to $p^0\enul+|\vec{p}|\ethr$ is denoted $\Ro_p$,
so 
\begin{equation}
   \Ro_pp = p^0\enul+|\vec{p}|\ethr \qquad\textrm{and}\qquad
   \Ro_p\tilde{p} = p^0\enul-|\vec{p}|\ethr\;\;.
\end{equation}
Since rotations only change the $3$-momentum, we shall use the same symbol if 
a rotation is restricted to three-dimensional space.

The physical PS of $\np$ particles is the $(3\np-4)$-dimensional
subspace of $\Real^{4\np}$, given by the restrictions that the energies of the 
particles are positive, the invariant masses squared
$\invs{p}{i}$ are fixed to given positive values $s_i$, and that
the sum 
\begin{equation}
   p_{(\np)}\df\sum_{i=1}^{\np}p_i 
\end{equation}
of the momenta is fixed to a given momentum $P$. The restrictions for the
separate momenta can be expressed with a `PS characteristic distribution'
\begin{equation}
   \vt_{s_i}(p) \df \delta(\invs{p}{}-s_i)\,\theta(p^0>0)\;\;, 
   \qquad\textrm{and}\qquad
   \vt(p) \df \vt_0(p) \;\;.
\end{equation}
The generic PS
integral, of a function $F$ of a set $\{p\}_\np\df\{p_1,\ldots,p_\np\}$ of
momenta, that has to be calculated is then given by 
\begin{equation}
   \int_{\Real^{4\np}}
   \Big(\prod_{i=1}^{\np}\dfp_i\vt_{s_i}(p)\Big)\, 
   \de^4(p_{(\np)}-P)\,F(\{p\}_\np) \;\;.
\end{equation}
We explicitly write down the number of degrees of freedom in the differentials
and the delta-distributions in order to keep track of the dimensions. Each 
momentum component carries the dimension of a mass.

\subsection{The algorithm}
\rambo\ was developed with the aim to generate the flat PS distribution of
$\np$ massless momenta as uniformly as possible, and such that the sum of the
momenta is equal to $\wcm\enul$ with $\scm$ a given squared energy. This
means that the system of momenta is in its center-of-mass frame (CMF), and that
the density is proportional to the `PS characteristic distribution' 
\begin{equation}
    \Theta_{\scm}(\{p\}_\np) \;\df\; 
   \delta^4(p_{(\np)}-\wcm\enul)\prod_{i=1}^{\np}\vt(p_i) \;\;.
\label{RamEq002}   
\end{equation}
The algorithm consists of the following
steps:
\begin{Alg}[{\tt RAMBO}]
\begin{enumerate}
\item generate $\np$ massless vectors $q_j$ with positive energy without 
      constraints but under some normalized density $f(q_j)$;
\item compute the sum $q_{(\np)}$ of the momenta $q_j$;
\item determine the Lorentz boost and scaling transform that
      bring $q_{(\np)}$ to $\wcm\enul$;
\item perform these transformations on the $q_j$, and call the result $p_j$.
\end{enumerate}
\end{Alg}
Trivially, the algorithm generates momenta that satisfy the various
$\de$-constraints, but it is not clear a priori that the momenta have the
correct distribution. To prove that they actually do, we apply the UAF.
It tells us that the generated density $\Phi_{\scm}$ is given by 
\begin{align}
   \Phi_{\scm}(\{p\}_\np) = 
   \int &\Big(\prod_{j=1}^{\np}d^4q_j\vt(q_j)f(q_j)\Big)\,
         d^4b\,\de^4\Big(b-\frac{q_{(\np)}}{\invm{q_{(\np)}}}\Big)\,
          dx\,\de\Big(x-\frac{\wcm}{\invm{q_{(\np)}}}\Big)\, \notag\\
        &\times \prod_{i=1}^{\np}\de^4(p_i-x\Bo_bq_i) \;\;.
\label{RamEq001}
\end{align}
To calculate the distribution yielded by this algorithm, the integral has to be
evaluated. First of all, some simple algebra using
$p_{(\np)}=x\Bo_bq_{(\np)}$, $q_{(\np)}=x^{-1}\Bo_b^{-1}p_{(\np)}$ and the
Lorentz and scaling properties of the Dirac $\de$-distributions leads to 
\begin{equation}
   \de^4\Big(b-\frac{q_{(\np)}}{\invm{q_{(\np)}}}\Big)\,
   \de\Big(x-\frac{\wcm}{\invm{q_{(\np)}}}\Big)
   \;=\; \frac{2\scm^2}{x}\,\de^4(p_{(\np)}-\wcm\enul)\,\de(\invs{b}{}-1) \;\;.
\end{equation}
Furthermore, since we may write
\begin{equation}
  d^4q_j\,\de(\invs{q}{j})\,\de^4(p_j-x\Bo_bq_j) 
  \;=\; \frac{1}{x^2}\,\de(\invs{p}{j}) 
\end{equation}
under the integral, the l.h.s.~of \eqn{RamEq001} becomes
\begin{equation}
   \int\Theta_{\scm}(\{p\}_\np)\,d^4b\,\de(\invs{b}{}-1)\,
           dx\frac{2\scm^2}{x^{2\np+1}}
	              \prod_{i=1}^{\np}f(\frac{1}{x}\Bo_b^{-1}p_i)\,
           \theta(\ip{\enul}{\Bo_b^{-1}p_j}>0) \;.
\end{equation}
In the standard \rambo\ algorithm, the following choice is made for $f$:
\begin{equation}
  f(q) \;=\; \frac{c^2}{2\pi}\,\exp(-cq^0) \;\;,
\end{equation}
where $c$ is a positive number with the dimension of an inverse mass.
Therefore, if we use that $p_{(\np)}=\wcm\enul$ and that $q^0=\ip{\enul}{q}$
for any $q$, then 
\begin{align}
   \prod_{i=1}^{\np}f(\frac{1}{x}\Bo_b^{-1}p_i)\,
                     \theta(\ip{\enul}{\Bo_b^{-1}p_i}>0)
   \;=\; &\left(\frac{c^2}{2\pi}\right)^{\np}
          \exp\left(-\frac{c}{x}\,\ip{\enul}{\Bo_b^{-1}p_{(\np)}}\right) \,
          \prod_{i=1}^{\np}\theta(\ip{\enul}{\Bo_b^{-1}p_i}>0)  \notag\\
   \;=\; &\left(\frac{c^2}{2\pi}\right)^{\np}\,
          \exp\left(-\frac{c\wcm}{x}\,b^0\right)\,\theta(b^0>0) \;\;. 
\end{align}
As a result of this, the variables $p_i$, $i=1,\ldots,\np$ only
appear in $\Theta_{\scm}$, as required. 
The remaining integral is calculated in the Appendix at the end of this
section, with the result that \rambo\ generates the density 
\begin{equation}
   \Phi_{\scm}(\{p\}_\np) 
   \;=\; \Theta_{\scm}(\{p\}_\np)\left(\frac{2}{\pi}\right)^{\np-1}
         \frac{\Gamma(\np)\Gamma(\np-1)}{\scm^{\np-2}} \;\;.		      
\end{equation}
Incidentally, we have computed here the volume of the
PS for $n$ massless particles:
\begin{equation}
   \int_{\Real^{4\np}}\dnp\,\Theta_{\scm}(\{p\}_\np) 
   \;=\; \left(\frac{\pi}{2}\right)^{\np-1}
                   \frac{\scm^{\np-2}}{\Gamma(\np)\Gamma(\np-1)}\;\;.
\end{equation}
Note, moreover, that $c$ does not appear in the final answer; this is
only natural since any change in $c$ will automatically be compensated
by a change in the computed value for $x$. Finally, it is important
to realize that the `original' PS has dimension $3\np$, while
the resulting one has dimension $3\np-4$: there are configurations
of the momenta $q_j$ that are different, but after boosting and
scaling end up as the same configuration of the $p_j$. It is this
reduction of the dimensionality that necessitates the integrals over
$b$ and $x$.

The first step of the algorithm consists of generating massless momenta with 
positive energy. To generate such momenta, we use that 
\begin{align}
   \dfp\vt(p)
   \;=\; d\vhi\,dz\,dp^0p^0\,
         \theta(p^0>0)\,\theta(0\leq\vhi\leq2\pi)\,\theta(-1\leq z\leq1) \;\;,
\end{align}
with $p = (\,p^0,p^0\Hat{n}(z,\vhi)\,)$, where 
\begin{equation}
   \Hat{n}_1(z,\vhi) \df \sqrt{1-z^2}\sin\vhi\;\;,\quad
   \Hat{n}_2(z,\vhi) \df \sqrt{1-z^2}\cos\vhi\;\;,\quad
   \Hat{n}_3(z,\vhi) \df z \;\;.
\label{DefHatn}   
\end{equation}
From this we can directly see that, to generate $p$ following a
density proportional to $\vt(p)f(p^0)$, one should
\begin{Alg}[{\tt MASSLESS MOMENTUM}]
\begin{enumerate}
\item generate $p^0$ in $[0,\infty)$ following a density $\sim p^0f(p^0)$;
\item generate $\vhi$ uniformly distributed in $[0,2\pi]$ and $z$ uniformly in 
      $[-1,1]$;
\item construct $\Hat{n}(z,\vhi)$ and put $\vec{p}\lar p^0\Hat{n}(z,\vhi)$.
\end{enumerate}
\end{Alg}
To generate $p^0$ following the density $p^0\exp(-p^0)$, one can
\begin{Alg}[{\tt 0-COMPONENT}]
\begin{enumerate}
\item generate $x_1$ and $x_2$ distributed uniformly in $[0,1]$;
\item put $p^0\lar -\log(x_1x_2)$, 
\end{enumerate}
\end{Alg}
since
\begin{equation}
   \int dx_1dx_2\,\theta(0\leq x_{1,2}\leq1)\,\de(\,p^0+\log(x_1x_2)\,)
   \;=\; \int_{e^{-p^0}}^1 dx\,\frac{e^{-p^0}}{x}
   \;=\; p^0e^{-p^0} \;.
\end{equation}

\subsection{Appendix}
We have to calculate the integral
\begin{equation}
   2\scm^2\left(\frac{c^2}{2\pi}\right)^n\int dxd^4b\,
   \de(\invs{b}{}-1)\,\theta(b^0>0)\,\frac{1}{x^{2n+1}}\,
   \exp\left(-\frac{c\wcm}{x}\,b^0\right)
   \;=\;\frac{2\Gamma(2n)\,B(n)}{(2\pi)^n\scm^{n-2}}\;\;, \notag
\end{equation}
where
\begin{equation}
   B(n) \;\df\; \int d^4b\,\de(\invs{b}{}-1)\,\theta(b^0>0)\,(b^0)^{-2n}
        \;=\;   2\pi\int_1^\infty db^0\,(b^0)^{-2n}\,\sqrt{(b^0)^2-1}  
	\;\;.\notag
\end{equation}
The `Euler substitution' $b^0\df\sfrac{1}{2}\,(v^{1/2}+v^{-1/2})$ 
casts the integral in the form
\begin{equation}
   B(n) \;=\; 
   2^{2n-2}\pi\int_1^\infty dv\,\frac{(v-1)^2v^{n-2}}{(v+1)^{2n}}  \;\;.\notag
\end{equation}
By the transformation $v\ra1/v$ it can easily be checked that the integral from
$1$ to $\infty$ is precisely equal to that from $0$ to $1$, so that we may 
write
\begin{equation}
   B(n) \;=\; \frac{2^{2n-2}\pi}{2}\int_0^\infty dv\,
              \frac{v^n-2v^{n-1}+v^{n-2}}{(1+v)^{2n}}
	\;=\; 4^{n-1}\pi\,\frac{\Gamma(n-1)\Gamma(n)}{\Gamma(2n)}\;\;,\notag
\end{equation}
where we have used, by writing $z\df1/(1+v)$, that
\begin{equation}
   \int_0^\infty dv\,v^p(1+v)^{-q} 
   \;=\; \int_0^1 dz\,z^{q-p-2}(1-z)^p 
   \;=\; \frac{\Gamma(q-p-1)\Gamma(p+1)}{\Gamma(q)} \;\;.\notag
\end{equation}

\clearemptydoublepage

\newcommand{\gl}{\textrm{g}}
\newcommand{\GeV}{\textrm{GeV}}
\newcommand{\Nge}{N_{\textrm{ge}}}
\newcommand{\Nac}{N_{\textrm{ac}}}
\newcommand{\ee}[1]{\!\times\!10^{#1}}
\newcommand{\emu}[1]{\!\times\!10^{-#1}}
\newcommand{\hour}{\textrm{h}}
\newcommand{\seco}{\textrm{s}}
\newcommand{\texa}{t_{\textrm{exa}}}

\chapter{Generating QCD-antennas\label{ChapSar}}

An algorithm to generate random momenta, distributed with a density that
contains the singular structure typically found in QCD-processes, is 
introduced. 
For the notation used we refer to \Sec{SecRambo}.

\vspace{\baselineskip}

\minitoc

\section{Introduction}
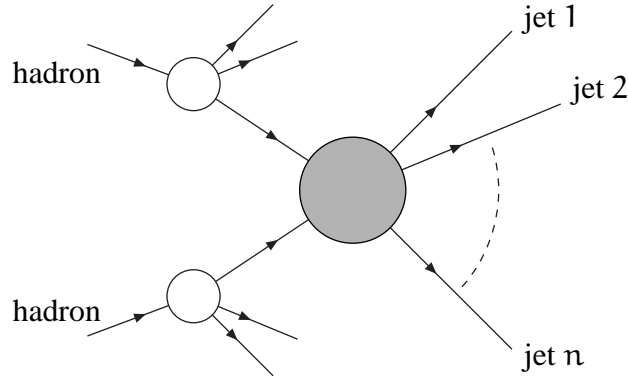
\begin{figure}[t]
\begin{center}
\begin{picture}(160,140)(0,13)
\ArrowLine(-20,135)(20,120)\Text(-15,125)[r]{hadron}
\ArrowLine(20,120)(80,80)
\ArrowLine(20,120)(50  ,150)
\ArrowLine(20,120)(59.2,136.2)
%
\ArrowLine(20,40)(80,80)
\ArrowLine(20,40)(50  ,10)
\ArrowLine(20,40)(59.2,23.8)
\ArrowLine(-20,25)(20,40)\Text(-15,35)[r]{hadron}
\ArrowLine(80,80)(140,140)\Text(145,145)[l]{jet $1$}
\ArrowLine(80,80)(158.4,112.5)\Text(163.4,118)[l]{jet $2$}
\DashCArc(80,80)(55,318.5,17){3}
\ArrowLine(80,80)(140,20)\Text(145,17)[l]{jet $\np$}
\GCirc(80,80){20}{0.7}
\BCirc(20,120){10}
\BCirc(20,40){10}
\end{picture}
\caption{A generic multi-jet event}
\label{SarFig01}
\end{center}
\end{figure}

In future experiments with hadron colliders, such as the LHC, many multi-jet
final states will occur, which have very high particle multiplicities. The
initial states will consist of two hadrons in the center-of-mass frame (CMF).
The processes involved in one transition (one {\em event}) are very
complicated, and are usually considered to consist of three steps. The generic
situation is depicted schematically in \fig{SarFig01}. Time can be considered to
flow from the left to the right in the picture. The hadrons start the
interaction with the emission of partons. The transition of a hadron into the
emitted parton and the leftover is represented by the white blobs. This is the
first step. In the second step, represented by the grey blob, the partons
interact, resulting in $\np$ new partons. In step three, these partons turn
into jets with high particle multiplicities.
The idea is that the contribution of the three steps more-or-less factorize 
in the transition amplitude of the whole event, and that the processes can 
be dealt with separately. In this chapter, we deal 
with step two, the grey blob. 

The multi-jet events that will occur in hadron colliders can be divided into
interesting events (IE) and very interesting events (VIE). The main difference
between the two classes is that the existing model of elementary particles, the
standard model, shall not have proven yet its capability of dealing with
the description of the VIE at the moment when they are analyzed. The IE shall
not manifest themselves as such a heavy test for the standard model. 
However, we still need to know the cross sections of the IE in order 
to compare the ratio of these and those of the VIE with the predictions of the 
standard model. 

\subsection{The problem}
Large part of the IE can be described by quantum chromo dynamics (QCD), the
formalism of quarks and gluons, with which multi-parton QCD-amplitudes are
calculated. It is well known \cite{Kuijf} that they contribute to the cross
section with a singular behavior in phase space (PS), given by the so-called
{\em antenna pole structure} (APS). In particular, for processes involving only
$n$ gluons the most important contribution comes from the sum of all
permutations in the momenta of 
\begin{equation}
   \frac{1}{(\ip{p_1}{p_2})(\ip{p_2}{p_3})(\ip{p_3}{p_4})\cdots
            (\ip{p_{\np-1}}{p_{\np}})(\ip{p_{\np}}{p_1})} \;\;,
\label{SarEq001}	    
\end{equation}
and the singular nature stems from the fact that the scalar products
$\ip{p_i}{p_j}$ can become very small. If functions, containing this kind of
kinematical structures, are integrated using the \rambo\ (\Sec{SecRambo}), which
generates the momenta distributed uniformly in PS, then a large number of
events is needed to reach a result to acceptable precision. As an illustration,
we present in the left table of \tab{SarTab01} the number of PS points needed
to integrate the single antenna of \eqn{SarEq001}, so not even the sum of its
permutations, to an expected error of $1\%$. 
\begin{table}
\begin{center}
\begin{tabular}{|c|c|c|}
  \hline
  \multicolumn{3}{|c|}{single antenna integrated to $1\%$ error}\\ \hline\hline
   \multirow{2}{18mm}{number of momenta}
  &\multirow{2}{20mm}{$\displaystyle\frac{\textrm{cut-off}}
                                         {\textrm{CM-energy}}$} 
  &\multirow{2}{18mm}{number of PS points} \\ 
  & & \\ \hline
  $3$ & $0.183$ & $10,069$ \\ \hline
  $4$ & $0.129$ & $26,401$ \\ \hline
  $5$ & $0.100$ & $58,799$ \\ \hline
  $6$ & $0.0816$ & $130,591$ \\ \hline
  $7$ & $0.0690$ & $240,436$ \\ \hline
  $8$ & $0.0598$ & $610,570$ \\ \hline
\end{tabular}
\qquad
\begin{tabular}{|c|c|c|}
  \hline
  \multicolumn{3}{|c|}{evaluation amplitude in $1$ PS point}\\ \hline\hline
   \multirow{2}{20mm}{number of final gluons}
  &\multicolumn{2}{c|}{cpu-time (seconds)}\\ \cline{2-3}
  & {\tt SPHEL} & exact \\ \hline
  $3$ & $2.83\emu{5}$ & $1.60\emu{1}$ \\ \hline
  $4$ & $9.76\emu{5}$ & $5.54\emu{1}$ \\ \hline
  $5$ & $4.88\emu{4}$ & $1.945$ \\ \hline
  $6$ & $3.26\emu{3}$ & $6.06$ \\ \hline
  $7$ & $2.57\emu{2}$ & $19.91$ \\ \hline
  $8$ &                    & $64.45$ \\ \hline
\end{tabular}
\caption{Typical number of PS points and computing times.}
\label{SarTab01}
\end{center}
\vspace{-20pt}
\end{table}
The antenna cannot be integrated over the whole of PS because of the
singularities, so these have to be cut out. This is done through the restriction
$\invs{(p_i+p_j)}{}\geq s_0$ for all $i,j=1,\ldots,\np$,\footnote{Remember that
$\invs{(p+q)}{}=2\ip{p}{q}$ since $p$ and $q$ are massless.} and in the table 
the ratio between $\sqrt{s_0}$ and the total energy $\wcm$ is given. These
numbers are based on the reasonable choice $s_0/s=0.2/[n(n-1)]$.

Performing MC integration with very many events is not a problem if the
evaluation of the integrand in each PS point is cheap in computing time.  This
is, for example, the case for algorithms to calculate the squared multi-parton
amplitudes based on the so called SPHEL-approximation, for which only the
kinematical structure of (\ref{SarEq001}) is implemented \cite{Kuijf}.
Nowadays, algorithms to calculate the {\em exact} matrix elements exist, which
are far more time-consuming \cite{DKP,CMMP}. As an illustration of what is
meant by `more time-consuming', we present the right table of \tab{SarTab01}
with the typical cpu-time needed for the evaluation in {\em one} PS point of
the integrand for processes of two gluons going to more gluons, both for the
SPHEL-approximation and the exact matrix elements \cite{Petros}. It is
expected, and observed, that the exact matrix elements reveal the same kind of
singularity structures as the APS, so that, according to the tables, the PS
integration for a process with $8$ final gluons would take in the order of $400$
days\ldots

\subsection{The solution}
The solution to this problem is importance sampling. Instead of \rambo, a PS
generator should be used which generates momenta with a density including the
APS. The following sections show the construction of such a PS generator,
called \sarge, which stands for {\tt S}taggered {\tt A}ntenna {\tt R}adiation
{\tt GE}nerator.

\section{The basic antenna}
As mentioned before, we want to generate momenta that represent radiated
partons with a density that has the antenna structure
$[\ipb{p_1}{p_2}\ipb{p_2}{p_3}\ipb{p_3}{p_4}\cdots,
  \ipb{p_{\np-1}}{p_{\np}}\ipb{p_{\np}}{p_1}]^{-1}$.
Naturally, the momenta can be viewed as coming from a splitting process: one
starts with two momenta, a third is radiated off creating a new pair of momenta
of which a fourth is radiated off and so on. In fact, models similar to this
are used in full-fledged Monte-Carlo generators like {\tt HERWIG}. Let us
therefore first try to generate a single massless momentum $k$, radiated from a
pair of given massless momenta $p_1$ and $p_2$. In order for the distribution
to have the correct infrared and collinear behavior, it should qualitatively be
proportional to $[\ipb{p_1}{k}\ipb{k}{p_2}]^{-1}$. Furthermore, we want the
density to be invariant under Lorentz transformations and scaling of the
momenta, keeping in mind that the momenta are three out of possibly more in a
CMF and that we have to perform these transformations in the end, like in
\rambo.
This motivates us to define the {\em basic antenna\/} structure as
\begin{equation}
   dA(p_1,p_2;k) 
   \;\df\; d^4k\vt(k)\,
           \frac{1}{\pi}\,\frac{\ipb{p_1}{p_2}}{\ipb{p_1}{k}\ipb{k}{p_2}}\;
           g\left(\frac{\ipb{p_1}{k}}{\ipb{p_1}{p_2}}\right)
           g\left(\frac{\ipb{k}{p_2}}{\ipb{p_1}{p_2}}\right)\;\;.
\label{SarEq002}
\end{equation}
Here, $g$ is a function that serves to regularize the infrared and collinear
singularities, as well as to ensure normalization over the whole space for $k$:
therefore, $g(\xi)$ has to vanish sufficiently fast for both $\xi\to0$ and
$\xi\to\infty$. To find out how $k$ could be generated, we evaluate $\int dA$
in the CMF of $p_1$ and $p_2$. Writing 
\begin{equation}
   E\df\sqrt{\ipb{p_1}{p_2}/2} \quad,\qquad
   p\df\Bo_{p_1+p_2}p_1     \quad,\qquad
   q\df\Bo_{p_1+p_2}k \;\;,
\end{equation}
we have
\begin{equation}
   \ipb{p_1}{p_2} = 2E^2              \;\;,\quad
   \ipb{p_1}{k} = Eq^0(1-z)  \;\;,\quad
   \ipb{k}{p_2} = Eq^0(1+z)  \;\;,
\end{equation}
where $z=\ip{\vec{p}}{\vec{q}}/(|\vec{p}||\vec{q}|)$. The azimuthal angle
of $\vec{q}$ is denoted $\vhi$, so that
$\vec{q}=|\vec{q}|\Ro_{p}^{-1}\Hat{n}(z,\vhi)$, with $\Hat{n}$ as in
(\ref{DefHatn}).  
We can write 
\begin{equation}
    d^4k\vt(k) \;=\; \sfrac{1}{2}\,q^0dq^0\,d\vhi\,dz
               \;=\; \sfrac{1}{2}\ipb{p_1}{p_2}\,d\vhi\,d\xi_1d\xi_2\;\;,
\end{equation}
where,  
\begin{equation}
   \xi_1 = \frac{\ipb{p_1}{k}}{\ipb{p_1}{p_2}} \quad\quad\textrm{and}\quad\quad
   \xi_2 = \frac{\ipb{k}{p_2}}{\ipb{p_1}{p_2}} \;\;,
\end{equation} 
so that $z=(\xi_2-\xi_1)/(\xi_2+\xi_1)$ and $q^0=E(\xi_2+\xi_1)$.
The integral over $dA$ takes on the particularly simple form
\begin{equation}
   \int dA(p_1,p_2;k)
   \;=\; \left(\int_0^\infty d\xi\,\frac{1}{\xi}\,g(\xi)\right)^2\;\;.
\end{equation}
The antenna $dA(p_1,p_2;k)$ will therefore correspond to a unitary algorithm
when we let the density $g$ be normalized by
\begin{equation}
  \int_0^\infty d\xi\,\frac{1}{\xi}\,g(\xi) = 1\;\;.
\end{equation}
Note that the normalization of $dA$ fixes the overall factor uniquely: in
particular the appearance of the numerator $\ipb{p_1}{p_2}$ is forced upon us
by the unitarity requirement.

For $g$ we want to take, at this point, the simplest possible function we can
think of, that has a sufficiently regularizing behavior. We introduce a
positive non-zero number $\xim$ and take
\begin{equation}
   g(\xi) \;\df\; \frac{1}{2\log\xim}\,\theta(\xim^{-1}\leq\xi\leq\xim) \;\;.
\label{SarEq006}   
\end{equation}
The number $\xim$ gives a cut-off for the quotients $\xi_1$ and $\xi_2$ of the
scalar products of the momenta, and not for the scalar products themselves.  It
is, however, possible to relate $\xim$ to the total energy $\wcm$ in the
CMF and a cut-off $s_0$ on the invariant masses, i.e., the requirement that 
\begin{equation}
   \invs{(p_i+p_j)}{} \geq s_0 \qquad \textrm{for all momenta $p_i\neq p_j$.}
   \label{SarEq014}
\end{equation}
This can be done by choosing 
\begin{equation}
   \xim \;\df\; \frac{\scm}{s_0} - \frac{(\np+1)(\np-2)}{2} \;\;.
\label{SarEq007}   
\end{equation}
With this choice, the invariant masses $\invs{(p_1+k)}{}$ and
$\invs{(k+p_2)}{}$ are regularized, but can still be smaller than $s_0$ so that
the whole of PS, cut by (\ref{SarEq014}), is covered. The $s_0$
can be derived from physical cuts $p_T$ on the transverse momenta and
$\theta_0$ on the angles between the outgoing momenta:
\begin{equation}
   s_0 \;=\;
   2p_T^2\cdot\min\left(1-\cos\theta_0\,,\,
                        \left(1+\sqrt{1-p_T^2/\scm}\right)^{-1}\right) \;\;.
\label{SarEq018}			
\end{equation}
With this choice, PS with the physical cuts is covered by PS with the cut of
(\ref{SarEq014}). To generate the physical PS, the method of crude Monte Carlo
(\Sec{CrudeMC}) can be used, i.e., if momenta of an event do not satisfy the
cuts, the whole event is rejected.  We end this section with the piece of the
PS algorithm that corresponds to the basic $dA(p_1,p_2;k)$:
\begin{Alg}[{\tt BASIC ANTENNA}]
\begin{enumerate}
\item given $\{p_1,p_2\}$, put $p\lar\Bo_{p_1+p_2}p_1$ 
      and put $E\lar\sqrt{\ipb{p_1}{p_2}/2}$\;;
\item generate two numbers $\xi_{1}$, $\xi_{2}$ independently, each from the
      density $g(\xi)/\xi$, and $\vhi$ uniformly in $[0,2\pi)$\;;
\item put $z\lar(\xi_2-\xi_1)/(\xi_2+\xi_1)$,\; $q^0\lar E(\xi_2+\xi_1)$ and 
      $\vec{q}\lar q^0\Ro_p^{-1}\Hat{n}(z,\vhi)$\;;
\item put $k\lar\Bo_{p_1+p_2}^{-1}q$\;.
\end{enumerate}
\end{Alg}

\section{A complete QCD antenna}
The straightforward way to generate $n$ momenta with the antenna structured 
density is by repeated use of the basic antenna. Let us denote 
\begin{equation}
      dA^i_{j,k}\df dA(q_j,q_k;q_i) \;\;,
\end{equation}
then 
\begin{equation}
   dA^2_{1,{\np}}dA^3_{2,{\np}}dA^4_{3,{\np}}\cdots 
            dA^{{\np}-1}_{{\np}-2,{\np}} 
   \;=\; \frac{\ipb{q_1}{q_{\np}}\,g_{\np}(\{q\}_{\np})}
              {\pi^{{\np}-2}\ipb{q_1}{q_2}\ipb{q_2}{q_3}\ipb{q_3}{q_4}\cdots
	                    \ipb{q_{\np-1}}{q_{\np}}}\,
          \prod_{i=2}^{{\np}-1}d^4q_i\vt(q_i) \;\;,\notag 
\end{equation}
where
\begin{equation}
   g_{\np}(\{q\}_{\np}) \;\df\; 
     g\left(\frac{\ipb{q_1}{q_2}}{\ipb{q_1}{q_{\np}}}\right)
     g\left(\frac{\ipb{q_2}{q_{\np}}}{\ipb{q_1}{q_{\np}}}\right)
     g\left(\frac{\ipb{q_2}{q_3}}{\ipb{q_2}{q_{\np}}}\right)
     g\left(\frac{\ipb{q_3}{q_{\np}}}{\ipb{q_2}{q_{\np}}}\right)\cdots
     g\left(\frac{\ipb{q_{\np-1}}{q_{\np}}}{\ipb{q_{\np-2}}{q_{\np}}}\right)
	   \;\;.
\end{equation}
So if we have two momenta $q_1$ and $q_{\np}$, then we can easily generate
$\np-2$ momenta $q_j$ with the antenna structure. Remember that this
differential PS volume is completely invariant under Lorentz
transformations and scaling transformations, so that it seems self-evident to 
force the set of generated momenta in the CMF with a given energy, using the 
same kind of transformation as in the case of \rambo. 
If the first two momenta are
generated with density $f(q_1,q_{\np})$, then the UAF tells us that generated 
density $\Ant(\{p\}_{\np})$ satisfies
\begin{align}
   \Ant(\{p\}_{\np}) \;=\; 
   &\int d^4q_1\vt(q_1)d^4q_{\np}\vt(q_{\np})\,f(q_1,q_{\np})\,
    dA^2_{1,{\np}}dA^3_{2,{\np}}dA^4_{3,{\np}}\cdots 
            dA^{{\np}-1}_{{\np}-2,{\np}}\notag\\
       &\times d^4b\,\de^4(b-q_{(\np)}/\invm{q_{(\np)}})\,
        dx\,\de(x-\wcm/\invm{q_{(\np)}})\,\prod_{i=1}^{\np}\de^4(p_i-x\Bo_bq_i) 
	\;.
\end{align}
If we apply the same manipulations as in the proof of the correctness of
\rambo, we obtain the equation 
\begin{align}
   \Ant(\{p\}_{\np}) \;&=\; 
   \Theta_{\scm}(\{p\}_\np)\,
   \frac{\ipb{p_1}{p_{\np}}\,g_{\np}(\{p\}_{\np})}
        {\pi^{\np-2}\ipb{p_1}{p_2}\ipb{p_2}{p_3}\ipb{p_3}{p_4}\cdots
	 \ipb{p_{\np-1}}{p_{\np}}}
     \notag\\
       &\times\;\int d^4b\,\de(\invs{b}{}-1)\,dx\,\frac{2\scm^2}{x^5}\,
                f(x^{-1}\Bo_b^{-1}p_1\,,\,x^{-1}\Bo_b^{-1}p_{\np})  \;\;.
\label{SarEq003}	       
\end{align}
Now we choose $f$ such that $q_1$ and $q_{\np}$ 
are generated back-to-back in their CMF with total energy $\wcm$, i.e.,
\begin{equation}
   f(q_1,q_{\np}) \;=\; \frac{2}{\pi}\,\de^4(q_1+q_{\np}-\wcm\enul) \;\;.
\end{equation}
If we evaluate the second line of \eqn{SarEq003} with this $f$, we arrive at 
\begin{multline}
   \frac{4\scm^2}{\pi}\int dx\,\frac{1}{x^5}\,d^4b\,\de(\invs{b}{}-1)\,
   \de^4(x^{-1}\Bo_b^{-1}(p_1+p_{\np})-\wcm\enul)  \\
   \;=\; \frac{4}{\pi}\int_0^\infty dx\,\frac{1}{x^5}\,
          \de\left(\frac{\invs{(p_1+p_{\np})}{}}{\scm x^2}-1\right)
    \;=\; \frac{\scm^2}{2\pi\ipb{p_1}{p_{\np}}^{2}} \;\;,	  
\end{multline}
so that the generated density is given by  
\begin{align}
  \Ant(\{p\}_n) \;=\; 
    \Theta_{\scm}(\{p\}_\np)\,
    \frac{\scm^2}
         {2\pi^{\np-2}}\,
    \frac{g_{\np}(\{p\}_{\np})}
         {\ipb{p_1}{p_2}\ipb{p_2}{p_3}\ipb{p_3}{p_4}\cdots
	  \ipb{p_{\np-1}}{p_{\np}}\ipb{p_{\np}}{p_1}}   \;\;.  
\label{SarEq010}		      
\end{align}
Note that, somewhat surprisingly, also the factor $\ipb{p_{\np}}{p_1}^{-1}$
comes out, thereby making the antenna even more symmetric. In fact, if the 
density $f(q_1,q_2)=c^4\exp(-cq_1^0-cq_2^0)/4\pi^2$ is taken instead of the one
we just used, the calculation can again be done exactly, with exactly the same 
result. The algorithm to generate $n$ momenta with the above antenna structure
is given by 
\begin{Alg}[{\tt QCD ANTENNA}]
\begin{enumerate}
\item generate massless momenta $q_1$ and $q_{\np}$;
\item generate $n-2$ momenta $q_j$  by the basic
      antennas $dA^2_{1,{\np}}dA^3_{2,{\np}}dA^4_{3,{\np}}\cdots 
                dA^{{\np}-1}_{{\np}-2,{\np}}$;
\item compute $q_{(\np)} = \sum_{j=1}^{\np}q_j$, and the
      boost and scaling transforms that bring $q_{(\np)}$ to $\wcm\enul$;
\item for $j=1,\ldots,{\np}$, boost and scale the $q_j$ accordingly, into the 
      $p_j$.
\end{enumerate}
\end{Alg}
Usually, the event generator is used to generate cut PS. 
If a generated event does not satisfy the physical cuts, it is rejected. In the
calculation of the weight coming with an event, the only contribution coming
from the functions $g$ is, therefore, their normalization. In total, this gives
a factor $1/(2\log\xim)^{2{\np}-4}$ in the density.

\section{Incoming momenta and symmetrization}
The density given by the algorithm above, is not quite what we want. First of
all, we want to include the incoming momenta $\ppin$ and $\pnip$ in the APS, so
that the density becomes proportional to $[\ipb{\ppin}{p_1}\ipb{p_1}{p_2}\cdots
\ipb{p_{\np-1}}{p_{\np}}\ipb{p_{\np}}{\pnip}]^{-1}$ instead of
$[\ipb{p_1}{p_2}\cdots\ipb{p_{\np-1}}{p_{\np}}\ipb{p_{\np}}{p_1}]^{-1}$. Then
we want the sum of all permutations of the momenta, including the incoming
ones.

\subsection{Generating incoming momenta\label{SecInc}}
The incoming momenta can be generated after the antenna has been generated. 
To show how, let us introduce the following ``regularized'' scalar product:
\begin{equation}
   \ipb{p}{q}_\de \;\df\;    \ipb{p}{q} + \de p^0q^0\;\;,
\end{equation}
where $\de$ is a small positive number. This regularization is not completely
Lorentz invariant, but that does not matter here. Important is that it is still
invariant under rotations, as we shall see. Using this regularization, we are
able to generate a momentum $k$ with a probability density
\begin{equation}
   \frac{1}{2\pi I_\de(p_1,p_2)}\,
   \frac{\vt(k)\,\de(k^0-1)}{\ipb{p_1}{k}_\de\ipb{\tilde{k}}{p_2}_\de} \;\;.
\label{SarEq004}       
\end{equation}
To show how, we calculate the normalization $I_\de(p_1,p_2)$. Using the 
Feynman-representation of $1/[\ipb{p_1}{k}_\de\ipb{\tilde{k}}{p_2}_\de]$, it 
is easy to see that
\begin{equation}
   I_\de(p_1,p_2)
   \;=\; \frac{1}{4\pi p_1^0p_2^0}\int dzd\vhi\int_0^1
         \frac{dx}{(1+\de-|\vec{p}_x|z)^2} \;\;,
\end{equation}
where $\vec{p}_x=x\Hat{p}_1+(x-1)\Hat{p}_2$. The integral over $z$ and $\vhi$
can now be performed, with the result that
\begin{equation}
   I_\de(p_1,p_2)
   \;=\; \frac{1}{p_1^0p_2^0}\int_0^1\frac{dx}{(1+\de)^2-|\vec{p}_x|^2}
   \;=\; \frac{1}{2\ipb{p_1}{\tilde{p}_2}}
         \int_0^1\frac{dx}{(x_+-x)(x-x_-)} \;\;,
\end{equation}
where $x_{\pm}$ are the solutions for $x$ of the equation 
$1+\de=|\vec{p}_x|$. Further evaluation finally leads to 
\begin{equation}
   I_\de(p_1,p_2) 
   \;=\; \frac{\ipb{p_1}{\tilde{p}_2}^{-1}}{x_+-x_-}\,
         \log\left|\frac{x_+}{x_-}\right| \;\;,\quad
   x_\pm 
   \;=\; \frac{1}{2}\pm\frac{1}{2}
         \sqrt{1+\frac{2p_1^0p_2^0(2\de+\de^2)}{\ipb{p_1}{\tilde{p}_2}}} \;\;.
\end{equation}
Notice that there is a smooth limit to the case in which $p_1$ and $p_2$ are 
back-to-back:
\begin{equation}
   I_\de(p,\tilde{p}) \;=\; \lim_{q\ra\tilde{p}}I_\de(p,q) 
   \;=\;\frac{1}{(p^0)^2(2\de+\de^2)} \;\;.
\end{equation}
The algorithm to generate $k$ can be derived by reading the 
evaluations of the integrals backwards.

Because $k$ and
$\tilde{k}$ are back-to-back, they can serve as the incoming momenta. To fix
them to $\enul+ \ethr$ and $\enul-\ethr$, the whole system of momenta can be
rotated. If we generate momenta with the density $\Ant$, use the first 
two momenta to generate the incoming momenta and rotate, we get a density
\begin{align}
  D_{\scm}(\{p\}_{\np})
   \;&=\;\int d^{4n}q\,\Ant(\{q\}_n)\;d^4k\,
         \frac{1}{2\pi I_\de(q_1,q_2)}
         \frac{\vt(k)\,\de(k^0-1)}
              {\ipb{q_1}{k}_\de\ipb{q_2}{\tilde{k}}_\de}\,   
         \prod_{i=1}^{\np}\de^4(p_i-\Ro_{k}q_i)      \notag\\
     &=\;  \Ant(\{p\}_n)\,I_\de(p_1,p_2)^{-1}\, 
           \int d^4k\vt(k)\,\de(k^0-1)\,
	   \frac{(2\pi)^{-1}}
	        {\ipb{p_1}{\Ro_{k}k}_\de\ipb{p_2}{\Ro_{k}\tilde{k}}_\de} \;\;, 
\end{align}
where we used the fact that the whole expression is invariant under rotations, 
and that these are orthogonal transformations.
The last line of the previous expression can be evaluated further with the 
result that
\begin{equation}
   D_{\scm}(\{p\}_{\np})
   \;=\; \Ant(\{p\}_{\np})\,\frac{I_\de(p_1,p_2)^{-1}}
	        {\ipb{p_1}{\ppin}_\de\ipb{\pnip}{p_2}_\de}
   \quad\textrm{with}\quad
   \ppin=\enul+\ethr \;,\;\; \pnip=\enul-\ethr \;.
\label{SarEq012}		
\end{equation}
The algorithm to generate the incoming momenta is given by
\begin{Alg}[{\tt INCOMING MOMENTA}]\label{SarAlg1}
\begin{enumerate}
\item given a pair $\{p_1,p_2\}$, calculate $x_+$ and $x_-$; 
\item generate $x$ in $[0,1]$ with density $\sim [(x_+-x)(x-x_-)]^{-1}$, 
      and put $\vec{p}_x\lar x\Hat{p}_1+(x-1)\Hat{p}_2$\;;
\item generate $\vhi$ uniformly in $[0,2\pi)$, $z$ in $[-1,1]$ with 
      density $\sim(1+\de-|\vec{p}_x|z)^{-2}$\;;
\item put $\vec{k}\lar\Ro_{p_x}^{-1}\Hat{n}(z,\vhi)$
      and $k^0\lar1$\;;       
\item rotate all momenta with $\Ro_{k}$\;;      
\item put $\ppin\lar\half\wcm(e_0+e_3)$ and 
          $\pnip\lar\half\wcm(e_0-e_3)$\;.      
\end{enumerate}
\end{Alg}
Notice that $I_\de(p_1,p_2)\ipb{p_1}{\ppin}_\de\ipb{\pnip}{p_2}_\de$ is
invariant under the scaling $p_1,p_2\ra cp_1, cp_2$ with a constant $c$, so
that scaling of $\ppin$ and $\pnip$ has no influence on the density.

The pair $(q_1,q_2)$ with which $k$ is generated is free to choose because we
want to symmetrize in the end anyway. We should only choose it such, that we
get rid of the factor $\ipb{q_1}{q_2}$ in the denominator of
$\Ant(\{q\}_{\np})$.

\subsection{Choosing the type of antenna with incoming momenta\label{SecCho}}
A density which is the sum over permutations can be obtained by generating
random permutations, and returning the generated momenta with permutated
labels. This, however, only makes sense for the outgoing momenta. The incoming
momenta are fixed, and should be returned separately from the outgoing momenta
by the event generator. Therefore, a part of the permutations has to be
generated explicitly. There are two kinds of terms in the sum: those in which
$\ipb{\ppin}{\pnip}$ appears, and those in which it does not.

\paragraph{Case 1: antenna with $\ipb{\ppin}{\pnip}$.}
To generate the first kind, we can choose a label $i$ at random with weight
$\ipb{p_i}{p_{i+1}}/\Sigma_1(\{p\}_{\np})$
where $\Sigma_1(\{p\}_{\np})$ is the sum of all scalar products in the 
antenna~\footnote{Read $i+1\!\mod\np$ when $i+1$ occurs in this section}:
\begin{equation}
   \Sigma_1(\{p\}_{\np}) \;\df\; \sum_{i=1}^n\ipb{p_i}{p_{i+1}}  \;\;.
\end{equation}
This is a proper weight, since all scalar products are positive. The total
density gets this extra factor then, so that $\ipb{p_i}{p_{i+1}}$ cancels. The
denominator of the weight factor does not give a problem, because its singular
structure is much softer than the one of the antenna. The pair
$\{p_i,p_{i+1}\}$ can then be used to generate the incoming momenta, as shown
above. So in this case, a density 
$\Ant(\{p\}_{\np})B_1(\{p\}_{\np})/\Sigma_1(\{p\}_{\np})$ is generated, where
\begin{equation}
   B_1(\{p\}_{\np}) \;\df\; 
   \sum_{i=1}^{\np}\frac{\ipb{p_i}{p_{i+1}}\,I_\de(p_i,p_{i+1})^{-1}}
        {\ipb{p_i}{\ppin}_\de\ipb{\pnip}{p_{i+1}}_\de} \;\;.
\end{equation}

\paragraph{Case 2: antenna without $\ipb{\ppin}{\pnip}$.}
To generate the second kind, we can choose two non-equal labels $i$ and
$j$ with weight $\ipb{p_i}{p_{i+1}}\ipb{p_j}{p_{j+1}}/\Sigma_2(\{p\}_{\np})$, 
where 
\begin{equation}
   \Sigma_2(\{p\}_{\np})
   \;\df\;\sum_{i\neq j}^{\np}\ipb{p_i}{p_{i+1}}\ipb{p_j}{p_{j+1}} \;\;.
\end{equation}
Next, a pair $(k,l)$ 
of labels has to be chosen from the set  of pairs
\begin{equation}
   \{(i,j)\}_+
   \;\df\; \{(i,j)\,,\,(i,{j+1})\,,\,({i+1},j)\,,\,({i+1},{j+1})\} \;\;.
\label{SarEq019}   
\end{equation}
If this is done with weight $I_\de(p_k,p_l)/\Sigma_{i,j}(\{p\}_{\np})$, where
\begin{equation}
   \Sigma_{i,j}(\{p\}_{\np})
   \;\df\;  \sum_{(k,l)\in\{(i,j)\}_+}I_\de(p_k,p_l)  \;\;,
\end{equation}
then the density $\Ant(\{p\}_{\np})B_2(\{p\}_{\np})/\Sigma_2(\{p\}_{\np})$ is
generated, where
\begin{align}
  B_2(\{p\}_{\np}) 
  \;&=\; \sum_{i\neq j}^\np\ipb{p_i}{p_{i+1}}\ipb{p_j}{p_{j+1}}
        \sum_{(k,l)\in\{(i,j)\}_+}\frac{I_\de(p_k,p_l)}
                                  {\Sigma_{i,j}(\{p\}_{\np})}\cdot
                             \frac{I_\de(p_k,p_l)^{-1}}
			          {\ipb{p_k}{\ppin}_\de\ipb{\pnip}{p_l}_\de}
				  \notag\\
  \;&=\; \sum_{i\neq j}^{\np}
      \frac{\ipb{p_i}{p_{i+1}}\ipb{p_j}{p_{j+1}}}
           {\ipb{p_i}{\ppin}_\de\ipb{p_{i+1}}{\ppin}_\de
	    \ipb{\pnip}{p_j}_\de\ipb{\pnip}{p_{j+1}}_\de}\cdot
	    \frac{\sum_{(k,l)\in\{(i,j)\}_+}\ipb{p_k}{\ppin}_\de
	                                    \ipb{\pnip}{p_l}_\de}
	         {\sum_{(k,l)\in\{(i,j)\}_+}I_\de(p_k,p_l)} \;\;.
\end{align}
Before all this, we first have to choose between the two cases, and the 
natural way to do this is with relative weights 
$\sfrac{1}{2}\scm\Sigma_1(\{p\}_{\np})$ and $\Sigma_2(\{p\}_{\np})$, 
so that the complete density is equal to 
\begin{align}
   S^{\scriptscriptstyle\textrm{QCD}}_{\scm}(\{p\}_{\np}) \;=\;
   \frac{1}{n!}\sum_{\textrm{perm.}}
   \Ant(\{p\}_{\np})\,
   \frac{\sfrac{1}{2}\scm B_1(\{p\}_{\np})+B_2(\{p\}_{\np})}
        {\sfrac{1}{2}\scm\Sigma_1(\{p\}_{\np})+\Sigma_2(\{p\}_{\np})} \;\;,
\label{SarEq009}		     
\end{align}
where the first sum is over all permutations of $(1,\ldots,n)$. One can, of 
course, try to optimize the weights for the two cases using the adaptive 
multichannel method (\Sec{SecAMC}). The result of using the sum of the two
densities is that 
the factors $\ipb{p_i}{p_{i+1}}$ in the numerator of $B_1(\{p\}_{\np})$ and
$\ipb{p_i}{p_{i+1}}\ipb{p_j}{p_{j+1}}$ in the numerator of $B_2(\{p\}_{\np})$
cancel with the same factors in the denominator of $\Ant(\{p\}_\np)$, so that
we get exactly the pole structure we want. The `unwanted' singularities in
$B_1(\{p\}_{\np}),B_2(\{p\}_{\np})$ and
$\Sigma_1(\{p\}_{\np}),\Sigma_2(\{p\}_{\np})$ are much softer than the ones
remaining in $\Ant(\{p\}_\np)$, and cause no trouble. 
The algorithm to generate the incoming momenta and the permutation is given by
\begin{Alg}[{\tt CHOOSE INCOMING POLE STRUCTURE}]
\begin{enumerate}
\item choose case 1 or 2 with relative weights 
      $\sfrac{1}{2}\scm\Sigma_1(\{p\}_{\np})$ and 
      $\Sigma_2(\{p\}_{\np})$\;;
\item in case 1, choose $i_1$ with relative weight 
      $\ipb{p_{i_1}}{p_{i_1+1}}$ and put $i_2\lar i_1+1$\;;
\item in case 2, choose $(i,j)$ with $(i\neq j)$ and relative weight 
      $\ipb{p_i}{p_{i+1}}\ipb{p_j}{p_{j+1}}$, and then \\
      choose $(i_1,i_2)$ 
      from $\{(i,j)\}_+$ with 
      relative weight $I_\de(p_{i_1},p_{i_2})$\;;
\item use $\{p_{i_1},p_{i_2}\}$ to generate the incoming momenta with 
      Algorithm \ref{SarAlg1};
\item generate a random permutation $\si\in S_{n}$ and put
      $p_i\leftarrow p_{\si(i)}$ for all $i=1,\ldots,n$. 
\end{enumerate}
\end{Alg}
An algorithm to generate the random permutations can be found in \cite{Knuth}.
An efficient 
algorithm to calculate a sum over permutations can be found in \cite{Kuijf1}.

\section{Improvements}
When doing calculations with this algorithm on a PS, cut such that
$\invs{(p_i+p_j)}{}>s_0$ for all $i\neq j$ and some reasonable $s_0>0$, we
notice that a very high percentage of the generated events does not pass the
cuts.  An important reason why this happens is that the cuts, generated by the
choices of $g$ (\eqn{SarEq006}) and $\xim$ (\eqn{SarEq007}), are implemented
only on quotients of scalar products that appear explicitly in the generation
of the QCD-antenna:
\begin{equation}
   \xi^i_1 \df \frac{\ipb{p_{i-1}}{p_i}}{\ipb{p_{i-1}}{p_\np}} 
   \qquad\textrm{and}\qquad
   \xi^i_2 \df \frac{\ipb{p_{i}}{p_{\np}}}{\ipb{p_{i-1}}{p_\np}}  \;\;,
   \quad i=2,3\ldots,\np-1\;\;.
\end{equation}
The total number of these $\xi$-variables is
\begin{equation}
   n_\xi \df 2\np-4 \;\;,
\end{equation}
and the cuts are implemented such that 
$\xim^{-1}\leq\xi^i_{1,2}\leq\xim$ for $i=2,3\ldots,\np-1$.
We show now how these cuts can be implemented on {\em all} quotients 
\begin{equation}
   \frac{\ipb{p_{i-1}}{p_{i}}}{\ipb{p_{j-1}}{p_{j}}} \;,\quad
   \frac{\ipb{p_{i-1}}{p_{i}}}{\ipb{p_{j}}{p_{\np}}} \quad\textrm{and}\quad
   \frac{\ipb{p_{i}}{p_{\np}}}{\ipb{p_{j}}{p_{\np}}} \;,
   \quad i,j=2,3,\ldots,\np-1 \;\;.
\label{SarEq013}   
\end{equation}
We define the $m$-dimensional convex polytope 
\begin{equation}
   \Pol_m\df\{(x_1,\ldots,x_m)\in[-1,1]^m\,\big|\;
                  |x_i-x_j|\leq1\;\forall\,i,j=1,\ldots,m\} \;\;,
\label{SarEq015}		  
\end{equation}
and replace the generation of the the $\xi$-variables by the following:
\begin{Alg}[{\tt IMPROVEMENT}]
\begin{enumerate}
\item generate $(x_1,x_2,\ldots,x_{n_\xi})$ distributed uniformly in 
      $\Pol_{n_\xi}$;
\item define $x_0\df0$ and put,  
      \begin{equation}
         \xi^i_1\lar e^{(x_{2i-3}-x_{2i-4})\log\xim}\;\;,\quad
	 \xi^i_2\lar e^{(x_{2i-2}-x_{2i-4})\log\xim}
	 \label{SarEq011}
      \end{equation}
      for all $i=2,\ldots,n-1$.
\end{enumerate}
\end{Alg}
Because all the variables $x_i$ are distributed uniformly such that
$|x_i-x_j|\leq1$, {\em all} quotients of (\ref{SarEq013}) are distributed such
that they are between $\xim^{-1}$ and $\xim$. In terms of the variables $x_i$,
this means that we generate the volume of $\Pol_{n_\xi}$, which is $n_\xi+1$,
instead of the volume of $[-1,1]^{n_\xi}$, which is $2^{n_\xi}$.  
In \Sec{PolSec}, 
we give the algorithm to generate variables distributed uniformly in
$\Pol_m$. We have to note here that this improvement only makes sense because
the algorithm to generate these variables is very efficient. The total density
changes such, that the function $g_n$ in \eqn{SarEq010} has to be replaced by
\begin{equation}
   g^{\Pol}_{n}(\xim;\{\xi\})
   \;\df\; \frac{1}{(n_\xi+1)(\log\xim)^{n_\xi}}\,
           \theta(\,(x_1,\ldots,x_{n_\xi})\in\Pol_{n_\xi}\,) \;\;,
\end{equation}
where the variables $x_i$ are functions of the variables $\xi^i_{1,2}$ as 
defined by $(\ref{SarEq011})$.
Because crude MC is used to restrict generated events to cut PS, again only the
normalization has to be calculated for the weight of an event.

\begin{figure}
\begin{center}
\begin{picture}(160,100)(0,0)
\GOval(80,50)(80,50)(90){1}
\GOval(80,50)(56,35)(90){0.7}
\BBoxc(80,50)(72,50)
\Line(133,75)(170,85)\Text(175,86)[l]{phase space}
\Line(90,40)(170,20)\Text(175,20)[l]{cut phase space}
\Line(35,50)(-20,50)\Text(-25,51)[r]{generated phase space}
\end{picture}
\caption{Schematic view on phase space.}
\label{SarFig02}
\end{center}
\end{figure}
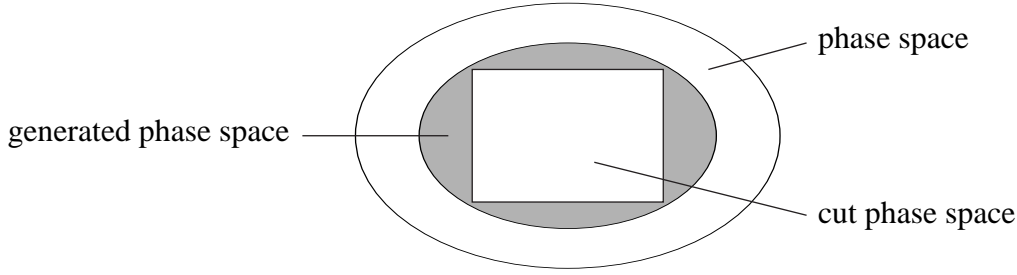

With this improvement, still a large number of events does not pass the cuts.
%
The situation with PS is depicted in \fig{SarFig02}.
Phase space contains generated phase space which contains cut phase space.  The
problem is that most events fall in the shaded area, which is the piece of
generated PS that is not contained in cut PS. To get a higher
percentage of accepted events, we use a random variable $\xi_v\in[0,\xim]$,
instead of the fixed number $\xim$, to generate the variables
$\xi^{i}_{1,2}$. This means that the size of the generated PS
becomes variable.  If this is done with a probability distribution such that
$\xi_v$ can, in principle, become equal to $\xim$, then whole of cut phase
space is still covered. We suggest the following, tunable, density:
\begin{equation}
   h_\al(\xi_v) 
   \;=\; \frac{\al n_\xi+1}{(\log\xim)^{\al n_\xi+1}}\cdot
         \frac{(\log\xi_v)^{\al n_\xi}}{\xi_v}\,\theta(1\leq\xi_v\leq\xim) 
	 \;\;,\quad \al\geq0 \;\;.
\end{equation}
If $\al=0$, then $\log\xi_v$ is distributed uniformly in $[0,\log\xim]$, and
for larger $\al$, the distribution peaks more and more towards $\xi_v=\xim$.
Furthermore, the variable is easy to generate and the total generated density
can be calculated exactly: 
$g^{\Pol}_{n}(\xim;\{\xi\})$ should be replaced by 
\begin{align}
   G^{\Pol}_{n}(\al,\xim;\{\xi\}) 
   \;\df&\;\int d\xi_v\,h_\al(\xi_v)\,g^{\Pol}_{n}(\xi_v;\{\xi\}) \notag\\
   \;=&\; \frac{1}{n_\xi+1}\cdot
          \frac{\al n_\xi+1}{(\log\xim)^{\al n_\xi+1}}
	  \int_{\log\xilow}^{\log\xim}dx\,x^{(\al-1)n_\xi} \;\;,
\label{SarEq016}
\end{align}
where $\xilow$ is the maximum of the ratios of scalar products in 
(\ref{SarEq013}).

\section{Results and conclusions}
We compare \sarge\ with \rambo\ in the integration of the {\tt SPHEL}-integrand
for processes of the kind $\gl\gl\ra \np\gl$, which is given by 
\begin{equation}
   \sum_{\textrm{perm.}}
   \frac{2\sum_{i\neq j}^{n+1}
   \ipb{p_i}{p_j}^4}{\ipb{p_1}{p_2}\ipb{p_2}{p_3}\ipb{p_3}{p_4}\cdots
            \ipb{p_{\np}}{p_{\np+1}}\ipb{p_{\np+1}}{p_{\np+2}}
	                            \ipb{p_{\np+2}}{p_1}} \;\;,
\end{equation}
where $p_1$ and $p_2$ are the incoming momenta, and the first sum is over all 
permutations of $(2,3,\ldots,n+2)$ except the cyclic permutations.
The results are presented in \tab{SarTab02}.
\begin{table}
\begin{center}
\begin{tabular}{|>{$}c<{$}|}
  \hline n \\
  \hline \tau_{\texttt{SPHEL}}(\seco) \\ 
  \hline \tau_{\textrm{exact}}(\seco) \\ \hline
\end{tabular}  
\begin{tabular}{|>{$}c<{$}|>{$}c<{$}|>{$}c<{$}|>{$}c<{$}|}
  \hline 4 &5 &6 &7 \\
  \hline 5.40\emu{5} &2.70\emu{4} &1.80\emu{3} &1.41\emu{2} \\ 
  \hline 3.07\emu{1} &1.08 &3.35 &10.92 \\ \hline
\end{tabular}  
\end{center}
\caption{cpu-times ($\tau_{\texttt{SPHEL}}$) in seconds needed to evaluate the 
         {\tt SPHEL}-integrand one time with a $300$-MHz UltraSPARC-IIi 
	 processor, and the cpu-times ($\tau_{\textrm{exact}}$) needed to 
	 evaluate the exact integrand, estimated with the help of 
	 \tab{SarTab01}.}  
\label{SarTab03}	 
\end{table}

\begin{table}
\begin{center}
\begin{tabular}{|c|}
   \hline \\ $\gl\gl\rightarrow4\gl$ \\ \\ $1\%$ error \\ \\ \hline
\end{tabular}
\begin{tabular}{|c|}
   \hline alg.     \\ 
   \hline $\sigma$ \\ 
   \hline $\Nge$   \\
   \hline $\Nac$   \\
   \hline $\tcpu(\hour)$  \\ 
   \hline $\texa(\hour)$  \\ \hline
\end{tabular}
\begin{tabular}{|>{$}c<{$}|>{$}c<{$}|>{$}c<{$}|>{$}c<{$}|}
  \hline \rambo     &\sarge,\al=0.0 &\sarge,\al=0.5 &\sarge,\al=10.0\\
  \hline 4.30\ee{8} &4.31\ee{8}     &4.37\ee{8}     &4.32\ee{8} \\
  \hline 4,736,672  &296,050        &278,702        &750,816 \\
  \hline 3,065,227  &111,320        &40,910         &23,373 \\
  \hline 0.198      &0.0254         &0.0172         &0.0348 \\
  \hline 262        &9.52           &3.51           &2.03 \\ \hline
\end{tabular}
\end{center}
\begin{center}
\begin{tabular}{|c|}
   \hline \\ $\gl\gl\rightarrow5\gl$ \\ \\ $1\%$ error \\ \\ \hline
\end{tabular}
\begin{tabular}{|c|}
   \hline alg.     \\ 
   \hline $\sigma$ \\ 
   \hline $\Nge$   \\
   \hline $\Nac$   \\
   \hline $\tcpu(\hour)$  \\ 
   \hline $\texa(\hour)$  \\ \hline
\end{tabular}
\begin{tabular}{|>{$}c<{$}|>{$}c<{$}|>{$}c<{$}|>{$}c<{$}|}
  \hline\rambo       &\sarge,\al=0.0 &\sarge,\al=0.5 &\sarge,\al=10.0\\
  \hline 3.78\ee{10} &3.81\ee{10}    &3.80\ee{10}    &3.81\ee{10} \\
  \hline 4,243,360   &715,585        &1,078,129      &6,119,125 \\
  \hline 1,712,518   &167,540        &36,385         &21,111 \\
  \hline 0.286       &0.133          &0.0758         &0.277 \\
  \hline 514         &51.6           &11.7           &9.10 \\ \hline
\end{tabular}
\end{center}
\begin{center}
\begin{tabular}{|c|}
   \hline \\ $\gl\gl\rightarrow6\gl$ \\ \\ $1\%$ error \\ \\ \hline
\end{tabular}
\begin{tabular}{|c|}
   \hline alg.     \\ 
   \hline $\sigma$ \\ 
   \hline $\Nge$   \\
   \hline $\Nac$   \\
   \hline $\tcpu(\hour)$  \\ 
   \hline $\texa(\hour)$  \\ \hline
\end{tabular}
\begin{tabular}{|>{$}c<{$}|>{$}c<{$}|>{$}c<{$}|>{$}c<{$}|}
  \hline \rambo      &\sarge,\al=0.0 &\sarge,\al=0.5 &\sarge,\al=10.0\\
  \hline 3.07\ee{12} &3.05\ee{12}    &3.13\ee{12}    &3.05\ee{12} \\
  \hline 3,423,981   &2,107,743      &6,136,375      &68,547,518 \\
  \hline 700,482     &276,344        &34,095         &17,973 \\
  \hline 0.685       &1.32           &0.471          &3.17 \\
  \hline 653         &258            &32.2           &19.9 \\ \hline
\end{tabular}
\end{center}
\begin{center}
\begin{tabular}{|c|}
   \hline \\ $\gl\gl\rightarrow7\gl$ \\ \\ $3\%$ error \\ \\ \hline
\end{tabular}
\begin{tabular}{|c|}
   \hline alg.     \\ 
   \hline $\sigma$ \\ 
   \hline $\Nge$   \\
   \hline $\Nac$   \\
   \hline $\tcpu(\hour)$  \\ 
   \hline $\texa(\hour)$  \\ \hline
\end{tabular}
\begin{tabular}{|>{$}c<{$}|>{$}c<{$}|>{$}c<{$}|>{$}c<{$}|}
  \hline \rambo      &\sarge,\al=0.0 &\sarge,\al=0.5 &\sarge,\al=10.0\\    
  \hline 2.32\ee{14} &2.16\ee{14}    &2.20\ee{14}    &2.28\ee{14} \\
  \hline 605,514     &710,602        &5,078,153      &125,471,887 \\
  \hline 49,915      &42,394         &3,256          &1,789 \\
  \hline 0.224       &1.86           &0.452          &6.74 \\
  \hline 152         &130            &10.3           &12.2 \\ \hline
\end{tabular}
\caption{Results for the integration of the {\tt SPHEL}-integrand.
         The CM-energy and the cuts used are 
	 $\wcm=1000$, $p_{T}=40$ and $\theta_0=30^\circ$. Presented are 
	 the finial result ($\si$), the number of generated ($\Nge$) and 
	 accepted ($\Nac$) events, the cpu-time $(\tcpu)$ in hours, and
	 the cpu-time ($\texa$) it would take to integrate the exact matrix 
	 element, estimated with the help of \tab{SarTab03}.
	 In the calculation of this table, adaptive multichanneling 
	 in the two cases of \Sec{SecCho} was used, and $\delta=0.01$ 
	 (\Sec{SecInc}).}
\label{SarTab02}	 
\end{center}
\end{table}

\begin{figure}
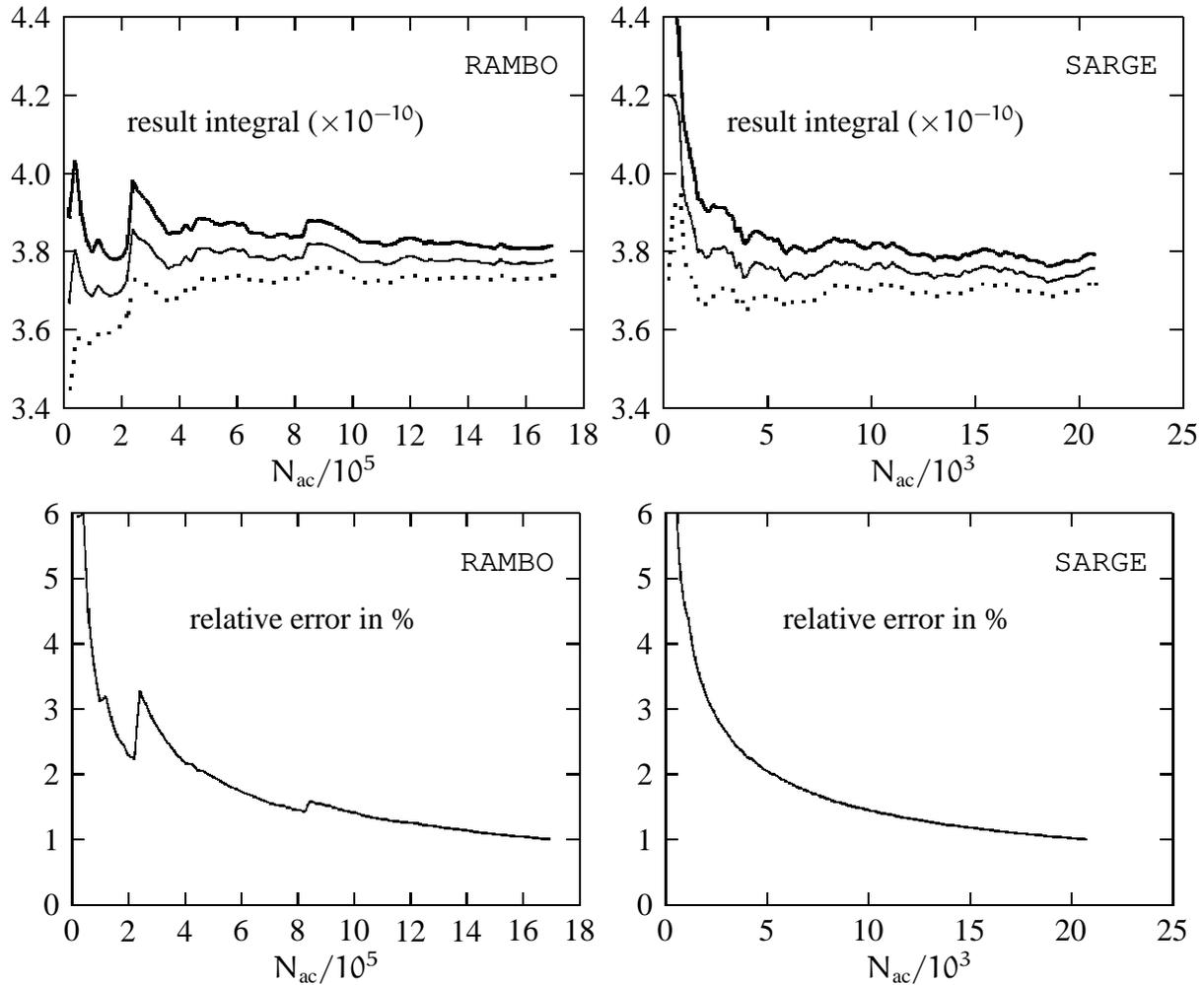

\begin{center}
\hspace{-45pt}
%
%
\setlength{\unitlength}{0.240900pt}
\ifx\plotpoint\undefined\newsavebox{\plotpoint}\fi
\sbox{\plotpoint}{\rule[-0.200pt]{0.400pt}{0.400pt}}%

\caption{The convergence of the MC-process in the integration of the 
         {\tt SPHEL}-integrand for $\np=5$, with $\wcm=1000$, $p_T=40$ and
	 $\theta_0=30^\circ$. The upper graphs show the 
	 integral itself as function of the number of 
	 accepted events, together with the estimated bounds on the expected
	 deviations. The lower graphs show the relative
	 error. \sarge\ was used with adaptive multi-channeling in the two 
	 cases of \Sec{SecCho}, with $\delta=0.01$ (\Sec{SecInc}) and without 
	 the variable $\xi_v$. The number of generated events was $6,699,944$, 
	 and the cpu-time was $0.308$ hours.}
\label{SarFig03}	 
\end{center}
\end{figure}
The calculations were done at a CM-energy $\wcm=1000$ with cuts $p_T=40$ on
each transverse momentum and $\theta_0=30^\circ$ on the angles between the
momenta. We present the results for $\np=4,5,6,7$, calculated with \rambo\ and
\sarge\ with different values for $\al$ (\eqn{SarEq016}). The value of $\si$ is
the estimate of the integral at an estimated error of $1\%$ for $\np=4,5,6$ and
$3\%$ for $\np=7$. These numbers are only printed to show that different
results are compatible. Remember that they are not the whole cross sections: 
flux factors, color factors, sums and averages over helicities, and coupling 
constants are not included.
The other data are the number of generated events
($\Nge$), the number of accepted events ($\Nac$) that passed the cuts, the
cpu-time consumed ($\tcpu$), and the cpu-time the calculation would have
consumed if the exact matrix element had been used ($\texa$), both in hours.
This final value is estimated with the help of \tab{SarTab03} and the formula
\begin{equation}
   \texa 
   \;=\; \tcpu + \Nac(\tau_{\textrm{exact}} - \tau_{\texttt{SPHEL}})\;\;,
\end{equation}
where $\tau_{\textrm{exact}}$ and $\tau_{\texttt{SPHEL}}$ are the cpu-times it 
takes to evaluate the squared matrix element once.
Remember that the integrand only has to be evaluated for accepted events. The
calculations have been performed with a single $300$-MHz UltraSPARC-IIi 
processor.

The first conclusion we can draw is that \sarge\ outperforms \rambo\ in
computing time for all processes. This is especially striking for lower number
of outgoing momenta, and this behavior has a simple explanation: we kept the
CM-energy and the cuts fixed, so that there is less energy to distribute over
the momenta if $\np$ is larger, and the cuts become relatively tighter. As a
result, \rambo\ gains on \sarge\ if $\np$ becomes larger. This effect would 
not appear if the energy, or the cuts, would scale with $\np$ like in 
\tab{SarTab01}. 
Another indication for this effect is the fact that the ratio 
$\Nac/\Nge$ for \rambo, 
which estimates the ratio of the 
volumes of cut PS and whole PS,  
decreases with $\np$.

Another conclusion that can be drawn is that \sarge\ performs better if $\al$
is larger. Notice that the limit of $\al\ra\infty$ is equivalent with dropping
the improvement of the algorithm using the variable $\xi_v$ (\eqn{SarEq016}).
Only if the integrand becomes too flat, as in the case of $\np=7$ with the 
energy and the cuts as given in the table, smaller values are preferable. Then, 
too many events do not pass the cuts if $\alpha$ is large.

As an extra illustration of the performance of \sarge, we present in
\fig{SarFig03} the evaluation of MC-integrals as function of the number of
accepted events. 
Depicted are the integral $\si$ with the bounds on the expected deviation
coming from the estimated expected error, and the relative error. Especially
the graphs with the relative error are illustrative, since they show that
it converges to zero more smoothly for \sarge\ then for \rambo. Notice the
spike for \rambo\ around $\Nac=25000$, where an event obviously hits a
singularity.

\section{Other pole structures}
The APS of (\ref{SarEq001}) is not the only pole structure occurring in the 
squared amplitudes of QCD-processes; not even in purely gluonic processes. 
For example, in the case of $\gl\gl\ra4\gl$, also permutations of 
\begin{equation}
   \frac{1}{\ipb{p_1}{p_3}\ipb{p_2}{p_4}\ipb{\ppin}{p_1}\ipb{\pnip}{p_2}
            \invs{(\ppin-p_1-p_2)}{}}
\label{SarEq017}	    
\end{equation}
occur \cite{Kuijf}. If one is able to generate momenta with this density, it
can be included in the whole density with the use of the adaptive multichannel
technique.
In the interpretation of the transition amplitude as a sum of
Feynman diagrams, this kind of pole structures typically come from $t$-channel
diagrams, which are of the type
\begin{equation}
\qquad
\parbox{110pt}{\begin{picture}(100,70)(0,0)
\ArrowLine(0,0)(50,20)     \Text(10.0,10.0)[rb]{$\pnip$}
\ArrowLine(50,20)(100,0)   \Text(90.0,10.0)[lb]{$Q_2$}
\Vertex(50,20){2}
\ArrowLine(50,50)(50,20) \Text(44,35)[cr]{$k$}
\Vertex(50,50){2}
\ArrowLine(0,70)(50,50)   \Text(10.0,62.0)[rt]{$\ppin$}
\ArrowLine(50,50)(100,70) \Text(90.0,62.0)[lt]{$Q_1$}
\end{picture}}
\qquad,\notag
\end{equation}
and where, for this case, $Q_1=p_1+p_3$ and $Q_2=p_2+p_4$, so that
$k=p_0-p_1-p_3$. The natural way to generate a density with this pole structure
is by generating $s_i\df\invs{Q}{i}$ with a density proportional to $1/s_i$, a
variable $t$ that plays the role of $\invs{(p_0-p_1-p_3)}{}$, construct with
this and some generated angles the momenta $Q_i$, and then split new momenta
from each of these. 
For $\np=4$, only two momenta have to split off each $Q_i$, and there is a
reasonable simple algorithm to generate these. 

We shall now just present the algorithm that generates the density (\ref{OPS}),
and then show its correctness using the UAF. If we mention the generation of
some random variable $x$ `with a density $f(x)$' in the following, we mean a
density that is proportional to $f(x)$, and we shall not always write down the
normalization explicitly.  Furthermore, $\scm$ denotes the square of the
CM-energy and $\la\df\la(\scm,s_1,s_2)$ the usual Mandelstam variable
\begin{equation}
   \la \;\df\; \scm^2 + s_1^2 + s_2^2 - 2\scm s_1 - 2\scm s_2 - 2s_1s_2  \;\;.
\end{equation}
Of course, a cut has to be implemented in order to generate momenta
following (\ref{SarEq017}), and we shall be able to put $\ipb{p_i}{p_j}>\half
s_0$ for the scalar products occurring in the denominator, where $s_0$ only has
to be larger than zero. To generate the momenta with density (\ref{SarEq017}),
one should
\begin{Alg}[{\tt T-CHANNEL}]
\begin{enumerate}
\item generate $s_1$ and $s_2$ between $s_0$ and $\scm$ 
      with density $1/s_1$ and $1/s_2$;
\item generate $t$ between $\scm-s_1-s_2\pm\sqrt{\la(\scm,s_1,s_2)}$ 
      with density $1/[t(t+2s_1)(t+2s_2)]$;   
\item put $z\lar(\scm-s_1-s_2-t)/\sqrt{\la}$ and 
      generate $\vhi$ uniformly in $[0,2\pi)$;
\item put $Q_1\lar(\sqrt{s_1+\la/(4s)},\sqrt{\la/(4s)}\,\Hat{n}(z,\vhi)\,)$ 
      and $Q_2\lar\wcm\enul-Q_1$;
\item for $i=1,2$, generate $z_i>1-4s_0/(t+2s_i)$ with density 
      $1/(1-z_i)$ and $\vhi_i$ uniformly in $[0,2\pi)$, and put
      $q_i\lar\half\sqrt{s_i}\,(1,\Hat{n}(z_i,\vhi_i)\,)$; 
\item for $i=1,2$, rotate $q_i$ to the  CMF of $Q_i$, then boost it to the 
      CMF of $Q_1+Q_2$ to obtain $p_i$, and put $p_{i+2}\lar Q_i-p_i$;
\end{enumerate}
\end{Alg}
As a final step, the incoming momenta can be put to
$\ppin\lar\half\wcm(\enul+\ethr)$ and $\pnip\lar\half\wcm(\enul-\ethr)$.
The variables $s_i$ and $z_i$ can easily be obtained by inversion
(\Sec{Inversion}). The variable $t$ can best be obtained by generating
$x\df\half\log(4s_1s_2)-\log t$ with the help of the rejection method
(\Sec{Rejection}).
In the UAF, the steps of the algorithm read as follows. Denoting 
\begin{equation}
   \ve_{1}\df\enul+\ethr
   \quad,\qquad
   \ve_{2}\df\enul-\ethr 
   \quad,\qquad
   h_\pm \df \scm-s_1-s_2\pm\sqrt{\la} \;\;,
\end{equation}
and
 \begin{align}
    \textrm{nrm}(\scm,s_1,s_2)
   \;\df&\; \int\frac{dt}{t(t+2s_1)(t+2s_2)}\,
              \theta(h_- < t < h_+) \notag\\
   \;=&\; \frac{1/4}{s_1-s_2}
            \left[  \frac{1}{s_2}\log\frac{1 + 2s_2/h_-}{1 + 2s_2/h_+} 
 	          - \frac{1}{s_1}\log\frac{1 + 2s_1/h_-}{1 + 2s_1/h_+}
	    \right] \;,
\end{align}
we have
\begin{align}
 1.&\;\; \int\frac{ds_1}{s_1}\frac{ds_2}{s_2}\,
         \frac{\theta(s_0<s_{1,2}<\scm)}{(\log\frac{\scm}{s_0})^2} \notag\\
 2.&\;\; \int\frac{dt}{t(t+2s_1)(t+2s_2)}\,
              \frac{\theta(h_- < t < h_+)}
	           {\textrm{nrm}(s,s_1,s_2)} \notag\\
 3.&\;\; \int dz\,\de\left(z-\frac{s-s_1-s_2-t}{\sqrt{\la}}\right)
               \,\frac{d\vhi}{2\pi} \notag\\
 4.&\;\; \int \dQ_1\,\de\left(Q_1^0-\sqrt{s_1+\sfrac{\la}{4s}}\right)
               \de^3\left(\vec{Q}_1-\sqrt{\sfrac{\la}{4s}}\,
	                                      \Hat{n}(z,\vhi)\right)
	       \dQ_2\,\de^4(Q_1+Q_2-\wcm\enul) \notag\\	
 5.&\;\; \int\prod_{i=1}^2 
          \frac{dz_i}{1-z_i}\,\frac{\theta(1-z_i>\frac{4s_0}{t+2s_i})}
			           {\log\frac{t+2s_i}{2s_0}}\;
          \frac{d\vhi_i}{2\pi}\; 
	  d^4q_i\,\de(q_i^0-\half\sqrt{s_i}\,)\,
                  \de^3(\,\vec{q}_i-q_i^0\Hat{n}(z_i,\vhi_i)\,) \notag \\
 6.&\;\; \int\prod_{i=1}^2 d^4b_i\,\de^4(b_i-\Bo_{Q_i}\ve_{i})\,
                       \de^4(p_i-\Bo_{Q_i}^{-1}\Ro_{b_i}^{-1}q_i)\,
                       \de^4(p_{i+2}+p_i-Q_i) \;\;. \notag
\end{align}
The various assignments imply the following identities. First of all, we have 
\begin{equation}
   \invs{(p_i+p_{i+2})}{} = \invs{Q}{i} = s_i  \;\;.
\end{equation}
Using that $4ss_1+\la=(s+s_1-s_2)^2$ we find
\begin{equation}
   \sqrt{4s}\,(\ip{\ve_1}{Q_1})
   = s+s_1-s_2-z\sqrt{\la} = t+2s_1 
\end{equation}
and the same for $(1\leftrightarrow2)$, so that
\begin{equation}
   t = 4(\ip{\ppin}{Q_1}) - 2\invs{(p_1+p_3)}{}
     = -2\invs{(\ppin-p_1-p_3)}{} \;\;.
\end{equation}
Denote $\Lo_{Q_i}\df\Ro_{b_i}\Bo_{Q_i}$, so that $q_i=\Lo_{Q_i}p_i$. Because
$\Lo_{Q_i}\ve_i\sim\ve_1$, we find that
\begin{equation}
   1-z_i = \frac{2(\ip{\ve_1}{q_i})}{\sqrt{s_i}}
         = 2\frac{(\ip{\ve_1}{\Lo_{Q_i}p_i})}
	         {(\ip{\ve_1}{\Lo_{Q_i}Q_i})}
	 = 2\frac{(\ip{\ve_i}{p_i})}{(\ip{\ve_i}{Q_i})}	\;\;, 
\end{equation}
so that
\begin{equation}
   (t+2s_1)(1-z_1) = 8(\ip{\ppin}{p_1}) 
   \quad\textrm{and}\quad
   (t+2s_2)(1-z_2) = 8(\ip{\pnip}{p_2}) \;\;.
\end{equation}
We can conclude so far that the algorithm generates the correct pole structure.
For the further evaluation of the integrals one can forget about the factors
$s_i$, $t$, $t+2s_i$ and $1-z_i$ in the denominators.
Using that
\begin{equation}
   d^4q_i\,\de(q_i^0-\half\sqrt{s_i}\,)\,
                  \de^3(\,\vec{q}_i-q_i^0\Hat{n}(z_i,\vhi_i)\,)
   \;=\; 2\,d^4q_i\vt(q_i)\,
         \de^3(\frac{2}{\sqrt{s_i}}\,\vec{q}_i-\Hat{n}(z_i,\vhi_i)\,)\;,
\end{equation}
and replacing step 4 by 
\begin{equation}
   \left(\prod_{i=1}^22\sqrt{s_1+\sfrac{\la}{4s}}\,
         \dQ_i\vt_{s_i}(Q_i)\right)
   \de(z(\vec{Q}_1)-z)\,\de(\vhi(\vec{Q}_1)-\vhi)\,
   \de^4(Q_1+Q_2-\wcm\enul)\;,
\end{equation}
the integrals can easily be performed backwards, i.e., in the order 
$q_i$, $\vhi_i$, $z_i$, $b_i$, $Q_i$, $\vhi$, $z$, $t$, $s_1$, $s_2$. 
The density finally is 
\begin{align}
   \Theta_{\scm}(\{p\}_4)\,
   &\frac{\theta(2\ipb{\ppin}{p_1}>s_0)\,\theta(2\ipb{\pnip}{p_2}>s_0)\,
         \theta(2\ipb{p_1}{p_3}>s_0)\,\theta(2\ipb{p_2}{p_4}>s_0)}
        {\ipb{\ppin}{p_1}\ipb{\pnip}{p_2}\ipb{p_1}{p_3}\ipb{p_2}{p_4}
	 [-\invs{(\ppin-p_1-p_3)}{}]} \notag\\
   &\times 
    \frac{\scm}{24(2\pi)^3}
    \left[\Big(\log\frac{s}{s_0}\Big)^2
          \log\Big(\frac{t+2s_1}{2s_0}\Big)\,\log\Big(\frac{t+2s_2}{2s_0}\Big)\,
	  \textrm{nrm}(\scm,s_1,s_2)\right]^{-1} \;,
\label{OPS}	  
\end{align}
where $s_i\df\invs{(p_i+p_{i+2})}{}$ and $t\df-2\invs{(\ppin-p_1-p_3)}{}$.

\section{Generating a uniform distribution inside a polytope\label{PolSec}}
We consider the $m$-dimensional convex polytope $\Pol_m$ defined in 
(\ref{SarEq015}).
The task is to generate $m$-dimensional points uniformly inside $\Pol_m$.  A
straightforward way is to generate points inside the hypercube $[-1,1]^m$ and
implement the other conditions by rejection, with an efficiency given by
$\Vol(\Pol_m)/2^m$, where $\Vol(\Pol_m)$ is the volume of the polytope. This
may, however, become slow if $\Vol(\Pol_m)$ does not increase fast with $m$.
Let us, therefore, compute $\Vol(\Pol_m)$. We distinguish positive and negative
$x_i$ values.  Define
\begin{equation}
   V_m(k) \;\df\; \int_{\Pol_m} dx_1\cdots dx_m\,\theta(x_{1,2,\ldots,k}\le0)
   \,\theta(x_{k+1,\ldots,m}\ge0)\;\;.
\end{equation}
We then have
\begin{equation}
  \Vol(\Pol_m) \;=\; \sum_{k=0}^m\frac{m!}{k!(m-k)!}\,V_m(k)\;\;.
\end{equation}
In the calculation of $V_m(k)$ we notice that the only
nontrivial constraints are of the type $x_i-x_j<1$,
with $i=k+1,k+2,\ldots,m$ and $j=1,2,\ldots,k$.
Writing $x_i=-y_i$ for $i=1,2,\ldots,k$, we therefore have
\begin{equation}
  V_m(k) \;=\; \int_0^1 dy_1dy_2\cdots dy_kdx_{k+1}dx_{k+2}\cdots dx_m\;
               \theta\left(\max_jx_j + \max_iy_i < 1\right)\;\;.
\end{equation}
Relabeling such that $\max_iy_i = y_1$ and $\max_jx_j = x_m$
then leads us to
\begin{align}
   V_m(k) \;&=\; k(m-k)\int_0^1dy_1\int_0^{y_1}dy_2\cdots dy_k
                 \int_0^1dx_m\int_0^{x_m}dx_{k+1}dx_{m-1}\theta(x_m+y_1<1)
		 \notag\\
          \;&=\; k(m-k)\int_0^1dy_1\,y_1^{k-1}\int_0^{1-y_1}
                 dx_m\,x_m^{m-k-1}
		 \notag\\
          \;&=\; k\int_0^1dy_1\,y_1^{k-1}(1-y_1)^{m-k}
          \;=\;  \frac{k!(m-k)!}{ m!}\;\;,
\end{align}
so that we find
\begin{equation}
   \Vol(\Pol_m) = m+1\;\;.
\end{equation}
Accordingly, the rejection algorithm will quickly become inefficient,
below 1\% for $n>10$.
The above calculation actually allows us to construct an optimal
algorithm by working backwards. In the following each $\rho_i$ stands for a
new call to the random number source.
\begin{Alg}[{\tt POLYTOPE}]
\begin{enumerate}
\item choose an integer $k$. Since $m!V_m(k)/k!(m-k)!=1$, it should be chosen
      uniformly in $[0,m]$, so
      $$
        k \lar \lfloor (m+1)\rho_0\rfloor\;\;.
      $$
\item if $k=0$ we simply have
      $$
        x_i \lar \rho_i\;\;i=1,m\;\;.
      $$
      If $k=m$ we use
      $$
        x_i \lar -\rho_i\;\;i=1,m\;\;.
      $$
\item for $0<k<m$, generate $y_1$ in $[0,1]$ according to the distribution
      $y_1^{k-1}(1-y_1)^{m-k}$.
      An efficient algorithm to do this is Cheng's rejection algorithm BA for
      beta random variates (cf.~\cite{Devroye})\footnote{There is an error on
      page 438 of \cite{Devroye}, where ``$V\lar\lambda^{-1}U_1(1-U_1)^{-1}$''
      should be replaced by ``$V\lar\lambda^{-1}\log[U_1(1-U_1)^{-1}]$''.}, but
      also the following works:
      $$
        v_1\lar -\log\left(\prod_{i=1}^k\rho_i\right)\;\;,\;\;
        v_2\lar -\log\left(\prod_{j=1}^{m-k+1}\rho_j\right)\;\;,\;\;
        y_1 \lar \frac{v_1}{ v_1+v_2}\;\;.
      $$
\item generate $x_m$ in $[0,1-y_1]$ according to the distribution
      $x_m^{m-k-1}$. The algorithm to do this is
      $$
        x_m \lar (1-y_1)\rho_m^{1/(m-k)}\;\;.
      $$
\item generate the $y_{2,\ldots,k}$ uniformly in $[0,y_1]$ and flip sign:
      $$
        x_1 \lar -y_1\;\;,\;\;x_i \lar -\rho_iy_1\;\;,i=2,3,\ldots,k\;\;.
      $$
\item generate the $x_{k+1,\ldots,m-1}$ uniformly in $[0,x_m]$:
      $$
        x_j \lar \rho_jx_m\;\;,j=k+1,k+2,\ldots,m-1\;\;.
      $$
\item Finally, perform a random permutation of the whole set of $x$ values.
\end{enumerate}
\end{Alg}

\subsection{Computational complexity}
The number usage $S$, that is, the expected number of calls to the random
number source $\rho$ per event can be derived easily. In the first place, 1
number is used to get $k$ for every event. In a fraction $2/(m+1)$ of the
cases, only $m$ calls are made. In the remaining cases, there are $k+(m-k+1) =
m+1$ calls to get $y_1$, and 1 call for all the other $x$ values. Finally, the
simplest permutation algorithm calls $m-1$ times \cite{Knuth}. The expected
number of calls is therefore
\begin{equation}
   S =1 + \frac{2m}{ m+1} + \frac{m-1}{m+1}(m+1 + (m-1) + (m-1)) =
   {3m^2 - m + 2\over m+1}\;\;.
\end{equation}
For large $m$ this comes to about $3m-1$ calls per event.
Using a more sophisticated permutation algorithm would use at least 1 call,
giving
\begin{equation}
  S = 1 + \frac{2m}{ m+1} + \frac{m-1}{m+1}(m+1 + (m-1) + (1)) = 2m\;\;.
\end{equation}
We observed that Cheng's rejection algorithm to obtain $y_1$ uses about
2 calls per event. Denoting this number by $C$ the expected number
of calls becomes
\begin{equation}
S = {2m^2 + (C-1)m - C + 3\over m+1} \sim 2m + C - 1
\end{equation}
for the simple permutation algorithm, while the more sophisticated one
would yield
\begin{equation}
S = {m^2 + (C+2)m - C +1\over m+1} \sim m + C + 2\;\;.
\end{equation}
We see that in all these cases the algorithm is uniformly efficient
in the sense that the needed number of calls is simply
proportional to the problem's complexity $m$, as $m$ becomes large.
An ideal algorithm would of course still need $m$ calls, while the
straightforward rejection algorithm rather has
$S = m2^m/(m+1) \sim 2^m$ expected calls per event.

In the testing of algorithms such as this one, it is useful to study
expectation values of, and correlations between, the various $x_i$.
Inserting either $x_i$ or $x_ix_j$ in the integral expression for
$V(P)$, we found after some algebra the following expectation
values:
\begin{equation}
 \Exp(x_i) = 0\;\;\;,\;\;\;
 \Exp(x_i^2) = \frac{m+3}{6(m+1)}\;\;\;,\;\;\;
 \Exp(x_ix_j) = \frac{m+3}{12(m+1)}\;\;(i\ne j)\;\;,
\end{equation}
so that the correlation coefficient 
between two different $x$'s is precisely 1/2
in all dimensions! This somewhat surprising fact allows for a simple
but powerful check on the correctness of the algorithm's implementation.

As an extra illustration of the efficiency, we present in \tab{PolTab} the
cpu-time ($\tcpu$) needed to generate $1000$ points in an $m$-dimensional
polytope, both with the algorithm just presented (\ouralg) and the
rejection method (\reject). In the latter, we just
\begin{enumerate}
\item put $x_i\lar 2\rho_i-1$ for $i=1,\ldots,m$;
\item reject $x$ if $|x_i-x_j|>1$ for any
      $i=1,\ldots,m-1$ and $j=i+1,\ldots,m$.
\end{enumerate}
The computations were done using a single $300$-MHz UltraSPARC-IIi processor,
and the random number generator used was {\tt RANLUX} on level 3.
\begin{table}
\[
\begin{array}{|c|c|c|}
  \hline
  \multicolumn{1}{|c|}{}&\multicolumn{2}{c|}{\tcpu ({\rm sec})}\\\hline
  m & \ouralg & \reject\\ \hline
    2 & 0.03  & 0.01  \\ \hline
    3 & 0.03  & 0.02  \\ \hline
    4 & 0.03  & 0.04  \\ \hline
    5 & 0.04  & 0.08  \\ \hline
    6 & 0.05  & 0.17  \\ \hline
    7 & 0.06  & 0.32  \\ \hline
    8 & 0.07  & 0.67  \\ \hline
    9 & 0.08  & 1.33  \\ \hline
   10 & 0.09  & 2.76  \\ \hline
\end{array}
\qquad
\begin{array}{|c|c|r|}
  \hline
  m & \ouralg & \reject\\ \hline
   11 & 0.09  &    5.15 \\ \hline
   12 & 0.10  &   10.94 \\ \hline
   13 & 0.11  &   21.71 \\ \hline
   14 & 0.12  &   44.06 \\ \hline
   15 & 0.13  &   87.90 \\ \hline
   16 & 0.14  &  169.65 \\ \hline
   17 & 0.15  &  336.67 \\ \hline
   18 & 0.16  &  671.46 \\ \hline
   19 & 0.17  & 1383.33 \\ \hline
   20 & 0.18  & 2744.82 \\ \hline
\end{array}
\]
\caption{The cpu-time (in seconds) needed to generate $1000$ points 
         in $\Pol_m$.}
\label{PolTab}	 
\end{table}
For $m=2$ and $m=3$, the rejection method is quicker, but from $m=4$ on, the
cpu-time clearly grows linearly for \ouralg\, and exponentially for the
rejection method.

\subsection{Extension}
Let us, finally, comment on one possible extension of this algorithm. 
Suppose that the points $x$ are distributed on the polytope $\Pol_m$,
but with an additional (unnormalized) density given by
\newcommand{\spi}[1]{\textrm{s}(#1)}
\newcommand{\cpi}[1]{\textrm{c}(#1)}
\begin{equation}
F(x) = \prod\limits_{i=1}^m\cos(\half\pi x_i)\;\;,
\end{equation}
so that the density is suppressed near the edges. It is then still
possible to compute $V_{m}(k)$ for this new density. 
Writing $\spi{x}\df\sin(\half\pi x)$ and $\cpi{x}\df\cos(\half\pi x)$, we have
\begin{align}
   V_{m}(k) 
   \;&=\; k(m-k)\int_0^1 dy_1\cpi{y_1}\int_0^{1-y_1}dx_m\cpi{x_m}
          \left(\int_0^{y_1}dy\,\cpi{y}\right)^{k-1}
          \left(\int_0^{x_m}dx\,\cpi{x}\right)^{m-k-1}             \notag\\
   \;&=\; k(m-k)\left({2\over\pi}\right)^m\int_0^1d\spi{y_1}\,\spi{y_1}^{k-1}
          \int_0^{\cpi{y_1}}d\spi{x_m}\,\spi{x_m}^{m-k-1}        \notag\\
   \;&=\; {2^{m-1}k\over\pi^m}\int_0^1ds\,s^{k/2-1}(1-s)^{(m-k)/2}
    \;=\; \left({2\over\pi}\right)^m
          {\Gamma(1+\frac{k}{2})\Gamma(1+\frac{m-k}{2})\over
           \Gamma(1+\frac{m}{2})}\;\;,
\end{align}
where we used $s\df\spi{y_1}^2$.
Therefore, a uniformly efficient algorithm can be constructed in this
case as well, along the following lines.
Using the $V_{m}(k)$, the relative weights for each $k$ can
be determined. Then $s$ is generated as a $\beta$-distribution.
The generation of the other $x_i$'s involves only manipulations with sine
and arcsine functions. Note that, for large $m$, the weighted volume 
$V(\Pol_m)$ is
\begin{equation}
V(\Pol_m) \;=\; \sum\limits_{k=0}^m
\left({2\over\pi}\right)^m
{\left({k\over2}\right)!\left({m-k\over2}\right)!\over\left({m\over2}\right)!}
{m!\over k!(m-k)!} 
\;\sim\; m\sqrt{{\pi\over8}}\left({8\over\pi^2}\right)^{m/2}\;\;,
\end{equation}
so that a straightforward rejection algorithm would have number usage
\begin{equation}
S \sim \sqrt{{8\over\pi}}\left({\pi^2\over2}\right)^{m/2}\;\;,
\end{equation}
and a correspondingly decreasing efficiency.

\clearemptydoublepage

\renewcommand{\baselinestretch}{1}
\lhead[\fancyplain{}{\small\bfseries\thepage}]{}
\rhead[\fancyplain{}{\small\bfseries Bibliography}]
{\fancyplain{}{\small\bfseries\thepage}}
\cfoot[]{\fancyplain{}{}}
\newcommand{\bibspace}{\vspace{-0.5\baselineskip}}

\clearemptydoublepage

\lhead[\fancyplain{}{\small\bfseries\thepage}]{}
\rhead[\fancyplain{}{\small\bfseries Summary}]
{\fancyplain{}{\small\bfseries\thepage}}
\cfoot[]{\fancyplain{}{}}
\thispagestyle{empty}
\newcommand{\leftlabel}[1]{\mbox{}\marginpar{\raggedleft\hspace{0pt}{\scriptsize{\it #1}}}}
\newcommand{\rightlabel}[1]{\marginpar{{\scriptsize{\it #1}}}}

\chapter*{Summary}
\addcontentsline{toc}{chapter}{Summary}



\noindent
Since particle physicists consider themselves scientists, they apply the
scientific method in their exploration of nature, which means that they perform
experiments and have a model that tries to reproduce the results of the
experiments. Furthermore, as the scientific method demands, they make
predictions derived from the model, which are believed to be testable in the
(near) future. The model gives the best description of nature on the most
fundamental level that is currently accessible by the experiments, and is
called the Standard Model.  It is based on the physical concept of quantum
mechanical particles and the mathematical construction of quantum field theory.

Quantum theories describe nature by the dynamics of states, and for a model of
quantum particles, these are states of particles. At one time, this state with
these particles is appropriate to a physical situation, and at another time,
that state with those particles is appropriate. An important piece of
information provided by the model comes from the {\em transition
probabilities}\rightlabel{transition}\rightlabel{probabilities}, which give the
probability to get, in a certain situation, from one certain particle state to
another certain particle state. These probabilities are interpreted as the
ratios of the number of times the different states should appear, starting from
the same state every time. In an experiment, one particular state is prepared
millions of times, and then it is counted how often it goes over into which
other state. These numbers are then compared with the probabilities predicted
by the model in order to check its validity. 

\vspace{\baselineskip}

Conceptually, the connection between model and experiments with the help of the
transition probabilities is easy. Practically, however, it is difficult. One of
the difficulties lies in the fact that, among other things, the 
{\em momenta}\rightlabel{momenta} of the particles belong to
the characteristics of a state. These include the
information in which directions the particles are moving, and it is possible
that states only differ in these directions. The types and numbers of particles
involved may be exactly the same; if the momenta are just slightly different,
the states are different. The problem is that it is impossible to interpret
the probabilities as described before, if the number of different states is so
large. If a state with a definite momentum configuration for the particles
appears at all, it will appear at most one time, and the model predicts
probability zero for the state to occur. One can only speak about a non-zero
probability for the direction of a particle, if it concerns a certain (small)
range of directions within which the particular direction is predicted to be.
The set of all possible momentum configurations is a continuum, called {\em
phase space}\rightlabel{phase space}, and transition probabilities are
probability densities over phase space. 

The solution to the problem is to consider states that only differ in momentum
configuration as equivalent. In an experiment, this just means that such states
should not be counted separately, but should be collected together. For the
model, this means that the average of the transition probabilities over the
momentum configurations, over phase space, has to be calculated, and has to be
multiplied by the total magnitude, the volume, of phase space. This is an
example of the mathematical procedure of integration. It asks for a measure on
phase space, which is given by the model of quantum particles. The
quantity of which the average has to be calculated, in
this case the probability density, is called the {\em
integrand}\leftlabel{integrand}. The integration problem is again difficult
to solve, but its solution is accessible, especially with the help of
computers. 

A popular method to integrate an integrand over phase space is the {\em Monte
Carlo}\leftlabel{Monte Carlo} method, and the idea behind it is, again, simple;
just do the same as the experimentalists. Take the average over a (finite)
number of momentum configurations, or phase space {\em
points}\leftlabel{points}, chosen at random, and hope that the result comes
close to the exact average. Probability theory tells us that, if the points are
chosen according to a uniform distribution, and their number becomes larger and
larger, then the result converges to the exact average. To understand what is
meant by `chosen according to a uniform distribution', it is helpful to
consider the process of choosing points in phase space as the delivery of
points by phase space itself. If the points are {\em distributed
uniformly}\leftlabel{uniform}\leftlabel{distribution}, 
then the probability for each
region of phase space to deliver the following point is proportional to the
volume of that region; at each instance in the process, all regions should get
a fair chance to deliver a point. The volumes are measured by the measure
mentioned before.
The number of points, needed to get a
result that is as close to the exact average as demanded, can be derived from a
formula for the expected deviation at each number of used points. This formula
is supplied by probability theory, and shows an expected deviation which
becomes smaller with larger number of used points.

The Monte Carlo method almost always works. There are some restrictions on the
integrand, but the number of degrees of freedom over which it has to be
averaged, the number of dimensions, does not matter. The only drawback of the
method is that it can be rather slow, because the number of phase space points
often has to be large. In those cases, it pays to `load the dice', and not to
give all regions a fair chance to deliver points. 

The reason why the Monte Carlo method works is that enough information about
the integrand is obtained to make a good estimate of its average. The points
get distributed uniformly over phase space, so that information is obtained
that is diverse enough for a trustworthy average. However, if it is known for
which regions of phase space the integrand shows its most diverse behavior, one
would like to use more points in that region, and less in the less interesting
regions. \leftlabel{importance}\leftlabel{sampling}This can be achieved by
giving a larger (than fair) probability to the interesting regions to deliver
points. These probabilities have to be known exactly, in order to compensate
for the cheating when the average is calculated: points coming from
uninteresting regions should get a higher weight, since less of them are used.
This improvement of the Monte Carlo method is called {\em importance sampling},
and the second part of this thesis deals with an explicit application to
specific kinds of transition probabilities.

\vspace{\baselineskip}

Calculations of the kind described above are usually done with the help of
computers, for which there are `standard' algorithms to deliver, or {\em
generate}, numbers between $0$ and $1$. These can be considered distributed `as
good as' randomly according to the uniform distribution. They are called {\em
(pseudo) random numbers}\rightlabel{random}\rightlabel{numbers}, where the
`pseudo' represents the `as good as' in the previous sentence. A computer
cannot do things at random, but it can run algorithms that deliver randomly
looking results, and for the purpose of Monte Carlo integration, this suffices.  
With the algorithms mentioned above, all other kind of random points in spaces,
needed for the application of Monte Carlo method, have to be constructed,
and in practice, this is almost always possible.  Every problem of calculating
an average with the Monte Carlo method has to be reduced to a problem of taking
the average of a (complicated) integrand over many degrees of freedom that all
range from $0$ to $1$. This is also the case for importance sampling, where
one just chooses a smart integrand. So eventually, one always applies the
ordinary Monte Carlo method on a space of variables between $0$ and $1$, called
a {\em hypercube}\rightlabel{hypercube}, using configurations of random
numbers, again called {\em points}. 

As noted before, the Monte Carlo method works because the points get
distributed uniformly over the hypercube. However, random numbers will not
necessarily deliver points that are distributed as uniformly as possible over
the hypercube, and there is room for improvement.  This sounds confusing, but
there is a difference between the uniform distribution in the probabilistic
sense, with fair probabilities for all regions of the hypercube to deliver a
point, and the uniformity of the distribution over the hypercube of a given set
of points; one only uses the same words. In the first case, there are fair
probabilities at each instance in the process, so that two following points can
still get close together, and this is something one would like to avoid
happening. If\rightlabel{Quasi}\rightlabel{Monte Carlo} there is a region where
a few points have shown up already, and another which is still empty, then it
is time that the latter region delivers a point. As a result of this kind of
`fudging', the points are not chosen independently anymore, but one might need
less of them for a good estimate of the average. This method is called the
{\em Quasi Monte Carlo} method, and the points are called {\em quasi random}.

The Quasi Monte Carlo method also has its drawbacks. First of all, it is easier
to let a computer choose the points with fair chances than distributed as
uniformly as possible. Secondly, the formula for the expected deviation of the
result only works for the normal Monte Carlo method. So the Quasi Monte Carlo
result may be better, you only do not know how much. There are formulas that
can be used, and they ask for the rate of non-uniformity, the {\em
discrepancy}\rightlabel{discrepancy}, of the set of used points. These
formulae, however, are very complicated.

One way to compare the normal and the Quasi Monte Carlo method is by
calculating the probability for a set of points, consisting of random
numbers, to have a certain discrepancy. If there is a large probability for the
discrepancy to be equally small, compared with a quasi random set of
points, then the two methods are equally good. If this probability is
small, then one better uses the Quasi Monte Carlo method. The first part of
this thesis is devoted to the calculation of such probability distributions.

\clearemptydoublepage
\selectlanguage{dutch}
\lhead[\fancyplain{}{\small\bfseries\thepage}]{}
\rhead[\fancyplain{}{\small\bfseries Samenvatting}]
{\fancyplain{}{\small\bfseries\thepage}}
\cfoot[]{\fancyplain{}{}}
\thispagestyle{empty}
\chapter*{Samenvatting}
\addcontentsline{toc}{chapter}{Samenvatting}



\noindent
Aangezien deeltjesfysici zichzelf als wetenschappers beschouwen, hanteren zij de
wetenschappelijke methode bij hun onderzoek van de natuur. Dit betekent dat ze
experimenten doen en een model hebben dat de resultaten van de experimenten
tracht te reproduceren. Bovendien doen ze voorspellingen aan de hand van het
model, waarvan geloofd wordt dat ze verifieerbaar zijn in de (nabije) toekomst,
zoals de wetenschappelijke methode verlangt. Dit model geeft de beste
beschrijving van de natuur op het meest fundamentele niveau dat toegankelijk is
met de huidige experimenten en wordt het Standaard Model genoemd. Het is 
gebaseerd op het fysische concept van quantummechanische deeltjes en het 
wiskundige formalisme van de quantumveldentheorie.

Quantumtheorie\"en beschrijven de natuur met behulp van de dynamica van
toestanden en voor een model van quantumdeeltjes zijn dit deeltjestoestanden. Op
het ene ogenblik is deze toestand met deze deeltjes van toepassing op een
fysische situatie en op een ander ogenblik die toestand met die deeltjes.
Belangrijke informatie, die door het model geleverd wordt, komt van de {\em
overgangswaarschijnlijkheden}\rightlabel{overgangs-}\rightlabel{waarschijn-}\rightlabel{lijkheden},
die de waarschijnlijkheid geven om, in een bepaalde situatie, van de ene
toestand over te gaan in de andere toestand. Deze waarschijnlijkheden worden
ge\"{\i}nterpreteerd als de verhoudingen van het aantal keren dat de
verschillende toestanden zouden verschijnen, wanneer er telkens met dezelfde
toestand gestart wordt. In een experiment wordt \'e\'en en dezelfde toestand
miljoenen keren geprepareerd en dan wordt er geteld hoe vaak deze over gaat in
welke andere toestanden. Deze getallen worden dan vergeleken met de door het
model voorspelde waarschijnlijkheden, zodat zijn geldigheid nagegaan kan
worden.

\vspace{\baselineskip}

Conceptueel is het verband tussen het model en de experimenten met behulp van
de overgangswaarschijnlijkheden gemakkelijk te leggen. Praktisch is het echter
moeilijk. E\'en van de moeilijkheden ligt in het feit dat de impulsen van de
deeltjes deel uit maken van de karakteristieken van een toestand.  Deze
bevatten o.a.\! de richtingen in welke de deeltjes zich bewegen en het is
mogelijk dat toestanden alleen verschillen in deze richtingen. De types en
aantallen deeltjes mogen precies hetzelfde zijn; als de impulsen verschillen,
dan verschillen de toestanden. Het probleem is dat het onmogelijk is om de
waarschijnlijkheden te interpreteren zoals zojuist beschreven, als het aantal
toestanden zo groot is. Als een toestand met een bepaalde impulsconfiguratie
zich \"uberhaupt voor doet, dan hoogstens \'e\'en keer en het model voorspelt
een kans gelijk aan nul dat hij zich voor doet. Men kan met betrekking tot de
richting van een deeltje alleen over een kans spreken die niet nul is, als het
een bepaald bereik van richtingen betreft waarbinnen de richting voorspeld
wordt te liggen. De verzameling van alle mogelijke impulsconfiguraties is een
contin\"uum dat de {\em faseruimte}\leftlabel{faseruimte} genoemd wordt, en de
overgangswaarschijnlijkheden zijn waarschijnlijkheidsdichtheden op de
faseruimte.

De oplossing voor het probleem is het equivalent beschouwen van toestanden die
alleen in impulsconfiguratie verschillen. In een experiment betekent dit
eenvoudigweg dat zulke toestanden niet apart geteld moeten worden, maar bij
elkaar genomen moeten worden. Voor het model betekent dit dat het gemiddelde
van de overgangswaarschijnlijkheden over de faseruimte genomen moet worden, en
vermenigvuldigd moet worden met de totale uitgebreidheid, het volume, van de
faseruimte. Dit is een voorbeeld van de wiskundige procedure van integratie.
Er is een maat op de faseruimte voor nodig, die geleverd wordt door het model 
van quantumdeeltjes. De grootheid van welke het gemiddelde uitgerekend moet 
worden, in dit geval de overgangswaarschijnlijkheid, wordt de 
{\em integrand}\leftlabel{integrand} genoemd. Het integratieprobleem is 
weerom moeilijk op te lossen, maar de oplossing is bereikbaar, in het bijzonder
met behulp van computers.

Een populaire methode om een integrand over de faseruimte te integreren is de
{\em Monte Carlo}\leftlabel{Monte Carlo} methode en de gedachte erachter is,
weerom, eenvoudig; doe maar hetzelfde als de experimentatoren. Neem het
gemiddelde over een (eindig) aantal willekeurig gekozen impulsconfiguraties,
ook wel {\em punten}\leftlabel{punten} in de faseruimte genoemd, en hoop dat
het resultaat dicht bij het exacte gemiddelde komt. De waarschijnlijkheidsleer
vertelt ons dat het resultaat naar het exacte gemiddelde convergeert als de
punten gekozen worden volgens een uniforme verdeling en hun aantal groter en
groter wordt. Om te begrijpen wat er bedoeld wordt met `gekozen worden volgens
een uniforme verdeling' is het nuttig om het proces van het kiezen van punten
in de faseruimte te zien als het leveren van punten door de faseruimte zelf. 
Als de punten {\em uniform verdeeld}\leftlabel{uniforme}\leftlabel{verdeling} zijn, dan is
voor ieder gebied van de faseruimte de waarschijnlijkheid om het volgende punt
te leveren evenredig aan het volume van dat gebied: op ieder moment van het
proces behoren alle gebieden een eerlijke kans te krijgen om een punt te
leveren. De volumes worden gemeten met de eerder genoemde maat. Het aantal
punten dat nodig is om een resultaat te verkrijgen dat dicht genoeg bij het
exacte gemiddelde ligt kan afgeleid worden van een formule voor de verwachte
afwijking na ieder aantal gebruikte punten. Deze formule komt uit de
waarschijnlijkheidsleer en laat een verwachte afwijking zien die afneemt met
het aantal gebruikte punten.

De Monte Carlo methode werkt bijna altijd. Er zijn een paar restricties op de 
integrand, maar het aantal vrijheidsgraden waarover het gemiddelde genomen 
moet worden, het aantal dimensies, maakt niet uit. Het enige nadeel van de 
methode is dat hij nogal traag kan zijn, omdat het aantal benodigde punten 
vaak groot is. In die gevallen loont het zich om `vals te spelen' en niet alle 
gebieden een eerlijke kans te geven.

De reden waarom de Monte Carlo methode werkt is dat er voldoende informatie
over de integrand wordt verkregen om een goede schatting van zijn gemiddelde te
doen. De punten worden gelijkmatig over de faseruimte verdeeld, zodat de
informatie afwisselend genoeg is voor een betrouwbaar gemiddelde. Echter, als
het bekend is in welke gebieden van de faseruimte de integrand zijn meest
afwisselende gedrag vertoont, dan zou men daar meer punten willen gebruiken dan
in de minder interessante gebieden. 
Dit\leftlabel{importance}\leftlabel{sampling} kan
bereikt worden door de interessante gebieden een grotere (dan eerlijke) kans te
geven om punten te leveren. Deze kansen moeten wel exact bekend zijn, zodat er
gecompenseerd kan worden voor het `vals spelen' wanneer het gemiddelde
uitgerekend wordt: punten uit de oninteressante gebieden moeten een hoger
gewicht krijgen, want er worden er minder van gebruikt. Deze verbetering van de
Monte Carlo methode wordt {\em importance sampling} genoemd en het tweede deel
van dit proefschrift behandelt een expliciete toepassing op een specifiek soort
overgangswaarschijnlijkheden.

\vspace{\baselineskip}

Het boven beschreven soort van berekeningen wordt gewoonlijk gedaan met behulp
van computers, waarvoor `standaard' algorithmes bestaan die getallen tussen $0$
en $1$ leveren. Deze kunnen beschouwd worden als `zo goed als' willekeurig
verdeeld volgens de uniforme verdeling. Ze worden {\em (pseudo)
toevalsgetallen}\rightlabel{toevals-}\rightlabel{getallen} genoemd, waar het
`pseudo' het `zo goed als' in de vorige zin representeert. Een computer kan
geen willekeurige dingen doen, maar hij kan wel algorithmes uitvoeren die
resultaten leveren die er willekeurig uit zien en voor het doel van Monte Carlo
integratie voldoen. Met de bovengenoemde algorithmes moeten alle andere soorten
van willekeurige punten in ruimtes, benodigd voor de toepassing van de Monte
Carlo methode, geconstrueerd worden en dit is in de praktijk bijna altijd
mogelijk. Ieder probleem van de berekening van een gemiddelde met behulp van de
Monte Carlo methode moet gereduceerd worden tot het nemen van het gemiddelde
van een (ingewikkelde) integrand over veel vrijheidsgraden, die allemaal lopen
van $0$ tot $1$. Dit is ook het geval voor importance sampling, waarbij men
enkel een slimme integrand kiest. Dus uiteindelijk past men altijd de gewone
Monte Carlo methode toe op een ruimte van variabelen die lopen van $0$ tot $1$,
een {\em hyperkubus}\rightlabel{hyperkubus} genoemd, waarbij configuraties van
toevalsgetallen, weerom {\em punten} genoemd, gebruikt worden.

Zoals zojuist beschreven werkt de Monte Carlo methode, omdat de punten
gelijkmatig over de hyperkubus verdeeld worden. Toevalsgetallen geven echter
niet noodzakelijk de meest gelijkmatige verdeling die mogelijk is en er is
ruimte voor verbetering \footnote{In het Engels kan hierover verwarring
ontstaan, omdat voor het woord `gelijkmatig' ook het woord `uniformly' gebruikt
wordt.}. Met de toevalsgetallen zijn er gelijke kansen op ieder ogenblik in het
proces en kan het gebeuren dat twee opeenvolgende punten vlak bij elkaar komen
te liggen, wat men zou willen proberen te verhinderen. 
Als\rightlabel{Quasi}\rightlabel{Monte Carlo} er een gebied is
waar reeds enige punten zijn verschenen en een ander waar er nog geen zijn, dan
wordt het tijd dat dit laatste gebied een punt levert. Als resultaat van dit 
`geknoei' worden de punten niet meer onafhankelijk van elkaar gekozen, maar 
zijn er mogelijk minder nodig voor een goede schatting van het gemiddelde. 
Deze methode wordt de {\em Quasi Monte Carlo} methode genoemd en de punten 
worden {\em quasi toevallig} genoemd.

De Quasi Monte Carlo methode heeft ook zijn nadelen. Ten eerste is het 
gemakkelijker om een computer punten te laten kiezen met gelijke kansen dan 
zo gelijkmatig mogelijk verdeeld. Ten tweede werkt de formule voor de 
verwachte afwijking alleen voor de gewone Monte Carlo methode. Dus het 
Quasi Monte Carlo resultaat mag dan wel beter zijn, je weet alleen niet 
hoeveel. Er zijn formules die wel gebruikt kunnen worden en ze vragen 
naar de mate van niet-gelijkmatigheid, de 
{\em discrepantie}\rightlabel{discrepantie}, van de verzameling van 
gebruikte punten. Deze formules zijn echter erg ingewikkeld.

Een manier om de normale en de Quasi Monte Carlo methode te vergelijken is 
door de waarschijnlijkheid uit te rekenen dat een verzameling van punten, 
bestaande uit toevalsgetallen, een bepaalde discrepantie vertoont. Als er
een grote kans is voor de discrepantie om even klein te zijn als voor een quasi 
toevallige verzameling, dan zijn de twee methoden even goed. Als deze kans
klein is, dan kan de Quasi Monte Carlo methode beter gebruikt worden. Het eerst
gedeelte van dit proefschrift is gewijd aan de berekening van zulke 
waarschijnlijkheidsverdelingen.

%
%
\selectlanguage{english}
\lhead[\fancyplain{}{\small\bfseries\thepage}]{}
\rhead[\fancyplain{}{\small\bfseries List of publications}]
{\fancyplain{}{\small\bfseries\thepage}}
\cfoot[]{\fancyplain{}{}}
\chapter*{List of publications}
\addcontentsline{toc}{chapter}{List of publications}


\begin{enumerate}
\item A.~van Hameren, R.~Kleiss and J.~Hoogland,
      {\it Gaussian limits for discrepancies: I. Asymptotic results},
      Comp.~Phys.~Comm.~107 (1997) 1-20.
\item A.~van Hameren, R.~Kleiss and J.~Hoogland,
      {\it Gaussian limits for discrepancies},\\
      Nucl.~Phys.~B (Proc.~Suppl.) 63A-C (1998) 988-990.      
\item A.~van Hameren and R.~Kleiss, 
      {\it Quantum field theory for discrepancies},\\
      Nucl.~Phys.~B 529 [PM] (1998) 737-762.
\item A.~van Hameren and R.~Kleiss,
      {\it Computer-aided analysis of Riemann sheet structures},
      Comp.~Phys.~Comm.~116 (1999) 311-318.
\item A.~van Hameren, R.~Kleiss and C.G.~Papadopoulos, 
      {\it Quantum field theory for discrepancies II: 1/N corrections using
      fermions}, Nucl.~Phys.~B 558 [PM] (1999) 604-620.
\item A.~van Hameren and R.~Kleiss, 
      {\it Scaling limits for the Lego discrepancy},\\
      Nucl.~Phys.~B 558 [PM] (1999) 621-636.
\item A.~van Hameren and R.~Kleiss,
      {\it A fast algorithm for generating a uniform distribution inside a 
           high-di\-men\-sional polytope}, 
      Comp.~Phys.~Comm.~133 (2000) 1-5.
\item A.~van Hameren, R.~Kleiss and P.~Draggiotis, 
      {\it SARGE: an algorithm for generating QCD-antennas},
      Phys.~Lett.~B 483 (2000) 124-130.
\item A.~van Hameren and R.~Kleiss, 
      {\it Generating QCD-antennas}, \\
      Eur.~Phys.~J.~C 17, (2000) 611-621.      
\end{enumerate}


\newpage
\selectlanguage{dutch}

\vspace*{4\baselineskip}

\noindent{\Huge\bf Curriculum vitae}
\addcontentsline{toc}{chapter}{Curriculum vitae}
\thispagestyle{empty}

\vspace{2\baselineskip}

\begin{center}
\begin{tabular}{ll}
 4-12-1973: &geboren te Horst.\\
  & \\
 1992: &gymnasium diploma behaald aan het Boschveld College te Venray.\\
  & \\
 1996: &doctoraal examen in de opleiding natuurkunde met lof afgelegd aan de\\
       &Katholieke Universiteit Nijmegen.\\
  & \\
 1997-2000: &promotie onderzoek verricht onder Prof. Dr. R.H.P. Kleiss aan 
             de Katholieke\\ & Universiteit Nijmegen.
\end{tabular}
\end{center}

\end{document}